\documentclass[12pt]{article}
\pdfoutput=1
\usepackage[a4paper]{geometry}
\usepackage[labelfont=bf]{caption}
\usepackage{subcaption}

\pdfoutput=1

\usepackage{jheppub,hyperref,float,array,adjustbox,mathtools,physics, xcolor}
\usepackage{graphicx,latexsym} 
\usepackage{amssymb,amsmath}
\usepackage{amsthm,lscape}
\usepackage{longtable,booktabs,setspace} 
\usepackage{url}
\usepackage{xcolor}

\definecolor{bittersweet}{rgb}{1.0, 0.44, 0.37}
\usepackage[bbgreekl]{mathbbol}

\usepackage{amsmath,amssymb,amsfonts,amsxtra, mathrsfs, makeidx,graphicx,amsthm,epsfig, bm,longtable,float, 
color,tikz,mathtools,xfrac,footnote,rotating, lscape, makecell, environ,mathtools, empheq, physics,cleveref,tensor,
slashed,subfiles,natbib,youngtab,multirow}

\usepackage{bbold}
\DeclareSymbolFontAlphabet{\mathbb}{AMSb}
\DeclareSymbolFontAlphabet{\mathbbl}{bbold}

\newcommand{\asymm}
{$
\begin{array}{c}
\vspace{-2.8mm}
\square \\
\square
\end{array}
$}

\newcommand{\casymm}
{$
\begin{array}{c}
\vspace{-2.8mm}
\overline
\square \\
\square
\end{array}
$}

\newcommand{\symm}
{$ \square\!\square$}

\newcommand{\asymmF}
{
	\begin{array}{c}
		\vspace{-2.8mm}
		\square \\
		\square
	\end{array}
}
 
\newcommand{\asymmBF}
{
	\begin{array}{c}
		\vspace{-2.8mm}
		\overline
		\square \\
		\square
	\end{array}
}

\newcommand{\symmF}
{ \square\!\square}

\newcommand{\symmBF}
{ \overline{\square\!\square}}

\title{
\begin{center}
Confinement for 3d $\mathcal{N}=2$ $SU(N)$ with a Symmetric tensor
\end{center}
}

\author[a]{Antonio Amariti,}	
\author[a]{Fabio Mantegazza}

\affiliation[a]{INFN, Sezione di Milano, Via Celoria 16, I-20133 Milano, Italy}

\emailAdd{antonio.amariti@mi.infn.it}
\emailAdd{fabio.mantegazza@mi.infn.it}

\abstract{
In this paper we study 3d $\mathcal{N}=2$ $SU(N)$ confining gauge theories with a matter field in the rank-two index symmetric representation. The models found here are obtained from the application of the duplication formula for hyperbolic gamma functions from \emph{parent} confining models, with antisymmetric fields and (anti)-fundamental matter by \emph{freezing} some of the mass parameters for the latter.
We find a series of identities that can give rise to candidate confining theories with $SU(N)$ gauge group, a symmetric tensor and in addition other charged matter fields, in general with a non-vanishing superpotential. 
We provide for each case further checks of the proposed dualities, by studying the Coulomb branch and by deconfining the tensorial matter by using other known 3d dualities.
From the final picture a refined classification  emerges, distinguishing the confining theories obtained in the paper in  classes with common properties.
}

\begin{document}
\maketitle
\flushbottom
\allowdisplaybreaks

\section{Introduction}

A quite useful and powerful  feature of supersymmetry is the notion of dualities
among different UV gauge theories, sharing the common IR description.
Such IR dualities have been proposed in  various dimensions and with a different amount of supersymmetry. Furthermore there are limiting cases of such dualities
where the gauge degrees of freedom disappear in the dual description leaving
a pure WZ model that reproduces the confining dynamics 
of the  gauge theory.

This is for example the case of 4d SQCD with $SU(N)$ gauge group and $N+1$ 
fundamental flavors and of 4d SQCD with $USp(2N)$ gauge group and $2N+4$ fundamentals. These two cases are indeed limiting cases of Seiberg \cite{Seiberg:1994pq} and Intriligator-Pouliot \cite{Intriligator:1995ne} dualities and they have been proven in \cite{Bajeot:2022kwt} to be the building blocks to reconstruct
the  classification of 4d $\mathcal{N}=1$ s-confining theories with a single $USp(2N)$ or $SU(N)$ gauge group, charged matter with one or two indices and vanishing superpotential found  in \cite{Csaki:1996sm,Csaki:1996zb}.

Upon compactification similar confining dualities can be worked out in three dimensions as well, even if in this case a full classification of s-confining gauge theories has not been found yet.
In three dimensions indeed there are various reasons that make such classification more intricate.

There are many ingredients that indeed one need to consider in this case.
First of all chiral matter contents that would be anomalous in 4d are allowed in the 3d picture (up to constraints imposed by parity anomaly \cite{Redlich:1983kn,Redlich:1983dv,Niemi:1983rq}). Furthermore the presence of real masses imply the existence of 
CS terms, and this enlarges the families of confining theories as well.
Another issue is related to the possibility of considering $U(1)$ factors in the gauge groups, and the associated topological symmetry is crucial in defining new types of confining dualities, following from the generalization of the notion of mirror symmetry \cite{Intriligator:1996ex} to the cases with four supercharges \cite{deBoer:1997ka}.
Related to this last point, another ingredient that plays a crucial role is the presence of a Coulomb branch in the moduli space of vacua, parameterized by monopole operator. 
This last aspect makes the analysis of the moduli space and the operator mapping quite non-trivial and require some care in the analysis.

It then follows that one may expect new type of confining dualities in three dimensions with respect to the ones classified in 4d in \cite{Csaki:1996sm,Csaki:1996zb}.

Motivated by this discussion here we focus on 3d $SU(N)$ gauge theories and we look for confining dualities with two index symmetric tensors.
While in 4d there are no examples (at least with vanishing superpotential)
of $SU(N)$ confining dualities with matter fields in the symmetric representation,  this possibility is not excluded in principle by the 3d dynamics.
Here we do not find any example of such type, however we find confining theories when a superpotential interaction between a symmetric and an antifundamental is turned on. Furthermore we find cases with vanishing superpotential and with a quantum deformed moduli space.

Our starting point are the confining dualities for 3d $\mathcal{N}=2$ $SU(N)$
gauge theories with antisymmetric tensor(s) and (anti)-fundamental matter studied 
in \cite{Nii:2019ebv}. While some of these models derive from dimensional reduction of 4d parents others are genuinely new 3d dualities. These confining  
dualities, translated in the language of localization on the squashed three sphere \cite{Hama:2011ea},  produce identities for hyperbolic integrals (the "basic" dualities for ABCD type 3d $\mathcal{N}=2$ SQCD were originally translated into the language of localization on $S^3$ in  \cite{Willett:2011gp,Benini:2011mf}) . 

Here is the crucial observation: by \emph{freezing}
 \footnote{Here we borrow the terminology of \cite{Kim:2023qwh,Hayashi:2023boy,Kim:2024vci} that has recently given a
brane interpretation for the duplication formula for theories with 8 supercharges.} some of the mass parameters and by using the duplication formula for the hyperbolic gamma functions we can transform the identities for the cases with an antisymmetric tensor  into confining dualities for a symmetric tensor\footnote{See \cite{Dolan:2008qi,Spiridonov:2010qv,Benini:2011mf,Amariti:2022wae} for similar applications of the duplication formula for models with four supercharges in four and three dimensions.}. In general the procedure produces also an antifundamental field interacting with the symmetric tensor through a non-vanishing superpotential, as we will clarify in the body of the paper.

The confining dualities conjectured by the identities obtained through such procedure require then further physical explanation and checks.
For this reason we explore the Coulomb branches and associate some of the exotic operator that arise from the analysis to dressed monopole operators, similarly to the ones studied in \cite{Intriligator:2013lca,Aharony:2013kma,Csaki:2014cwa,Amariti:2015kha,Nii:2019wjz,Nii:2019ebv,Benvenuti:2021nwt,Amariti:2022wae}.

Furthermore we derive the dualities from tensor deconfinement.
The procedure consists of finding an equivalent description of the model with tensors in terms of a theory with  multiple gauge groups. The new gauge groups introduced by such description are actually confining and their confinement leads to the theory we started with. This idea of \emph{deconfining} two-index tensor fields was originally proposed by  in \cite{Berkooz:1995km} and then refined in \cite{Pouliot:1995me,Luty:1996cg}. The application of this mechanism  has been recently deeply investigated in 4d and 3d \cite{Pasquetti:2019uop,Bottini:2022vpy,Benvenuti:2020gvy,Etxebarria:2021lmq,Bajeot:2022lah,Amariti:2022wae}.
After deconfining a tensor it is possible to sequentially applying  other known dualities, until one reaches the expected confining WZ model.
Actually in some case the procedure requires to alternate sequential confinements \cite{Benvenuti:2017kud} and deconfinements \cite{Pasquetti:2019uop}. Furthermore the procedure can be applied to the analysis of the three sphere partition function as well, providing an alternative proof of the identities that we started with.

The paper is organized as follows.
In Section \ref{review} we survey the main aspects of three dimensional gauge theories with four supercharges, focusing on the analysis of the Coulomb branch and the relation with the monopole operators. This section is  helpful to set notations and to explain the computation of the global charges of the monopoles that play a crucial role in our analysis.
Then in Section \ref{classification} we give an overview of the strategy used in the paper in order to claim the existence of 3d confining dualities with $SU(N)$ gauge group and symmetric matter. We conclude the section by classifying the gauge theories analyzed in the paper, that are specified, with the help of a quiver description, by their matter content.
Then in Section \ref{sec4} we start our case by case analysis, studying the first family of confining gauge theories. We distinguish three cases for such family, and in each case we start from a mother confining duality, i.e. a confining duality with an antisymmetric.
Then, considering the three sphere partition function, we observe that it is possible, by freezing the mass parameters, to claim the existence of an identity that represents a confining duality with a symmetric tensors. Then we interpret the field content of the
confining duality by studying the quantum charges of the monopole operators, matching the results with the one obtained from the integral identity.
The analysis continues with the derivation of the dualities through tensor deconfinement, and for each case we conclude by a consistency check of the procedure at the level of localization. Reversing the interpretation, the last step is an independent proof of the identity obtained by the application of the duplication formula.
The same logic applies to the second family studied in Section \ref{sec:famII}.
The main aspects of the dualities studied in this section consists in the fact that in the proof of the dualities from tensor deconfinement we have used the results of 
 Section \ref{sec4}. In these sense the confining dualities of the second family descend from the ones of the first family.
The analysis of the third family in Section \ref{sec:famIII} requires a slightly different analysis. Indeed in such case we need a the application of tensor deconfinement in a recursive way.  In this way we arrive at a low rank at confining dualities that coincide with the ones analyzed in Section \ref{sec4} and/or \ref{sec:famII}.
 In Section \ref{unitarity} we comment on the unitarity of the dualities proposed in the paper. We observe that in general the exact R-symmetry extracted from F-maximization does not satisfy the unitarity bound, and discuss in the simplest cases the procedure necessary to  claim the existence of a conformal window.
 In Section \ref{conclusions} we give some conclusive remarks on the results obtained in the paper and discuss a series of further developments.
 We have also added a series of appendices that contain some review of the tools used in the body of the paper and also new results on the partition functions that have been necessary to prove some of the identities.
  In Appendix \ref{appPZ} we reviewed basic aspects of the three sphere partition functions, mostly to set the general  notations used in the paper.
 In Appendix \ref{integralsfromdecforAS} we have derived some of the identities for the cases with antisymmetric matter that could not be derived from dimensional reduction of the superconformal index.
 In Appendix \ref{Appendix_Branching_Rules} we have summarized the branching rules used in the paper to compute the quantum charges of the dressed monopole operators.
 In Appendix \ref{3dreview} we have reviewed the dualities used to deconfine tensorial matter in the paper. Some of the results on the partition function for the $O_{+}(N)$ cases are new and they are exposed in Appendix \ref{apppO}.

\section{General aspects of $3d$ $\mathcal{N}=2$ field theory}
\label{review}

Here we collect  general topics of $3d$ $\mathcal{N}=2$ gauge theories, 
summarizing some aspects explained in full generality in 
\cite{deBoer:1997kr,Aharony:1997bx,Aharony:2013dha,Intriligator:2013lca,Csaki:2014cwa}. The matter content of these kind of theories can be handily obtained by dimensional reduction of $4d$ $\mathcal{N}=1$ gauge theories. Thus the chiral superfield can be written as
\begin{equation}
	Q = \phi + i \bar{\theta} \gamma^i \theta \partial_i \phi + \frac{1}{4} \theta^2 \bar{\theta}^2 \partial^2 \phi + \theta \psi -i \theta^2 \bar{\theta} \gamma^i \partial_i \psi + \theta^2 F,
\end{equation}
where $\phi$ is a scalar field, $\psi$ is a two-component complex Weyl fermion, $F$ is a complex auxiliary field and $\gamma^i$ are elements of the $3d$ Clifford algebra. Note that under the $3d$ Lorentz group the fermion $\psi$ decomposes into two real independent Majorana fermions. Similarly, the vector super fields $V$ can be written as
\begin{equation}
	V = -i \theta \bar{\theta} \sigma - \theta \gamma^i \bar{\theta} A_i + i \theta^2 \bar{\theta} \lambda - i \bar{\theta}^2 \theta \lambda + \frac{1}{2} \theta^2 \bar{\theta}^2 D,
\end{equation}
where $A_i$\footnote{Hereafter we use the latin indices $i$ to denote vectors of the $3d$ Lorentz group.} is the $3$ dimensional vector field, $\sigma$ is the real gauge scalar descending from $A_3$, $\lambda$ are the gauginos and $D$ is a real auxiliary field. We can see that the main difference between this $3d$ vector multiplet and its $4d$ counterpart is the presence of the scalar $\sigma$, which may acquire a vacuum expectation value (VEV), as we will review in the following. First of all, let us write down the generic action for a $3d$ $\mathcal{N}=2$ gauge theory
\begin{equation}
	S_{3d, \mathcal{N}=2} = S_{YM} + S_{\text{chiral}},
\end{equation}
where $S_{YM}$ is the action describing the kinetic gauge action and $S_{\text{chiral}}$ describes the matter content's dynamics. They read
\begin{equation}
	\begin{split}	
		S_{YM} & =  \frac{1}{g^2} \int d^2 \theta d^3 x (\text{Tr} W_{\alpha} W^{\alpha} + \text{c.c.}) \\
		& = \frac{1}{g^2} \int d^3 x \text{Tr} \left( \frac{1}{4} F_{ij} F^{ij} + \mathcal{D}_i \sigma \mathcal{D}^i \sigma + D^2 + \lambda^{\dagger} \gamma^i \mathcal{D}_i \lambda \right),
	\end{split}
\end{equation}
where $W_{\alpha} = -\frac{1}{4} \bar{D}^2 e^{-V} D_{\alpha} e^V$. The action for a chiral matter field $Q$ reads
\begin{equation}
	S_{\text{chiral}} = \int d^3 x d^4 \theta K(Q,Q^{\dagger}) + \left( \int d^3 x d^2 \theta W(Q) +\text{c.c.} \right),
\end{equation}
where $W(Q)$ is the chiral superpotential and $K(Q,Q^{\dagger})$ is the real K\"ahler potential, written as $K(Q,Q^{\dagger}) = Q e^V Q^{\dagger}$. This term gives rise to an action of the form
\begin{equation}
	\int \!\!\! d^3 \! x d^4 \theta Q e^V \! Q^{\dagger} \! = \! \! \int \!\!\! d^3 \! x |\mathcal{D}_i \phi|^2 \! - \phi^{\dagger} \sigma \phi + \phi^{\dagger} D \phi + i \psi^{\dagger} \gamma^i \mathcal{D}_i \psi - i \psi^{\dagger} \sigma \psi + i \phi^{\dagger} \lambda \phi - i \phi^{\dagger} \lambda^{\dagger} \phi + |F|^2.
\end{equation}
There are two additional terms that one can add to $S_{3d, \mathcal{N}=2}$. Because the number of dimensions is odd, we can add a Chern-Simons (CS) term of the form
\begin{equation}
	S_{\text{CS}} = \int d^3 x \text{TR} \left[ \epsilon^{ijk} \left( A_i \partial_j A_k + i \frac{2}{3} A_i A_j A_k \right) +2 \mathcal{D} \sigma - \lambda^{\dagger} \lambda \right].
\end{equation}
Note that for Abelian gauge groups we can also have mixed CS terms. Indeed, if $A^{(a)}$ is the gauge field for the gauge group $U(1)_a$ and $A^{(b)}$ for the gauge group $U(1)_b$, we can write a term of the form
\begin{equation}
	S_{\text{CS, mixed}} = \int d^3 x \epsilon^{ijk} A_i^{(a)} \partial_j A_k^{(b)} + D^{(a)} \sigma^{(b)} + D^{(b)} \sigma^{(a)}.
\end{equation}
We will comment below on the radiative birth of these term and on the role they play. The second term we can add to the $3$-dimensional supersymmetric action has a topological origin. Indeed, $3d$ Abelian gauge theories posses an additional topological global symmetry with respect to the global symmetries obeyed by the superpotential $W$. This follows from the conservation of the current $\mathcal{J}^i$, 
\begin{equation}
	\mathcal{J}^i = \frac{1}{2} \epsilon^{ijk} F_{jk},
\end{equation}
thanks to the Bianchi identity. In supersymmetry, any conserved current can be written as a component of a linear superfield $\Sigma$ satisfying $D^2 \Sigma = \bar{D}^2 \Sigma = 0$,
\begin{equation}
	\Sigma = - \frac{i}{2} \epsilon^{\alpha \beta} \bar{D}_{\alpha} D_{\beta} V = \sigma + \theta \bar{\lambda} + \bar{\theta} \lambda + \frac{1}{2} \theta \gamma^i \bar{\theta} \mathcal{J}_i + i \theta \bar{\theta} D + \frac{i}{2} \bar{\theta}^2 \theta \gamma^i \partial_i \lambda - \frac{i}{2} \theta^2 \bar{\theta} \gamma^i \partial_i \bar{\lambda} + \frac{1}{4} \theta^2 \bar{\theta}^2 \partial^2 \sigma.
\end{equation}
The $S_{\text{YM}}$ part of the action can be rewritten in terms of $\Sigma$
\begin{equation}
	S_{\text{YM}} = \int d^3 x d^4 \theta f(\Sigma),
\end{equation}
where $f$ is a real function. Using this we can write a dual version of $S_{\text{YM}}$ where the gauge multiplet is described by a chiral superfield. For the free case this is easily done. We rewrite the path integral over the linear superfield $\Sigma$ as a path integral over a general real superfield $\Sigma'$, with the addition of Lagrange multiplier chiral superfields $\Phi$ and $\Phi^{\dagger}$, which equations of motion enforce the condition $D^2 \Sigma' = \bar{D}^2 \Sigma' = 0$:
\begin{equation}
	S_{\text{YM}} = \int d^3 x d^4 \theta \left( f(\Sigma') - (\Phi + \Phi^{\dagger}) \frac{\Sigma'}{2 \pi} \right).
	\label{Eq:SYM_Sigma}
\end{equation}
By performing the path integral over $\Sigma'$ instead of $\Phi$, we obtain the relation
\begin{equation}
	\Phi + \Phi^{\dagger} = 2 \pi \frac{\partial f(\Sigma')}{\partial \Sigma'}.
	\label{Eq:Sigma_rel}
\end{equation}
A few comments about this procedure are necessary, In general the action for the linear superfield $\Sigma$ can receive radiative corrections when we integrate out massive fields, modifying the relation \eqref{Eq:Sigma_rel}. This points out that we should only rely on this equation when all the matter fields are integrated out at a scale where the gauge coupling is small. By plugging the condition \eqref{Eq:Sigma_rel} in \eqref{Eq:SYM_Sigma} we obtain an action of the form
\begin{equation}
	S_{\text{YM}} = \int d^3 x d^4 \theta K(\Phi + \Phi^{\dagger}),
\end{equation}
where $K$ is the Legendre transform of $f$. By using the tree-level form $f(\Sigma) = \Sigma^2 / e^2$ we get that the scalar component of $\Phi$, $\phi = \phi_R + i \phi_I$, can be related to the gauge degrees of freedom by the relationships
\begin{equation}
	\begin{split}
		& \phi_R = \frac{2\pi}{e^2} \sigma, \\
		& \partial_i \phi_I = - \frac{\pi}{e^2} \mathcal{J}_i.			
	\end{split}
\end{equation}
The field $\phi_I$ is known as the dual photon, with which we can rewrite the Hodge dual of the field strength tensor, $\star F$, $\left[ \star F \right]_i = \left[ d \gamma \right]_i$. Note that the equation
\begin{equation}
	\mathcal{J}^i_e = \partial_j F^{ij} = \epsilon^{ijk} \partial_j \mathcal{J}_k = - \frac{\pi} {e^2} \epsilon^{ijk} \partial_j \partial_k \phi_I
\end{equation}
relates $\phi_I$ with the total electric charge $Q_e$ through
\begin{equation}
	Q_e = - \frac{2 \pi}{e^2} \oint_{r \rightarrow \infty} \partial_{\theta} \phi_I = - \frac{2 \pi}{e^2} \left( \phi_I(r \rightarrow \infty, \theta = 2\pi) - \phi_I (r \rightarrow \infty, \theta = 0) \right).
\end{equation}
So $\phi_I$ winds by an amount $- \frac{e^2}{2 \pi} Q_e$ as it circles the charge. Since the charge $Q_e$ is quantized, $\phi_I$ is periodic under integer shifts of $\frac{e^2}{2 \pi}$ and takes values on $S^1$. Since also $\phi_I$ can take VEV, we see from this condition that the topology of the Coulomb branch for a $U(1)$ gauge theory is $\mathbb{R} \times S^1$.

\subsection{Moduli space}
Thanks to the machinery presented above we can map out the classical moduli space of $3d$ $SU(N)$ gauge theories \cite{deBoer:1997kr}. We start by analyzing theories without CS terms. The moduli space of a $3d$ $\mathcal{N}=2$ theories is richer than the one of $4d$ ones. Indeed, it contains two different kind of branches, the Higgs branch, present also in $4d$, and the Coulomb branch. In the first one all the $\sigma$'s are set to zero and we give VEV to the matter fields in such a way that the D-term potential vanishes. This branch can be parametrized by the VEVs of all the gauge singlet composite operators that can be formed out of the matter fields up to classical constraints. The Coulomb branch is peculiar to the $3d$ case and corresponds to setting all the matter fields to zero and giving VEV's to the $\sigma$ fields. Without any loss of generality we can use the gauge transformations to diagonalize the $\sigma$'s and write the general VEV as $\langle \sigma \rangle = \sigma^i H_i$, where $H_i$ are elements of the Cartan subalgebra. By writing so, we give mass to all the matter fields and we break the gauge group to its maximal Abelian subgroup, which for $SU(N)$ is $U(1)^{N-1}$. In addition, to get rid of the leftover gauge redundancy we impose additional constraint which restrict $\sigma$ to a Weyl chamber. It reads
\begin{equation}
	\langle \sigma \rangle = \text{diag} (\sigma_1, \sigma_2,..., \sigma_N),
\end{equation}
with the ordering $\sigma_1 > \sigma_2 > ... > \sigma_{N-1} > \sigma_N$ and the $SU$ constraint $\sum_i \sigma_i = 0$. From this analysis we can conclude that in absence of matter the Coulomb branch has topology $\mathbb{R} \times S^1$ and it is described by the $N-1$ coordinates 
\begin{equation}
	Y_i \sim e^{\Phi_i},
	\label{Eq:Y_i}
\end{equation}
where the $\Phi_i$ are the chiral superfield for the unbroken $U(1)$'s, containing in their lowest component the dual photon and the adjoint scalar $\sigma$. It can be shown that in the full UV theory, the $Y_i$ operators are realized as ``disorder operators" which impose a unit magnetic flux on all Euclidian $\mathbf{S}^2$ spheres enclosing a given point $x$ in the path integral. For this reason, the $Y_i$'s are often called monopole operators. In principle we can choose the $Y_i$'s to correspond to whichever basis of low-energy $U(1)$'s that we
care to use. As we will soon see, it will prove convenient to choose our $Y_i$'s so that their lowest (scalar) component is classically equivalent to
\begin{equation}
	Y_i = \exp(\frac{2 \pi}{g^2} (\sigma_i - \sigma_{i+1}) + i (a_i - a_{i+1})), \  i = 1,..., N-1.
	\label{Eq:Y_i_Monopoles}
\end{equation}	
The presence of matter and non-perturbative effects change drastically this classical picture. In particular, in presence of matter fields we can have region in which the Higgs branch pinches the Coulomb branch. Indeed, the VEV for the scalar field $\sigma$ gives an effective mass to the matter fields and this mass can vanish in some cases. For fields in the fundamental representation it happens when one or more of the $\sigma_i$ vanish. For a field in the antisymmetric representation it happens whenever $\sigma_i + \sigma_j = 0$, $i \neq j$. In these regions of the Coulomb branch we can turn on VEV for the massless matter fields and the Higgs branch intersects the Coulomb one. We refer to these regions of intersections as ``pinches" in the coulomb branch. So the Coulomb branch splits in several regions, each of which is separated by the Higgs branch, where a description in terms of \eqref{Eq:Y_i} has to be substituted by a combination of them. However, this description remains valid in the bulk of the Coulomb branch, that is to say far away from the pinches, where all matter fields can be integrated out at a scale in which the gauge interactions are perturbative. Besides matter fields, also non-perturbative corrections modify the classical picture.  In $3d$ $\mathcal{N}=2$ $SU(N)$ gauge theories there exist instantons, classical solutions to the Euclidian path integral, which can potentially generate effective superpotentials for the low-energy degrees of freedom. These instantons are mathematically similar to $4d$ 't Hooft-Polyakov monopoles, with the scalar field $\sigma$ in the $3d$ gauge multiplet playing the role of the adjoint Higgs scalar in $4d$. For this reason, these instantonic field configurations are also referred to as ``monopoles" \cite{Weinberg:1979zt,Weinberg:1982ev}. There are $N-1$ such monopoles, each of which corresponds to a particular embedding of $SU(2)$ within the $SU(N)$ gauge group. There is then a one-to-one association between these monopole instantons and the monopole operators as defined by eq. \eqref{Eq:Y_i_Monopoles}. Then, if a charged fermion under a $U(1)$ symmetry has zero modes in this instanton/monopole background, also $Y_i$ acquires charge under this $U(1)$. Since the zero modes counting changes between the various Coulomb branch, also the charge of $Y_i$ changes. This discontinuity requires two independent operators, $Y_i$ and $\tilde{Y}_i$, to describe each side of the Coulomb branch. In a pure $SU(N)$ theory the monopole instanton configurations generate the low-energy superpotential \cite{Affleck:1982as}
\begin{equation}
	W_{\text{monopoles}} = \sum_i Y_i^{-1},
	\label{Eq:W_Monopole}
\end{equation}
which lifts entirely the Coulomb branch. This picture changes in presence of matter. These fields may have zero modes, according to Callias index theorem \cite{Callias:1977kg}. These zero modes can suppress some of the monopole contributions to the potential \eqref{Eq:W_Monopole}, leaving parts of the Coulomb branch unlifted \cite{Aharony:1997bx}. This will not always happen: each instanton/monopole is associated to a $SU(2)$ embedded in $SU(N)$ and this subgroup can be broken by the VEV of the matter field, according with the decomposition of its representation under this subgroup. This VEV set the monopole scale, which for the monopole corresponding to the operator $Y_i$ is
\begin{equation}
	\frac{1}{\rho_i} = \frac{1}{2} |\sigma_i- \sigma_{i+1}|.
\end{equation}
We have to compare this scale with the induced mass of some of the fermions which transform under the $SU(2)$ embedding that correspond to the monopole. If this mass is larger than the monopole scale, these fermions will be integrated out before they can effect the effective superpotential \eqref{Eq:W_Monopole}. Vice versa, the fermions contribute to the zero mode counting, with a number of zero mode depending on the representation. The first scenario will always realize in the bulk of the Coulomb branch, that is to say that every configuration in the bulk of the Coulomb branch can be taken to a configuration where a superpotential is generated without having to go through a pinch. Even for these, it is often possible to make the mass of some matter fields arbitrarily larger than some of the monopole scales, so that only a very restricted part of the pinch regions will actually survive. So, for a theory with only fields in the fundamental and antifundamental representation, the non-perturbative effects leave unlifted the part of Coulomb branch corresponding to the non vanishing semi-classical configuration \cite{Seiberg:1996nz}
\begin{equation}
	\langle \sigma \rangle = \text{diag}(\sigma,0,...,0, -\sigma).
\end{equation}
This direction is parametrized by the low-energy monopole
\begin{equation}
	Y = \prod_{i=1}^{N-1} Y_i = \exp(\frac{2 \pi}{g} (\sigma_1 - \sigma_N) + i (a_1-a_N)).
\end{equation}
In presence of antisymmetric matter field, due to the different effective real masses, we have additional surviving directions (indeed one region has a zero mode structure similar to the one with only fundamental and antifundamental fields and so is still well parameterized by $Y$) corresponding to
\begin{equation}
	\langle \sigma \rangle_m = \text{diag} \left( \underbrace{ \sigma,...,\sigma}_{m},\underbrace{0,...,0}_{N-2m}, \underbrace{-\sigma,...,-\sigma}_{m} \right) .
	\label{Eq:Sigma_m}
\end{equation}
These directions are parameterized by $\tilde{Y}_m$ operators, defined as follows
\begin{equation}
	\begin{split}
		& \tilde{Y}_2 = \sqrt{Y_1 Y_2^2 Y_3^2...Y_{N-2}^2 Y_{N-1}}, \\
		& \tilde{Y}_3 = \sqrt{Y_1 Y_2^2 Y_3^3...Y_{N-3}^3 Y_{N-2}^2 Y_{N-1}},\\
	\end{split}
\end{equation}
and so on.

\subsection{One-loop CS terms}

Until now, we developed all the analysis above without caring for non-vanishing CS term $k \neq 0$. But, we know that the low-energy behavior of $3d$ $\mathcal{N}=2$ theories depends on these terms, so let us explain how these kind of terms are dynamically generated and how they influence our treatment.
We know that in the $3d$ IR two-point correlation function between conserved current arise conformally invariant contact terms. The UV description of these contact terms is given by CS interactions for the background gauge fields coupling to the conserved current \cite{Witten:2003ya,Closset:2012vg,Closset:2012vp}. The Lagrangian description of these terms has the following form
\begin{equation}
	\mathcal{L}_{\text{CS}} \sim k \text{TR}(\epsilon^{\mu \nu \rho } A_{\mu} \partial_{\nu} A_{\rho}) = k \text{TR}(\epsilon^{\mu \nu \rho } A_{\mu} F_{\nu \rho}).
\end{equation}
This additional interaction modifies the Gauss' law (i.e. the $A_0$ equation of motion), introducing a ``magnetic'' contribution to the electric charge when $k \neq 0$
\begin{equation}
	- \frac{1}{e^2} \partial_i F_{0i} = \rho_{\text{matter}} - \frac{k}{2 \pi} F_{12},
\end{equation}
where $\rho_{\text{matter}} = \frac{\delta \mathcal{L}_{\text{matter}}}{\delta A_0}$ is the contribution from matter fields. In fact, by imposing Coulomb gauge to preserve the canonical quantization (i.e. setting $A_0=0$), we make $A_0$ become a non-dynamical field. As a consequence we have to impose Gauss' law as a constraint and this implies that any charged field under an $U(1)_J$ symmetry acquires an electric charge
\begin{equation}
	q_{\text{elec}}^{\text{CS}} = -k q_J.
	\label{Eq:q_el}
\end{equation}
The story does not end here, in fact this effect is also present in theories with no tree-level CS term, that is if $k=0$. This because CS terms are generated perturbately when UV divergences are regulated in a gauge-invariant way \cite{Avdeev:1991za,Redlich:1983kn}.  Let us see how this works. As we saw earlier a non-vanishing VEV for the $\sigma$'s generate a real mass for the matter fields $m_i(\sigma) = n_i \langle \sigma \rangle$, where $n_i$ is the $U(1)$ charge of the fields. So, if all real masses are set to zero, when we explore directions of the Coulomb branch when $\sigma \neq 0$, we have to integrate out the heavy fields in order to get an effective action for the light degrees of freedom. 
Considering the vacuum polarization diagram which generates one-loop Chern-Simons terms for the gauge field $A_i$, where in the loop circulate the heavy fermion, we have this effective action 
\begin{equation}
	S_{\text{eff}}[A] = \frac{1}{2} n_i^2 \int \frac{d^3 q}{(2 \pi)^3} \frac{d^3 p}{(2 \pi)^3} A_l(-p) A_m(-q) [-i \Pi^{lm}(p)],
	\label{Eq:S_eff}
\end{equation}
where (we suppress the index $i$ in $m_i(\sigma)$)
\begin{equation}
	\Pi^{lm}(p) = -i e^2 \int \frac{d^3 k}{(2 \pi)^3} \text{Tr} \left[ \gamma^l \frac{-i}{(p+k) - m(\sigma)} \gamma^m \frac{-i}{k-m(\sigma)} \right].
\end{equation}
For large $m$,
\begin{equation}
	\Pi^{lm} \rightarrow -i \epsilon^{lmk} p_k \text{sign}[m(\sigma)] + ...
	\label{Eq:PI}
\end{equation}
Where the dots represent divergent term which need to be regularized. Substituting \eqref{Eq:PI} in \eqref{Eq:S_eff}, we can see that each charged heavy fermion, with charge $n_i$, generates a one-loop CS term with coefficient $k_{\text{eff,i}} = n_i^2 \text{sign}(n_i \sigma)$. By integrating out all the massive fermions along the Coulomb branch we obtain
\begin{equation}
	k_{\text{eff}}(\sigma) = k + \frac{1}{2} \sum_i n_i^2 \text{sign} (n_i \sigma).
	\label{Eq:k_eff}
\end{equation}
Remembering \eqref{Eq:q_el}, from \eqref{Eq:k_eff} we read the induced one-loop electric charge for the $U(1)_{\mathcal{J}}$ charged field
\begin{equation}
	q_{\text{elec}}^{\text{CS}} = - k_{\text{eff}} q_{\mathcal{J}} = \left( k + \frac{1}{2} \sum_i n_i^2 \text{sign} (n_i \sigma) \right) q_{\mathcal{J}}.
\end{equation}
This formula can be easily generalized to the case with non-vanishing real masses $m_{\mathbb{R},i}$, it reads
\begin{equation}
	q_{\text{elec}}^{\text{CS}} = \left( k + \frac{1}{2} \sum_i n_i^2 \text{sign} (m_i( \sigma)) \right) q_{\mathcal{J}} = \left( k + \frac{1}{2} \sum_i n_i^2 \text{sign} (m_{\mathbb{R},i} + n_i \sigma)) \right) q_{\mathcal{J}}.
\end{equation}
The same reasoning is true for the generation of effective Abelian charges. One-loop $U(1)$ charges are generated by mixed Chern-Simons term. Let us see this by weakly gauging a  $U(1)$ symmetry. The weakly gauged $U(1)$ vector bosons $a_i$ interact with the heavy fermions charged under the same $U(1)$, with charge $\tilde{n}_i$. Integrating out these massive fermionic fields we encounter the vacuum polarization diagram which generates one-loop mixed Chern-Simons term involving both $a_i$ and $A_j$. These diagrams induce one-loop mixed CS terms of the form
\begin{equation}
	\mathcal{L}_{\text{CS}}^{\text{mixed}} \sim k_{\text{eff}}^{\text{mix}} (\sigma) \text{Tr}(\epsilon^{ijk} a_i F_{jk}),
\end{equation}
with
\begin{equation}
	k_{\text{eff}}^{\text{mix}} (\sigma) = \frac{1}{2} \sum_{i} n_i \tilde{n}_i \text{sign} (n_i \sigma).
\end{equation}
Analogously with the previous case, the non-vanishing mixed CS term will modify Gauss' law for the weakly gauged $U(1)$ (i.e. the $a_0$ equation of motion), which in turns will induce a $U(1)$ charge for fields which have a non-vanishing $q_{\mathcal{J}}$. This reads
\begin{equation}
	\tilde{q} = \frac{1}{2} \left( \sum_{i} n_i \tilde{n}_i \text{sign} (n_i \sigma) \right) q_{\mathcal{J}}.
\end{equation}
In this way the Abelian charges mix in a non trivial way, as a result, the $Y$'s operators, despite the fact that they are not charged classically, acquire Abelian charges quantum mechanically. In this way, as we will review in the next paragraph, they acquire also charge under the gauge group. But, in order to parametrize the Coulomb branch we need gauge-invariant operator. We construct them by ``dressing" the bare monopole operators with massless matter fields.

\subsection{Coulomb branch analysis}
\label{genmondisc}
In this subsection we present how the analysis for the CB operators analysis has been performed. In general we will consider the following operators corresponding to different directions of the Coulomb branch. $Y$ will refer to the direction $\text{diag}(\sigma,0,...,0,\sigma)$. This direction gives this braking pattern
\begin{equation}
	SU(N) \rightarrow SU(N-2) \times U(1)_1 \times U(1)_2,
	\label{Eq:SU(N-2)_Breaking}
\end{equation} 
where 
\begin{equation}
	\begin{split}
		& U(1)_1 \sim \text{diag}(1, \underbrace{0,...,0}_{N-2},-1), \\
		& U(1)_2 \sim \text{diag} (N-2,\underbrace{-2,...,-2}_{N-2}, N-2).
	\end{split}
\end{equation}
$\tilde{Y}_m$ will refer to the direction \eqref{Eq:Sigma_m} and breaks $SU(N)$ as follows
\begin{equation}
	SU(N) \rightarrow SU(m)^t \times SU(N-2m) \times SU(m)^b \times U(1)_{\tilde{1}} \times U(1)_{\tilde{2}},
\end{equation}
where 
\begin{equation}
	\begin{split}
		& U(1)_{\tilde{1}} \sim \text{diag}(\underbrace{1,...,1}_{m}, \underbrace{0,...,0}_{N-2m}, \underbrace{-1,...,-1}_{m}), \\
		& U(1)_{\tilde{2}} \sim \text{diag}(\underbrace{N-2m,...,N-2m}_{m}, \underbrace{-2m,...,-2m}_{N-2m}, \underbrace{N-2m,...,N-2m}_{m}).
	\end{split}
\end{equation}
Finally, only for $SU(2N)$ gauge groups, $\hat{Y}$ refers to the direction $\text{diag}(\sigma, ..., \sigma, -\sigma,..., -\sigma)$. The induced breaking pattern is
\begin{equation}
	SU(2N) \rightarrow SU(N)^t \times SU(N)^b \times U(1)_{\hat{1}},
\end{equation}
where
\begin{equation}
	U(1)_{\hat{1}} \sim \text{diag}(\underbrace{1,...,1}_{N}, \underbrace{-1,...,-1}_{N}).
\end{equation}
The branching rules for the case of equation (\ref{Eq:SU(N-2)_Breaking}) are explictly reported in Appendix \ref{Appendix_Branching_Rules}. In chiral field theories, when the number of fundamental and antifundamental fields is different, the bare Coulomb branch operators are not gauge invariant, since the mixed CS terms between the various $U(1)$'s of the theories are usually non-vanishing. For example, for the breaking pattern \eqref{Eq:SU(N-2)_Breaking} the mixed CS term between $U(1)_1$ and $U(1)_2$ in a theory with $F$ fundamental fields and $\bar{F}$ antifundamental fields is
\begin{equation}
	k_{\text{eff}}^{U(1)_1,U(1)_2} = (N-2)(F-\bar{F}),
\end{equation}
and the effective $U(1)_2$ charge for the monopole is $-k_{\text{eff}}^{U(1)_1,U(1)_2}$. The problem is solved by dressing these monopole operators by matter fields to reestablish gauge invariance.

\section{Generalities and classification}
\label{classification}

In this section we expose the general structure of the analysis performed in the rest of the paper.
We initiate our investigation by considering the classification of 3d dualities with $SU(N)$ gauge groups and antisymmetric 
tensors.
Such classification is based on the analysis of the quantum
structure of the moduli space of vacua. The non trivial role played by the (dressed) monopole operators
is indeed crucial in order to provide the correct matching of the gauge invariant operators that build the confining phase.
Furthermore, in some cases such pure 3d confining dualities can be obtained by reducing the 4d confining dualities classified in \cite{Csaki:1996sm,Csaki:1996zb} and then by removing the KK monopole superpotential by real mass flows. We will comment on the 4d origin of some of these dualities below.
Our main claim here is that the analysis of the matching of the 3d partition functions for such dualities 
hints the existence of 3d confining dualities for $SU(N)$ gauge groups with two index symmetric tensors.

Such proposal follows from  the application of the duplication formula for the hyperbolic Gamma functions (\ref{duplication}),
operated by \emph{freezing} the mass parameters of three or four $SU(N)$ fundamental fields.
The procedure indeed converts, in the integrand of the partition function, the one loop determinant for an antisymmetric into the one of a symmetric tensor.
Let's distinguish the two physically relevant patterns, associated to the possibility of \emph{freezing} the masses of  three or four fundamentals respectively.
\begin{itemize}
\item First we consider an antisymmetric with mass parameter $\tau$ and four fundamentals with mass parameters $\mu_{1,2,3,4}$. Then we freeze such masses as 
\begin{equation}
\label{freezing1}
\mu_1 = \frac{\tau}{2},\quad
\mu_2 = \frac{\tau}{2}+\frac{\omega_1}{2},\quad
\mu_3 = \frac{\tau}{2}+\frac{\omega_2}{2},\quad
\mu_4 = \frac{\tau}{2}+\omega.
\end{equation}
Using the duplication formula (\ref{duplication}), we obtain 
\begin{equation}
\prod_{a=1}^{4} \Gamma_h(\sigma_i + \mu_a)
=
\Gamma_h(2\sigma_i+\tau),
\end{equation}
which converts the one loop determinant of a two index $SU(N)$ antisymmetric field to the one of a two index $SU(N)$ symmetric one (for this reason we will keep referring to the parameter $\tau$ as $\tau_S$ in the following).
\item  Then we consider an antisymmetric with mass parameter $\tau$ and three fundamentals with mass parameters $\mu_{1,2,3}$. 
Here we freeze such masses as 
\begin{equation}
\label{freezing2}
\mu_1 = \frac{\tau}{2},\quad
\mu_2 = \frac{\tau}{2}+\frac{\omega_1}{2},\quad
\mu_3 = \frac{\tau}{2}+\frac{\omega_2}{2}.\quad
\end{equation}
In this case, by applying the  duplication formula (\ref{duplication}),  we obtain 
\begin{equation}
\prod_{a=1}^{4} \Gamma_h(\sigma_i + \mu_a)
=
\Gamma_h(2\sigma_i+\tau) \Gamma_h\left(\omega-\frac{\tau}{2}-\sigma_i\right).
\end{equation}
We interpret this result again as converting an $SU(N)$ antisymmetric field into an $SU(N)$ symmetric one, but in addition a further $SU(N)$ antifundamental is generated in the process.
The mass parameter of the latter is constrained and this is compatible with a cubic superpotential between the symmetric and the antifundamental. For this reason we will denote the symmetric field as $S$ and the antifundamental arising from this procedure as $ \widetilde{Q}_S$. The cubic superpotential has then the form
\begin{equation}
\label{WSQSQS}
W = S \widetilde{Q}_S^2.
\end{equation}
\end{itemize}

We then focus on explicit 3d confining theories with $SU(N)$ gauge groups and two index tensor antisymmetric matter fields.
In absence of CS terms and tree level superpotential such cases have been classified in \cite{Nii:2019ebv}. 
In order to apply the two freezings (\ref{freezing1}) and/or (\ref{freezing2}) we look for the cases classified in  \cite{Nii:2019ebv}  with at least three fundamentals in addition to the antisymmetric.
There are six possibilities for generic $SU(N)$ (for fixed and low ranks there are also other sporadic cases that we do not study here).
These possibilities correspond to 
\begin{itemize}
\item $SU(N)$ with \asymm, $(N-1)$ $\overline \square$ and 3 $\square$;
\item $SU(N)$ with \asymm, $(N-2)$ $\overline \square$ and 4 $\square$;
\item $SU(N)$ with \asymm, \casymm, 1 $\overline \square$  and  3 $\square$;
\item $SU(N)$ with\asymm, \casymm and 4 $\square$;
\item $SU(N)$ with 2 \asymm, 1 $\overline \square$ and 3 $\square$;
\item $SU(2N)$ with 2 \asymm and 4 $\square$.
\end{itemize}
By freezing the mass parameters of the fundamentals as in  (\ref{freezing1}) and (\ref{freezing2}) 
these cases give rise to 9 candidate confining theories with two index symmetric tensors.
In three cases there is a vanishing tree level superpotential, while in the remaining six cases we have a tree level superpotential of the form (\ref{WSQSQS}). 

\begin{figure}
\begin{center}
  \includegraphics[width=12cm]{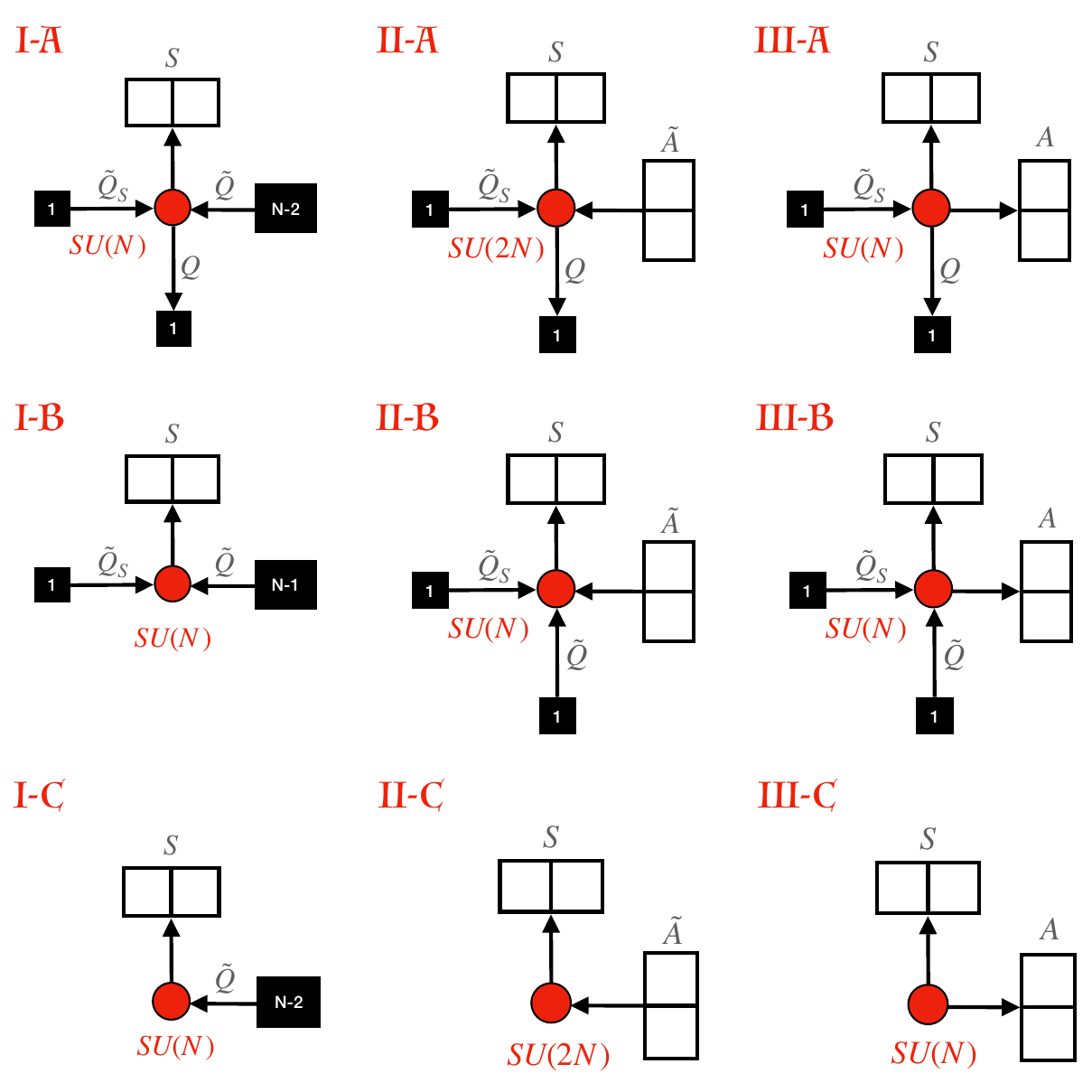}
  \end{center}
  \caption{
  In this figure we have summarized the nine models discussed in this paper. The black squares represent the $SU(N)$ gauge group while the 
  black boxes refer to the flavors. 
  The symmetric and the antisymmetric matter fields are represented in the quivers by boxes, 
  while the orientation of the arrows is associated to the conjugation of the representation. 
 We have named the columns as I, II and III and the rows are denoted with A, B, C. The reason for such refined classification will be clear in the body of the paper. Essentially the models identified by the same letter I,II or III have the same field content in terms of tensors while the model identified by the same letter A, B or C are related by a similar  analysis in the proof of the duality through tensor deconfinement. In this sense the case II-A and III-A can be proven using the results found in the deconfinement of case I-A and the same observation holds for the other cases.
 }
    \label{clasification}
\end{figure}
In Figure \ref{clasification} we have represented these nine cases with the help of a quiver description.
We organized the nine cases studied below in three columns, (labeled I, II and III) and three raws (labeled A,B and C).
The label of the column distinguish the model by their two-index tensor matter content. In the family I there is just a symmetric, in family II a symmetric and a conjugate antisymmetric and in case III there are a symmetric and an antisymmetric. 
The labels of the raws, on the other hand,  has been decided a posteriori, by proving the dualities by deconfining the two-index tensors in terms of orthogonal and symplectic gauge group. 
We chose such ordering because, in the analysis below, we have observed that the cases labeled by II and III can be studied from the knowledge of the cases denoted by I in the same raw.

Once these nine conjectured dualities have been established we need to check their validity.
First we can read the operator mapping from the matching of the partition functions with the pure WZ theory.
Indeed the singlets appearing in the RHS of the identities must correspond to gauge singlets of the electric theory. Such gauge singlets can be of mesonic and baryonic type, and in such cases it is straightforward to interpret them   across the duality.
However, a second typology of singlets is crucial in our analysis. Such singlets correspond to monopole operators of the electric theory, and, in presence of a chiral matter content, they are not gauge invariant.
We then need to find the gauge singlets built up by dressing such operators by  charged matter fields.
Once we have obtained the singlets associated to the dressed monopoles we can compare with the RHS of the identities obtained on the three sphere partition function by applying the duplication formula.
This analysis provides a quite robust check of the conjectured confining dualities.
Another check that we perform in each case consists of proving the confining dualities (and the relations from localization) by deconfining the two index tensors, in terms of  orthogonal and symplectic gauge groups.
This is a strong check of the confinement of the models discussed here, because it provides a derivation by using only other known dualities. Essentially  we observe that  there are few building blocks that can be used to prove the confinement of all the models discussed here,  namely the confining duality for 3d SQCD with ABCD gauge algebras. This is consistent with the various results discussed in the literature on tensor deconfinement in 4d and 3d.
Furthermore, proving the dualities in terms of such building blocks gives also alternative proofs of the integral identities that we have derived from the duplication formula and that we have used as a starting point to conjecture the existence of the confining dualities for $SU(N)$ with symmetric matter.


\section{Family I}
\label{sec4}

The first family studied in this paper corresponds to $SU(N)$ gauge theories with a  two index 
symmetric tensor.
Following the classification of \cite{Nii:2019ebv} we have found two cases where the duplication formula can be used in order to give rise to two 3d confining dualities with symmetric tensors, corresponding to
\begin{itemize}
\item $SU(N)$ with \asymm, $(N-1)$ $\overline \square$ and $3$ $ \square$;
\item $SU(2N)$ with  \asymm, $(N-2)$ $\overline \square$ and $4$ $. \square$\end{itemize}
These model can be obtained by reducing the corresponding 4d confining duality found in \cite{Csaki:1996sm,Csaki:1996zb}
and then by performing an opportune real mass flow.

In the following  we apply the duplication formula on the integral identities representing the realization of these dualities  on the squashed three sphere. The identities we start with can obtained by  using the prescription for the 4d/3d reduction of 4d supersymmetric index of \cite{Aharony:2013dha}. The 4d matching of the superconformal index follows from the analysis of \cite{spiridonov2003theta}. Furthermore, a physical proof of such matchings can be given by applying the discussion of \cite{Bajeot:2022kwt}, using the integral identities for the elementary dualities in the deconfinement of the antisymmetric tensor.
In both the cases at hand here the 3d identities obtained for these confining dualities are in agreement with the proposal of \cite{Nii:2019ebv}.

Using these identities we will argue here that three new confining dualities arise in presence of a symmetric 
tensor.
One case originates from the confining duality with three fundamentals and $N-1$ antifundamentals, while the other two cases arise from the confining duality with four fundamentals and $N-2$ antifundamentals.

In each case we will investigate the consequence of the matching of the partition function after the application of the duplication formula to the cases with the antisymmetric.
We explicitly study the Coulomb branch, showing the precise mapping of the dressed monopoles of the gauge theory with respect to the singlets of the dual phase. 
Furthermore, we provide for each case an explicit derivation of the confining dualities by deconfining the symmetric tensors using an $SO(K)$ or $O_{+}(K)$  gauge group. When applied to the three sphere partition function this approach furnishes an alternative derivation of the integral identities.

The discussion is organized following the classification of Figure \ref{clasification},
where we distinguish the three theories in this first family in three cases, named A,B and C.
This classification will be useful in the following sections as it will become clear 
when we will prove the dualities through tensor deconfinement in family II and III.

\subsection{Case I-A: $S \oplus Q \oplus \tilde Q_S \oplus (N-2) \tilde Q$}
\label{SQQtQtS}

We start our analysis by considering $SU(N)$ with an antisymmetric tensor $A$, $N-2$ antifundamentals  $\tilde Q$ and four fundamentals $Q$.
In this case the theory has $W=0$ and it confines and the analysis for odd and even $N$ must be taken separately.
The theory confines as discussed in \cite{Nii:2019ebv} and in the following we will review the details of the dual WZ models, distinguishing the situation in the case of $N=2n$ and $N=2n+1$.

\subsubsection*{Applying the duplication formula}
In the even case $N=2n$ the elementary degrees of freedom are
\begin{eqnarray}
\label{singlets1nii}
M \equiv Q \tilde Q, \quad
T \equiv A^n, \quad
B_{n-1} \equiv A^{n-1} Q^2, \quad
B_{n-2} \equiv A^{n-2} Q^4, \quad
\widetilde B_1 = A \tilde Q^2.
\end{eqnarray}
In addition there are two dressed monopoles of the electric theory that act as singlets in the dual phase. Such operators combine the non-gauge invariant bare monopole operator $Y_{SU(2n-2)}^{bare}$ as
\begin{equation}
\label{singlets2nii}
Y_A^{dressed} \equiv  Y_{SU(2n-2)}^{bare} A ,\quad 
Y_{\tilde Q^{2n-2}}^{dressed} \equiv  Y_{SU(2n-2)}^{bare} \tilde Q^{2n-2}.
\end{equation}
Here and in the rest of the paper we will refer to the $SU(N)$ bare monopole 
$Y_{SU(N-2)}^{bare}$ as the one corresponding to the Coulomb branch coordinate 
$(\sigma,0,\dots,0,\sigma)$. It is in general not gauge invariant and its charge is obtained along the lines of the discussion in sub-section \ref{genmondisc}. Furthermore, the branching rules necessary to compute the monopole charges are summarized in Appendix \ref{Appendix_Branching_Rules}.
The confining superpotential is \cite{Nii:2019ebv} 
\begin{equation}
W =Y_A^{dressed}  
(\widetilde B_1 ^{n-2} M^2 B_{n-1} +T  \widetilde B_1 ^{n-3} M^4)
+
Y_{\tilde Q^{2n-2}}^{dressed} 
(B_{n-1}^2+T B_{n-2} ).
\end{equation}
Observe that this confining superpotential can be derived also from the dimensional reduction of the 4d confining gauge theory for $SU(2N)$ 
with and antisymmetric, $2N$ antifundamentals and $4$ antifundamentals,
by first considering the effective theory on $S^1$ and then by a real mass flow.
The same observation holds in the following for all the models with a 4d origin.

This confining duality then follows from the 4d parent and, at the level of the partition function, we can reduce the corresponding identity between the superconformal indices obtaining the 3d identity 
\begin{eqnarray}
\label{idIAAS}
&&
Z_{SU(2n)}\left(\vec \mu;,\vec \nu;-;-;\tau_A;- \right)
=
\prod_{a=1}^{4} \prod_{b=1}^{2n-2} \Gamma_h(\mu_a + \nu_b) 
\prod_{1\leq a<b \leq 4}  \Gamma_h((n-1)\tau_A + \mu_a + \mu_b)
\nonumber \\
&\times&
\Gamma_h( n \tau_A)
 \Gamma_h \left((n-2)\tau_A + \sum_{a=1}^{4} \mu_a  \right)
 \prod_{1\leq a<b \leq 2n-2}  \Gamma_h(\tau_A + \nu_a + \nu_b)
  \\
&\times&
\Gamma_h(2 \omega - \sum_{b=1}^{2n-2} \nu_b  - \sum_{a=1}^{4} \mu_a-(2n-3) \tau_A)
\Gamma_h(2 \omega  - \sum_{a=1}^{4} \mu_a-(2n-2) \tau_A),
\nonumber
\end{eqnarray}
where we refer the reader to appendix \ref{appPZ} for the notations relative to the partition function used in the paper.

In the odd case $N=2n+1$ the elementary degrees of freedom are
\begin{eqnarray}
M \equiv Q \tilde Q, \quad
B_{n} \equiv A^{n} Q, \quad
B_{n-1} \equiv A^{n-1} Q^3, \quad
\widetilde B_1 = A \tilde Q^2.
\end{eqnarray}
In addition there are two dressed monopoles of the electric theory that act as singlets in the dual phase. Such operators combine the non-gauge invariant bare monopole operator $Y_{SU(2n-1)}^{bare}$ as
\begin{equation}
Y_A^{dressed} \equiv  Y_{SU(2n-1)}^{bare} A ,\quad 
Y_{\tilde Q^{2n-1}}^{dressed} \equiv  Y_{SU(2n-1)}^{bare} \tilde Q^{2n-1}.
\end{equation}
The confining superpotential is 
\begin{equation}
W =Y_A^{dressed}  
(\widetilde B_1 ^{n-1} M^2 B_{n-1} + \widetilde B_1 ^{n-2} M^3 B_n)
+
Y_{\tilde Q^{2n-1}}^{dressed} 
(B_{n} B_{n-1} ).
\end{equation}
This confining duality follows from the 4d parent and at the level of the partition function it corresponds to the identity 
\begin{eqnarray}
\label{idIAAS2}
&&Z_{SU(2n+1)}\left(\vec \mu;,\vec \nu;-;-;\tau_A ;-\right)
=
\prod_{a=1}^{4} \prod_{b=1}^{2n-1} \Gamma_h(\mu_a + \nu_b) 
\prod_{a=1}^3  \Gamma_h(n\tau_A + \mu_a )
\nonumber \\
&\times&
 \prod_{1\leq a<b<c \leq 4}  \Gamma_h((n-1)\tau_A + \mu_a + \mu_b+\mu_c)
 \prod_{1\leq a<b \leq 2n-1}  \Gamma_h(\tau_A + \nu_a + \nu_b)
  \\
&\times&
\Gamma_h(2 \omega - \sum_{b=1}^{2n-1} \nu_b  - \sum_{a=1}^{4} \mu_a-(2n-2) \tau_A)
 \Gamma_h(2 \omega  - \sum_{a=1}^{4} \mu_a-(2n-1) \tau_A).
\nonumber
\end{eqnarray}
Starting from the relations (\ref{idIAAS}) and (\ref{idIAAS2}) we are now going to 
derive two new identities holding in presence of an $SU(N)$ symmetric tensor.
In both cases we freeze the mass parameters for the fundamentals as in  
(\ref{freezing2}) and then we apply the duplication formula for the hyperbolic gamma functions.
We work in the even and in the odd case separately and we further change the name of the parameter $\tau_A$ into $\tau_S$.
We obtain  a unified relation holding for both $N=2n$ and $N=2n+1$. 
The relation is
\begin{eqnarray}
\label{ConfZVI}
&&
Z_{SU(N)}\left(\mu;\omega-\frac{\tau_S}{2},\vec \nu;\tau_S;-;-;- \right)
=
\prod_{b=1}^{ N-2} \Gamma_h \left(\omega-\frac{\tau_S}{2}-\nu_b \right) \Gamma_h(\nu_b+\mu)
\nonumber \\
&&
\Gamma_h(N \tau_S) \Gamma_h((N-1) \tau_S+2\mu)
\prod_{1\leq b\leq c\leq N-2} (\nu_b+\nu_c+\tau_S)
 \\
&&
\Gamma_h\left(\omega-\left(N-\frac{3}{2} \right)\tau_S -\mu - \sum_{b=1}^{N-2} \nu_b\right)
\Gamma_h\left(\omega-\left(N-\frac{1}{2} \right) \tau_S-\mu\right). \nonumber
\end{eqnarray}

\subsubsection*{Field theory interpretation}

Once the identity (\ref{ConfZVI}) is established 
we focus on its field theory interpretation.
On the electric side we have an $SU(N)$ gauge theory with a two index symmetric tensor $S$, one fundamental that we denote as $Q$,
$(N-2)$ antifundamentals $\widetilde Q$ and a further  antifundamental that we denote as $Q_S$ interacting with the symmetric tensor through (\ref{WSQSQS}).
The charges of these fields under the global symmetries are summarized in Table \ref{Tab:IAelectric}.
 \begin{table}[H]
	\centering
		\begin{tabular}{c |c| c c c c c}
Field & $SU(N)$ & $SU(N-2)$ & $U(1)_S$ & $U(1)_Q$  & $U(1)_{\tilde{Q}}$& $U(1)_R$ \\ 
\hline
$S$ & \symm & $1$&$1$&$0$&$0$&$0$\\
$Q$ &$\square$&$1$&$0$&$1$&$0$&$0$\\
$\tilde Q$ &$\overline \square$&$\square$&$0$&$0$&$1$&$0$\\
$\tilde Q_S$ &$\overline \square$&$1$&$-\frac{1}{2}$&$0$&$0$&$1$\\
 \end{tabular}
		\caption{Charged fields in the electric side of the case I-A. }
		\label{Tab:IAelectric}
\end{table}
From the identity (\ref{ConfZVI}) we expect that the theory confines in terms of its gauge invariant degrees of freedom. 
There are two types of gauge invariant that describe the moduli space. The first type of gauge invariants are the singlets  build from the charged superfields  in Table \ref{Tab:IAelectric}, after imposing the chiral ring constraints from the superpotential (\ref{WSQSQS}). The second type of gauge invariant operators involves the monopoles, with possible dressings.
We have found that the singlets appearing in the RHS of (\ref{ConfZVI})
can be summarized as  in Table \ref{Tab:IA}.
 \begin{table}[H]
	\centering
		\begin{tabular}{c|c| c c c c c}
			$\Phi$& & $SU(N\!-\!2)\!\!\!$ & $U(1)_S$ & $\!U(1)_Q\!$  & $\! U(1)_{\tilde{Q}}\!$& $\!U(1)_R$ \\ 
			\hline
			$\Phi_1$ & $\!\!Y^{bare}_{SU(N-2)} S^{N-1} \tilde{Q}^{N-3} Q\!\!$ & $\overline{\square}$ & $-\frac{1}{2}$ & $0$ & $1$ &$1$\\
			$\Phi_2$& $Q \tilde{Q}$ & $\square$ & $0$ & $1$ & $1$ & $0$ \\
			$\Phi_3$& $\text{det} S$& $1$ & $N$ & $0$ & $0$ & $0$ \\
			$\Phi_4$& $S^{N-1} Q^2$ & $1$ & $N-1$ & $2$ & $0$ & $0$ \\
			$\Phi_5$& $S \tilde{Q}^2$ & \symm & $1$ & $0$ & $2$ &$0$ \\
			$\Phi_6$& $Y^{bare}_{SU(N-2)} S$& $1$ & $\frac{3}{2} - N$ & $-1$ & $2-N$ & $1$ \\
			$\Phi_7$& $Y^{bare}_{SU(N-2)} \tilde{Q}^{N-2}$ & $1$ & $\frac{1}{2} - N$ & $-1$ & $0$ &$1$ \\ \hline 
			 & $Y^{bare}_{SU(N-2)}$ & $1$ & $\frac{1}{2}-N $ & $ -1$ & $2-N$ & $1$ \\
		\end{tabular}
		\caption{Gauge invariant combinations of the Case I-A reproducing the singlets  appearing  the RHS of the identity  (\ref{ConfZVI}). We reported also the charges of the bare monopole operator, which is not gauge invariant and it has $U(1)_2$-charge $-2(N-2)$.}  
		\label{Tab:IA}
\end{table}

In the last line of the table we have reported the global charges of the bare monopole $Y^{bare}_{SU(N-2)}$ computed along the lines of the discussion in subsection \ref{genmondisc} and using the branching rules presented in appendix \ref{Appendix_Branching_Rules}.
In this case the bare monopole is not gauge invariant and we need to dress it by the combinations of fields specified in the table. It turns out that there are three singlets, denoted as $\Phi_{1,6,7}$ corresponding to gauge invariant combination 
of the fields dressing the monopoles. 

From the combinations in Table \ref{Tab:IA}  we notice  that the most general superpotential compatible with the global symmetries is given by 
\begin{equation}
\label{claimWIA}
W_{conf} = \Phi _4 \Phi _6^2 \det \Phi _5+\Phi _1 \Phi _2 \Phi _3 \Phi _7
+\Phi _3 \Phi _4  \Phi _7^2+\Phi _5  \Phi _1^2 + \Phi_2^2 \Phi_3 \Phi_5^{N-3} \Phi_6^2.
\end{equation}
Below we show that such superpotential is generated from the deconfinement of the symmetric tensor and the following (confining) dualities.

\subsubsection*{Tensor deconfinement}
\begin{figure}[H]
\begin{center}
\includegraphics[width=12cm]{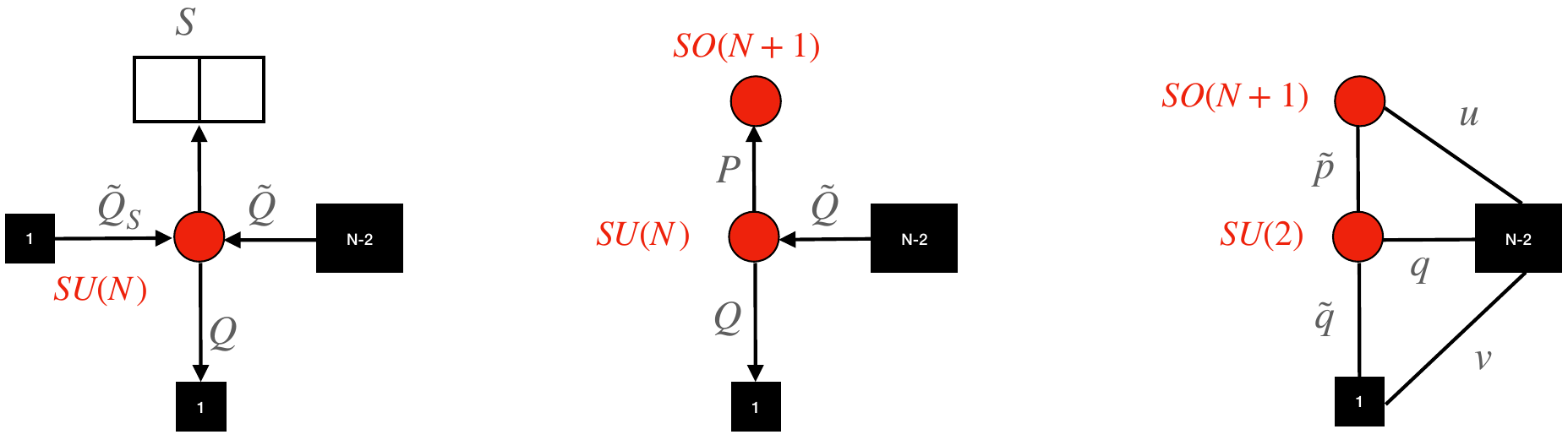}
\end{center}
\caption{On the left we provide the quiver for the electric model in Case I-A. In the central part of the figure we show the auxiliary quiver where the symmetric
is traded with an $SO(N+1)$ gauge group with a new bifundamental $P$. While a field $\sigma$ has to be considered in the quiver on the left, with the interaction in formula (\ref{IAstarting}), no further singlets need to be considered in the auxiliary quiver in the central picture. The $SU(N) \rightarrow SU(2)$  duality gives rise to the third quiver on the right.}
\label{IIAalt}
\end{figure}
Here we provide a derivation of the confining duality  by deconfining the symmetric tensor $S$. Such operation corresponds to find 
an auxiliary confining gauge group giving rise to a quiver where the original symmetric tensor is traded with a bifundamental.
In this case, being the two-index tensor in a symmetric representation we expect that the auxiliary gauge group is orthogonal.
Then by confining or dualiizing the other(s) node of the quiver using other (known) dualities one can arrive to the final WZ model  with superpotential (\ref{claimWIA}).
In general the procedure is not unique, i.e. there can be various quivers that can be used to deconfine a given tensor.
In some cases further flips may be required, in order to simplify the structure of the interactions in the various dual phases.
Here for example we provide two different deconfinements for the case at hand and we comment in the discussion what are the assumptions on the dynamical generation of the superpotential terms in the various phases of the process.

We start with the first derivation by simplifying the model, modifying the electric theory with an extra singlet $\sigma$, interacting through the superpotential
\begin{equation}
\label{IAstarting}
W = S \tilde Q_S^2 + \sigma^2 \det S.
\end{equation}
Then we deconfine the symmetric tensor using the quiver in Figure \ref{IIAalt}
with vanishing superpotential
\begin{equation}
W =0.
\end{equation}

This is a standard deconfinement that we review in appendix \ref{apppSO}. At this stage we have an $SU(N) \times SO(N+1)$ quiver and we observe that the $SU(N)$ gauge theory  can be dualized using the (chiral) duality
discussed in \cite{Nii:2018bgf} and reviewed in appendix \ref{Niidual}. The $SU(N)$ gauge group in this case is dual to  an $SU(2)$ gauge group, with superpotential
\begin{equation}
W = u \tilde p q + v \tilde q q, 
\end{equation}
where the mesons are $u=P\tilde Q$ and 
$v=Q\tilde Q$.
While the duality map is canonical also for the baryon, it is more intricate for the anti-baryons, indeed, they do not exist in the electric phase while they are allowed in the dual.
There is a non trivial mapping between these last and the dressed monopoles of the electric theory \cite{Nii:2018bgf}. We will discuss further details of this mapping in appendix \ref{Niidual}.

We then proceed observing that the $SO(N+1)$ gauge theory is confining and that the confining superpotential is
\begin{equation}
\label{superpotIAbis}
W =S_{\tilde p \tilde p} q_0^2 + S_{\tilde p u} q_0 q_1 +
S_{uu} q_1^2 + \rho^2 
\det\left(
\begin{array}{cc}
S_{\tilde p \tilde p} & S_{\tilde p u} \\
S_{u\tilde p}&S_{uu}
\end{array}
\right)
+
S_{\tilde p u} q
+ v \tilde q q, 
\end{equation}
where $q_0$ and $q_1$ are the baryon monopoles
\begin{equation}
q_0 = Y_{SO(N+1)}^- \epsilon \cdot (\tilde p u^{N-2}),
\quad
q_1 = Y_{SO(N+1)}^- \epsilon \cdot (\tilde p^2 u^{N-3})
\end{equation}
and $\rho$ corresponds to the monopole $Y_{SO(2N+1)}^+$.
The massive fields $S_{\tilde p u}$ and $q$ can be integrated out and they give rise to a quartic interaction $q_0 q_1  v \tilde q$ and a leftover interaction coming from the derivative of determinant of the type $\rho^2 S_{\tilde p \tilde p} v^2 \tilde q^2  S_{uu}^{N-3}$.
The last step consists of confining the leftover $SU(2)=USp(2)$ gauge theory, corresponding to the third quiver in Figure \ref{IIAalt}.
This theory has one adjoint $S_{\tilde p \tilde p}$, one fundamental $q_0$ and another fundamental $\tilde q$. The confining duality is reviewed in appendix \ref{app:BLMAdj}.

There are four $SO(N+1)$ singlets that interact in the WZ model, corresponding to the fundamental monopoles $Y_0$, the dressed monopole $Y_1 = Y_0  Tr( S_{\tilde p \tilde p})$, the meson $M = \tilde q^2 S_{\tilde p \tilde p}$ and the singlet $\phi = Tr S^2_{\tilde p \tilde p}$. 
Furthermore, the operator $q_0  \tilde q$ is mapped to $Y_0 M$.
All in all the superpotential of the WZ model becomes
\begin{equation}\label{spotSU2}
W=Y_0^2  M \phi
+
Y_1^2 M
+
S_{uu} q_1^2
+
\rho^2 \phi \det(S_{uu} )
+
q_1 v  Y_0 M 
+
\rho^2  M  v^2 S_{uu}^{N-3} .
\end{equation}
Using then the dictionary
\begin{eqnarray}
&&
\Phi_1 \leftrightarrow q_1 ,\quad 
\Phi_2 \leftrightarrow v   ,\quad 
\Phi_3 \leftrightarrow  M  ,\quad 
\Phi_4 \leftrightarrow \phi, \nonumber \\
&&
\Phi_5 \leftrightarrow S_{uu} ,\quad 
\Phi_6 \leftrightarrow \rho  ,\quad 
\Phi_7 \leftrightarrow  Y_0  ,\quad 
\sigma  \leftrightarrow Y_1,
\end{eqnarray}
we arrive at the final superpotential 
\begin{equation}
\label{claimmeWIA}
W_{conf} = \Phi _4 \Phi _6^2 \Phi _5^{N-2}+\Phi _1 \Phi _2 \Phi _3 \Phi _7
+\Phi _3 \Phi _4  \Phi _7^2+\Phi _5  \Phi _1^2 + \sigma^2 \Phi_3 + \Phi_2^2 \Phi_3 \Phi_5^{N-3} \Phi_6^2.
\end{equation}
Observe that we can also remove the singlet $\sigma$ from both (\ref{IAstarting}) and (\ref{claimmeWIA})  consistently.

\subsubsection*{Tensor deconfinement again:  an alternative derivation}
\begin{figure}[H]
\begin{center}
\includegraphics[width=8cm]{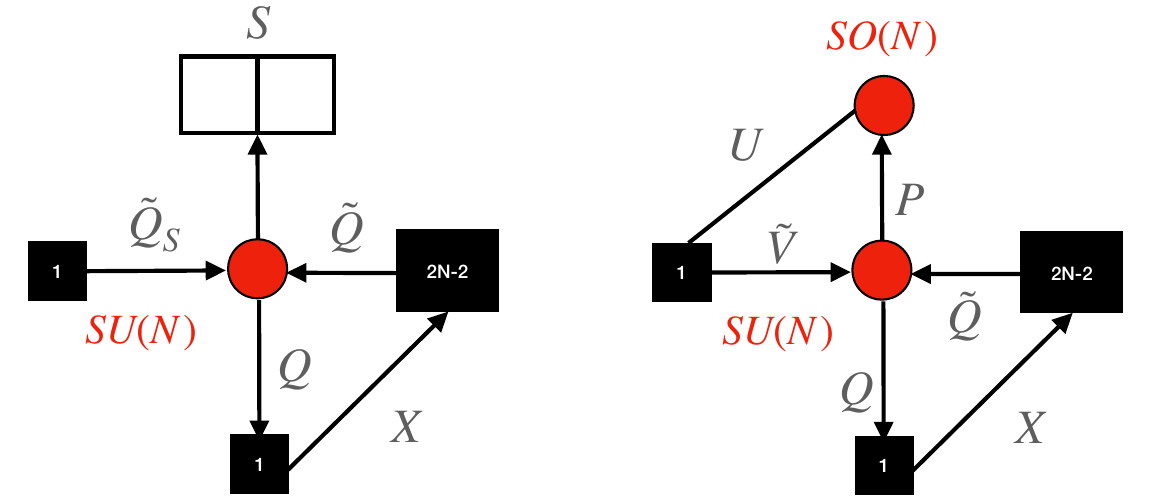}
\end{center}
\caption{On the left we provide the quiver for the electric model in Case I-A. On the right of the picture we show the auxiliary quiver where the symmetric
is traded with an $SO(N)$ gauge group with a new bifundamental $P$. 
In this case we have added the flipper $X$ on the electric side, flipping the mesonic component $Q \tilde Q$ and we have added the flipper $V$ in the
auxiliary quiver. There are also two further flippers denoted as $\gamma$ and $\alpha$ in the superpotential (\ref{WIAdecnuovo}), but  we have not depicted them here.
}
\label{quivdecI-A}
\end{figure}

Here we discuss an alternative derivation of the confining duality.
Again we want to prove the duality read from the identity (\ref{ConfZVI})
by deconfining the symmetric tensor using another confining duality for $SO(N)$ with vectors, reviewed in appendix \ref{apppSO}.
In addition we flip the meson $Q \tilde Q$ of the electric theory through the superpotential 
\begin{equation}
W_{I-A}^{ele} = S Q_S^2 + X Q \tilde Q.
\end{equation}
Next we deconfine the symmetric tensors and we obtain the $SO(N) \times SU(N)$ quiver 
in figure (\ref{quivdecI-A}) with superpotential
\begin{equation}
\label{WIAdecnuovo}
W = Y_{SO(N)}^+ + P U \tilde V + \alpha U^2 + Q \tilde Q X + \gamma \, \epsilon_N \cdot P^N.
\end{equation}
The next step consists of confining the $SU(N)$ gauge node with $N-1$ antifundamentals and $N+1$
fundamentals.

Such a duality has been derived in \cite{Nii:2018bgf} and it is reviewed in appendix \ref{Niiconf}. The theory is confining and it confines in terms of mesons, baryons and dressed monopoles, with superpotential (\ref{dualconf3dMbbt}). Details about this 
confining duality are given in Appendix \ref{Niiconf}.
In our case these fields correspond to the following gauge invariant combinations
\begin{equation}
\left(
\begin{array}{cc}
M_{\tilde Q P} & M_{\tilde V P} \\
M_{\tilde Q Q} & M_{\tilde V  Q}
\end{array}
\right),
\quad
\left(
\begin{array}{c}
B_0 \\
B_S
\end{array}
\right)
= \epsilon_N \cdot \left(
\begin{array}{c}
P^{N-1} Q \\
P^N
\end{array}
\right),
\quad
\left(
\begin{array}{c}
y_1 \\
y_2
\end{array}
\right)
= \left(
\begin{array}{c}
Y_d \tilde Q^{N-2}\\
Y_d \tilde Q^{N-3} \tilde V
\end{array}
\right).
\end{equation}
The superpotential for this confined theory becomes
\begin{eqnarray}
\label{422}
W &=& M_{\tilde Q P} B_0 y_1 + M_{\tilde V P} B_0 y_2+
M_{\tilde Q Q} B_S y_1+ M_{\tilde V Q} B_S y_2 + \nonumber 
\\
&+&
 U M_{\tilde V P} + \alpha U^2 + M_{\tilde Q Q}  X + \gamma B_S
+ M_{\tilde V Q} Y_{SO(N) \,\epsilon \cdot (M_{\tilde Q P} ^{N-3} B_0)}^-,
\end{eqnarray}
where we claim that the last term is dynamically generated, such that $M_{\tilde V Q}$ flips the baryon monopole of the $SO(N)$ group.
Then we integrate out the massive fields and we are left with 
\begin{eqnarray}
\label{quasi}
W &=& M_{\tilde Q P} B_0 y_1 + \alpha (B_0 y_2 )^2 
 + M_{\tilde VQ} Y_{SO(N) \,\epsilon \cdot (M_{\tilde Q P} ^{N-3} B_0)}^-.
\end{eqnarray}
The final step consists of confining the $SO(N)$ gauge theory with $N-1$ vectors and superpotential (\ref{quasi}).
The dual theory has superpotential 
\begin{equation}
W = \mathcal{S} q^2 + \sigma^2 \det 
\mathcal{S}
 + 
S_{01} y_1 + \alpha S_{11} y_2^2 +
 M_{\tilde V Q}  q_1,
 \end{equation}
 where
\begin{equation}
\mathcal{S}
=
\left(
\begin{array}{cc}
S_{00}& S_{01} \\
S_{01} & S_{11}
\end{array}
\right)
=
\left(
\begin{array}{cc}
M_{\tilde Q P} ^2 &  M_{\tilde Q P} B_0 \\
M_{\tilde Q P} B_0  & B_0^2
\end{array}
\right), \quad
q = 
\left(\begin{array}{c}
q_0\\
q_1
\end{array}
\right)=
\left(\begin{array}{c}
Y_{SO(N) \, \epsilon \cdot (M_{\tilde Q P}^{N-2})}^-
\\
Y_{SO(N) \, \epsilon \cdot (M_{\tilde Q P} ^{N-3} B_0 )}^-
\end{array}
\right)
\end{equation}
 and after integrating out the massive fields we obtain
 \begin{equation}
 \label{WfinaleIA}
W = S_{00} q_0^2 + \sigma^2 S_{11} \det S_{00} + \alpha S_{11} Y_2^2 .
 \end{equation}
 By following the operator map across the various steps we can map the 
 gauge invariant operators obtained at this final stage with the fields $\Phi_I$ described above.
 We have the following dictionary
 \begin{eqnarray}
\Phi_1  \leftrightarrow q_{0},\quad
\Phi_3  \leftrightarrow   \alpha, \quad
\Phi_4  \leftrightarrow   S_{11},\quad
\Phi_5 \leftrightarrow S_{00}, \quad
\Phi_6  \leftrightarrow  \sigma, \quad
\Phi_7  \leftrightarrow  Y_2.
\end{eqnarray}
By substituting this map in (\ref{WfinaleIA}) we obtain the superpotential claimed in (\ref{claimWIA}), where the second and last terms are actually missing because the singlet $\Phi_2$ has been flipped by the field $X$ in the electric description.

\subsubsection*{A proof of (\ref{ConfZVI}) from field theory I}

Here we give a proof of  (\ref{ConfZVI}) using the first deconfinement discussed above.
The even and the odd case require different treatment and  here we study only the even case
 $SU(2n)$. The analysis in the odd case is analogous, and the only differences stay in the 
 Weyl factor and in the contribution of the zero weights for the vectors.
 We start considering the partition function of the $SU(2n)$ gauge theory by multiplying it by the hyperbolic gamma function corresponding to the singlet $\sigma$ in (\ref{IAstarting}).
We have
\begin{equation}
\label{decaaa}
\Gamma_h(\omega - n \tau_S) Z_{SU(2n)}(\mu;\vec \nu,\omega - \frac{\tau_S}{2};\tau_S;-;-;-).
\end{equation}
The partition function of the deconfined theory, corresponding to the auxiliary quiver in the right side of figure \ref{IIAalt} can be obtained using formula (\ref{IAstarting}). 
The partition function is
\begin{eqnarray}
\label{ZIAfirstprima}
Z =
&&\!\!\!
\frac{1 }{(2n)! \cdot n! \cdot 2^{n}} 
\int \prod_{i=1}^{2n} d \sigma_i \prod_{\alpha=1}^{n} d \rho_\alpha
\prod_{i=1}^{2n} \left(\Gamma_h(\mu + \sigma_i) \prod_{b=1}^{2n-2}  \Gamma_h(\nu_b - \sigma_i)\right)
\nonumber\\
\times&&
\frac{ \delta(\sum_{i=1}^{2n} \sigma_i)
\prod_{i=1}^{2n} \left(\Gamma_h 
\left(\sigma_i  +\frac{\tau_S}{2} \right)
\prod_{\alpha=1}^{n} 
\Gamma_h \left(\sigma_i \pm \rho_\alpha +\frac{\tau_S}{2} \right)\right)}
{\prod_{i<j} \Gamma(\pm(\sigma_i - \sigma_j))
\prod_{\alpha} \Gamma_h(\pm \rho_\alpha )
\prod_{\alpha<\beta} \Gamma_h(\pm \rho_\alpha \pm \rho_\beta))}.
 \end{eqnarray}
Then we dualized the $SU(2n)$ theory using formula (\ref{chiralpqstarduality}) on the partition function. This gives an $SU(2) \times SO(2n+1)$ quiver.
The partition function for this quiver is
\begin{eqnarray}
\label{ZIAfirstseconda}
&&Z =\!
\frac{\prod_{b=1}^{2n-2}\Gamma_h\left(\nu_b+\mu\right) }{ n! \cdot 2^{n+1}} \!\!
\int \!\! d \sigma \! \prod_{\alpha=1}^{n} \! d \rho_\alpha
\frac{ \Gamma_h \left(\pm \sigma +X-\frac{\tau_S}{2} \right)
\prod_{\alpha=1}^{n} 
\Gamma_h \left(\pm \sigma \pm \rho_\alpha +X-\frac{\tau_S}{2} \right)}
{\Gamma(\pm 2\sigma )
\prod_{\alpha} \Gamma_h(\pm \rho_\alpha )
\prod_{\alpha<\beta} \Gamma_h(\pm \rho_\alpha \pm \rho_\beta))}\nonumber\\
&&\times
\Gamma_h(X \!-\!\mu \pm  \sigma)\!\! \prod_{b=1}^{2n-2} \!\! \left(\!\Gamma_h(2\omega \!- \! X \!-\!\nu_b \! \pm \! \sigma)
\Gamma_h \! \! \left(\nu_b \! + \! \frac{\tau_S}{2} \right) \! \prod_{\alpha=1}^{n} 
\Gamma_h \! \left(\nu_b+\frac{\tau_S}{2} \pm \rho_\alpha \right)\! \right),
 \end{eqnarray}
with
$
X =\frac{(2n+1) \tau_S +2\mu}{4}
$.
Then we confine the $SO(2n+1)$ gauge group using the identity (\ref{SOelemag}).
The partition function for the remaining $SU(2)=USp(2)$ gauge theory is
\begin{eqnarray}
\label{ZIAfirstterza}
Z =
&&
\!\!\!\!\!\!\!\!
\prod_{1\leq b \leq c \leq 2n-2} \!\!\!\!\!\! \!\!\Gamma_h(\nu_b \!+\!\nu_c \!+\! \tau_S)
 \Gamma_h \! \left(\! \omega\!-\!\left(2n\!-\!\frac{3}{2}\right)\!\tau_S\!-\!\mu\! -\!\!\sum_{b=1}^{2n-2}\nu_b \!\right)
 \!\!\! 
\prod_{b=1}^{2n-2}
\!\!
\Gamma_h\!\left(\nu_b+\mu,\omega\!-\!\nu_b\!-\!\frac{\tau_S}{2} \right)
\nonumber\\
&&\!\!\!\!\!\!\!\!
\frac{1}{2} \!\!
\int \!\! d \sigma
\frac{\Gamma_h\left(\frac{(2N+1)\tau-2\mu}{4} \pm  \sigma,\omega-\frac{(2n-1)\tau+2\mu}{4} \pm \sigma \right) }
{\Gamma(\pm 2\sigma )} \Gamma_h\left(\frac{(2N-1)\tau_S}{2}+\mu\pm 2 \sigma \right),
 \end{eqnarray}
where in the integral we see the contributions of the fundamentals $q_0$ and $\tilde q$ and of the adjoint $S_{\tilde p \tilde p}$, with mass parameters consistent with the superpotential (\ref{superpotIAbis}).
The last step consists of dualizing the $USp(2)$ node using the identity (\ref{AR22}) with $N=1$.
The final partition function is 
\begin{eqnarray}
\label{finalIdeconf1}
Z =
&&
 \Gamma_h\left(\omega-\left(2n-\frac{3}{2}\right)\tau_S-\mu -\sum_{b=1}^{2n-2}\nu_b\right)
\prod_{1\leq b \leq c \leq 2n-2} \Gamma_h(\nu_b + \nu_c + \tau_S)
\nonumber\\
&&
\prod_{b=1}^{2n-2}\Gamma_h\left(\nu_b+\mu,\omega-\nu_b-\frac{\tau_S}{2} \right)
\Gamma_h\left(\omega-\mu -\left(2 n-\frac{1}{2}\right) \tau _S\right)
\nonumber\\
&&
\Gamma_h(2n \tau_S) \Gamma_h((2n-1)\tau_S+2\mu) \Gamma_h(\omega-n \tau_S).
 \end{eqnarray}
We conclude comparing (\ref{decaaa}) and (\ref{finalIdeconf1}), observing that 
once we simplify the terms $ \Gamma_h(\omega-n \tau_S)$ we have obtained the relation 
(\ref{ConfZVI})  for $N=2n$.

We conclude commenting on the derivation in the odd $SU(2n+1)$ case.
The main difference is in formulas (\ref{ZIAfirstprima}) and (\ref{ZIAfirstseconda}), due to the different contributions of the weights  of the $SO(N)$ vector in the numerator and of the adjoint representation in the denominator.
Such difference disappears anyway in (\ref{ZIAfirstterza}) and indeed the final result (\ref{ConfZVI}) 
is independent on the parity of the rank of the $SU(N)$ gauge group.

\subsubsection*{A proof of (\ref{ConfZVI}) from field theory II}

Here we give a proof of  (\ref{ConfZVI}) using the second deconfinement discussed above.
 Again we restrict to the even $SU(2n)$
case, even if the derivation for $SU(2n+1)$ is  analogous.
In this case the deconfined partition function becomes
\begin{eqnarray}
&&
Z \!=\!
\frac{\Gamma_h(2n \tau_S)\Gamma_h(2\omega-n \tau_S) }{n! (2n)! 2^{n}} 
\prod_{b=1}^{2n-2} \!\! \Gamma_h(2\omega\!-\!\nu_b \!-\!\mu) \!\!
\int \prod_{i=1}^{2n} d \sigma_i  \delta\Big(\sum_{i=1}^{2n} \sigma_i\Big)\!\! \prod_{\alpha=1}^{n} d \rho_\alpha
\Gamma_h (m\pm \rho_\alpha) 
\nonumber\\
&&\!\!\!
\frac{
\prod_{i=1}^{2n} \left(\Gamma_h(\mu + \sigma_i)  \Gamma_h\left(2\omega-m-\frac{\tau_S}{2} -\sigma_i\right)\prod_{\alpha=1}^{n} 
\Gamma_h \left(\sigma_i \pm \rho_\alpha +\frac{\tau_S}{2} \right)
\prod_{b=1}^{2n-2}  \Gamma_h(\nu_b - \sigma_i)\right)}
{\prod_{i<j} \Gamma(\pm(\sigma_i - \sigma_j))
\prod_{\alpha<\beta} \Gamma_h(\pm \rho_\alpha \pm \rho_\beta))}, \nonumber\\
 \end{eqnarray}
 with the balancing condition 
 \begin{equation}
 \label{bc1a}
m=\omega-n \tau_S.
 \end{equation}
The $SU(2n)$ gauge theory has $2n+1$ fundamentals and $2n-1$ antifundamentals and it can be dualized 
into a confining theory. On the partition function this duality corresponds to the identity (\ref{chiralpqstarconfining}).
By applying such identity we obtain 
\begin{eqnarray}
Z =
&&
\Gamma_h(2n \tau_S)
\Gamma_h\left(\omega-\left(2n-\frac{1}{2} \right) \tau_S -\mu \right)
 \nonumber\\
 \times &&
\Gamma_h\left( \omega +\mu + \left(N-\frac{1}{2} \right) \tau_S \right)
\Gamma_h\left( 2\omega-N \tau_S-\mu -\sum_{b=1}^{2n-2} \nu_b\right) \nonumber\\
\times &&
\int \prod_{i=1}^{2n}
\prod_{\alpha=1}^{n} d \rho_\alpha
\frac{ 
\prod_{\alpha=1}^{n} \Gamma_h\left( \left(n-\frac{1}{2} \right) \tau_S +\mu \pm \rho_\alpha \right)
\prod_{b=1}^{2n-2}\Gamma_h \left(\nu_b +\frac{\tau_S}{2} \pm \rho_{\alpha}\right)}
{\prod_{\alpha<\beta} \Gamma_h(\pm \rho_\alpha \pm \rho_\beta))}, \nonumber\\
\end{eqnarray}
where we have simplified the expression by using the balancing condition (\ref{bc1a}) and the duplication formula (\ref{duplication}).

The $SO(2n)$ gauge theory has now $2n-1$ vectors and it can be dualized into singlets thanks to the confining duality reviewed in appendix \ref{apppSO}.
On the partition function this confining duality corresponds to formula (\ref{SOelemag}) and the final expression coincides exactly with (\ref{ConfZVI}), where the contribution of the singlet $\Phi_2$ is traded with the contribution of the flipper $X$ appearing in the RHS of the equality.
\subsection{Case I-B: $S \oplus \tilde Q_S \oplus (N-1) \tilde Q$}
\label{subsec:I-B}
Here we consider an  $SU(N)$ gauge theory with an antisymmetric tensor $A$, $N-1$ antifundamentals  $\tilde Q$ and three fundamentals $Q$.
In this case the theory confines and the analysis for odd and even $N$ must be taken separately.

\subsubsection*{Applying the duplication formula}
In the even case $N=2n$ the elementary degrees of freedom are
\begin{eqnarray}
M \equiv Q \tilde Q, \quad
T \equiv A^n, \quad
B_{n-1} \equiv A^{n-1} Q^2, \quad
\widetilde B_1 = A \tilde Q^2.
\end{eqnarray}
In addition  we need to consider the bare monopole of the electric theory that acts as  a singlet in the dual phase. Such operators corresponds to $Y_{SU(2n-2)}^{bare}$. 
The confining superpotential is 
\begin{equation}
W= Y_{SU(2n-2)}^{bare} ( M B_{N-1} \widetilde B_1^{N-1} + T M^3 \widetilde B_1^{N-2}).
\end{equation}
This confining duality follows from the 4d parent and at the level of the partition function it corresponds to the identity 
\begin{eqnarray}
&&
Z_{SU(2n)}\left(\vec \mu;,\vec v;-;-;\tau_A;- \right)
=
\prod_{a=1}^{3} \prod_{b=1}^{2n-2} \Gamma_h(\mu_a + \nu_b) 
\prod_{1\leq a<b \leq 3}  \Gamma_h((n-1)\tau_A + \mu_a + \mu_b)
\nonumber \\
&\times&
\Gamma_h( n \tau_A)\!\!\!\! \!\!\!
 \prod_{1\leq a<b \leq 2n-1}  
 \!\!\!\! \!\!\!
 \Gamma_h(\tau_A + \nu_a + \nu_b) \,
\Gamma_h(2 \omega - \sum_{b=1}^{2n-1} \nu_b  - \sum_{a=1}^{3} \mu_a-2(n-1) \tau_A).
\end{eqnarray}

In the odd case $N=2n+1$ the elementary degrees of freedom are
\begin{eqnarray}
M \equiv Q \tilde Q, \quad
B_{n}\equiv A^n Q, \quad
B_{n-1} \equiv A^{n-1} Q^3, \quad
\widetilde B_1 = A \tilde Q^2.
\end{eqnarray}
In addition  we need to consider the bare monopole of the electric theory that acts as  a singlet in the dual phase. Such operators corresponds to $Y_{SU(2n-1)}^{bare}$.
The confining superpotential is 
\begin{equation}
W =Y_{SU(2n-1)}^{bare}(\widetilde B_1^n B_{n-1} + \widetilde B_1 ^{n-1} M^2 B_{n}).
\end{equation}
This confining duality follows from the 4d parent and at the level of the partition function it corresponds to the identity 
\begin{eqnarray}
&&Z_{SU(2n+1)}\left(\vec \mu;,\vec v;-;-;\tau_A;-\right)
=
\prod_{a=1}^{3} \prod_{b=1}^{2n} \Gamma_h(\mu_a + \nu_b) 
\prod_{a=1}^3  \Gamma_h(n\tau_A + \mu_a )
\nonumber \\
&\times&
 \prod_{1\leq a<b<c \leq 3}  \Gamma_h((n-1)\tau_A + \mu_a + \mu_b+\mu_c)
 \prod_{1\leq a<b \leq 2n}  \Gamma_h(\tau_A + \nu_a + \nu_b)
  \\
&\times&
\Gamma_h(2 \omega - \sum_{b=1}^{2n} \nu_b  - \sum_{a=1}^{3} \mu_a-(2n-1) \tau_A).
\nonumber
\end{eqnarray}
Starting from these relations we apply the duplication formula by freezing the masses of the fundamentals as in (\ref{freezing2}).
We work in the even and in the odd case separately and we further change the name of the parameter $\tau_A$ into $\tau_S$.
We obtain a unified relation holding for both $N=2n$ and $N=2n+1$. The relation is
\begin{eqnarray}
\label{ZIB}
&&
Z_{SU(N)}\left(-;\omega-\frac{\tau_S}{2},\vec{\nu};\tau_S;-;-;-\right)=
\Gamma_h(N \tau_S) 
\prod_{b=1}^{N-1} \Gamma_h \left(\omega-\frac{\tau_S}{2} -\nu_b\right) 
\nonumber \\
&&
\Gamma_h\left(\omega-\left(N-\frac{1}{2}\right)\tau_S-\sum_{b=1}^{N-1} \nu_b \right)
\prod_{1\leq b\leq c\leq N-1} \Gamma_h(\tau_S + \nu_b+\nu_c).
\end{eqnarray}

\subsubsection*{Field theory interpretation}

We then move to the field theory interpretation of the relation (\ref{ZIB}).
On the electric side we have an $SU(N)$ gauge theory with a two index symmetric tensors $S$, 
$(N-1)$ antifundamentals $\widetilde Q$ and an antifundamental  $Q_S$ interacting with the symmetric tensor through (\ref{WSQSQS}).
The local and global  charges of these fields are summarized in Table \ref{Tab:IBele}.
\begin{table}[H]
	\centering
\begin{tabular}{c|c|cccc}
    &SU(N)&$U(1)_{\tilde Q}$ &$U(1)_S$ & $SU(N-1)$ &$U(1)_R$ \\
\hline
$S$ &\symm&0&1&0&0 \\
$\tilde Q_S$ &$\overline \square$&0&$-1/2$&0&1 \\
$\tilde Q$ &$\overline {\square}$&1&$0$&$\square$&0 \\
	\end{tabular}
	\caption{Charged fields in the electric side of the case I-B}
	\label{Tab:IBele}
\end{table}
On the dual side we have tow following gauge invariant combinations
\begin{table}[H]
	\centering
	\begin{tabular}{c | c c c c}
		&  $SU(N-1)$ & $U(1)_S$ & $U(1)_{\tilde{Q}}$& $U(1)_R$ \\ 
		\hline
		$\Phi_2\equiv \text{det} S$ & $1$ & $N$ & $0$ & $0$ \\
		$\Phi_3\equiv S \tilde{Q}^2$ & $\symmF$ & $1$ & $2$ & $0$ \\
		$\Phi_4= Y^{bare}_{SU(N-2)} $ & $1$ & $1/2 - N$ & $1-N$ & $1$ \\
		$\Phi_1\equiv Y^{bare}_{SU(N-2)} \tilde{Q}^{N-2} S^{N-1}$ & $\overline{\square}$ & $-1/2$ & $-1$ & $1$ \\
	\end{tabular}
	\caption{Gauge invariant combinations of the Case I-B reproducing the singlets appearing in the RHS of the identity \eqref{ZIB}. In this case the bare monopole is gauge invariant.}
	\label{Tab:IB}
\end{table}
and the superpotential compatible with such charges is given by
\begin{equation}
\label{WIB}
W = \Phi_1^2 \Phi_3 +\Phi_4^2\Phi_2  \det \Phi_3.
\end{equation}

\subsubsection*{Tensor deconfinement}

Here we show that the confining duality following from the field theoretical interpretation of the relation (\ref{ZIB}) is a consequence of other 3d dualities. We start by deconfining the symmetric tensor using an $SO(N)$ confining duality and then proceed by dualizing the unitary gauge node and confining back the orthogonal one. In this way we will be able to obtain the confined description discussed above in terms of ordinary dualities that do not involve any tensor field.

\begin{figure}[H]
\begin{center}
  \includegraphics[width=8cm]{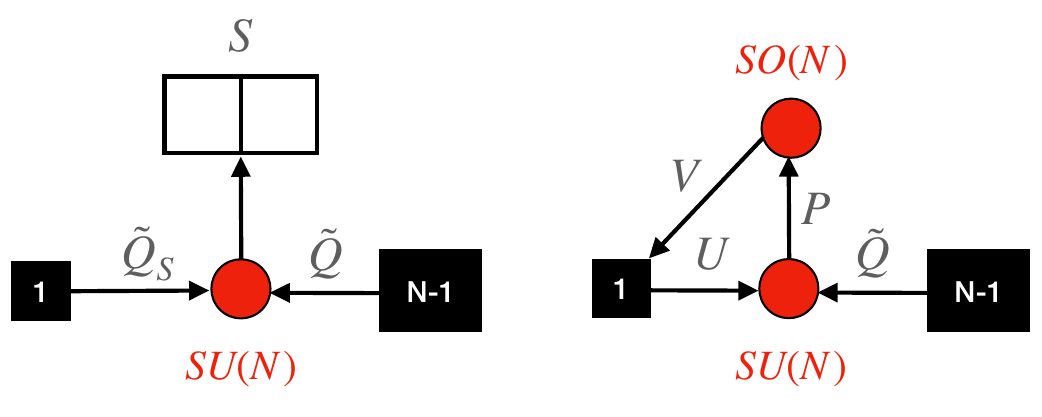}
  \end{center}
 \caption{On the left we provide the quiver for the electric model in Case I-B. On the right of the picture we show the auxiliary quiver where the symmetric
is traded with an $SO(N)$ gauge group with a new bifundamental $P$ and a new vector $V$. 
In this case we have added the flipper $U$ in the
auxiliary quiver. There are also two further flippers denoted as $\gamma$ and $\alpha$ in the superpotential (\ref{spotdec}), but  we have not depicted them here.} 
  \label{deecib}
\end{figure}

 We start by considering the model depicted in Figure \ref{deecib} with superpotential 
 \begin{equation}
 \label{spotdec}
 W =Y_{SO(N)}^+ + VUP +\alpha V^2+\gamma \, \epsilon_{N} \cdot P^{N}.
 \end{equation}
 
 By confining the $SO(N)$ node this model coincides with the original one. On the other hand it is possible to confine the $SU(N)$ gauge group has well, because it has $N$ non-chiral flavors. This  confining theory can indeed be regarded as a limiting case of
3d $SU(N)$  Aharony duality, as shown in \cite{Aharony:2013dha}.  The dual superpotential is
 \begin{eqnarray}
 \label{Wconf1}
W =Y_{SO(N)}^+ +V M_{UP}+ a V^2+\gamma \tilde b+Y_{SU(N)} (b \tilde b -\det M),
  \end{eqnarray}
where the last terms enforces the classical constraints on the moduli space for the confining $SU(N)$ gauge node.
The fields $V$, $M_{UP}$, $\gamma$ and $\tilde b$ are massive in (\ref{Wconf1}) and they can be integrated out. This procedure removes the linear monopole superpotential as well and, by inspection on the quantum charges,  we claim that there is a further interaction where the $SU(N)$ baryon $b$ flips the monopole $Y_{SO(N)}^+$. Summarizing, the superpotential for this phase is 
\begin{eqnarray}
W =b \, Y_{SO(N)}^+ + a \, Y_{SU(N)}^2 (\epsilon_N \cdot M_{\tilde Q P}^{N-1})^2.
  \end{eqnarray}
The $SO(N)$ model is now confining, and its confined degrees of freedom are the symmetric tensor $S_M$  built from the $N-1$ vectors  $M_{\tilde Q P}$ and the baryon monopole $q=Y^-_{SO(N)\, \epsilon \cdot M_{\tilde Q P}^{N-2}}$. These fields interact through the superpotential term  $W \subset S_M q^2$. On the other hand, in this last confining duality the baryon flipping the $Y_{SO(N)}^+$ monopole disappears.
Considering now the identification 
\begin{eqnarray}
\Phi_1 \leftrightarrow q\, \quad
\Phi_2 \leftrightarrow a \, \quad
\Phi_3 \leftrightarrow S_M \, \quad
\Phi_4 \leftrightarrow Y_{SU},
\end{eqnarray}
 we can see that the deconfinement procedure discussed here reproduces the superpotential
 (\ref{WIB}).

 \subsubsection*{Tensor deconfinement 2: an alternative derivation}
 
\begin{figure}[H]
\begin{center}
  \includegraphics[width=8cm]{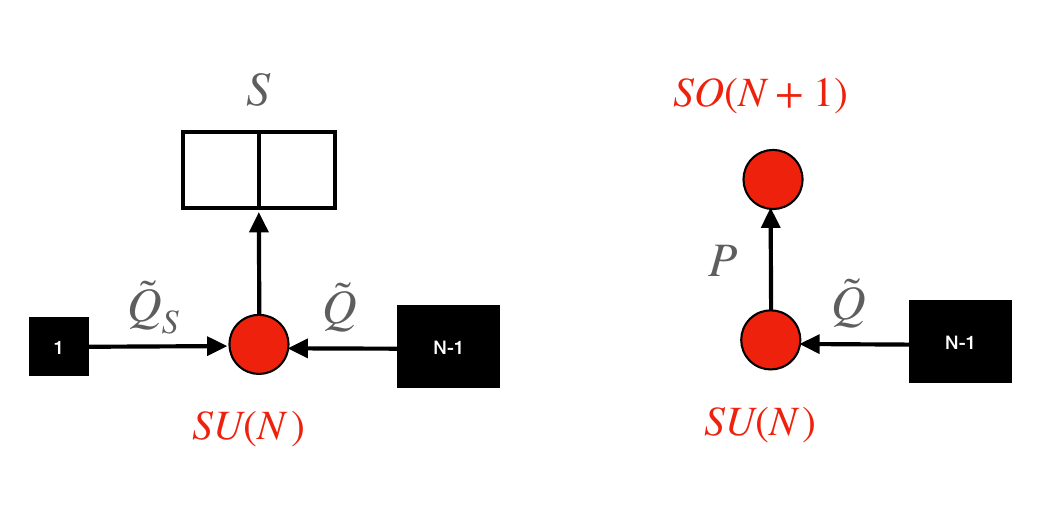}
  \end{center}
 \caption{On the left we provide the quiver for the electric model in Case I-B. On the right of the picture we show the auxiliary quiver where the symmetric
is traded with an $SO(N+1)$ gauge group with a new bifundamental $P$. In this case  
no further singlets need to be considered in the quiver on the left while
a singlet $\sigma$ has to be considered in the auxiliary quiver on the right, with the interaction given in formula (\ref{flippedIIB}) flipping the monopole 
$Y_{SO(N+1)}^+$.}
  \label{IIBbis}
\end{figure}

 We can also study the deconfinement in an alternative way, that is more similar to the 
 procedures that we will consider in the rest of the paper. We consider the deconfined quiver in  Figure \ref{IIBbis},
 where in the deconfinement we have introduced a singlet $\sigma$  and the corresponding interaction flipping the monopole $Y_{SO(N+1)}^+$ 
 \begin{equation}
 \label{flippedIIB}
 W = \sigma Y_{SO(N+1)}^+.
  \end{equation}
 The $SU(N)$ with $N+1$ fundamentals and $N-1$ antifundamentals  confines as reviewed in appendix  (\ref{Niiconf}). After its confinement we have 
 an $SO(N+1)$ gauge theory with superpotential
  \begin{equation}
 \label{WdualIIBbis}
  W = M B Y_{d}+\sigma Y^-_{SO(N+1) \epsilon_{N+1}\cdot(M^{N-2}B)},
  \end{equation}
 where the $SU(N)$ gauge invariant combinations appearing in (\ref{WdualIIBbis})
 correspond to the rectangular meson $M = P \tilde Q$, the baryon $B = \epsilon_{N} P^N$ and  the minimal dressed monopole $Y_{d} =Y \epsilon_{N} \cdot \tilde Q^{N-2}$, where $Y = Y_1\dots Y_{N-1}$ is the product of the fundamental monopoles that breaks the gauge group to $SU(N-2)$. This last combination is not gauge invariant and it needs indeed to be dressed by the combination $\epsilon_{N} \cdot \tilde Q^{N-2}$, such that $Y_{d}$ transforms in an  antifundamental  representation of the flavor symmetry $SU(N-1)$.
 %
 %
 Furthermore we claim that the flipper $\sigma$ in this phase flips the 
 baryon monopole $Y^-_{SO(N+1) \, \epsilon \cdot(M^{N-2}B)}$ of the $SO(N+1)$ gauge group.
 The $SO(N+1)$ gauge group can be then reconfined giving rise to the superpotential
   \begin{equation}
 \label{WconflIIBbis}
 W=S q^2 + \rho^2 \det S+S_{01} Y_{d}+\sigma q_0,
   \end{equation}
      where 
  \begin{equation}
  S = 
  \left(
  \begin{array}{cc}
  S_{00} & S_{01} \\
  S_{10} & S_{11}
  \end{array}
  \right)
  ,\quad
q = 
  \left(
  \begin{array}{c}
q_0 \\
q_1
  \end{array}
  \right)
  \end{equation}    
      and
  the components of the symmetric meson are $S_{00} = B^2$,
    $S_{01} = B M $ and $S_{11} = M^2$ while the fields $q_0$ and $q_1$ are
    the baryon monopoles
    \begin{equation}
    q_0=Y^-_{\epsilon \cdot (M^{N-1})}
    ,\quad
    q_1=Y^-_{\epsilon \cdot (M^{N-2}B)}.
    \end{equation}
    There is a further singlet $\rho$ that corresponds to the monopole $Y_{SO(N+1)}^+$.
    By integrating out the massive fields we are left with the superpotential
      \begin{equation}
W  =    S_{11} q_1^2  +\rho^2 S_{00} \det S_{11}.
       \end{equation}
 That coincides with (\ref{WIB}) once we use the dictionary      
       \begin{equation}
       \Phi_1=q_1,\quad
       \Phi_2=S_{00},\quad
       \Phi_3=S_{11},\quad
       \Phi_4\leftrightarrow \rho.
       \end{equation}

 \subsubsection*{A proof of  (\ref{ZIB}) from field theory I}

Here we prove the identity  (\ref{ZIB})  using the first deconfinement discussed above.
In the analysis of the partition function we need to distinguish the even $SU(2n)$ and the odd $SU(2n+1)$. 
 Here we discuss explicitly only the $SU(2n)$ case but we observe that the distinction between the two cases is not immanent and indeed we can write in the last step a single identity for  both the cases in terms of the same singlets. 
 \begin{eqnarray}
Z =
&&\!\!\!
\frac{\Gamma_h(2n \tau_S)\Gamma_h(2\omega-n \tau_S) }{(2n)! n! 2^{n}} 
\!\!
\int \prod_{i=1}^{2n} d \sigma_i \prod_{\alpha=1}^{n} d \rho_\alpha
\frac{ \delta(\sum_{i=1}^{2n} \sigma_i)
\prod_{i=1}^{2n} \prod_{\alpha=1}^{n} 
\Gamma_h \left(\sigma_i \pm \rho_\alpha +\frac{\tau_S}{2} \right)}
{\prod_{i<j} \Gamma(\pm(\sigma_i - \sigma_j))
\prod_{\alpha<\beta} \Gamma_h(\pm \rho_\alpha \pm \rho_\beta))}\nonumber\\
\times&&
\prod_{i=1}^{2n} \prod_{b=1}^{2n-1}  \Gamma_h(\nu_b - \sigma_i)
\prod_{\alpha=1}^{n}
\Gamma_h (m\pm \rho_\alpha)  \prod_{i=1}^{2n} \Gamma_h\left(2\omega-m-\frac{\tau_S}{2} -\sigma_i\right),
 \end{eqnarray}
 with the balancing condition    $m+n \tau_S = \omega$. 
 The partition function becomes
 \begin{eqnarray}
 \label{almostIB}
Z =
&&
\frac{1}{n! 2^{n}} 
\Gamma_h\left(\!\omega+\left(n-\frac{1}{2}\right)\tau_S\!+\!\sum_{b=1}^{2n-1} \nu_b,\omega-\left(2n-\frac{1}{2}\right)\tau_S\!-\!\sum_{b=1}^{2n-1} \nu_b\!\right) 
\nonumber\\
\times
&&
\Gamma_h(2n \tau_S) 
\int \prod_{\alpha=1}^{n} d \rho_\alpha
\frac{ 
 \prod_{\alpha=1}^{n}  \prod_{b=1}^{2n-1} 
\Gamma_h\left(\frac{\tau_S}{2} +\nu_b \ \pm \rho_\alpha \right)}
{
\prod_{\alpha<\beta} \Gamma_h(\pm \rho_\alpha \pm \rho_\beta))}.
 \end{eqnarray}
We  conclude the derivation  by plugging  (after using the inversion relation on the first term) the relation
(\ref{SOelemag}) into (\ref{almostIB}) obtaining the expected relation (\ref{ZIB}).

 \subsubsection*{A proof of  (\ref{ZIB}) from field theory II}

Here we prove the identity  (\ref{ZIB})  using the second deconfinement discussed above.
Again the analysis for the $SU(2n)$ and the $SU(2n+1)$ is different, even if the final result can be formulated with a unique formula. Here we choose to study the  $SU(2n+1)$ explicitly
and we refer to the difference in the  $SU(2n)$ in the bulk of the discussion.

In this case, once we deconfine the symmetric tensor $S$, with mass parameter $\tau_S$ we obtain an auxiliary $SO(2n+2)$ gauge group, with an $SO(2n+2) \times SU(2n+1)$ bifundamental 
$P$ with mass parameter $\frac{\tau_S}{2}$ and a singlet $\sigma$ with mass parameter 
$\left(\omega+\left(n +\frac{1}{2} \right)\tau_S \right)$ 
 flipping the $Y_{SO(2n+2)}^+$ monopole. The partition function for the $SO(2n+2) \times SU(2n+1)$ quiver is
\begin{eqnarray}
&&
Z = \frac{\Gamma_h(\omega+\left(n +\frac{1}{2} \right) \tau_S)}{(2n+1)! (n+1)! 2^{n+1}}
\int \prod_{j=1}^{2n+1} d \sigma_i \prod_{\alpha=1}^{n+1} d \rho_\alpha \nonumber \\
&&
 \frac{
\prod_{j=1}^{2n+1} 
\left(
\prod_{b=1}^{2n}  \Gamma_h\left( \nu_b - \sigma_j\right) 
\prod_{\alpha=1}^{n+1}\Gamma_h\left(\pm \rho_\alpha + \sigma_j + \frac{\tau_S}{2} \right)
\right)}
{\prod_{i<j} \Gamma_h(\pm(\sigma_i-\sigma_j)) \prod_{\alpha<\beta} \Gamma_h(\pm \rho_\alpha \pm \rho_\beta)}.
\end{eqnarray}
Then we confine the $SU(2n+1)$ gauge theory using the identity (\ref{chiralpqstarconfining}).
In this way the $SO(2n+2)$ gauge theory has $2n$ fundamentals from the $SU(2n+1)$ mesons,
with mass parameter $\nu_b + \frac{\tau_S}{2}$ and an extra fundamental corresponding to the 
$SU(2n+1)$ baryon with mass parameter $ \frac{2n+1}{2} \tau_S $. There is also another singlet 
corresponding to the dressed monopole $Y_d$, with mass parameter $(2\omega - \nu_b - (n+1)\tau_S) $. The partition function becomes
\begin{eqnarray}
&&
Z=
\frac{\Gamma_h\left(\omega+\left(n +\frac{1}{2} \right) \tau_S\right)}{(n+1)! 2^n}
\prod_{b=1}^{2n} \Gamma_h(2\omega - \nu_b - (n+1)\tau_S) \nonumber \\
&&
\int 
\frac{\prod_{\alpha=1}^{n+1} d \rho_\alpha \Gamma_h\left( \pm \rho_\alpha + \frac{2n+1}{2} \tau_S \right)
 \prod_{b=1}^{2n} \Gamma_h\left(\pm \rho_\alpha  + \nu_b+\frac{\tau_S}{2} \right)}{\prod_{\alpha<\beta} \Gamma_h(\pm \rho_\alpha \pm \rho_\beta)}.
\end{eqnarray}
Observe that this partition function is compatible with the claim that the second interaction in 
(\ref{WdualIIBbis}) is dynamically generated in the process. The last step corresponds to the application of the identity (\ref{SOelemag}) for $K=2n+1$, where actually the first term on the LHS of (\ref{SOelemag}) needs to be moved on the numerator of the RHS using the inversion relation (\ref{inversion}).
We obtain the final identity 
\begin{eqnarray}
Z=
&&\Gamma_h\left(\omega+\left(n +\frac{1}{2} \right) \tau_S\right)
\prod_{b=1}^{2n} \Gamma_h(2\omega - \nu_b - (n+1)\tau_S) \nonumber \\
\times &&
\prod_{b=1}^{2n}\Gamma_h \left(\omega-\nu_b-\frac{\tau_S}{2} \right)
\Gamma_h\left(\omega-\left(n +\frac{1}{2} \right) \right)
\Gamma_h \left(\omega-\sum_{b=1}^{2n} \nu_b-\left(2n+\frac{1}{2}\right)\tau_S\right)
\nonumber \\
\times &&
\prod_{b\leq c}\Gamma_h \left(\nu_b + \nu_c +\tau_S\right)
\Gamma_h \left((2 n+1) \tau_S \right)
\prod_{b=1}^{2n}\Gamma_h \left(\nu_b + \left(n+1\right) \tau_S  \right) 
\nonumber \\
=&&
\prod_{b=1}^{2n}\Gamma_h \left(\omega-\nu_b-\frac{\tau_S}{2} \right)
\Gamma_h \left(\omega-\sum_{b=1}^{2n} \nu_b-\left(2n+\frac{1}{2}\right)\tau_S\right)
\nonumber \\
\times &&
\prod_{b\leq c}\Gamma_h \left(\nu_b + \nu_c +\tau_S\right)
\Gamma_h \left((2 n+1) \tau_S \right).
\end{eqnarray}
This last formula, obtained by applying the inversion relation (\ref{inversion}) on the first equality, coincides with  (\ref{ZIB}) for $N=2n+1$. This concludes the derivation of the identity using the tensor deconfinement procedure discussed at field theory  level above.

 \subsection{Case I-C:  $S \oplus (N-2) \tilde Q$}
 \label{SQt}

We conclude this section  by considering again the $SU(N)$ theory with an antisymmetric tensor $A$, $N-2$ antifundamentals  $\tilde Q$ and four fundamentals $Q$ discussed in subsection (\ref{SQQtQtS}).

\subsubsection*{Applying the duplication formula}

The main difference with the case considered in subsection (\ref{SQQtQtS}) is that in this case we are freezing all the four mass parameters for the fundamentals
as in (\ref{freezing1}).
We obtain the following identity 
\begin{eqnarray}
\label{ZIC}
Z_{SU(N)}\left(-;\vec \nu;\tau_S;-;-;-\right)\!=\!
\Gamma_h\!\! \left(\!-\!\left(N\!-\!1\right)\tau_S\!-\!\sum_{b=1}^{N-2} \nu_b \! \right)
\Gamma_h(N \tau_S) 
\!\!\! \!\!\! \!\!\!  \prod_{1\leq b\leq c\leq N-2} \!\!\!\! \!\!\!\!  \Gamma_h(\tau_S + \nu_b+\nu_c).
 \nonumber \\
\end{eqnarray}

\subsubsection*{Field theory interpretation}

We interpret the identity (\ref{ZIC}) as the fact that an $SU(N)$ gauge theory with a symmetric and $N-2$ antifundamentals is confining with a quantum deformed moduli space. Indeed, by looking at the charges we observe that in this case it is not possible to write a consistent superpotential with the singlets that one can extract from the RHS of (\ref{ZIC}), but we can consistently write a neutral  combination  of the singlets.
The charged fields for this duality are summarized in table \ref{Tab:ICele}
\begin{table}[H]
	\centering
\begin{tabular}{c|c|cccc}
    &$SU(N)$&$U(1)_{\tilde Q}$ &$U(1)_S$ & $SU(N-2)$ &$U(1)_R$ \\
\hline
$S$ &\symm&$0$&$1$&$0$&$0$ \\
$\tilde Q$ &$\overline \square$&1&$0$&$\square$&$0$ \\
	\end{tabular}
	\caption{Charged matter fields for gauge theory studied in case I-C}
	\label{Tab:ICele}
\end{table}
while the singlets on the RHS of (\ref{ZIC}) correspond to the gauge invariant combinations
\begin{table}[H]
	\centering
	\begin{tabular}{l|c c c c}
		& $SU(N-2)$ & $U(1)_S$ & $U(1)_{\tilde{Q}}$& $U(1)_R$ \\ 
		\hline
		$\Phi_1 \equiv \text{det} S$ & $1$ & $N$ & $0$ & $0$ \\
		$\Phi_2 \equiv S \tilde{Q}^2$ & $\symmF$ & $1$ & $2$ & $0$ \\
		$\Phi_3 \equiv Y^{bare}_{SU(N-2)} S$ & $1$ & $1-N$ & $2-N$ & $0$ \\
		\hline
		$Y^{bare}_{SU(N-2)}$ & $1$ & $-N$ & $2-N$ & $0$ \\
	\end{tabular}
	\caption{Gauge invariant combinations of the Case I-C reproducing the singlets appearing in the RHS of the identity \eqref{ZIC}. The bare monopole is not gauge invariant and it has $U(1)_2$-charge $-2(N-2)$.}
	\label{Tab:IC}
\end{table}

We observe that the combination $\Phi_3^2 \Phi_1 \det \Phi_2 $ is neutral and it suggests that there is no superpotential for such model.
In the following we will see that 
this behavior can be obtained  from tensor deconfinement.

Observe that this case can be obtained directly from the case I-A by adding an holomorphic mass term $W = m  Q \tilde Q_S$ to the model studied in case I-A. 
On the gauge theory side we can integrate out the massive fields reproducing the field content and the superpotential of case I-C.
The massive deformation  is associated in the dual to the composite operator $W_{dual} \subset \Phi_1 \Phi_2 + \Phi_4 \Phi_7$.
Adding this superpotential to (\ref{claimWIA}) and integrating out the massive fields we arrive to a theory with the singlets $\Phi_{3,5,6}^{(I-A)}$ where we added the superscript (I-A) to the singlets in order to avoid clutter.
Translating the singlets into the ones of case 
I-C we have $\Phi_3^{(I-A)}  = \Phi_1^{(I-C)} $, $\Phi_6^{(I-A)}  = \Phi_3^{(I-C)}  $, and $\Phi_5^{(I-A)}  = \Phi_2^{(I-A)} $.
We conclude observing that the massive deformation $W = m  Q \tilde Q_S$ imposes at the level of the partition function the same constraint on the real mass $\mu$ of the field $Q$ imposed by (\ref{freezing1}), consistent with the  claim that the identity (\ref{ConfZVI}) 
becomes the identity (\ref{ZIC}).

\subsubsection*{Tensor deconfinement}

Here we show that the confining duality following from the field theoretical 
 interpretation of the relation (\ref{ZIC}) is a consequence of other 3d dualities.
 
\begin{figure}[H]
\begin{center}
  \includegraphics[width=6cm]{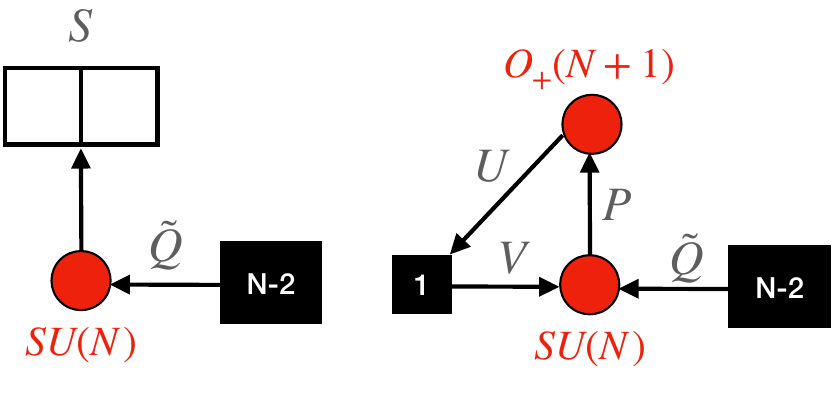}
  \end{center}
 \caption{On the left we provide the quiver for the electric model in Case I-C. On the right of the picture we show the auxiliary quiver where the symmetric
is traded with an $O_+(N+1)$ gauge group with a new bifundamental $P$ and a new vector $U$. 
}
  \label{quivdecI-C}
\end{figure}

 We start by deconfining the symmetric tensor using an $O_+(N+1)$ confining duality and then proceed by dualizing the unitary gauge node and confining back the orthogonal one. In this way we will be able to obtain the confined description discussed above in terms of ordinary dualities that do not involve any tensor field.
After deconfining the symmetric tensor we have the $O_+(N+1) \times SU(N)$ quiver 
in Figure \ref{quivdecI-C}, with superpotential
\begin{equation}
\label{wdecic}
W_{dec} =  \alpha U^2 + PUV +Y_{O(N+1)}^+ +  \alpha Y_d,
\end{equation}
where the crucial observation is that the last interaction is dynamically generated by the deconfinement, where $Y_d$ is a dressed monopole of the $SU(N)$ node 
corresponding to $Y_{SU(N-2)}^{bare}$ dressed by the antifundamentals $\tilde Q^{N-2}$, in formula $Y_d =Y_{SU(N-2)}^{bare} \tilde Q^{N-2}$. The next step consists of dualizing the $SU(N)$ gauge node with $N-1$ antifundamentals and $N+1$ fundamentals. This can be done using the confining duality of appendix (\ref{Niiconf}).
Integrating out the massive fields the dual superpotential becomes
\begin{equation}
\label{interactionICdec}
W_{dual} = B M_{P \tilde Q} Y_0.
\end{equation}
This superpotential term is  compatible with the quantum charges of the fields obtained in the duality map.
On the other hand, the linear monopole deformation for the $O_+(N+1)$ gauge group is lifted here because of the holomorphic mass term  $U M_{PV}$.
Observe that $Y_0$ corresponds to the dressed monopole $Y_0\equiv Y_{SU(N-2)}^{bare} \tilde Q^{N-3} V$. 
By integrating out the massive fields we are left with an $O_{+}(N+1)$ gauge theory with $N-1$ vectors ($M_{P \tilde Q}$ and $B$ ) and interaction (\ref{interactionICdec}) . This theory confines with a quantum deformed moduli space
\footnote{The discussion in  \cite{Aharony:2011ci} shows that even if the moduli space is unlifted  there isn't any singularity in the moduli space 
 and  the quantum moduli space is a smooth  version of the classical one. Here, by using these results in the deconfinement analysis we claim that
 the same conclusions apply to cases I-C, II-C and III-C.
 This reflects also in the fact that the identity at the level of the partition function does not have any singular behavior, i.e. there are no delta functions, because the singlets in the dual theory can take any value, and the origin of the field space, where the global symmetry is fully realized, is allowed. } 
\cite{Aharony:2011ci}, where the combination $Y_{O_+(2n+1)}^2 \det S$ is uncharged and  where 
\begin{equation}
S = 
\left(
\begin{array}{cc}
S_{00} & S_{01} \\
S_{01} &S_{11}
\end{array}
\right)= 
\left(
\begin{array}{cc}
M_{P \tilde Q}^2& M_{P \tilde Q} B \\
M_{P \tilde Q} B& B^2
\end{array}
\right).
\end{equation}
 In this case due to the leftover superpotential we further have  to set $S_{01}=0$ such that uncharged combination is $ Y_{O_+(2n+1)}^2 S_{11} \det S_{00}$.
The three singlets appearing in this combination are associated to the three singlets in Table \ref{Tab:IC} as
\begin{equation}
\Phi_1 \leftrightarrow S_{11},\quad
\Phi_2 \leftrightarrow S_{00},\quad
\Phi_3 \leftrightarrow Y_{O_+(2n+1)}.
\end{equation}

 \subsubsection*{A proof of (\ref{ZIC}) from field theory}
 
 Here we derive the identity (\ref{ZIC}) by applying the chain of deconfinements and confining dualities discussed on the field theory side. The analysis requires a different 
 treatment for the even $SU(2n)$ and the odd $SU(2n+1)$ case. Here we show the explicit computation in the even case, leaving the calculation for the odd case to the interested reader.

The partition function corresponding to the model obtained after deconfining the symmetric tensor is
\begin{eqnarray}
Z =
&&
\frac{\Gamma_h(m)\Gamma_h(2\omega-2m) }{(2n)!^2 2^{n+1}} 
\int \prod_{i=1}^{2n} d \sigma_i \prod_{\alpha=1}^{n} d \rho_\alpha
\frac{ \delta(\sum_{i=1}^{2n} \sigma_i)
\prod_{i=1}^{2n} \prod_{\alpha=1}^{n} 
\Gamma_h \left(\sigma_i \pm \rho_\alpha +\frac{\tau_S}{2} \right)}
{\prod_{i<j} \Gamma(\pm(\sigma_i - \sigma_j))
\prod_{\alpha<\beta} \Gamma_h(\pm \rho_\alpha \pm \rho_\beta))}\nonumber\\
\times&&
\frac{
\prod_{i=1}^{2n} 
\Gamma_h \left(\sigma_i  +\frac{\tau_S}{2}  \right)  \Gamma_h\left(2\omega-m-\frac{\tau_S}{2} -\sigma_i\right)  \prod_{b=1}^{2n-2}  \Gamma_h(\nu_b - \sigma_i) }
{
\prod_{\alpha=1}^n \Gamma_h(\pm \rho_\alpha)} \prod_{\alpha=1}^{n}
\Gamma_h (m\pm \rho_\alpha), \nonumber\\
 \end{eqnarray}
with the balancing condition $n\tau_S + m = 0$.
Then we confine the $SU(2n)$ gauge node. At the level of the partition function such confinement corresponds to apply the formula (\ref{chiralpqstarconfining}). We obtain
\begin{eqnarray}
Z =
&&
\frac{\Gamma_h (n \tau_S) }{(2n)! 2^{n+1}} 
\prod_{b=1}^{2n-2} \Gamma_h \left(\nu_b +\frac{\tau_S}{2},2\omega-\left(n+\frac{1}{2} \right)\tau_S +\nu_b\right)
\nonumber\\
\times &&
\int \prod_{\alpha=1}^{n} d \rho_\alpha
\frac{
\prod_{\alpha=1}^{n} \Gamma_h(n\tau_S\pm \rho_\alpha )}
{
\prod_{\alpha<\beta} \Gamma_h(\pm \rho_\alpha \pm \rho_\beta))}
\prod_{\alpha=1}^{n} \frac{ \prod_{b=1}^{2n-2} \Gamma_h \left(\pm \rho_{\alpha} +\nu_b +\frac{\tau_S}{2} \right)
 }
{
 \Gamma_h(\pm \rho_\alpha)}. \nonumber\\
 \end{eqnarray}
The $O_+(2n+1)$ gauge group has now $2n-1$ vectors and we can use the relation (\ref{Opluswobc}) in the appendix (after applying the inversion relation (\ref{inversion}))
and eventually obtaining the expected relation (\ref{ZIC}).


\section{Family II}
\label{sec:famII}

The second family studied in this paper corresponds to $SU(N)$  with a  two index 
symmetric  and  a  two index conjugated
antisymmetric tensor.
Following the classification of \cite{Nii:2019ebv} we have found two cases where the duplication formula can be used in order to give rise to 3d confining dualities.
These two cases correspond to 
\begin{itemize}
\item $SU(N)$ with an antisymmetric tensor, a conjugate antisymmetric tensor, three fundamentals and one antifundamental;
\item $SU(2n)$ with an antisymmetric tensor, a conjugate antisymmetric tensor and four fundamentals.
\end{itemize}
The model with three fundamentals and one antifundamental can be obtained by reducing the corresponding 4d confining duality found in \cite{Csaki:1996sm,Csaki:1996zb} and then by performing an opportune real mass flow.
 The integral identity for such case can be  obtained  by  using the prescription for the 4d/3d reduction of the 4d supersymmetric index of \cite{Aharony:2013dha}. The 4d matching of the superconformal index follows from the analysis of \cite{spiridonov2003theta}. Furthermore, a physical proof of such matching
can be given by applying the discussion of \cite{Bajeot:2022kwt}, using the 
integral identities for the elementary dualities  in the deconfinement of the antisymmetric tensor.
In  the second case, the one with  four fundamentals,  the confining duality cannot be derived from 4d, and we need to find the integral identity by tensor deconfinement. We will discuss this derivation in appendix \ref{appA1}.
The integral identities for the two confining dualities are in agreement with the proposal of \cite{Nii:2019ebv}.

Using these identities we argue here that three new confining dualities arise in presence of a symmetric and a conjugate antisymmetric two index tensor.
One case originates from the confining duality with three fundamentals and one antifundamental, while the other two cases arise from the confining duality with four fundamentals.
In each case we will investigate the consequence of the matching of the partition function after the application of the duplication formula to the cases with the antisymmetric flavor.
We explicitly study the Coulomb branch, showing the precise mapping of the dressed monopoles 
of the gauge theory with respect to the singlets of the dual phase. 
Furthermore we study  the explicit derivation of the confining dualities by deconfining the antisymmetric tensors using a confining duality for SQCD with a symplectic gauge groups. In each case we observe that the confining dualities of this section follow from the ones studied in the previous one.
We conclude showing that, when applied to the three sphere partition function, this approach furnishes an alternative derivation of the integral identities we started with.

\subsection{Case II-A: $S \oplus \tilde A \oplus Q \oplus \tilde Q_S$}
\label{subsec:IIA}

Here we consider an $SU(2n)$ gauge theory with an antisymmetric tensor A, 
a conjugate antisymmetric $\tilde A$ and four fundamentals $Q$. In this case the theory confines but the confining duality does not derive from a 4d parent.
For this reason in appendix \ref{appA1} we will discuss the derivation of the integral identity associated to the duality using the tensor deconfinement.
\subsubsection*{Applying the duplication formula}

The confining duality in this case has been proposed in \cite{Nii:2019ebv}.
The gauge invariant operators associated to such duality are
\begin{eqnarray}
&&
M_{m=0,\dots,n-2} = A (A \tilde A)^m  Q^2, \quad
B_n = A^n, \quad
B_{n-1} = A^{n-1} Q^2 \nonumber \\
&&
B_{n-2} = A^{n-2} Q^4, \quad
\tilde B_n = \tilde A^n, \quad
T_{\ell=1,\dots,n-1} = (A \tilde A)^\ell.
\end{eqnarray}
In addition we need to consider the dressed monopole of the electric theory
 that act as singlets in the confining phase.  They are
\begin{eqnarray}
&&
Y_{a=0,\dots,n-1}^{dressed} = Y_{SU(2n-2)}^{bare} A^2 (A \tilde A)^a,\nonumber \\
&&
Y_{\tilde A}^{dressed}= Y_{SU(2n-2)}^{bare} (\tilde A^{n-1})^2,  \\
&&Y_{\tilde A A}^{dressed} =Y_{SU(2n-2)}^{bare} \tilde A^{n-1} A. \nonumber 
\end{eqnarray}
The superpotential for this confining duality has not been showed in \cite{Nii:2019ebv}.
However in appendix \ref{appA1} we will discuss, by deconfining  the conjugated antisymmetric field, 
a constructive procedure to build such superpotential.

The integral identity for such confining duality has been derived in formula
(\ref{finalidasa4fond}).
Next we apply the duplication formula on this identity by freezing the masses 
as in (\ref{freezing2}) arriving at the relation 
\begin{eqnarray}
\label{confIIA}
&&
Z_{SU(2n)} \left(\mu;\omega-\frac{\tau_S}{2};\tau_S;-;-;\tau_{\tilde A}\right) =\Gamma_h(2 n\tau_S) \Gamma_h(n \tau_{\tilde A})
\prod_{m=0}^{n-2} \Gamma_h(2(m+1)(\tau_S +\tau_{\tilde A}) \nonumber \\
&&
\prod_{m=0}^{n-2} 
\Gamma_h( (2m+1)\tau_S + 2(m+1)\tau_{\tilde A} +2\mu)
\Gamma_h((2n-1)\tau_S + 2 \mu)\nonumber \\
&&
\Gamma_h\left(\omega-\mu-\left( 2n-\frac{3}{2}\right)\tau_S-(n-1)\tau_A\right)
\Gamma_h\left(\omega-\left(2n-\frac{1}{2} \right) \tau_S-\mu\right)
 \nonumber\\
&& 
\prod_{m=0}^{2n-3} \Gamma_h\left(\omega-\mu-\left(m+\frac{1}{2}\right) \tau_S -(m+1)\tau_{\tilde A}\right).
\end{eqnarray}

\subsubsection*{Field theory interpretation}

We then move to the field theory interpretation of the relation (\ref{confIIA}).
On the electric side we have an $SU(2n)$ gauge theory with a two index symmetric tensors $S$, a two index conjugate antisymmetric tensors $\tilde A$, a fundamental $Q$ and an antifundamental $Q_S$ interacting with the symmetric tensor through (\ref{WSQSQS}).
The field content of the gauge theory and the relative charges with respect to the gauge and flavor symmetries are summarized in Table \ref{Tab:eleIIA}.
\begin{table}[H]
	\centering
	\begin{tabular}{l|c|c c c c}
		&$SU(2n)$ & $U(1)_S$ & $U(1)_A$ & $U(1)_{Q}$& $U(1)_R$ \\ 
		\hline
		$ S$ & \symm & $1$ & $0$ & $0$& $0$ \\
		$A$ & $\begin{array}{c}
\vspace{-2.57mm}
\overline
\square \\
\square
\end{array}
$ & $0$ & $1$ & $0$& $0$ \\
		$Q$ & $\square$ & $0$ & $0$ & $1$ & $0$\\
		$\tilde Q_S$ & $\overline \square $ & $-\frac{1}{2}$ & $0$ & $0$& $1$ \\
	\end{tabular}
	\caption{Charged matter fields for the $SU(2n)$ gauge theory studied in  Case II-A.}
	\label{Tab:eleIIA}
\end{table}
Then from RHS of the relation (\ref{confIIA}) we can read the charges of the gauge singlets
that are expected in the confined case. 
 The global charges of these singlets  are summarized in Table \ref{Tab:IIA}.
\begin{table}[H]
	\centering
	\begin{tabular}{c|c|c c c c}
		& & $U(1)_S$ & $U(1)_A$ & $U(1)_{Q}$& $U(1)_R$ \\ 
		\hline
		$\Psi_1$ &$\text{det} S$ & $2n$ & $0$ & $0$ & $0$ \\
		$\Psi_2$&$\tilde{A}^n$ & $0$ & $n$ & $0$ & $0$ \\
		$\Psi_3$&$(S \tilde{A})^{2j +2}$ & $2(j+1)$ & $2(j+1)$ & $0$ & $0$ \\
		$\Psi_4$&$(S \tilde{A})^{2j +1} \tilde{A} Q^2$ & $2j+1$ & $2(j+1)$ & $2$ & $0$ \\
		$\Psi_5$&$S^{2n-1} Q^2 $ & $2n-1$ & $0$ & $2$ & $0$ \\
		$\Psi_6$&$Y^{bare}_{SU(2n-2)} \tilde{A}^{2n-2}$ & $1/2-2n$ & $0$ & $-1$ & $1$ \\
		$\Psi_7$&$Y^{bare}_{SU(2n-2)} S \tilde{A}^{n-1}$ & $3/2 - 2n$ & $1-n$ & $-1$ & $1$ \\
		$\Psi_8$&$Y^{bare}_{SU(2n-2)} S^2 (S \tilde{A})^{2n-3-m}$ & $-(1/2+m)$ & $-(m+1)$ & $-1$ & $1$ \\
		\hline
		&$Y^{bare}_{SU(2n-2)}$ & $1/2-2n$ & $2-2n$ & $-1$ & $1$ \\
	\end{tabular}
	\caption{Gauge invariant combinations of the Case II-A reproducing the singlets appearing in the RHS of the identity \eqref{confIIA}. In the table $j=0,...,n-2$ and $m= 0,..., 2n-3$, and the bare monopole operator has $U(1)_2$-charge $-2(2n-2)$.}
	\label{Tab:IIA}
\end{table}

By looking at the Table \ref{Tab:IIA} 
we have found that the most generic superpotential compatible with the charges is
	\begin{equation}
	\label{WfabioIIA}
		\begin{split}
		W = & \Psi_6^2 \Psi_5 \Psi_1 + 
		\Psi_7^2 \Psi_{4}^{(n-3-j)} \Psi_{3}^{(j)} \Psi_1 +  \Psi_7^2 \Psi_5 \Psi_{3}^{(n - 2)} + \Psi_{8}^{(2 j - k)} \Psi_{8}^{(k)} \Psi_{4}^{(j)} \\ &+ \Psi_{8}^{(2 j - k + 2 l + 2)} \Psi_{8}^{(k)} \Psi_{4}^{(j)} \Psi_{3}^{(l)} + \Psi_{8}^{(m)} \Psi_{8}^{(2 n - m - 2)} \Psi_5  \Psi_2^2 \\ & + \Psi_{8}^{(m)} \Psi_{8}^{(2 n + 2 j - m)} \Psi_{3}^{(j)}  \Psi_5  \Psi_2^2 + \Psi_6  \Psi_{8}^{(2 j + 1)} \Psi_{4}^{(j)} \Psi_1 + \Psi_7 \Psi_{8}^{(2+2j)} \Psi_{4}^{(j)} \Psi_1 \Psi_2 \\ & + \Psi_7  \Psi_{8}^{(0)}  \Psi_5  \Psi_2 + \Psi_7 \Psi_{8}^{(2+2j)} \Psi_5 \Psi_{3}^{(j)} \Psi_2,
		\end{split}
	\end{equation} 
where the sums over the various indices are understood. In the following we will not reproduce such generic structure from the tensor deconfinement because we will flip the fields $\Psi_3^{(0,\dots,n-2)}$.

\subsubsection*{Tensor deconfinement}

\begin{figure}[H]
\begin{center}
  \includegraphics[width=8cm]{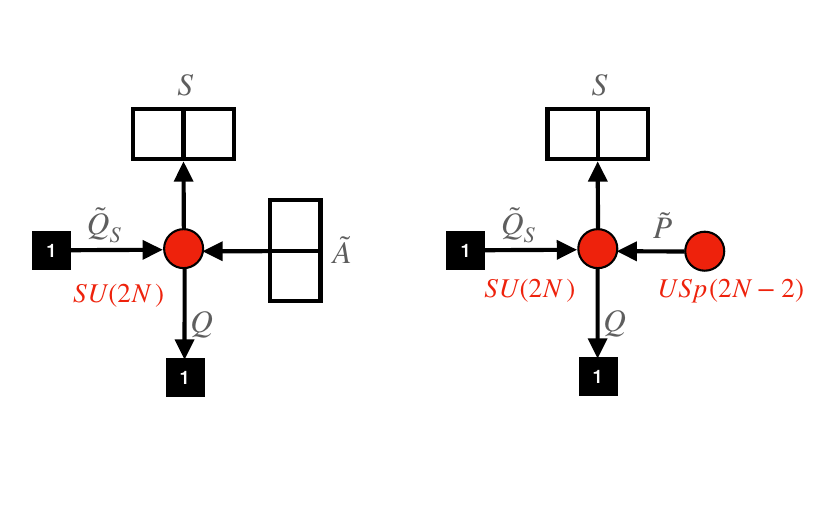}
  \end{center}
 \caption{On the left we provide the quiver for the electric model in Case II-A. On the right of the picture we show the auxiliary quiver where the conjugate antisymmetric
is traded with an $USp(2n-2)$ gauge group with a new bifundamental $\tilde P$. In this case  a further singlet $\sigma$ has to be considered in the auxiliary quiver on the right, flipping the fundamental monopole $Y_{USp(2n-2)}$  with the interaction given in formula (\ref{decIIA}).}
  \label{quivIIAdec}
\end{figure}

The duality can be obtained from the one derived in Case I-A, by deconfining the conjugated antisymmetric in terms of an $USp(2n-2)$ gauge theory with a flipped monopole. 
The deconfined theory is a quiver with $USp(2n-2) \times SU(2n)$ gauge group, plotted in Figure \ref{quivIIAdec}.
This deconfined theory has superpotential 
\begin{equation}
\label{decIIA}
W = S Q_S^2 +
 \sigma Y_{USp(2n-2)}.
\end{equation}
The $SU(2n)$ gauge theory is now confining, because it corresponds to the 
I-A model studied in subsection (\ref{SQQtQtS}).
Actually in this case the difference with the model I-A consists in the fact that the 
$SU(2n-2)$ flavor symmetry has been  gauged as $USp(2n-2)$, indeed we have $2n-2$ antifundamentals $\tilde P$ that are in the fundamental of such gauged symmetry.

The $SU(2n)$ node confines giving rise to an $USp(2n-2)$ gauge theory with an adjoint and two vectors. 
Using the terminology of subsection (\ref{SQQtQtS}) the adjoint is referred to as 
$\Phi_5$ while the fundamentals are $\Phi_1$ and $\Phi_2$.
Using the superpotential (\ref{claimWIA}) for the confining duality I-A,  the superpotential of the $USp(2n-2)$ theory is
 \begin{equation}
 \label{IIAonetolast}
W \subset \sigma  \Phi _6 \Phi _4 \hat Y_{USp(2n-2)}^{(2n-3)} + \Phi _4 \Phi _6^2 \det \Phi _5+\Tr (\Phi _1 \Phi _2) \Phi _3 \Phi _7
+\Phi _3 \Phi _4  \Phi _7^2+\Tr (\Phi _5  \Phi _1^2),
\end{equation}
where we have included the interaction of the flipper $\sigma$ with the singlets
$\Phi_6$ and $\Phi_4$ and the fundamental monopole $\hat Y_{USp(2n-2)}$ dressed by $\Phi_5^{2n-3}$. Observe that we have denoted the fundamental monopole of the $USp(2n)$ theory with the adjoint  with an hat to distinguish it  from the one appearing in (\ref{decIIA}).

The final step consists of confining the $USp(2n-2)$ gauge theory. Such theory is indeed confining as proved in \cite{Benvenuti:2021nwt}. We review the basic aspects of such confining duality  in appendix \ref{app:BLMAdj}.
The confining duality was studied from adjoint deconfinement and it was shown that the WZ model had a very simple form in terms of cubic interactions by flipping on the electric side the tower of operators corresponding to the (even) traces of the adjoint. 
Here in our analysis such traces are associated to the operators $(S \tilde A)^{2k}$ with $k=1,\dots n-1$ in the $SU(2n)$ electric theory.
This operator corresponds here to $ \Tr (\Phi_5)^{2k}$. Flipping this field corresponds then to add the extra superpotential interaction 
\begin{equation}
\label{WIIAfliop}
W \subset \sum_{k=1}^{n-1} \beta_k \Tr (\Phi_5)^{2k}.
\end{equation}
The final superpotential of the $USp(2n-2)$ theory is then given by adding 
(\ref{WIIAfliop}) to (\ref{IIAonetolast}) and by observing that the F-terms of the flippers set to zero the operator $\det \Phi_5$ as well.
We can now dualize this theory obtaining the final WZ model.
Let us define the dressed mesons of the dual phase as $M_j = \Phi_2^2 \Phi_5^{2j+1}$ 
It is important to stress that the product $Tr (\Phi_1 \Phi_2)$ is a composite operator \cite{Benvenuti:2021nwt} and it is dual to the combination $\sum_{j=0}^{n-2} M_{j} \mathcal{M}_{2(n-2-j)}$, where the last  term are the dressed $USp(2n-2)$ monopoles acting as singlets in this phase.  Defining the fundamental monopole as $\mathcal{M} \equiv \hat Y_{USp(2n-2)}$, the dressed ones are $\mathcal{M}_\alpha \equiv \mathcal{M} \Phi_5^{\alpha}$, for $\alpha=0,\dots,2n-3$.
The superpotential then becomes
 \begin{equation}
 \label{IIAfinalfromdec}
W = \sigma  \Phi _6 \Phi _4  \mathcal{M}_{2n-3} +\sum_{j=0}^{n-2} M_{j} \mathcal{M}_{2(n-2-j)}\Phi _3 \Phi _7 +
\Phi _3 \Phi _4  \Phi _7^2 + \sum_{\alpha,\beta,k} \mathcal{M}_\alpha  \mathcal{M}_\beta  M_k \delta_{\alpha+\beta+2k-4n-2 }.
\end{equation}
Then we can read the operator map between the singlets $\Psi_i$ in table \ref{Tab:IIA} 
and the singlets in (\ref{IIAfinalfromdec}).
The flipped terms $\Phi_5^{2j+2}$ correspond to the singlets $\Psi_3^{(j)}$ and they disappear.
The mapping for the other singlets is
\begin{equation}
\label{mapIIA}
\Phi_4 \leftrightarrow \Psi_5,
\,
\Phi_3 \leftrightarrow \Psi_1,
\,
\Phi_6 \leftrightarrow \Psi_7,
\,
\Phi_7 \leftrightarrow \Psi_6,
\,
\sigma \leftrightarrow \Psi_2,
\,
M_j = \psi_4^{(j)},
\,
\mathcal{M}_\alpha =\psi_8^{(2n-3-\alpha)}.
\end{equation}
In this way we have reproduced most of the terms in  the superpotential (\ref{WfabioIIA})  after flipping $\Psi_3^{(j)}$.
Actually there are two contributions that we have not reproduced using the constructions spelled out above. Such terms are  
$\Psi_{8,m} \Psi_{8,2 n - m - 2} \Psi_5  \Psi_2^2$ and $ \Psi_7 \Psi_{8,2+2j} \Psi_{4,j} \Psi_1 \Psi_2 $.
Both such terms involve the flipper $\sigma$ generated in (\ref{decIIA}) and their existence should be dynamical in (\ref{IIAonetolast}). It would be interesting to further investigate on the stability of such terms and we leave such analysis to future investigations.
Summarizing, using the dictionary (\ref{mapIIA})  we have obtained the superpotential
\begin{eqnarray}
W = 
&&\Psi_2  \Psi _7 \Psi _5  \Psi_8^{(0)} +
\Psi_1 \Psi _6 \sum_{j=0}^{n-2} \Psi_4^{(j)} 
\Psi_8^{(2j+1)}+\Psi _1 \Psi _5  \Psi _6^2 \nonumber \\
+&&
 \sum_{\alpha,\beta,k} \Psi_8^{(2n-3-\alpha)}  \Psi_8^{(2n-3-\beta)}  \Psi_4^{(k)} \delta_{\alpha+\beta+2k-4n+6 }.
\end{eqnarray}
For future reference we also present the superpotential for the $SU(4)$ case, after 
flipping the field $\Psi_2$. The superpotential in such case becomes
\begin{eqnarray}
\label{IIASU(4)}
W = 
\Psi_1 \Psi _6 \Psi_4^{(0)} 
\Psi_8^{(1)}+\Psi _1 \Psi _5  \Psi _6^2 + \Psi_8^{(0)}  \Psi_8^{(0)}  \Psi_4^{(0)}. \end{eqnarray}
Observe that in this case the singlet $\Psi_7$ does not appear in the superpotential and it 
must be thought as a free field. This is indeed compatible with F-maximization, where the exact $R$-charges of the singlets in this duality are $R=1/2$ for all the singlets except the field $R_{\Psi_8^{(0)}}=\frac{3}{4}$. In this case the exact R-charge is indeed given by the combination $U(1)_R + \frac{1}{16} (2 U(1)_S + 2U(1)_A + U(1)_Q)$. We refer the reader to Section \ref{unitarity} for a more general discussion on the unitarity of the dualities discussed in the paper.  
 
\subsubsection*{A proof of (\ref{confIIA}) from field theory}

We conclude the analysis of the II-A case by reproducing the identity (\ref{confIIA}) by applying the deconfinement procedure 
explained above at the level of the three sphere partition function.

In the first step, once we deconfine the conjugate antisymmetric, we have a flipper with mass parameter   $n \tau_{\tilde A}$ and 
an $SU(2n) \times USp(2n-2)$ bifundamental with mass parameter $n \frac{\tau_{\tilde A}}{2}$. Then we confine the $SU(2n)$ node using the identity (\ref{ConfZVI}) for even gauge rank, obtaining the partition function
\begin{eqnarray}
\label{almostIIA}
&&
 \Gamma_h(m_\sigma,m_{\Phi_3},m_{\Phi_4},m_{\Phi_6},m_{\Phi_7})
Z_{USp(2n-2)} \left(m_{\Phi_1}, m_{\Phi_2};-;m_{\Phi_5} \right),
 \end{eqnarray}
where the mass parameters are 
\begin{eqnarray}
m_{\sigma} &=&n \tau_{\tilde A},\quad 
m_{\Phi_1} = \omega-\frac{\tau_{\tilde A}+\tau_S}{2}, \nonumber \\
m_{\Phi_2}&=&\frac{\tau_{\tilde A}}{2} +\mu, \quad
m_{\Phi_3}=2n \tau_S,\quad\nonumber \\
m_{\Phi_4}&=&(2n-1) \tau_S+2\mu,\quad
m_{\Phi_5}=\tau_{\tilde A}+\tau_S\nonumber \\
m_{\Phi_6}&=&\omega-\left(2n-\frac{3}{2} \right)\tau_S -\mu - (n-1) \tau_{\tilde A} 
\nonumber \\
m_{\Phi_7}&=&\omega-\left(2n-\frac{1}{2} \right) \tau_S-\mu. \quad
\end{eqnarray}
The final step consists of confining the leftover $USp(2n-2)$ gauge theory with an adjoint and two fundamentals. In this case we plug the relation (\ref{AR22}) into (\ref{almostIIA}) and obtain the expected identity (\ref{confIIA}).
In this way we have proved the identity, obtained above from the duplication formula, by applying the tensor deconfinement and the subsequent confining dualities explained on the field theory side.

\subsection{Case II-B: $S \oplus \tilde A \oplus \tilde Q_S \oplus \tilde Q$}

Here we consider an $SU(N)$ gauge theory with an antisymmetric tensor A, 
a conjugate antisymmetric $\tilde A$, one antifundamentals $\tilde Q$ and three fundamentals $Q$. This is a  3d s-confining duality with 4d origin. It can be indeed derived from the 4d case of $SU(N)$ with \asymm, \casymm, $3$ $\square$ and $3$ $\overline \square$. The 3d duality is obtained by reducing the confining duality on $S^1$ and by removing the KK monopole superpotential triggering a real mass flow for two antifundamentals, assigning them an opposite mass. 
Following these steps on the matching of the 4d SCI, proven in \cite{spiridonov2003theta}, one obtains the matching of the 3d partition functions necessary for the  application of the duplication formula after \emph{freezing} the mass parameters of the antifundamentals.
The final 3d dualities have been extensively studied in \cite{Nii:2019ebv}, where it was observed that the even and the odd case need a separate discussion.

\subsubsection*{Applying the duplication formula}

Here we first review the 3d confining duality distinguishing the even and the odd case.
For even $N=2n$ the gauge invariant combinations, build from the charged chiral fields, that appear  
in the dual WZ model are
\begin{eqnarray}
&&
M_{k=0,\dots,n-1} = Q (A \tilde A)^k \tilde Q, \quad
H_{m=0,\dots,n-2} = \tilde A(A \tilde A)^m Q^2, \quad
T_{j=1,\dots,n-1} = (A \tilde A)^j,
\nonumber \\
&&
\tilde B_n = \tilde A^n,\quad
B_n = A^n, \quad
B_{n-1} = A^{n-1} Q^2.
\end{eqnarray}
Furthermore, there are two types of dressed monopoles
\begin{eqnarray}
Y_{a=0,\dots,n-2}^{dressed} =Y_{SU(2n-2)}^{bare} A (A \tilde A)^a,
\quad 
Y_{\tilde A}^{dressed}   = Y_{SU(2n-2)}^{bare}  \tilde A^{n-1}.
\end{eqnarray}

As discussed above, in this case the matching of the partition functions for the two phases of this  
confining duality can be proven starting from the 4d SCI and the final 3d identity is
\begin{eqnarray}
&&
Z_{SU(2n)}(\vec \mu;\nu;-;-;\tau_A,\tau_{\tilde A})=
\prod_{k=0}^{n-1} \prod_{a=1}^3 \Gamma_h(\mu_a+\nu+k(\tau_A+\tau_{\tilde A}) )
\Gamma_h(n \tau_A) \Gamma_h(n \tau_{\tilde A})
\nonumber \\
&&
\prod_{1\leq a \leq b \leq 3} 
\left(\Gamma_h((n-1) \tau_A + \mu_a+\mu_b)
\prod_{m=0}^{n-2} 
\Gamma_h(\mu_a+\mu_b+m(\tau_A+\tau_{\tilde A}) +\tau_{\tilde A})
\right)
\nonumber \\
&&
\prod_{j=1}^{n-1} \Gamma_h(j(\tau_A+\tau_{\tilde A}))
\Gamma_h (2\omega - \nu - \sum_{a=1}^{3} \mu_a-(N-1) (2 \tau_A+
\tau_{\tilde A} ) 
\nonumber \\
&&
\prod_{j=0}^{n-2}
\Gamma_h \left(2\omega - \nu - \sum_{a=1}^{3} \mu_a +(2-2n+j)
(\tau_A+\tau_{\tilde A})+\tau_A\right).
\end{eqnarray}
Starting from this identity we apply the duplication formula using (\ref{freezing2}) 
on the parameters $\mu_a$, substituting also the parameter $\tau_A$ with $\tau_S$.
We obtain a new identity
\begin{eqnarray}
\label{eq:ZSAQQts}
&&
Z_{SU(2n)} \left(-;\omega-\frac{\tau_S}{2},\nu;\tau_S;-;-;\tau_{\tilde A} \right)=
 \Gamma_h\left(\left(\frac{1}{2}-2n\right)\tau_S +(1-n) \tau_A+\omega-\nu\right)
 \nonumber \\
&&
\Gamma_h(2 n \tau_S) \Gamma_h(n\tau_{\tilde A}) 
\prod_{k=0}^{n-1} \Gamma_h((2k+1)\tau_S +2k\tau_{\tilde A} + 2 \nu)
\prod_{k=0}^{n-2}\Gamma_h(2(k+1) (\tau_S +\tau_{\tilde A}))
 \nonumber \\
 &&
\prod_{k=0}^{2n-2} \Gamma_h\left(\omega-\left(k+\frac{1}{2}\right)\tau_S-k \tau_{\tilde A}-\nu\right),
  \nonumber \\
\end{eqnarray}
that corresponds to an $SU(2n)$ gauge theory with a symmetric $S$, a conjugated antisymmetric $\tilde A$ and two antifundamentals $\tilde Q$ and $\tilde Q_S$, where the last is compatible with the superpotential interaction (\ref{WSQSQS}).

For odd $N=2n+1$ the gauge invariant combinations, build from the charged chiral fields, that appear  
in the dual WZ model are
\begin{eqnarray}
&&
M_{k=0,\dots,n-1} = Q (A \tilde A)^k \tilde Q, \quad
H_{m=0,\dots,n-2} = \tilde A(A \tilde A)^m Q^2, \quad
B_n = A^n Q,
\nonumber \\
&&
\tilde B_n = \tilde A^n \tilde Q , \quad
B_{n-1} = A^{n-1} Q^3,\quad 
T_{j=1,\dots,n} = (A \tilde A)^j.
\end{eqnarray}
Furthermore, there are two types of dressed monopoles
\begin{eqnarray}
Y_{a=0,\dots,n-1}^{dressed} =Y_{SU(2n-1)}^{bare} A (A \tilde A)^a,
\quad 
Y_{\tilde A}^{dressed}   = Y_{SU(2n-1)}^{bare}  \tilde A^{n-1} \tilde Q.
\end{eqnarray}
In this case the matching of the partition functions for the two phases of this  
confining duality can be proven starting from 4d and the 3d identity is
\begin{eqnarray}
&&
Z_{SU(2n+1)}(\vec \mu;\nu;-;-;\tau_A;\tau_{\tilde A})=
\prod_{k=0}^{n-1} \prod_{a=1}^3 \Gamma_h(\mu_a+\nu+k(\tau_A+\tau_{\tilde A}) )
\nonumber \\
&&
\Gamma_h(n \tau_{\tilde A}+\nu)
\Gamma_h\left((n-1) \tau_A + \sum_{a=1}^3 \mu_a\right)
\prod_{m=0}^{n-2} 
\Gamma_h(\mu_a+\mu_b+m(\tau_A+\tau_{\tilde A}) +\tau_{\tilde A})
\nonumber \\
&&
\prod_{a=1}^3\Gamma_h(n \tau_A+\mu_a) 
\prod_{j=1}^{n} \Gamma_h(j(\tau_A+\tau_{\tilde A}))
\Gamma_h \left(2\omega - \nu - \sum_{a=1}^{3} \mu_a-(2n-1) \tau_A-n
\tau_{\tilde A} \right) 
\nonumber \\
&&
\prod_{j=0}^{n-1}
\Gamma_h \left(2\omega - \nu - \sum_{a=1}^{3} \mu_a +(1-2n+j)
(\tau_A+\tau_{\tilde A})+\tau_A\right).
\end{eqnarray}
Starting from this identity we apply the duplication formula using (\ref{freezing2}) 
on the parameters $\mu_a$, substituting also the parameter $\tau_A$ with $\tau_S$.
We obtain a new identity
\begin{eqnarray}
\label{eq:ZSAQQts2}
&&
Z_{SU(2n+1)} \left(-;\omega-\frac{\tau_S}{2},\nu;\tau_S;-;-;\tau_{\tilde A} \right)=
\Gamma_h((2 n+1) \tau_S)
\Gamma_h(n\tau_{\tilde A}+\nu)
  \nonumber \\
&&
\prod_{k=0}^{n-1} \Gamma_h((2k+1)\tau_S + 2k\tau_{\tilde A} + 2 \nu)
\prod_{k=0}^{n-1}\Gamma_h(2(k+1) (\tau_S +\tau_{\tilde A}))
   \\
&&
\prod_{j=0}^{2n-1}\Gamma_h\left(\omega-\left(j+\frac{1}{2} \right )\tau_S -j \tau_{\tilde A} -\nu\right) \Gamma_h\left(\omega-\left(2n+\frac{1}{2} \right) \tau_S-n\tau_{\tilde A}\right), \nonumber
\end{eqnarray}
that corresponds to an $SU(2n+1)$ gauge theory with a symmetric $S$, a conjugated antisymmetric $\tilde A$ and two antifundamentals $\tilde Q$ and $\tilde Q_S$, where the last is compatible with the superpotential interaction (\ref{WSQSQS}).

\subsubsection*{Field theory interpretation}
 
 Here we provide a field theory interpretation of the identities (\ref{eq:ZSAQQts})
 and (\ref{eq:ZSAQQts2}).
From these identities we expect that an $SU(N)$ gauge theory with a symmetric $S$, a conjugated antisymmetric $\tilde A$, an antifundamental $\tilde Q$ and an antifundamental $\tilde Q_S$, with superpotential (\ref{WSQSQS}), is confining.
The charges of the matter fields of the $SU(N)$ gauge theory are summarized in Table \ref{Tab:IIB_ele}.
\begin{table}[H]
	\centering
\begin{tabular}{c|ccccc}
    &SU(N)&$U(1)_{\tilde Q}$ &$U(1)_S$ & $U(1)_A$ &$U(1)_R$ \\
\hline
$S$ &\symm&0&1&0&0 \\
$\tilde A$ &
\casymm
&0& 0&1&0\\
$\tilde Q_S$ &$\overline {\square}$&0&$-1/2$&0&1 \\
$\tilde Q$ &$\overline {\square}$&1&$0$&0&0 \\
	\end{tabular}
	\caption{Charged matter fields of the $SU(N)$ gauge theory for case II-B.}
	\label{Tab:IIB_ele}
\end{table}

 The dual spectrum is made out gauge singlets built up from the matter fields and from (dressed) monopole operators. 
Here we have verified that it is indeed the case, by computing the quantum charges of the gauge invariant combinations as summarized in Tables \ref{Tab:IIB_even} and \ref{Tab:IIB_odd}.
The charges of these singlets match indeed with the ones that can be extracted from the  identities 
(\ref{eq:ZSAQQts}) and (\ref{eq:ZSAQQts2}).
Then in the dual confining phase we need to distinguish the charges of the singlets in the WZ model.
When considering $SU(2n)$ we have 
\begin{table}[H]
	\centering
	\begin{tabular}{l|c c c c}
		& $U(1)_S$ & $U(1)_A$ & $U(1)_{\tilde{Q}}$& $U(1)_R$ \\ 
		\hline
		$\Psi_1= \text{det} S$ & $2n$ & $0$ & $0$ & $0$ \\
		$\Psi_2= \tilde{A}^n$ & $0$ & $n$ & $0$ & $0$ \\
		$\Psi_3^{(j)}= (S)^{2j+1}  ( \tilde{A}^{2j} \tilde{Q}^2)$ & $2j+1$ & $2j$ & $2$ & $0$ \\
		$\Psi_4^{(k)}= (S \tilde{A})^{2k+2}$ & $2(k+1)$ & $2(k+1)$ & $0$ & $0$ \\
		$\Psi_5= Y^{bare}_{SU(2n-2)} \tilde{A}^{n-1}$ & $1/2 - 2n$ & $1-n$ & $-1$ & $1$ \\
		$\Psi_6^{(m)}= Y^{bare}_{SU(2n-2)} S (S \tilde{A})^{2n-2-m}$ & $-1/2-m$ & $-m$ & $-1$ & $1$ \\
		\hline
		$Y^{bare}_{SU(2n-2)}$ & $1/2 - 2n$ & $ 2-2n$ & $-1$ & $1$ \\
	\end{tabular}
	\caption{Gauge invariant combinations of the Case II-B even reproducing the singlets appearing in the RHS of the identity \eqref{eq:ZSAQQts}. In the table $j=0,...,n-1$, $k= 0,...,n-2$ and $m= 0,...,2n-2$, and the bare monopole has $U(1)_2$-charge $-(2n-2)$.}
	\label{Tab:IIB_even}
\end{table}
By looking at the Table \ref{Tab:IIB_even} 
we have found that the most generic superpotential compatible with the charges is
	\begin{equation}
	\label{WfabioIIBeven}
		\begin{split}
			W= & \Psi_5^2 \Psi_{3}^{(n - 1)} \Psi_1 + \Psi_5^2 \Psi_{3}^{(j)} \Psi_1 \Psi_{4}^{(n - j - 2)} +	 \Psi_{6}^{(m)} \Psi_{6}^{(2 j - m + 2 k + 2)} \Psi_{3}^{(j)} \Psi_{4}^{(k)} \\ & +	
			\Psi_{6}^{(2 j + 1)} \Psi_5 \Psi_{3}^{(j)} \Psi_1 \Psi_2 + \Psi_{6}^{(2 j + 3 + 2 k)} \Psi_5 \Psi_{3}^{(j)} \Psi_{4}^{(k)} \Psi_1 \Psi_2 +\Psi_{6}^{(m)} \Psi_{6}^{(2 j - m)} \Psi_{3}^{(j)},
		\end{split}
\end{equation}
where the sums over the various indices are understood. In the following we will not reproduce such generic structure from the tensor deconfinement because we will flip the fields $\Psi_4^{(0,\dots,n-2)}$.

A similar analysis holds for $SU(2n+1)$, where the singlets in the confining phase are summarized in Table \ref{Tab:IIB_odd}.
\begin{table}[H]
	\centering
	\begin{tabular}{l|c c c c}
		& $U(1)_S$ & $U(1)_A$ & $U(1)_{\tilde{Q}}$& $U(1)_R$ \\ 
		\hline
		$\Psi_1 \equiv \text{det} S$ & $2n+1$ & $0$ & $0$ & $0$ \\
		$\Psi_2 \equiv \tilde{A}^n \tilde{Q}$ & $0$ & $n$ & $1$ & $0$ \\
		$\Psi_3^{(j)} \equiv (S)^{2j+1} (\tilde{A}^{2j} \tilde{Q}^2)$ & $2j+1$ & $2j$ & $2$ & $0$ \\
		$\Psi_4^{(j)} \equiv (S \tilde{A})^{2j+2}$ & $2(j+1)$ & $2(j+1)$ & $0$ & $0$ \\
		$\Psi_5 \equiv Y^{bare}_{SU(2n-1)} \tilde{Q} \tilde{A}^{n-1}$ & $-(1/2+2n)$ & $-n$ & $0$ & $1$ \\
		$\Psi_6^{(m)} \equiv Y^{bare}_{SU(2n-1)} S (S \tilde{A})^{2n-1-m}$ & $-(m+1/2)$ & $-m$ & $-1$ & $1$ \\
		\hline
		$Y^{bare}_{SU(2n-1)} $ & $-(1/2+2n)$ & $1-2n$ & $-1$ & $1$ \\
	\end{tabular}
	\caption{Gauge invariant combinations of the Case II-B odd reproducing the singlets appearing in the RHS of the identity \eqref{eq:ZSAQQts2}. In the table $j=0,...,n-1$ and $m= 0,...,2n-1$, and the bare monopole has $U(1)_2$-charge $-(2n-1)$.}
	\label{Tab:IIB_odd}
\end{table}
By looking at the Table \ref{Tab:IIB_odd} 
we have found that the most generic superpotential compatible with the charges is
\begin{equation}
	\label{WfabioIIBodd}
	\begin{split}
	W = & \Psi_5^2 \Psi_1 \Psi_{4}^{(n - 1)} + \Psi_{6}^{(m)} \Psi_{6}^{(2 n - m)} \Psi_2^2 \Psi_1 + \Psi_{6}^{(m)} \Psi_{6}^{( 2 n - m + 2 j + 2)} \Psi_{4}^{(j)} \Psi_2^2 \Psi_1 + \Psi_{6}^{(m)} \Psi_{6}^{(2 j - m)} \Psi_{3}^{(j)} \\ & + \Psi_{6}^{(m)} \Psi_{6}^{(2 + 2 j + 2 k - m)} \Psi_{3}^{(j)} \Psi_{4}^{(k)} +	\Psi_5 \Psi_{6}^{(0)} \Psi_2 \Psi_1 + \Psi_5 \Psi_{6}^{(2 j + 2)} \Psi_{4}^{(j)} \Psi_2 \Psi_1,
	\end{split}
\end{equation}
where the sums over the various indices are understood. In the following we will not reproduce such generic structure from the tensor deconfinement because we will flip the fields $\Psi_4^{(0,\dots,n-1)}$.

\subsubsection*{Tensor deconfinement}

Here we will give a proof of the confinement of the model in both the $SU(2n)$ and $SU(2n+1)$ case by deconfining the antisymmetric tensor. In this way we will be able to reformulate the models in terms of other theories that have been already proved to be confining.
After deconfining the conjugated antisymmetric indeed we will see that the $SU(2n)$ and the $SU(2n+1)$ gauge theory correspond to the models studied in case I-B, with a (partially) gauged flavor symmetry. Such gauge nodes are confining and leave us in both cases with a symplectic gauge group with two fundamentals and one adjoint interacting with just one of the fundamentals through a cubic superpotential term.
Such theory is confining as proven in \cite{Benvenuti:2021nwt} and this leaves us with the final WZ model representing the confinement of the model with the symmetric tensor.
\begin{figure}[H]
\begin{center}
  \includegraphics[width=8cm]{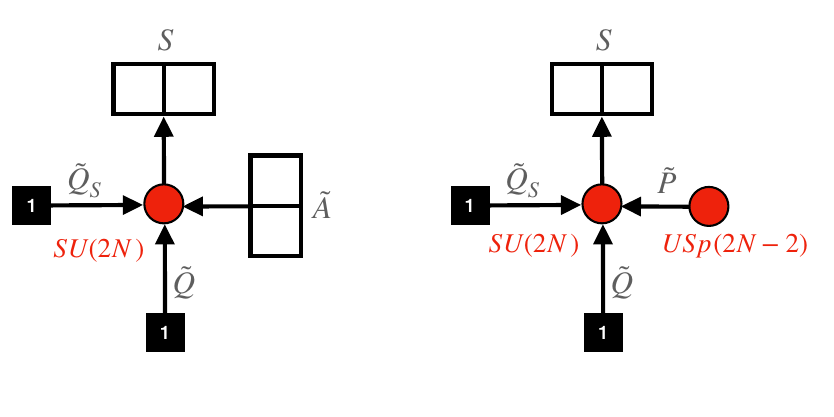}
  \end{center}
 \caption{On the left we provide the quiver for the electric model in Case II-B for an $SU(2n)$ gauge group. On the right of the picture we show the auxiliary quiver where the conjugate antisymmetric
is traded with an $USp(2n-2)$ gauge group with a new bifundamental $\tilde P$. In this case  a further singlet $\sigma$ has to be considered in the auxiliary quiver on the right, flipping the fundamental monopole $Y_{USp(2n-2)}$  with the interaction given in formula (\ref{IIBdecWW}).}
  \label{decIIBevenfig}
\end{figure}

In the even case the duality can be obtained from the one derived in Case I-B, by deconfining the conjugated antisymmetric in terms of an $USp(2n-2)$ gauge theory with monopole superpotential. The quiver of the deconfined theory is given in Figure \ref{decIIBevenfig}.
The deconfined theory has superpotential 
\begin{equation}
\label{IIBdecWW}
W = Q_S^2 S + \sigma Y_{USp(2n-2)},
\end{equation}
where $ Y_{USp(2n-2)}$ is the fundamental monopole of $USp(2n-2)$.
Then we observe  that the $SU(2n)$ gauge theory is a type I-B theory and it confines, as shown in sub-section (\ref{subsec:I-B}). 
After confining this node we are left with an $USp(2n-2)$ gauge theory with two fundamentals and one adjoint. 

Actually in this case the difference with the model I-B consists in the fact that the 
$SU(2n-1)$ flavor symmetry has been partially gauged, indeed we have $2n-2$ antifundamentals $\tilde P$ that are in the fundamental of such gauged symmetry, while there is an extra antifundamental $\tilde Q$. Such gauging has two effects in the confinement of the $SU(2n)$ node. 

First of all we must split the fields in Table \ref{Tab:IB} accordingly.
The fields involved in such splitting are the $SU(2n-1)$ symmetric field $\Phi_3$ and the $SU(2n-1)$ antifundamental $\Phi_1$.
The symmetric is split into an $USp(2n-2)$ symmetric (adjoint) $\Phi_3^S$, an
$USp(2n-2)$ fundamental $\Phi_3^F$ and an  $USp(2n-2)$ singlet $\Phi_3^s$.
The $SU(2n-1)$ antifundamental $\Phi_1$ is split into an  $USp(2n-2)$
fundamental $\Phi_1^F$ and an  $USp(2n-2)$ singlet  $\Phi_1^s$. 
The superpotential (\ref{WIB}) is modified accordingly.
Furthermore, an interaction involving the flipper $\sigma$, the singlets of this phase and the fundamental $USp(2n-2)$ monopole $\hat Y_{USp(2n-2)}$ dressed by $(\Phi_3^S)^{2n-3}$,
\begin{equation}
\label{WIIBmon}
W \subset \sigma \Phi_2 \Phi_3^s \Phi_4 \hat Y_{USp(2n-2)}^{(2n-3)},
\end{equation}
is allowed. 
The $USp(2n-2)$ theory is confining as proved in \cite{Benvenuti:2021nwt}.
The confining duality was studied from adjoint deconfinement and it was shown that the WZ model had a very simple form in terms of cubic interactions by flipping on the electric side the tower of operators corresponding to the (even) traces of the adjoint.
Here in our analysis such traces are associated to the operators $(S \tilde A)^{2k}$ with $k=1,\dots n-1$ in the $SU(2n)$ electric theory. 
This operator corresponds here to $\Phi_3^S$. Flipping this field corresponds then to add the extra superpotential interaction 
\begin{equation}
\label{WIIBfliop}
W \subset \sum_{k=1}^{n-1} \beta_k  \Tr  (\Phi_3^S)^{2k}.
\end{equation}
Then the $USp(2n-2)$ gauge theory has superpotential 
\begin{eqnarray}
\label{WIIBonetolast}
W &=& \sum_{k=1}^{n-1} \beta_k \Tr (\Phi_3^S)^{2k} 
+
 \sigma \Phi_2 \Phi_3^s \Phi_4 \hat Y_{USp(2n-2)}^{(2n-3)}
+
 \Tr (\Phi_1^F   \Phi_1^F  \Phi_3^S) 
  \nonumber \\ &+&
   \Phi_1^s   \Phi_1^s  \Phi_3^s+
     \Phi_1^s    \Tr (\Phi_1^F  \Phi_3^F)
     +
     \Phi_2 \Phi_4^2 \det \Phi_3^S,
     \end{eqnarray}
where the traces and the determinants are associated to the $USp(2n-2)$ indices.
The final superpotential of this $USp(2n-2)$ theory is actually simplified with respect to 
(\ref{WIIBonetolast}) by observing that the F-terms of the flippers $\beta_k$ set to zero the operator $\det \Phi_3^S$ as well.
We can now dualize this theory obtaining the final WZ model.
Let us define the dressed mesons of the dual phase as $M_{j} = (\Phi_3^S)^{2j+1} (\Phi_1^F)^2$.
It is important to stress that the product $ \Tr (\Phi_1^F  \Phi_3^F)$ is a composite operator \cite{Benvenuti:2021nwt} and it is dual to the combination $\sum_{j=1}^{n-1} M_{j-1} \mathcal{M}_{2(n-1-j)}$, where the last  term are the dressed $USp(2n-2)$ monopoles acting as singlets in this phase.  Defining the fundamental monopole as $\mathcal{M} \equiv \hat Y_{USp(2n-2)}$, the dressed ones are $
\mathcal{M}_\alpha \equiv  \mathcal{M} (\Phi_3^S)^{\alpha}$ 
 for $\alpha=0,\dots,2n-3$.
The superpotential then becomes
 \begin{eqnarray}
 \label{IIBfinalfromdec}
W &=&\sigma \Phi_2 \Phi_3^s \Phi_4 \mathcal{M}_{2n-3}
 +
   \Phi_1^s   \Phi_1^s  \Phi_3^s
   +   \Phi_1^s   \sum_{j=1}^{n-1} M_{j-1} \mathcal{M}_{2(n-1-j)} 
   \nonumber \\
&+&    \sum_{\alpha,\beta,k} \mathcal{M}_\alpha  \mathcal{M}_\beta  M_k \delta_{\alpha+\beta+2k-2(2n-3)}.
\end{eqnarray}
Then we can read the operator map between the singlets $\Psi_i$ in table \ref{Tab:IIB_even} 
and the singlets in (\ref{IIBfinalfromdec}).
The flipped terms $(\Phi_3^S)^{2j+2}$ correspond to the singlets $\Psi_4^{(j)}$ and they disappear.
The mapping for the other singlets is
\begin{eqnarray}
\label{mapIIB}
&&
\sigma \leftrightarrow \Psi_2,
\quad
\Phi_2 \leftrightarrow \Psi_1,
\quad
\Phi_1^s \leftrightarrow \Psi_6^{(0)}, \quad
\Phi_3^s \leftrightarrow \Psi_3^{(0)},
\nonumber \\
&&
\Phi_4\leftrightarrow \Psi_5,
\quad
\mathcal{M}_\alpha\leftrightarrow \Psi_6^{(2n-2-\alpha)},
\quad
M_j\leftrightarrow \Psi_3^{(j+1)},
\end{eqnarray}
with $\alpha=0,\dots,2n-3$ and $j=0,\dots,n-2$.
Using the map (\ref{mapIIB}), after some rearrangement, the superpotential (\ref{IIBfinalfromdec}) becomes
\begin{equation}
\label{WfinalIIBbello}
W= \Psi_1 \Psi_2 \Psi_3^{(0)} \Psi_5 \Psi_6^{(1)}
 +
   \sum_{\alpha,\beta=0}^{2n-2} \sum_{k=0}^{n-1} \Psi_6^{(2n-2-\alpha)}  \Psi_6^{(2n-2-\beta)}  \Psi_3^{(k)}\delta_{\alpha+\beta+2k-2(2n-2) }.
\end{equation}

In this way we have reproduced most of the terms in  the superpotential (\ref{WfabioIIBeven})  after flipping $\Psi_4^{(j)}$.
Actually there are two contributions that we have not reproduced using the constructions spelled out above. Such terms are  
$\Psi_{6}^{(2 j + 1)} \Psi_5 \Psi_{3}^{(j)} \Psi_1 \Psi_2$  with $j>0$ and $\Psi_5^2 \Psi_{3}^{(n - 1)} \Psi_1$.
The first term involves the flipper $\sigma$ generated in (\ref{IIBdecWW}) and its existence should be dynamical in (\ref{WIIBonetolast}).
On the other hand, we have not find an interpretation of the  term  $\Psi_5^2 \Psi_{3}^{(n - 1)} \Psi_1$ in our construction. It would be interesting to further investigate on the stability of such term and we leave such analysis to future investigations.

\begin{figure}[H]
\begin{center}
  \includegraphics[width=10cm]{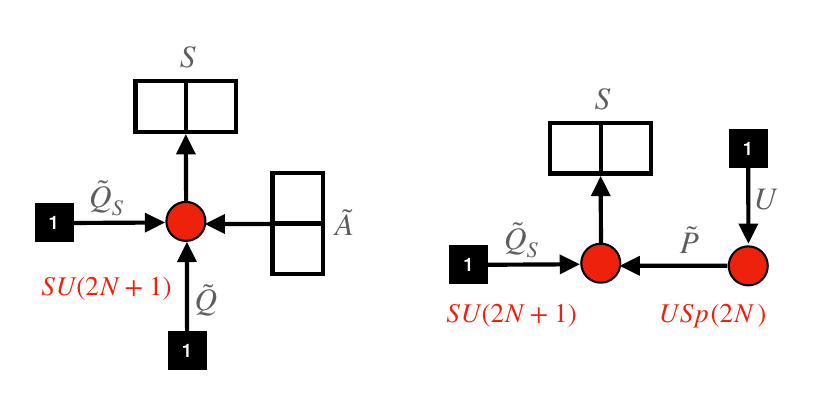}
  \end{center}
 \caption{On the left we provide the quiver for the electric model in Case II-B for an $SU(2n+1)$ gauge group. On the right of the picture we show the auxiliary quiver where the conjugate antisymmetric field is traded with an $USp(2n)$ gauge group with a new bifundamental $\tilde P$. In this case  a further singlet $\sigma$ has to be considered in the auxiliary quiver on the right, flipping the fundamental monopole $Y_{USp(2n)}$. Furthermore, there is a new fundamental $U$ charged under $USP(2n)$ (where the arrow refers to the $U(1)$ charge of the field) such that the antifundamental  
$\tilde Q$ in the quiver on the left corresponds to the composite $U \tilde P$ in the quiver on the right.}
  \label{decIIBodd}
\end{figure}
The situation is similar  in the odd case. For $SU(2n+1)$  we deconfine  the conjugate antisymmetric tensor with an $USp(2n)$ gauge group flipping the fundamental monopole with  the quiver in Figure \ref{decIIBodd}.
The superpotential of this phase is again given by (\ref{IIBdecWW}).
The $SU(2n+1)$ theory is of type I-B ant it confines as discussed in subsection (\ref{subsec:I-B}).
In this case the flavor symmetry $SU(2n)$ under which the fundamentals $\tilde P$ are charged is gauged as $USp(2n)$.
Confining the $SU(2n+1)$ gauge node we are left with an $USp(2n)$ gauge theory with two fundamentals $\Phi_1$ and $U$ and one adjoint $\Phi_3$.
Following the same discussion of the even case, the superpotential of this theory is 
\begin{equation}
\label{WIIBonetolastodd}
W = \Tr(\Phi_1^2 \Phi_3) + \det \Phi_3\Phi_4^2\Phi_2 + \sum_{k=1}^{n}
\beta_k Tr \Phi_3^{k}+ \sigma \hat Y_{USp(2n)}^{(2n-1)} \Phi_2 \Phi_4,
\end{equation}
where the traces and the determinants are associated to the $USp(2n)$ indices.

Using the confining duality of \cite{Benvenuti:2021nwt} for this $USp(2n)$ theory we arrive at the final WZ model. The final superpotential of this $USp(2n)$ theory is actually simplified with respect to (\ref{WIIBonetolastodd}) by observing that the F-terms of the flippers $\beta_k$ set to zero the operator $\det \Phi_3$ as well.
We can now dualize this theory obtaining the final WZ model.
Let us define the dressed mesons of the dual phase as $M_{j} = (\Phi_3)^{2j+1} U^2$. Defining the fundamental monopole as $\mathcal{M} \equiv \hat Y_{USp(2n)}$, the dressed ones are $ \mathcal{M}_\alpha \equiv  \mathcal{M} (\Phi_3)^{\alpha}$ for $\alpha=0,\dots,2n-1$.
The superpotential then becomes
 \begin{equation}
 \label{IIBfinalfromdecodd}
W =\sigma \Phi_2  \Phi_4 \mathcal{M}_{2n-1}
 +
 \sum_{\alpha,\beta,k} \mathcal{M}_\alpha  \mathcal{M}_\beta  M_k \delta_{\alpha+\beta+2k-2(2n-1) }.
\end{equation}

Then we can read the operator map between the singlets $\Psi_i$ in Table \ref{Tab:IIB_odd} 
and the singlets in (\ref{IIBfinalfromdecodd}).
The flipped terms Tr$(\Phi_3)^{2j+2}$ correspond to the singlets $\Psi_4^{(j)}$ and they disappear.
The mapping for the other singlets is
\begin{eqnarray}
\label{mapIIBodd}
&&
\sigma \leftrightarrow \Psi_2,
\quad
\Phi_2 \leftrightarrow \Psi_1,
\quad
\Phi_4\leftrightarrow \Psi_5,
\quad
\mathcal{M}_\alpha\leftrightarrow \Psi_6^{(2n-1-\alpha)},
\quad
M_j\leftrightarrow \Psi_3^{(j)},
\end{eqnarray}
with $\alpha=0,\dots,2n-1$ and $j=0,\dots,n-2$.
In this way we have reproduced most of the terms in  the superpotential (\ref{WfabioIIBodd})  after flipping $\Psi_4^{(j)}$.
Actually there is one contribution that we have not reproduced using the constructions spelled out above. Such term is  
$\Psi_{6}^{(m)} \Psi_{6}^{(2 n - m)} \Psi_2^2 \Psi_1 $.
This term involves the flipper $\sigma$ and its existence should be dynamical in (\ref{WIIBonetolastodd}). It would be interesting to further investigate on the stability of such terms and we leave such analysis to future investigations.

\subsubsection*{A proof (\ref{eq:ZSAQQts}) and  (\ref{eq:ZSAQQts2}) from field theory}

We conclude the analysis by reproducing the identities  (\ref{eq:ZSAQQts})  and (\ref{eq:ZSAQQts2}) for the confining duality II-B by applying the deconfinement just explained at the level of the three sphere partition function, both in the even and in the odd case.

Let's start from the case $N=2n$. In this case, after deconfining the antisymmetric tensor, we have to introduce the hyperbolic gamma function of the flipper $\sigma$, with mass parameter $n \tau_{\tilde A}$. Furthermore, the $SU(2n)\times USp(2n-2)$ bifundamental $\tilde P$ has mass parameter $\frac{\tau_{\tilde A}}{2}$.
We then dualize the $SU(2n)$ node using the identity  (\ref{ZIB})  derived for the I-B Case for even $N=2n$, obtaining the following partition function
\begin{eqnarray}
\label{almostIIBeven}
&&
\Gamma_h(  m_{\sigma},m_{\Phi_1^s},m_{\Phi_2},m_{\Phi_3^s},m_{\Phi_4})
)Z_{USp(2n-2)} \left( m_{\Phi_1^F}, m_{\Phi_3^F};-;m_{\Phi_3^S} )
 \right),
  \end{eqnarray}
where the charged fundamentals and the adjoint have mass parameters 
  \begin{equation}
  m_{\Phi_1^F} = \tau_S +\frac{\tau_{\tilde A}}{2}+\nu, \quad
m_{\Phi_3^F} = \omega-\frac{\tau_{\tilde A}+\tau_S}{2}, \quad 
m_{\Phi_3^S} = \tau_{\tilde A}+\tau_S
\end{equation}
  respectively, and the singlets have mass parameters
  \begin{eqnarray}
 && m_{\sigma} = n \tau_{\tilde A}, \quad
  m_{\Phi_1^s} = \omega-\frac{\tau_S}{2} -\nu,\quad
m_{\Phi_2}= 2 n \tau_S, \quad
m_{\Phi_3^s}=\tau_S +2\nu , 
\nonumber \\
&&
m_{\Phi_4} = \omega-\left(2n-\frac{1}{2}\right)\tau_S-\nu - (n-1) \tau_{\tilde A}.
  \end{eqnarray}
 The last step consists of applying the identity   (\ref{AR22})  for the integral corresponding to the leftover  $USP(2n-2)$ gauge node with an adjoint and two fundamentals. Actually before this step we need to flip the tower of singlets Tr$(S \tilde A)^{2k}$, that corresponds to add to (\ref{almostIIBeven}) the contribution of the singlets $\beta_k$ that have mass parameters $m_{\beta_k}= 2\omega - 2 k m_{\Phi_3^S} $.
We then plug  (\ref{AR22})  into  (\ref{almostIIBeven}),
 after we have added the contributions of the flippers $\beta_k$.
In this way we arrive at (\ref{eq:ZSAQQts}), obtaining an independent proof of such identity, derived above by using the duplication formula.

The derivation is similar in the case of odd rank, i.e. $N=2n+1$. In this case the flipper $\sigma$ has mass parameter $\nu+n\tau_{\tilde A}$ 
while the $SU(2n+1)\times USp(2n-2)$ bifundamental $\tilde P$ has still mass
parameter $\frac{\tau_{\tilde A}}{2}$. 
We then dualize the $SU(2n+1)$ node using the identity  (\ref{ConfZVI})  derived for the I-A Case for odd $N=2n+1$, obtaining the following partition function
\begin{eqnarray}
\label{almostIIBodd}
\Gamma_h(m_\sigma, m_{\Phi_2},m_{\Phi_4}) Z_{USp(2n)}(m_U,m_{\Phi_1};-;m_{\Phi_3}),
\end{eqnarray}
with
\begin{eqnarray}
&&
m_\sigma=\nu + n \tau_{\tilde A}, \quad
m_U=\nu-\frac{\tau_A}{2}, \quad
m_{\Phi_1}=\omega-\frac{\tau_{\tilde A}+\tau_S}{2}, \quad
m_{\Phi_2} = (2n+1) \tau_S, \nonumber \\
&&
m_{\Phi_3} = \tau_{\tilde A}+\tau_S, \quad
m_{\Phi_4}=\omega-n \tilde \tau_{\tilde A}-\left(2n+\frac{1}{2}\right)\tau_S.
\end{eqnarray}
We need also to flip the tower of singlets Tr$(S \tilde A)^{2k}$, that corresponds to add to (\ref{almostIIBodd}) the contribution of the singlets $\beta_k$ that have mass parameters $m_{\beta_k}= 2\omega - 2 k m_{\Phi_3} $.
We then plug (\ref{AR22})  into (\ref{almostIIBodd}) , after we have added the contributions of the flippers $\beta_k$, arriving at (\ref{eq:ZSAQQts2}), i.e.
 proving  the identity for the confining duality in the odd case and matching it with the one derived  by using the duplication formula above.

\subsection{Case II-C: $S \oplus \tilde A$}
\label{IIC}

We conclude this section by considering again the $SU(2n)$ theory with an antisymmetric tensor $A$, a conjugated antisymmetric tensor $\tilde A$ and  four fundamentals $Q$ discussed in subsection \ref{subsec:IIA}.

\subsubsection*{Applying the duplication formula}
The main difference with the case considered in subsection \ref{subsec:IIA} is that in this case we are freezing all the four mass parameters for the fundamentals as in formula (\ref{freezing1}).
We obtain the following identity
\begin{eqnarray}
\label{IICfinal}
&&
Z_{SU(2n)} (-;-;\tau_S;\tau_{\tilde A}) =\Gamma_h(2 n\tau_S) \Gamma_h(n \tau_{\tilde A})
\prod_{m=0}^{n-2} \Gamma_h(2(m+1)(\tau_S +\tau_{\tilde A}) \nonumber \\
&&
\Gamma_h(-(2n-1)\tau_S-(n-1)\tau_A)
\prod_{m=0}^{n-2} \Gamma_h(-(2m+1)(\tau_S +\tau_{\tilde A}).
\end{eqnarray}

\subsubsection*{Field theory interpretation}

From (\ref{IICfinal}) we have an $SU(2n)$ gauge theory with a two index symmetric tensor $S$ and a  conjugated antisymmetric tensor $\tilde A$.
The gauge singlets that we expect in the dual phase are summarized in Table 
\ref{Tab:IIC_even}.
\begin{table}[h]
	\centering
	\begin{tabular}{c |c c c}
		& $U(1)_S$ & $U(1)_A$ & $U(1)_R$ \\ 
		\hline
		$\Psi_1 \equiv \text{det} S$ & $2n$ & $0$ & $0$ \\
		$\Psi_2 \equiv \tilde{A}^n$ & $0$ & $n$ & $0$ \\
		$\Psi_3 \equiv (S \tilde{A})^{2j+2}$ & $2j+2$ & $2j+2$ & $0$ \\
		$\Psi_4\equiv Y^{bare}_{SU(2n-2)} S \tilde{A}^{n-1}$ & $1-2n$ & $1-n$ & $0$ \\
		$\Psi_5 \equiv Y^{bare}_{SU(2n-2)} S^2 (S \tilde{A})^{2n-2m-3}$ & $-(2m+1)$ & $-(2m+1)$ & $0$ \\
		\hline
		$Y^{bare}_{SU(2n-2)}$ & $-2n$ & $2-2n$ & $0$ \\
	\end{tabular}
	\caption{Gauge invariant combinations of the Case II-C reproducing the singlets appearing in the RHS of the identity \eqref{IICfinal}. In the table with $j=0,...,n-2$, $m=0,...,n-2$ and the bare monopole has $U(1)_2$-charge $-2(2n-2)$.}
	\label{Tab:IIC_even}
\end{table}
Similarly to the discussion in subsection \ref{subsec:IIA} in this case we observe that from the quantum charges of these fields we cannot write a superpotential for these singlets. 
For example the combinations $\Psi _1 \Psi _2 \Psi _4 \Psi _3^{(j)} \Psi _5^{(j+1)} $, $  \Psi _3^{(j_1)}  \Psi _3^{(j_2)} \Psi _5^{(j_3)}\Psi _5^{(j_4)} \delta _{j_1+j_2-j_3-j_4+1}$ and $ \Psi _4 ^2 \Psi_1 \Psi _3^{(n-2)}$ 
are neutral. Then we expect that the model confines with a quantum deformed moduli space.
In the following we will support this expectation by studying the dynamics using tensor deconfinement. In order to simplify the analysis we flip the Pfaffian of the antisymmetric tensor in the gauge theory, removing the singlet $\Psi_2$ from the singlets appearing in the confining phase.

Similarly to the analysis of case I-C also in this case  adding an holomorphic mass $W = m Q \tilde Q_S$ to case II-A leads to the case II-C. The deformation is not associated to a composite operator in the dual but to the combination $W_{dual} \subset \Psi_5 \Psi_6 + \sum_{j=0}^{n-2} \Psi_4^{(j)} \Psi_8^{(2j)}$.
Again we expect that the analysis of the equations of motions of the massive fields leads to the deformed moduli for the II-C theory.
Furthermore we observe that this construction is consistent with the expectations from the analysis of the partition function where the mass term $W = m Q \tilde Q_S$ corresponds to transform (\ref{freezing2}) into (\ref{freezing1}).

\subsubsection*{Tensor deconfinement}

If we consider the theory with  superpotential $W = \sigma\, \text{Pf} (A)$
we can deconfine the conjugated antisymmetric tensor in terms of an $USp(2n-2)$ gauge theory with vanishing
superpotential.
The $SU(2n)$ gauge theory is a type I-C theory and it confines, as shown in sub-section (\ref{SQt}). 
Again, differently from the case I-C, here the $2n-2$ antifundamentals  of the $SU(2n-2)$ flavor symmetry become a single $SU(2n) \times USp(2n-2)$ bifundamental for the deconfined quiver, denoted here as $\tilde P$.
The $SU(2n)$ gauge theory confines with a quantum deformed moduli space, with the 
uncharged combination $\Phi_3^2 \Phi_1 \det \Phi_2$.
After confining the $SU(2n)$ node we are left with an $USp(2n-2)$ gauge theory  with the adjoint $\Phi_2$. 
Such model has been discussed in appendix \ref{app:BLMAdj}, where we have aimed that this model, without any flipper, confines with a quantum deformed moduli space. 
The uncharged combinations $ \Psi _4 ^2 \Psi_1 \Psi _3^{(n-2)}$ follows from the one obtained above, where 
we have used the relation $\det \Phi_2 \sim  Tr \Phi _2^{n-2} \sim  \Psi _3^{(n-2)}$ holding ignoring multi traces, and we have the further
map $\Phi_3 \leftrightarrow \Psi _4 $ and $\Phi_1 \leftrightarrow \Psi _1$.
The other combination $  \Psi _3^{(j_1)}  \Psi _3^{(j_2)} \Psi _5^{(j_3)}\Psi _5^{(j_4)} \delta _{j_1+j_2-j_3-j_4+1}$ 
follows from the confinement of the $Usp(2n-2)$ gauge node as discussed in the appendix  \ref{app:BLMAdj}.

\subsubsection*{A proof of (\ref{IICfinal}) from field theory}

At the level of the partition function 
we have consider   the original model with the flipped Pfaffian. Then we deconfine the 
antisymmetric using (\ref{USpwomon}) and use the relation (\ref{ZIC}), obtaining
\begin{eqnarray}
\label{IICafertdecdual}
\Gamma_h(2 \omega - n \tau_{\tilde A}) 
Z_{SU(2N)} (-;-;\tau_S;-;-;\tau_{\tilde A})
&=&
\Gamma_h\left(\left(1-2n\right)\tau_S-(n-1) \tau_{\tilde A} \right)
 \\
&\times &
\Gamma_h(2n \tau_S) 
Z_{USp(2n-2)}(-;-; \tau_S +\tau_{\tilde A}).
\nonumber 
\end{eqnarray}
Plugging the identity (\ref{AM24}) in (\ref{IICafertdecdual}) with $n=n-1$ and 
$\tau= \tau_S +\tau_{\tilde A}$ we arrive at the expected relation (\ref{IICfinal}).

\section{Family III}
\label{sec:famIII}

The last family studied in this paper corresponds to $SU(N)$ gauge theories with a  index 
symmetric tensors an antisymmetric tensor.
Following the classification of \cite{Nii:2019ebv} we have found two cases where the duplication formula can be used in order to give rise two 3d confining dualities with symmetric tensors.

These two cases correspond to 
\begin{itemize}
\item $SU(N)$ with two antisymmetric tensors, three fundamentals and one antifundamental;
\item $SU(2N)$ with two antisymmetric tensors and four fundamentals.
\end{itemize}
For $N=2$ the first model can be obtained by reducing the corresponding 4d confining duality found in \cite{Csaki:1996sm,Csaki:1996zb}.
For higher ranks similar constructions are not possible and the  confining dualities are genuinely three dimensional.

In appendix \ref{casesIIIas} we will give a proof of such confining dualities using tensor deconfinement and in this way we will construct the integral identity between the three sphere partition functions of the dual phases.
Using these identities we will argue here that three new confining dualities arise in presence of a symmetric and an antisymmetric tensor.
One case originates from the confining duality with three fundamentals and one antifundamental, while the other two cases arise from the confining duality with four fundamentals.

In each case we will investigate the consequence of the matching of the partition function after the application of the duplication formula to the cases with the two antisymmetric tensors.
We explicitly study the Coulomb branch, showing the precise mapping of the dressed monopoles 
of the gauge theory with respect to the singlets of the dual phase. 
Furthermore we provide for each case an explicit derivation of the confining dualities by deconfining the symmetric and the antisymmetric tensors using of orthogonal and symplectic gauge groups. 

In this case we show that the derivation requires to iterate the procedure, i.e. alternating dualities and tensor deconfinements. We isolate the opportune recursive patterns that allow us to reduce the rank of the gauge groups, until we reach an $SU(4)$ gauge theory with a symmetric and a conjugated antisymmetric tensor.
The last step then requires to use the confining dualities proven in section \ref{sec4} for such tensors to obtain the confining theories we are looking for.
When applied to the three sphere partition function this approach provides a  derivation of the integral identities obtained from the duplication formula.

\subsection{Case III-A: $S \oplus A \oplus \tilde Q_S \oplus Q$}
\label{SAQsQ}
The first model discussed in this section  corresponds  to an $SU(N)$ gauge theory with two antisymmetric tensors $A_{1,2}$ and four fundamentals.

\subsubsection*{Applying the duplication formula}

Here we first review the 3d confining duality distinguishing the even and the odd case.
For even $N=2n$ the gauge invariant combinations, built from the charged chiral fields, that appear in the dual WZ model are
\begin{eqnarray}
\label{fourfund2ASevenmes}
&&
T_n = A^n, \quad
T_{n-1}=A^{n-1} Q^2, \quad
T_{n-2}= A^{n-2} Q^4.
\end{eqnarray}
Furthermore, there is a dressed monopole
\begin{eqnarray}
\label{fourfund2ASevenmon}
Y_{SU(2n-2)}^{dressed} =Y_{SU(2n-2)}^{bare} A^{2n-2},
\end{eqnarray}
The superpotential for such confining phase is
\begin{equation}
W = Y_{SU(2n-2)}^{dressed}  (T_n T_{n-2} + T_{n-1}^2).
\end{equation}
 In this case there is an $SU(2)$ symmetry that rotates the two antisymmetric and then the singlets  $T_j$ are in the $n-j$ symmetric representation of such flavor symmetry group, while the dressed monopole
is in the $2n-2$ symmetric representation of such $SU(2)$.
The integral identity representing this confining duality is
\begin{eqnarray}
\label{4fund2ASeven}
Z_{SU(2n)} (\vec \mu;-;-;-;\tau_{A},\tau_B;-)
&=&
\prod_{j=0}^{n} \Gamma_h(j \tau_A + (n-j) \tau_B)
\\
&\times&
\prod_{j=0}^{n-1}  \prod_{1\leq a <b \leq 4}\Gamma_h(j \tau_A + (n-j-1) \tau_B + \mu_a +\mu_b)
\nonumber \\
&\times&
\prod_{j=0}^{n-2} \Gamma_h \left(j \tau_A + (n-j-2) \tau_B + \sum_{a=1}^{4} \mu_a \right)
\nonumber \\
&\times& \!
 \prod_{j=0}^{2n-2} \Gamma_h \left(\!
2\omega-j \tau_A \! -(2n-j-2) \tau_B -\! \sum_{a=1}^{4} \mu_a \! \right).
\nonumber 
\end{eqnarray}
The relation (\ref{4fund2ASeven}) has been derived in appendix  \ref{casesIIIas}
and the next step consist of freezing the mass parameters appearing in the arguments of the fundamentals as in (\ref{freezing2}), where we rename $\tau_B = \tau_S$, explicitly breaking the $SU(2)$ global symmetry between the two antisymmetric tensors. In this way we
convert one of the two antisymmetric tensors into a rank-two symmetric $S$ tensor and an antifundamental $Q_S$.
By applying the duplication formula 
this assignation of parameters leaves on the electric side an $SU(N)$ gauge theory with one rank-two symmetric tensor $S$, one rank-two antisymmetric tensor $A$, one antifundamental $Q_S$ and one fundamental $Q$. 
The mass parameter of $Q_S$ is compatible with the superpotential interaction (\ref{WSQSQS})  and we claim that this superpotential is necessary for the confinement.
In this way we arrive at the identity 
\begin{eqnarray}
\label{eq:Z2NAtSQQts}
&&Z_{SU(2n)}\left(\mu;\omega-\frac{\tau_S}{2};\tau_S;-;\tau_A;-\right) 
=
\Gamma_h( n \tau_A) \Gamma_h \left(\omega-\frac{\tau_S}{2}- (n-1) \tau_A - \mu \right)
\nonumber  \\
&&
\prod_{j=0}^{n-1} \Gamma_h(2(j+1)\tau_S + 2(n-1-j) \tau_A)
\prod_{j=0}^{n-1}\Gamma_h((2j+1)\tau_S + 2(n-1-j) \tau_A+2 \mu)
\nonumber
\\
&&
\prod_{j=0}^{2n-2}\Gamma_h\left(
\omega+\left(j+\frac{1}{2}-2n\right) \tau_S-j\tau_A-\mu
\right).
\end{eqnarray}

For odd $N=2n+1$ the gauge invariant combinations, built from the charged chiral fields, that appear in the dual WZ model are
\begin{eqnarray}
\label{singletsoddfour1}
&&
T_n = A^n Q, \quad
T_{n-1}=A^{n-1} Q^3. \quad
\end{eqnarray}
Furthermore, there is a dressed monopole
\begin{eqnarray}
\label{singletsoddfour2}
Y_{SU(2n-1)}^{dressed} =Y_{SU(2n-1)}^{bare} A^{2n-1}.
\end{eqnarray}
The superpotential for such confining phase is
\begin{equation}
W = Y_{SU(2n-1)}^{dressed}  T_n  T_{n-1}.
\end{equation}
Again with an $SU(2)$ symmetry rotating the two antisymmetric tensors such that the singlets  $T_j$ are in the $n-j$ symmetric representation of such flavor symmetry group, while the dressed monopole
is in the $2n-1$ symmetric representation of such $SU(2)$.
The integral identity representing this confining duality is
\begin{eqnarray}
\label{4fund2ASodd}
Z_{SU(2n+1)} (\mu;-;-;&-&;\tau_{A},\tau_B;-)
=
\prod_{j=0}^{n}  \prod_{a=1}^{4} \Gamma_h(j \tau_A + (n-j) \tau_B + \mu_a)
\\
&\times&
\prod_{j=0}^{n-1}  \prod_{1\leq a <b<c \leq 4}\Gamma_h(j \tau_A + (n-j-1) \tau_B + \mu_a +\mu_b +\mu_c)
\nonumber \\
&\times&
\prod_{j=0}^{2n-1}  \Gamma_h \left(
2\omega-j \tau_A + (2n-j-1) \tau_B -\sum_{a=1}^{4} \mu_a \right).
\nonumber 
\end{eqnarray}

The relation (\ref{4fund2ASodd}) has been derived in appendix  \ref{casesIIIas}
and the next step consist of freezing the mass parameters appearing in the arguments of the fundamentals as in (\ref{freezing2}), where we rename $\tau_B = \tau_S$, explicitly breaking the $SU(2)$ global symmetry between the two antisymmetric tensors. In this way we
convert one of the two antisymmetric tensors into a rank-two symmetric tensor  $S$ and an antifundamental $Q_S$.
By applying the duplication formula 
this assignation of parameters leaves on the electric side an $SU(2n+1)$ gauge theory with one rank-two symmetric tensor $S$, one rank-two antisymmetric tensor $A$, one antifundamental $Q_S$ and one fundamental $Q$. 
The mass parameter of $Q_S$ is compatible with the superpotential interaction (\ref{WSQSQS})  and we claim that this superpotential is necessary for the confinement.
In this way we arrive at the identity 
\begin{eqnarray}
\label{eq:Z2NAtSQQts2}
&&Z_{SU(2n+1)}\left(\mu;\omega-\frac{\tau_S}{2};\tau_S;-;\tau_A;-\right) 
=
\Gamma_h( n \tau_A+\mu)  \prod_{j=0}^{n} \Gamma_h((2j+1)\tau_S + 2(n-j) \tau_A)
 \nonumber
\\
&&
\Gamma_h \left(\omega-\frac{\tau_S}{2}- n \tau_A  \right) 
\prod_{j=0}^{n-1}\Gamma_h(2(j+1)\tau_S +2 (n-1-j) \tau_A+2 \mu)
 \nonumber
\\
&&
\prod_{j=0}^{2n-1}
\Gamma_h\left(\omega+\left(j-\frac{1}{2}-2n\right) \tau_S-j\tau_A-\mu
\right).
\end{eqnarray}

\subsubsection*{Field theory interpretation}

In order to provide a field theory interpretation of the identities (\ref{eq:Z2NAtSQQts}) and 
(\ref{eq:Z2NAtSQQts}) we organize the fields in the electric side by assigning their charges 
with respect to the global symmetries. Such charges are summarized in Table
\ref{Tab:IIIA_elec}.
\begin{table}[H]
	\centering
\begin{tabular}{c|c|cccc}
    &$SU(N)$& $U(1)_S$ & $U(1)_A$ & $U(1)_Q$ & $U(1)_R$ \\
\hline
$S$ &\symm&1&0&0 &0\\
$A$ & \asymm& 0&1&0&0\\
$\tilde Q_S$ &$\overline{\square}$&$-1/2$&0& 0&1 \\
$ Q$ &$ \square$&$0$&0&1&0 \\
\end{tabular}
	\caption{Charges for the matter fields in the electric of case III-A.}
	\label{Tab:IIIA_elec}
\end{table}
Then we study the gauge invariant combinations arising from these fields and the superpotential (\ref{WSQSQS}).
The analysis is different for $N=2n$ and $N=2n+1$.
In the case of $SU(2n)$ we have found the  combinations in Table \ref{Tab:IIIA_even}.
\begin{table}[H]
	\centering
	\begin{tabular}{l|c c c c}
		& $U(1)_S$ & $U(1)_A$ & $U(1)_Q$ & $U(1)_R$ \\ 
		\hline
		$\Xi_1  \equiv A^n$ & $0$ & $N$ & $0$ & $0$ \\
		$\Xi_2^{(j)}\equiv  S^{2j+2} A^{2n-2j-2}$ & $2j+2$ & $2n-2j-2$ & $0$ & $0$ \\
		$\Xi_3^{(j)}\equiv  S^{2j+1} A^{2n-2j-2} Q^2$ & $2j+1$ & $2n-2j-2$ & $2$ & $0$ \\
		$\Xi_4\equiv  Y^{bare}_{SU(2n-2)} S^{2n-1} A^{n-1}$ & $-1/2$ & $1-n$ & $-1$ & $1$ \\
		$\Xi_5^{(k)}\equiv  Y^{bare}_{SU(2n-2)} S^k A^{2n-2-k}$ & $k+1/2-2n$ & $-k$ & $-1$ & $1$ \\
		\hline
		$Y^{bare}_{SU(2n-2)}$ & $1/2-2n$ & $2-2n$ & $-1$ & $1$ \\
	\end{tabular}
	\caption{Gauge invariant combinations of the Case III-A even reproducing the singlets appearing in the RHS of the identity \eqref{eq:Z2NAtSQQts}. In the table $j=0,..., n-1$, $ k=0,...,2n-2$ and the bare monopole has $U(1)_2$-charge $-(2n-2)^2$.}
	\label{Tab:IIIA_even}
\end{table}
The last two terms corresponds gauge invariant dressings of the bare  monopole $Y^{bare}_{SU(2n-2)}$. This monopole indeed is not gauge invariant because it has charge 
$-(2n-2)^2$ under $U(1)_2$. 
On the other hand, the two gauge invariant combinations $Y^{bare}_{SU(2n-2)}\\ S^{2n-1} A^{n-1}$ and $Y^{bare}_{SU(2n-2)} S^k A^{2n-2-k}$ are gauge invariant and they must be considered as singlets in the dual WZ model.
We observe that three sphere partition functions for the fields in Table \ref{Tab:IIIA_even}
is given by the RHS of (\ref{eq:Z2NAtSQQts}), corroborating the validity of the
conjectured confining duality.

In the case of $SU(2n+1)$ we have found the gauge invariant  combinations in Table \ref{Tab:IIIA_odd}.
\begin{table}[H]
	\centering
	\begin{tabular}{l|c c c c}
		& $U(1)_S$ & $U(1)_A$ & $U(1)_Q$ & $U(1)_R$ \\ 
		\hline
		$\Xi_1  \equiv A^n Q$ & $0$ & $n$ & $1$ & $0$ \\
		$\Xi_2^{(j)}\equiv S^{2j+1} A^{2n-2j}$ & $2j+1$ & $2n-2j$ & $0$ & $0$ \\
		$\Xi_3^{(\ell)}\equiv S^{2\ell+2} A^{2n-2\ell-2} Q^2$ & $2\ell+2$ & $2n-2\ell-2$ & $2$ & $0$ \\
		$\Xi_4\equiv Y^{bare}_{SU(2n-1)}S^{2n} A^{n-1} Q$ & $-1/2$ & $-n$ & $0$ & $1$ \\
		$\Xi_5^{(k)}\equiv  Y^{bare}_{SU(2n-1)} S^k A^{2n-1-k}$ & $k-1/2-2n$ & $-k$ & $-1$ & $1$ \\
		\hline
		$Y^{bare}_{SU(2n-1)}$ & $-1/2-2n$ & $1-2n$ & $-1$ & $1$ \\
	\end{tabular}
	\caption{Gauge invariant combinations of the Case III-A odd reproducing the singlets appearing in the RHS of the identity \eqref{eq:Z2NAtSQQts2}. In the table $j=0,..., n$, $\ell=0,..., n-1$ and $ k=0,...,2n-1$, and the bare monopole has $U(1)_2$-charge $-(2n-1)^2$.}
	\label{Tab:IIIA_odd}
\end{table}
The last two terms in the table corresponds gauge invariant dressings of the bare  $Y^{bare}_{SU(2n-1)}$. This monopole indeed is not gauge invariant because it has charge 
$-(2n-2)^2$ under $U(1)_2$.
On the other hand, the two gauge invariant combinations $Y^{bare}_{SU(2n-1)}S^{2n} A^{n-1} Q$ and $Y^{bare}_{SU(2n-1)} S^k A^{2n-1-k}$ are gauge invariant and they must be considered as singlets in the dual WZ model.
We observe that three sphere partition functions for the fields in Table \ref{Tab:IIIA_odd}
is given by the RHS of (\ref{eq:Z2NAtSQQts2}), corroborating the validity of the
conjectured confining duality.
Here we do not discuss the generic form of the superpotential but we will comment about
it in the next paragraphs.

\subsubsection*{Proof from tensor deconfinement}

In this case we can prove the duality by deconfining the 
symmetric and the antisymmetric tensor, by using a confining duality for an orthogonal and a symplectic confining duality, and then by dualizing the unitary node.
The discussion for $SU(N=2n)$ and  $SU(N=2n-1)$ need to be taken separately.

Let's start discussing the $SU(2n)$ case. 
After the deconfinement of the two two-index tensors we have a quiver with three gauge nodes. The $SU(2n)$ gauge node has a chiral matter content and its dual phase has gauge group $SU(2n-1)$. At this point we confine the symplectic gauge node  and then deconfine
the antisymmetric tensor   back, using a confining dualities with a linear monopole superpotential.
Then we apply another chiral duality on $SU(2n-1)$, obtaining an $SU(2n-2)$ gauge group.
The last step consists of confining the symplectic and the orthogonal node, arriving to the same field content of the initial theory, but with gauge rank  decreased of two unities and 
with four extra singlets interacting through a non trivial superpotential.
\begin{figure}[H]
\begin{center}
  \includegraphics[width=15cm]{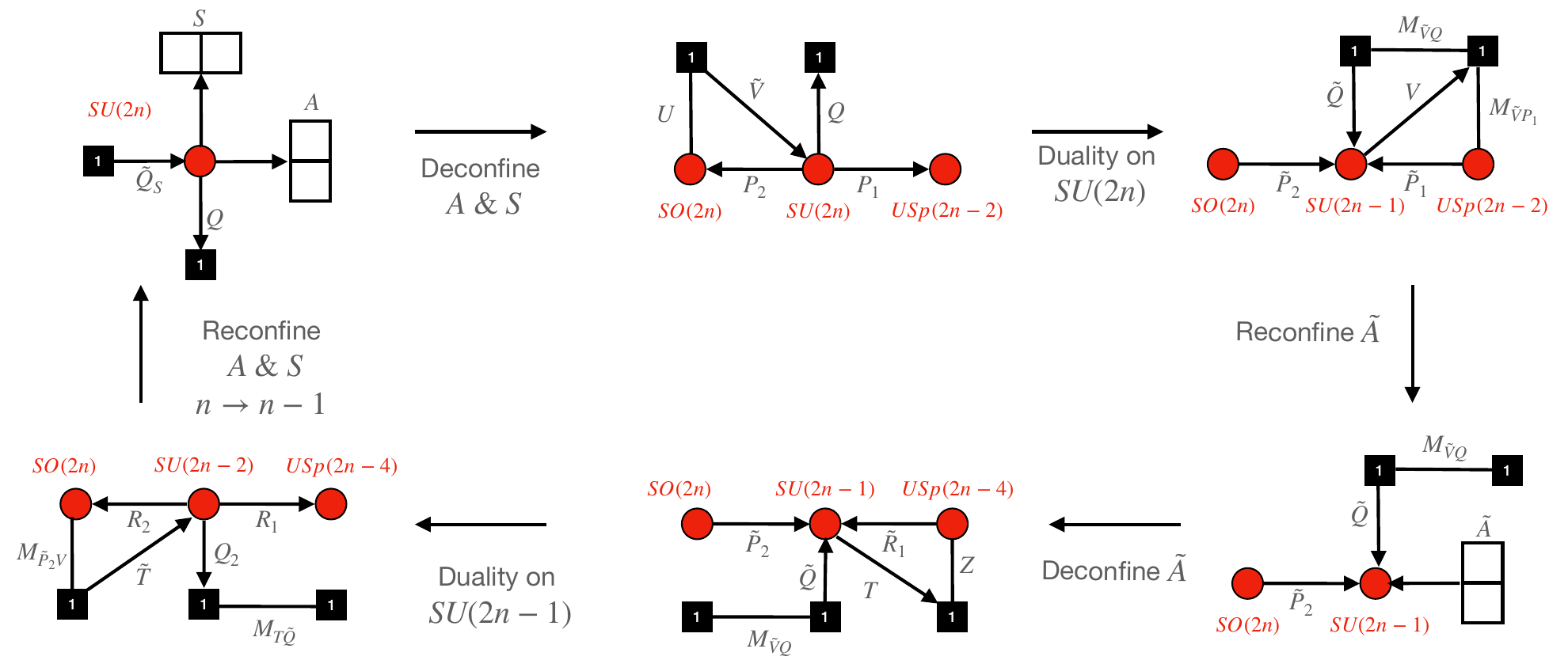}
  \end{center}
  \caption{In this figure we illustrate the various steps that we used in order to prove the confinement of case III-A for $SU(2n)$. Following the chain of  deconfinements and
  dualities in this figure we show how an $SU(2n)$ model is associated to an $SU(2n-2)$  
  model with the same charged matter content in addition to further singlets that are clarified in the discussion. Using this relation we can iterate the discussion until we reach an $SU(4)$ gauge theory. The last step, i.e. the confinement of the $SU(4)$ gauge theory follows from the confinement discussed above in case II-A. }
  \label{evenquiverIIIA}
\end{figure}
The process just described  is illustrated in figure \ref{evenquiverIIIA} where we did not 
represented explicitly all the singlets appearing in the various steps because we will be more clear and precise below.

This procedure can be iterated once the duality map between the gauge invariant combinations appearing in the original $SU(2n)$ theory and the one with $SU(2n-2)$
is established.
In this way we can proceed through a cascading behavior until we reach an $SU(4)$ gauge node, where the antisymmetric tensor is self conjugated. The field content is such case is the one of case IIA studied in subsection (\ref{subsec:IIA}). Using this confining duality we arrive to the WZ model dual to the $SU(2n)$ theory we started with.

In the following we will follow the steps just surveyed, studying the superpotential and the operator mapping at each step. In this way we will arrive to the precise formulation of the confining duality. Furthermore the discussion will be helpful for the analysis of the partition function in the next paragraph.

In order to simplify the analysis we study the model with the quiver 
in the  first figure in Figure \ref{evenquiverIIIA}  by flipping the Pfaffian operator associated to the antisymmetric tensor $A$. The superpotential for this theory is
\begin{equation}
\label{WIIIA-first}
W= \sigma \, \text{Pf } A + Q_S^2 S,
\end{equation}
where we omit to represent the flipper $\sigma$ in the Figure \ref{evenquiverIIIA}.

Then we deconfine $S$ and $A$ through an $SO(2n)$ and an $USp(2n-2)$ gauge group. 
The quiver for this phase is the second one in Figure  \ref{evenquiverIIIA}. 
The orthogonal gauge group is equipped also by  a linear monopole superpotential for the fundamental monopole.  The superpotential of this deconfined theory is
\begin{equation}
\label{WIIIA-second}
W = \tilde V P_2 U+\alpha U ^2 + \gamma \, \epsilon_{2n} \cdot P_2^{2n}+Y_{SO(2n)}^+,
\end{equation}
where we have not represented in the quiver the singlet $\gamma$ flipping the baryonic term  $\epsilon_{2n} \cdot P_2^{2n}$.
We also claim from the charge mapping that the singlet $\alpha$ corresponds to the operator $\det S$ and it is not represented in the figure as well. This observation will be crucial below.

The next step consists of a duality on $SU(2n)$ with $(4n-1)$ fundamentals  and one antifundamental. The dual theory, discussed in appendix \ref{SUSUappD1}, corresponds to $SU(2n-1)$   with $4n-1$ antifundamentals and one fundamental. 
The dual superpotential can be read from (\ref{WIIIA-second}) and from (\ref{spotchidu}).
By integrating out the massive field such superpotential is
\begin{equation}
\label{WIIIA-third}
W =  V \tilde P_1 M_{\tilde V P_1}+ V \tilde Q M_{\tilde V Q}+(V \tilde P_2)^2 \alpha+\gamma \, \epsilon \cdot (\tilde Q \tilde P_1^{2n-2}).
\end{equation}
Observe that in the superpotential (\ref{WIIIA-third}) the linear monopole superpotential for
$Y_{SO(2n)}^+$ is lifted because of the holomorphic mass term induced by  the deformation $\tilde V P_2 U$ in (\ref{WIIIA-second}). Furthermore the $SU(2n)$ baryon $ \epsilon \cdot P_2^{2n}$  in (\ref{WIIIA-second}) corresponds, through the duality map, to the 
$SU(2n-1)$  baryon $\epsilon\cdot (\tilde Q \tilde P_1^{2n-2}) $.

We then observe that the symplectic gauge group $USp(2n-2)$ with $2n$ fundamentals
is confining and its confinement gives rise to a conjugated  antisymmetric tensor denoted in the fourth quiver in Figure  \ref{evenquiverIIIA} as $\tilde A$. Integrating out the massive fields the superpotential for this phase is
\begin{equation}
\label{WIIIA-fourth}
W = \gamma \, \epsilon \cdot (\tilde Q \tilde A^{n-1} ) +\alpha \chi ^2  \tilde P_2^2  \tilde A^{2 n-2}+\tilde Q \chi  A^{n-1} M_{\tilde V Q},
\end{equation}
where the singlet $\chi$, that is not represented in the quiver, corresponds to the $USp(2n-2)$ fundamental monopole acting as a singlet in the superpotential (\ref{WIIIA-fourth}).

Then we deconfine again the conjugated antisymmetric tensor $\tilde A$, but this time we use a different confining duality, involving a deformation for the fundamental monopole of the $USp(2n-4)$ gauge group.
The quiver associated to this phase is  the fifth one in figure \ref{evenquiverIIIA}.
Observe that in this case we have named the bifundamental of $SU(2n-1)\times USp(2n-4)$  as $\tilde R_1$, in order to not generate an overlap in the terminology. 
The superpotential for this phase is obtained by looking at the operator map and we claim that in this phase it becomes
\begin{equation}
\label{WIIIA-fifth}
W = \gamma \tilde Q T+\alpha \chi ^2 \tilde P_2^2 T^2+  \chi M_{\tilde V Q} \tilde Q T+Z   \tilde R_1 T+ Y_{USp(2n-4)}.
\end{equation}
The $SU(2n-1)$ gauge theory has $4n-3$ antifundamentals and one fundamental. The dual theory (up to an overall conjugation), discussed in appendix \ref{SUSUappD1}, corresponds to $SU(2n-2)$   with $4n-3$ fundamentals and one antifundamental. 
The dual superpotential can be read from (\ref{WIIIA-fifth}) and from (\ref{spotchidu}).
Actually in this case we claim that there is also an extra interaction between the singlet 
$M_{\tilde V Q} $ and the baryon monopole $Y^-_{SO(2n) \, \epsilon \cdot(P_2^{2n-2})}$, this is the same type of interaction discussed in (\ref{422}).
By integrating out the massive field the superpotential for this phase is
\begin{equation}
\label{WIIIA-sixth}
W = M_{\tilde P_2  T}  P_2 \tilde T+ \alpha \chi^2 M_{\tilde P_2  T}^2+M_{\tilde V Q} Y^-_{SO(2n) \, \epsilon_{2n-2}P_2^{2n-2}}.
\end{equation}
At this point of the discussion we can confine together the $SO(2n)$ gauge group, with $2n-1$ vectors and the $USp(2n-4)$ gauge group with $2n-2$ fundamentals. The confinement in this case is standard and it gives rise to a symmetric tensor $S_2$, an antifundamental $\tilde Q_{S_2}$ (dual to the baryon monopole $Y_{SO(2n)}^- \epsilon \cdot(R_2^{2n-3} M_{\tilde P_2 T})$) and 
an antisymmetric tensors $A_2$. Furthermore, there are three  singlets arising in this phase. First of all there are two singlets $\Sigma_2$ and $\sigma_2$ corresponding to the monopole operators $Y_{SO(2n)}^+$ and $Y_{USp(2n-4)}$ acting as singlets in this phase.
There is also a singlet denoted as $\varphi_2 \equiv M_{\tilde P_2  V}^2 $.
The quiver for the charged matter fields in this phase is the same as the first one in Figure \ref{evenquiverIIIA}, with $n \rightarrow n-1$. The superpotential is
\begin{equation}
\label{WIIIA-seventh}
W= \sigma_2 \, \text{Pf } A_2 + \tilde Q_{S_2}^2 S_2+\alpha_2 \chi_2^2 \varphi_2+\Sigma_2^2 \varphi_2 \det S_2,
\end{equation}
where we renamed $\alpha \rightarrow \alpha_2$ and $\chi \rightarrow \chi_2$ for future aims.

At this point of the discussion we need to iterate the procedure, going from $SU(2n-2)$ to $SU(2n-4)$ and so on, until we reach $SU(4)$. At each step we generate three new singlets, generalizing the ones denoted as $\alpha_2$, $\chi_2$ and  $\Sigma_2$ in (\ref{WIIIA-seventh}). At each step, the extra flipper, named $\sigma_2$ in (\ref{WIIIA-seventh}), is actually removed in the second quiver in Figure \ref{evenquiverIIIA} and then a new flipper $\sigma_{j+1}$ is generated for the $SU(2(n-j))$ gauge group. 

When going from $SU(2n)$ to $SU(2n-2)$ we observe also that the duality map relates the operator $\det S_2 \sim S_2^{2n-2}$ to the operator $S^{2n-2} A^2$.
In general the relation becomes $S_{j+1}^{2(n-j)} = S^{2(n-j)} A^{2j}$ for $SU(2(n-j))$.
Furthermore, this implies the relations $A_{j+1}^{(n-j)} = A^{n}$.
Looking at the relation between the $j$-th step and the $(j+1)$-th one we have 
$S_{j+1}^{2(n-j)} = S_{j}^{2(n-j)} A_{j}^{2}$ and $A_{j+1}^{(n-j)} = A_j^{(n-j+1)}$.
We also have $Q_{j+1}^{2(n-j)} = Q_{j}^{2(n-j)} A_j^{j}$.
Then, after $j=n-2$ steps we obtain an $SU(4)$ gauge theory with superpotential
\begin{eqnarray}
\label{Witerato}
W &=&\sigma_{n-1} \, \text{Pf } A_{n-1}+  \tilde Q_{S_{n-1}}^2 S_{n-1}+ \sum_{j=1}^{n-2} \alpha_{j+1} \chi_{j+1}^2 \varphi_{j+1} \nonumber \\
&+&  \sum_{j=1}^{n-3} \Sigma_{j+1}^2 \varphi_{j+1} \alpha_{j+2}+\Sigma_{n-1}^2 \varphi_{n-1} \det S_{n-1},
\end{eqnarray}
where the singlets $\alpha_{j+2}$ correspond to the operators $\det S_{j+2}$ and their interactions with the combinations $\Sigma_{j+1}^2 \varphi_{j+1}$ are generated at each step from the last term in (\ref{WIIIA-seventh}). 

Before dualizing the $SU(4)$ gauge theory we  give the explicit map between the singlets
obtained at this level with the ones appearing in Table \ref{Tab:IIIA_even}.
We have
\begin{eqnarray}
\label{map1}
\alpha_{n-j} \leftrightarrow \Xi_2^{(j)}, \quad
\varphi_{n-j} \leftrightarrow \Xi_3^{(j)}, \quad
\chi_{\ell+1}\leftrightarrow \Xi_5^{(2\ell)}, \quad
\Sigma_{\ell+1}\leftrightarrow \Xi_5^{(2\ell+1)},
\end{eqnarray}
with $j=2,\dots,n-1$ and $\ell=0,\dots,n-3$.
At this point we can dualize the $SU(4)$ gauge theory and the singlets of this phase are 
denoted as the ones in Table \ref{Tab:IIA} with $n=2$.
Such fields can be related to the ones of Table \ref{Tab:IIIA_even} with the explicit mapping
\begin{eqnarray}
\label{map2}
&&
\Psi_1 \leftrightarrow \Xi_2^{(1)},\quad
\Psi_2 \leftrightarrow \Xi_1 ,\quad
\Psi_3^{(0)} \leftrightarrow \Xi_2^{(0)},\quad
\Psi_4^{(0)} \leftrightarrow \Xi_3^{(0)},\quad
\Psi_5 \leftrightarrow \Xi_3^{(1)}\nonumber \\
&&
\Psi_6  \leftrightarrow \Xi_5^{(2n-4)},\quad
\Psi_7  \leftrightarrow \Xi_5^{(2n-3)},\quad
\Psi_8^{(0)}  \leftrightarrow \Xi_4,\quad
\Psi_8^{(1)}  \leftrightarrow \Xi_5^{(2n-2)}.
\end{eqnarray}
Looking at the superpotential (\ref{Witerato}) we see that the coupling of the Pfaffian of $A_{n-1}$ with the singlet $\sigma_{n-1}$ becomes a mass term in the last step for the singlet $\Xi_1$. This is indeed consistent with the flip in (\ref{WIIIA-first}), which implies that the singlet $\Xi_1$ is set to zero in the chiral ring by the F-term $F_{\sigma}$.

The last comment regards the final superpotential. Using the map in (\ref{map1}) and in (\ref{map2}) we can read the interactions in the final superpotential from (\ref{Witerato}) and (\ref{IIASU(4)}). Actually not all the fields from the $SU(4)$ gauge theory appear in the final description. Indeed, the fields $\Xi_5^{(2n-3)}$ and $\Xi_2^{(0)}$ do not appear in 
(\ref{IIASU(4)}) and in general one has to ensure that they are free field in this duality.
In general this in not consistent with a unitary duality, as one can check through  
F-maximization. Flipping such fields is then necessary in order to satisfy the unitary bound.
This correspond to have a superpotential with only quartic couplings and $R=1/2$ for all the singlets in the chiral ring.

\begin{figure}[H]
\begin{center}
  \includegraphics[width=9cm]{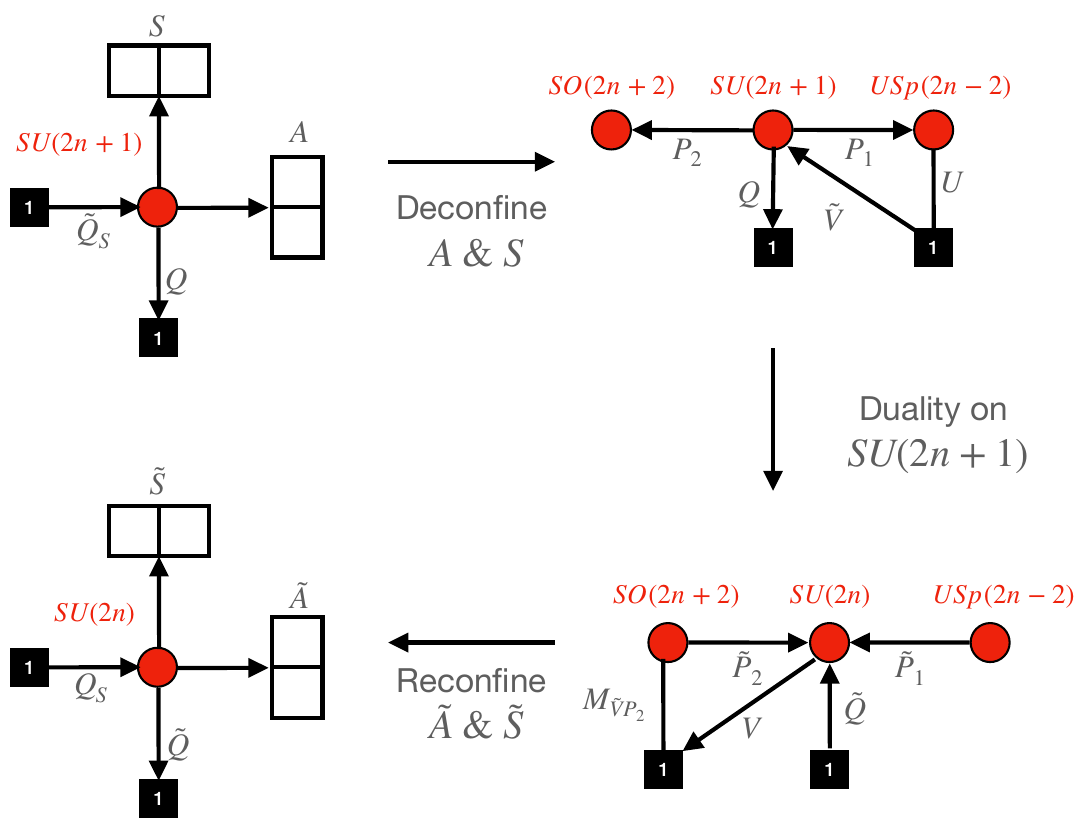}
  \end{center}
  \caption{In this figure we illustrate the various steps that we used in order to prove the confinement of case III-A for $SU(2n+1)$. Following the chain of  deconfinements and
  dualities in this figure we show how an $SU(2n)$ model is associated to an $SU(2n-1)$  
  model. Once we reach this model we observe that, up to conjugation and to a further superpotential term, this coincides with the $SU(2n)$ case discussed above. This proves the confining dynamics of this theory.}
  \label{oddquiverIIIA}
\end{figure}

A similar analysis holds for the $SU(2n+1)$ case. In this case the analysis is simplified because we can use the results obtained for the even case. We consider the first quiver in Figure \ref{oddquiverIIIA}.
As discussed in the even case the situation is simplified by flipping the pfaffian operator for the antisymmetric $A$. The superpotential for this phase is then 
\begin{equation}
\label{decoddIIIA-I}
W = \tilde Q_S^2 S + \sigma_0 \epsilon_{2n+1}(A^n Q) + \zeta^2 \det S.
\end{equation}
Then we deconfine the two tensor $A$ and $S$ using an $USp(2n-2)$ and an $SO(2n+2)$ gauge node respectively. In the symplectic case we turn on a linear monopole superpotential. Furthermore, following the operator map the flipper $\sigma_0$ interact with the combination $(\tilde V P_1)^2 $. The superpotential for the second quiver in Figure \ref{oddquiverIIIA} is then
\begin{equation}
\label{decoddIIIA-II}
W = Y_{USp(2n-2)}+ P_1 U \tilde V+\sigma_0 \tilde V Q.
\end{equation}
The $SU(2n+1)$ gauge group has $4n-1$ fundamentals and one antifundamental, then it can be dualized following the duality map reviewed in Appendix \ref{SUSUappD1}.
The dual gauge group becomes $SU(2n)$ with $4n-1$ antifundamentals and one fundamental. 
The dual quiver arising from this duality is the third one in  Figure \ref{oddquiverIIIA}.
Integrating out the massive fields the superpotential for this theory is
\begin{equation}
\label{decoddIIIA-III}
W =M_{\tilde V P_2} V \tilde P_2. 
\end{equation}
Then we observe that the $SO(2n+2)$  and  the $USp(2n-2)$ gauge groups have 
$2n+1$ vectors and $2n$ fundamentals respectively, i.e. they are confining.
Once we confine these groups we obtain a conjugated symmetric $\tilde S$ and a conjugated antisymmetric $\tilde A$. From the confinement of the orthogonal gauge group there are also a baryon monopole $Q_{\tilde S} = 
Y_{SO(2n+2) \epsilon \cdot (\tilde P_2^{2n-1} M_{\tilde V P_2})}^-$ in the fundamental representation of the $SU(2n)$ gauge group and a baryon monopole $q_1=Y_{SO(2n+2) \epsilon \cdot(\tilde P_2^{2n-2})}^-$, that is a singlet of the leftover $SU(2n)$ gauge group.
Furthermore, there are three other singlets that arise from this duality and that are 
massless in the superpotential, denoted as $\Sigma_1 = Y_{SO(2n+2)}^+$,
$\sigma_1 = Y_{USp(2n-2)}$ and $\varphi_1 = 
 M_{\tilde V P_2}^2$. The superpotential for this model is then
\begin{equation}
\label{decoddIIIA-IV}
W = Q_{\tilde S}^2 \tilde S+\sigma_1 \text{Pf}\tilde A+\Sigma_1^2 \varphi_1 \det \tilde S + \varphi_1 q_1^2.
\end{equation}

At this point of the discussion we observe that this model corresponds (up to an overall charge conjugation that is irrelevant for the discussion) to the model studied above, i.e.
case III-A with an $SU(2n)$ gauge group. The only difference stays in the last term in the superpotential (\ref{decoddIIIA-IV}), where the operator $\det \tilde S$ is dual to $S^2 (\epsilon \cdot (Q A^{n-1}))^2$.
It follows that we can arrive to the $SU(4)$ gauge theory as discussed above using the
superpotential (\ref{Witerato}) and adding the contribution of the last term in 
(\ref{decoddIIIA-IV}). We have
\begin{eqnarray}
\label{Witerato2}
W &=&\sigma_{n-1} \, \text{Pf } \tilde A_{n-1}+  Q_{\tilde S_{n-1}}^2 \tilde S_{n-1}+ \sum_{j=1}^{n-2} \alpha_{j+1} \chi_{j+1}^2 \varphi_{j+1} + \varphi_{1}  q_{1}^2  \nonumber \\
&+&  \sum_{j=0}^{n-3} \Sigma_{j+1}^2 \varphi_{j+1} \alpha_{j+2}+\Sigma_{n-1}^2 \varphi_{n-1} \det \tilde S_{n-1}, 
\end{eqnarray}
where the sum in the first term in the second line of (\ref{Witerato2}) now starts from $j=0$.
Similarly with what we have done in the even case we can map the singlets in Table \ref{Tab:IIIA_odd} with the ones in (\ref{Witerato2}). The missing singlets  are then obtained from the confinement of  $SU(4)$ gauge theory from case II-A. Again the superpotential generated from (\ref{Witerato2}) and (\ref{IIASU(4)}) does not include all the singlets of Table \ref{Tab:IIIA_odd} and similar comments about the unitarity of the duality at hand hold also in this case.

\subsubsection*{Proving the identities (\ref{eq:Z2NAtSQQts}) and (\ref{eq:Z2NAtSQQts2}) from field theory}

Here we conclude our analysis on the case III-A by proving the relations 
(\ref{eq:Z2NAtSQQts}) and (\ref{eq:Z2NAtSQQts2})  applying the 
iterative procedure discussed above at the level of the partition function.
For this reason we refer to the quiver in Figure \ref{evenquiverIIIA} and at each step we discuss the main aspects of the identities generated by the various dualities.

We start by fixing the parameters in the first quiver as $\tau_S$ for the symmetric, $\tau_A$ for the antisymmetric, $\mu$ for the fundamental and $\omega-\frac{\tau_S}{2}$
for the antifundamental. The flipper $\sigma$ on the other hand has mass parameter
$m_{\sigma} = 2 \omega - n \tau_A$.
Then we deconfine the two tensors using the relations (\ref{USpfondmon}) and
\footnote{Actually the relation (\ref{confsO}) needs to be modified accordingly due to the presence of the flippers $\gamma$ and $\alpha$ and the cubic superpotential interaction $P_2 U \tilde V$.} 
(\ref{confsO}) and the new bifundamentals $P_1$ and $P_2$ in the second quiver 
have mass parameter $\frac{\tau_A}{2}$ and $\frac{\tau_S}{2}$ respectively. The 
new fields  $\tilde V$ and $U$ have mass parameters
\begin{equation}
m_{\tilde V}= \omega+n \tau_S-\frac{\tau_S}{2},
\quad 
m_U = \omega-n \tau_S,
\end{equation}
where the mass for the $SO(2n)$ vector $U$ is determined by the balancing condition and signals the presence of the linear monopole superpotential $Y_{SO(2n)}^+$ in 
(\ref{WIIIA-second}).
The singlets $\gamma$ and $\alpha$ have mass parameters
\begin{equation}
\label{malpha}
m_{\gamma}= 2\omega-n \tau_S,
\quad 
m_\alpha = 2n \tau_S.
\end{equation}
Observe that this is consistent with the identification $\alpha = \det S$ claimed in the field theory discussion. 

Then we consider the third quiver in Figure \ref{evenquiverIIIA},  where the duality is obtained from (\ref{chiralpqstarduality}) with $N=2n$, $F=4n-1$ and $\tilde F=1$. 
In this case the antifundamentals have mass parameters 
\begin{equation}
m_{\tilde P_2} = X-\frac{\tau_S}{2},\quad 
m_{\tilde P_1} =X- \frac{\tau_A}{2},\quad 
m_{\tilde Q} =X- \mu,
\end{equation}
while the fundamental $V$ have mass parameter
$m_{V}= \omega-n \tau_S+\frac{\tau_S}{2}-X$.
The duality map is obtained by computing the quantity 
\begin{equation}  
X\equiv\frac{n \tau_S+(n-1)\tau_A+\mu}{2n-1}.
\end{equation}
The two mesons that survive at this step have mass parameter
\begin{equation}  
m_{M_{\tilde V Q}}=\omega+n \tau_S-\frac{\tau_S}{2}+\mu, \quad
m_{M_{\tilde V P_1}}=\omega+n \tau_S-\frac{\tau_S}{2}+\frac{\tau_A}{2},
\end{equation}
while the meson $m_{M_{\tilde V  P_2}}$ has mass parameter  $m_{M_{\tilde V P_2}}=2 \omega-m_U$ and it disappears (together with $U$) because of the inversion formula 
(\ref{inversion}).

The next step, corresponding to the fourth quiver in Figure \ref{evenquiverIIIA}, consists of confining the $USp(2n-2)$ gauge node through the identity (\ref{USpwomon}). This steps generates the conjugated antisymmetric $\tilde A$, with mass parameter $2X-\tau_{A}$ and the monopole $\chi$ with mass parameter 
\begin{equation}
\label{mchi}
m_\chi=\omega-2n \tau_S +\frac{\tau_S}{2}-\mu.
\end{equation}
On the other hand, the mass parameter of the $SU(2n-1)$ antifundamental built from $\tilde P_1 \cdot M_{\tilde V P_1}$ is eliminated together with the one of the $SU(2n-1)$ fundamental  $V$ using the duplication formula.

We then deconfine the conjugated antisymmetric $\tilde A$ using an $USp(2n-4)$ gauge group turning on a linear   superpotential for the fundamental monopole $Y_{USp(2n-4)}$, obtaining the fifth quiver  in Figure \ref{evenquiverIIIA}. The effect of this deformation is absorbed in the mass parameter of the auxiliary $USp(2n-4)$ fundamental $Z$, through the relation $m_{Z}= 2\omega- (2n-1) m_{\tilde R_1}$, where the bifundamental 
$\tilde R_1$ has mass parameter $X-\frac{\tau_A}{2}$. The mass parameter for the $SU(2n-1)$ fundamental $T$ is $m_T =(n-1) (2X-\tau_A)$ consistently with the superpotential term $\tilde R_1 Z T $ in (\ref{WIIIA-fifth}).

The following duality, on $SU(2n-1)$ requires to define the auxiliary quantity
\begin{equation}  
\label{Yaux}
Y=\frac{(n-2) \left(2 X-\tau _A\right)+n \left(2 X-\tau _S\right)+(X-\mu )}{2 (n-1)}=\frac{X-\mu }{2 n-2}-\tau _S+2 X,
\end{equation}
in order to read the duality map on the $(4n-3)$ $SU(2n-1)$ antifundamentals $\{\tilde R_1,\tilde P_2,\tilde Q\}$ and the fundamental $T$.
The mass parameter of the dual  $(4n-3)$ $SU(2n-2)$ fundamentals, denoted as $\{ R_1, R_2, Q_2\}$ in 
the sixth quiver  in Figure \ref{evenquiverIIIA} are 
\begin{eqnarray}
\label{impo}
m_{R_2} &=& Y-m_{\tilde P_2}= 
\frac{1}{2} \left(\frac{\tau _A}{n-1}+\tau _S\right),\nonumber \\
m_{R_1}&=&Y-m_{\tilde R_1}= 
\frac{n \tau _A}{2 (n-1)}, \\
m_{Q_2}&=&Y-m_{\tilde Q}= 
\frac{\tau _A}{2 (n-1)}+\mu, \nonumber
\end{eqnarray}
while the antifundamental $\tilde T$ has $m_T =2\omega-m_T-Y =\frac{\tau _A}{2-2 n}-\mu -n \tau _S+2 \omega$.
There are also three mesons $M_{T \tilde R_1 }$,  $M_{T \tilde P_2 }$ and  $M_{T \tilde Q }$,

Two of them are massive, $M_{T \tilde Q }$ and $M_{T \tilde R_1 }$  because of the interactions in  (\ref{WIIIA-fifth})
with $\gamma$ and $Z$ respectively. This emerges in the analysis of the partition function because 
$m_\gamma +m_{M_{T \tilde Q }}=2 \omega$ and  $m_Z+m_{M_{T \tilde R_1}}=2 \omega$.
The other meson, $M_{T \tilde P_2 }$, has mass parameter  $m_{M_{T \tilde P_2 }} =m_T+m_{ \tilde P_2}$.

The last step of this first iteration corresponds to the confinement of $USp(2n-4)$ with $2n-2$ fundamentals and 
$SO(2n)$ with $2n-1$ fundamentals. On the partition function we use the relations (\ref{USpwomon}) and (\ref{SOelemag}) respectively.

These confining dualities generate the new symmetric and antisymmetric $S_2$ and $A_2$, and their mass parameters 
can be read from (\ref{impo}), corresponding to $m_{S_2} = 2 m_{R_2} $ and $m_{A_2} = 2 m_{R_1}$.
The new antifundamental $Q_{S_2}$ corresponds to the baryon monopole   $Y_{SO(2n)}^- \epsilon \cdot(R_2^{2n-3} M_{\tilde P_2 T})$ has mass parameter $\omega - \frac{m_{S_2}}{2}$, consistently with the superpotential interaction $S_2 \tilde Q_{S_2}^2$.

There are also two singlets corresponding to the fundamental monopoles of the $SO(2n)$ and of the $USp(2n-4)$ gauge groups. These fields have beed denoted as $\Sigma_2$ and $\sigma_2$ above. Their mass parameters are
\begin{equation}
\label{mSigma}
m_{\Sigma_2}= \omega -\tau _A-\mu -2 n \tau _S+\frac{3 \tau _S}{2} , \quad 
m_{\sigma_2} = 2\omega-(2n-1)m_{R_1} = 2 \omega-n \tau_A.
\end{equation}
Observe that the singlet $\sigma$ flipping the Pfaffian in the superpotential 
(\ref{WIIIA-first}) has the same mass parameter of the singlet $\sigma_2$ in the superpotential (\ref{WIIIA-seventh}). This mapping extends through the iterative procedure,
until we arrive at the superpotential (\ref{Witerato}), where again the singlet $\sigma_{n-1}$ has the same mass parameter of $\sigma$.
There is another baryon monopole, corresponding to the singlet $Y_{SO(2n) \,\epsilon \cdot(R_2^{2n-2})}^-$,
with mass parameter 
\begin{equation}
\omega- m_{M_{\tilde P_2 T}} = \omega-\left(\frac{1}{2}-n\right) \tau _S-\mu
\end{equation}
and, as expected from the field theory analysis, it interacts through a holomorphic  mass term with the meson $M_{\tilde V Q}$.
The other singlet emerging at this step  is $\varphi_2=M_{\tilde P_2  T}^2$, with mass parameter
\begin{eqnarray}
\label{massaphi}
m_{\varphi_2}=2m_{M_{\tilde P_2  T}} =2 \mu +(2 n-1) \tau _S.
\end{eqnarray}
Summarizing we have obtained an $SU(2n-2)$ gauge theory with the same charged field content of the $SU(2n)$ case we started with and, in addition some singlets.
In this case the rules of the duality applied on the parameters $\tau_S, \tau_A$ and $\mu$ give
\begin{eqnarray}
\label{daiterare}
\tau_S^{(1)} = \tau_S &\rightarrow& \tau_S^{(2)} = \frac{\tau_A^{(1)}}{n-1}+\tau_S^{(1)}, 
\nonumber \\
\tau_A^{(1)} = \tau_A  &\rightarrow& \tau_A^{(2)} =  \frac{n }{n-1}\tau_A^{(1)},
\nonumber \\
\mu^{(1)} = \mu   &\rightarrow& \mu^{(2)} =  \frac{\tau_A^{(1)}}{2 (n-1)}+\mu^{(1)}.
\end{eqnarray}
The matching at the level of the  partition function is
\begin{eqnarray}
\label{daiterarepari}
&&
\Gamma_h(2\omega-n \tau_A^{(1)})
Z_{SU(2n)} \left(\mu^{(1)},\omega-\frac{\tau_S^{(1)}}{2};-;\tau_S^{(1)},\tau_A^{(1)} \right) =
\nonumber \\
&&
\Gamma_h\left(2 n \tau _S,\omega-\mu +\left(\frac{1}{2}-2 n\right) \tau _S,\omega-\tau _A-\mu -2 n \tau _S+\frac{3 \tau _S}{2},2 \mu +(2 n-1) \tau _S\right)
\nonumber \\
&&
\Gamma_h(2\omega-(n-1)\tau_A^{(2)} ) 
Z_{SU(2N-2)} \left(\mu^{(2)},\omega-\frac{\tau_S^{(2)}}{2};-;\tau_S^{(2)},\tau_A^{(2)} \right),
\nonumber \\
\end{eqnarray}
where $\Gamma_h(2\omega-(n-1)\tau_A^{(2)} ) =\Gamma_h(2\omega-n \tau_A^{(1)})$.
In the second lines we have collected the masses of the singlets $\alpha=\alpha_2$, $\chi=\chi_2$, $\Sigma_2$ and $\varphi_2$ that can be read from (\ref{malpha}), (\ref{mchi}), (\ref{mSigma}) and (\ref{massaphi}) respectively.

Then we iterate the relation (\ref{daiterarepari}) using the  relations 
\begin{eqnarray}
&&
\tau_S^{(j+1)} = \tau_S^{(1)}+\frac{j}{n-j}\tau_A^{(1)},  \nonumber \\
&&
\tau_A^{(j+1)} =\frac{n}{n-j}\tau_A^{(1)},
 \nonumber \\
&&
\mu^{(j+1)} = \mu^{(1)} +\frac{j}{2(n-j)}\tau_A^{(1)}, 
\end{eqnarray}
that can be derived by iterating (\ref{daiterare}).

At each step we generate new singlets $\{\alpha_{j+1},\varphi_{j+1},\chi_{j+1},\Sigma_{j+1} \}$  and their contribution to the partition function is summarized as
\begin{eqnarray}
\label{almost}
&&
\prod_{j=0}^{n-3} \Gamma_h \left( 2 j \tau _A+2(n-j) \tau _S, 2 \mu+2 j \tau _A +(2 n-1-2j) \tau _S\right) \nonumber \\
&&
\prod_{j=0}^{2n-5}   \Gamma_h \left( \omega- j \tau _A-\mu +\left(\frac{1}{2}+ j-2 n\right) \tau _S\right).
\end{eqnarray}

This construction is reliable as long as we reach $SU(4)$ (i.e. after iterating the procedure $n-2$ times), where the deconfinement of the antisymmetric tensor is not possible along the analysis performed above.
Anyway we can use the duality discussed in case II-B, because in this case the antisymmetric is self conjugate.

This boils  down to consider $SU(4)$ with one symmetric with mass parameter $\tau _S+\frac{n-2}{2}\tau_A$, a conjugate antisymmetric with mass parameter $\frac{n \tau _A}{2}$, a fundamental with mass parameter $\mu+\frac{1}{4} (n-2) \tau _A$ and an antifundamental with mass parameter $\omega-\frac{\tau _S}{2}-\frac{n-2}{4}\tau_A$.
By plugging these values into the RHS of (\ref{eq:Z2NAtSQQts}) for an $SU(4)$ gauge theory, we obtain the  missing singlets that we must add to (\ref{almost}) in order to prove the RHS of (\ref{eq:Z2NAtSQQts}). Actually, we arrive at the identity  (\ref{eq:Z2NAtSQQts}) with the flipper of the pfaffian $\sigma$ in the RHS of the equality.
It can be \emph{moved} on the LHS by applying the inversion relation (\ref{inversion}).
This concludes the proof of the confining duality for the $SU(2N)$ case from tensor deconfinement.

Then we move to the case of $SU(2n+1)$.
We start again by defining $\tau_S$, $\tau_A$ and $\mu$ as the mass parameters of the symmetric $S$, the  antisymmetric $A$ and the fundamental $\mu$. 
We follow the steps summarized in Figure \ref{oddquiverIIIA} on the partition function.
Proceeding as in the even case we arrive at the following relation between the $SU(2n+1)$ and the $SU(2n)$ partition function 
\begin{eqnarray}
\label{evodd}
&&
\Gamma_h(m_{\sigma_0},m_{\zeta})
Z_{SU(2n+1)} (\mu;\tau_S-\frac{\omega}{2};\tau_S;-;\tau_A;-) = \Gamma_h(m_{\sigma_1}, m_{\varphi_1} ,m_{\Sigma_1}, m_{q_1}) 
\nonumber \\
&&  Z_{SU(2n)} (\omega-X+\frac{\tau_S}{2};X-\mu;-;2X-\tau_S;-;2X-\tau_A), 
\end{eqnarray}
where 
\begin{equation}
\label{Xevodd}
X=\frac{(n+1)\tau_S + (n-1)\tau_A +\mu}{2n}
\end{equation}
and
\begin{eqnarray}
\label{masseeee}
&&m_{\sigma_0}=2\omega -n \tau_A-\mu, \quad\,
m_{\zeta} =\omega-(2n+1)\frac{\tau_S}{2}, \quad\,
m_q =\omega-n \tau _A-\frac{\tau _S}{2}, \nonumber \\
&&m_{\sigma_1} = 2\omega-(n+1)\tau_S+\tau_A-\mu, \quad\,
m_{\varphi_1} = 2n\tau_A+\tau_S, \nonumber \\ 
&& m_{\Sigma_1} = \omega-\mu-(2n-1)\tau_A-\frac{3}{2}\tau_S.  
\end{eqnarray}
Observe that in the RHS of (\ref{evodd})  the fields in the partition function are conjugated with respect to the ones appearing in the III-A case for even rank studied above. Due to the absence of CS levels in this case parity is unbroken and this partition function is equivalent to the one obtained by conjugating all the representations,
i.e. $Z_{SU(2n)} (X-\mu;\omega-X+\frac{\tau_S}{2};2X-\tau_S;-;2X-\tau_A;-)$.
We conclude the proof  of (\ref{eq:Z2NAtSQQts2}) by plugging the identity (\ref{eq:Z2NAtSQQts}) on the RHS of (\ref{evodd}).
\subsection{Case III-B: $S \oplus  A \oplus \tilde Q_S \oplus \tilde Q$}

Here we first review the 3d confining duality distinguishing the even and the odd case.
For even $N=2n$ the gauge invariant combinations, built from the charged chiral fields, that appear in the dual WZ model are
\begin{equation}
M \!=\! Q \tilde Q, \,\,
T_n \!=\! A^n, \,\,
T_{n-1}\!=\!A^{n-1} Q^2, \,\,
P_1 \!=\! A^{N-1} (A \tilde Q) Q, \,\,
P_3 \!=\! A^{N-2} (A \tilde Q) Q^3.
\end{equation}
Furthermore, there is a dressed monopole
\begin{eqnarray}
Y_{SU(2n-2)}^{dressed} =Y_{SU(2n-2)}^{bare} A^{2n-3}.
\end{eqnarray}
The superpotential for this confining phase is
\begin{equation}
W =Y_{SU(2n-2)}^{dressed}  (M T_n T_{n-1} + T_{n} P_3 +  T_{n-1} P_1  ).
\end{equation}
In this case there is an $SU(2)$ symmetry that rotates the two antisymmetric and then the singlets  $T_j$ are in the $n-j$ symmetric representation of such flavor symmetry group. The singlets  $P_1$ and $P_3$  are in the $n-1$ and $n-2$ symmetric representation of such $SU(2)$,
  while the dressed monopole
is in the $2n-3$ symmetric representation.
The integral identity representing this confining duality is
\begin{eqnarray}
\label{3fund2ASeven}
Z_{SU(2n)} (\vec \mu;\nu;-;-;\tau_{A},\tau_B;-)
&=&
\prod_{a=1}^{3}\Gamma_h(\mu_a +\nu)
\prod_{j=0}^{n} \Gamma_h(j \tau_A + (n-j) \tau_B)
\\
&\times&
\prod_{j=0}^{n-1}  \prod_{1\leq a <b \leq 4}\Gamma_h(j \tau_A + (n-j-1) \tau_B + \mu_a +\mu_b)
\nonumber \\
&\times& 
\prod _{j=0}^{n-2} 
\prod_{a=1}^{3}
\Gamma _h\left(\tau _A (n-1-j)+(j+1) \tau _B+\mu _a+\nu \right)
\nonumber\\
&\times&
\prod _{j=0}^{n-3} \Gamma _h\left(\tau _A (n-2-j)+(j+1) \tau _B+\sum_{a=1}^{3} \mu_a+\nu \right)
\nonumber\\
&\times&
\prod_{j=0}^{2n-3} \Gamma_h\! \left(
2\omega-(j+1) \tau_A \!-\! (2n\!-\!j\!-2) \tau_B \!-\!\sum_{a=1}^{3} \mu_a\!-\!\nu \right).
\nonumber 
\end{eqnarray}
The relation (\ref{3fund2ASeven}) has been derived in appendix  \ref{casesIIIas}
and the next step consist of freezing the mass parameters appearing in the arguments of the fundamentals as in (\ref{freezing2}), where we rename $\tau_B = \tau_S$, explicitly breaking the $SU(2)$ global symmetry between the two antisymmetric tensors. In this way we
convert one of the two antisymmetrics into a rank-two symmetric $S$ tensor and an antifundamental $\tilde Q_S$.
By applying the duplication formula, 
this assignation of parameters leaves on the electric side an $SU(2n)$ gauge theory with one rank-two symmetric tensor $S$, one rank-two antisymmetric tensor $A$, one antifundamental $\tilde Q_S$ and one antifundamental $\tilde Q$. 
The mass parameter of $\tilde Q_S$ is compatible with the superpotential interaction (\ref{WSQSQS})  and we claim that this superpotential is necessary for the confinement.
In this way we arrive at the identity 
\begin{eqnarray}
\label{IIIBevenfinal}
&&Z_{SU(2n)}\left(-;\omega-\frac{\tau_S}{2},\nu;\tau_S;-;\tau_A;-\right) 
= \Gamma_h( n \tau_A)
\Gamma_h( \tau_S + 2\nu) 
{\Gamma_h \left( \omega-\frac{\tau_S}{2} -\nu \right)}
\nonumber
 \\
&&
\prod_{j=0}^{n-1} \Gamma_h(2(j+1)\tau_S + 2(n\!-\!1\!-j) \tau_A)
\prod_{j=0}^{n-2} \Gamma_h((2j+3)\tau_S + 2(n\!-\!j\!-1) \tau_A+2\nu)
 \\
&&
\Gamma_h\left(\omega-\frac{3\tau_S}{2}-(n-1)\tau_A-\nu\right)
\prod_{j=0}^{2n-3} \Gamma_h\left( \left(j+\frac{1}{2}-2n\right)\tau_S - (j+1) \tau_A-\nu+\omega\right).
\nonumber 
\end{eqnarray}

For odd $N=2n+1$ the gauge invariant combinations, built from the charged chiral fields, that appear  
in the dual WZ model are
\begin{equation}
M = Q \tilde Q, \,\,
T_n = A^n Q,  \,\,
T_{n-1}=A^{n-1} Q^3,  \,\,
P_1 = A^{n} (A \tilde Q),  \,\,
P_2 = A^{n-1} (A \tilde Q) Q^2.
\end{equation}
Furthermore, there is a dressed monopole
\begin{eqnarray}
Y_{SU(2n-1)}^{dressed} =Y_{SU(2n-1)}^{bare} A^{2n-2}.
\end{eqnarray}
The superpotential for this confining phase is
\begin{equation}
W =Y_{SU(2n-1)}^{dressed}  (M T_n^2 + T_{n-1} P_1 +  T_n P_1  ).
\end{equation}
In this case there is an $SU(2)$ symmetry that rotates the two antisymmetric and then the singlets  $T_j$ are in the $n-j$ symmetric representation of such flavor symmetry group. The singlets  $P_1$ and $P_2$  are in the $n-1$ and $n-2$ symmetric representation of such $SU(2)$,
  while the dressed monopole
is in the $2n-2$ symmetric representation.
The integral identity representing this confining duality is
\begin{eqnarray}
\label{3fund2ASodd}
Z_{SU(2n+1)} (\vec \mu;\nu;-;&-&;\tau_{A},\tau_B;-)
=
\prod_{a=1}^{3}\Gamma_h(\mu_a +\nu)
\prod_{j=0}^{n} \Gamma_h(j \tau_A + (n-j) \tau_B+\mu_a) \nonumber
\\
&\times&
\prod_{j=0}^{n-1} \Gamma_h\left(j \tau_A + (n-j-1) \tau_B+\sum_{a=1}^3 \mu_a \right)
\nonumber
\\
&\times&
\prod_{j=0}^{n-1} \Gamma_h\left( (n-j)\tau _A+(j+1) \tau _B+\nu \right)
\\
&\times&
\prod_{j=0}^{n-2} \prod_{1\leq a<b\leq 3}
\Gamma_h\left((n-1-j)\tau _A +(j+1) \tau _B+\nu+\mu_a +\mu_b \right)
\nonumber
\\
&\times&
\prod_{j=0}^{2n-2} \Gamma_h \left(
2\omega-(j+1) \tau_A - (2n-j-1) \tau_B -\sum_{a=1}^{3} \mu_a \right).
\nonumber 
\end{eqnarray}
The relation (\ref{3fund2ASodd}) has been derived in appendix  \ref{casesIIIas}
and the next step consists of freezing the mass parameters appearing in the arguments of the fundamentals as in (\ref{freezing2}), where we rename $\tau_B = \tau_S$, explicitly breaking the $SU(2)$ global symmetry between the two antisymmetric tensors. In this way we
convert one of the two antisymmetrics into a rank-two symmetric $S$ tensor and an antifundamental $\tilde Q_S$.
By applying the duplication formula 
this assignation of parameters leaves on the electric side an $SU(2n+1)$ gauge theory with one rank-two symmetric tensor $S$, one rank-two antisymmetric tensor $A$, one antifundamental $\tilde Q_S$ and one antifundamental $\tilde Q$. 
The mass parameter of $\tilde Q_S$ is compatible with the superpotential interaction (\ref{WSQSQS})  and we claim that this superpotential is necessary for the confinement.
In this way we arrive at the identity 

\begin{eqnarray}
\label{IIIBoddfinal}
&&Z_{SU(2n+1)}\left(-;\omega-\frac{\tau_S}{2},\nu;\tau_S;-;\tau_A;-\right) 
= 
\Gamma_h \left( \omega-\frac{\tau_S}{2} -\nu \right)
\Gamma_h( \tau_S+n \tau_A +\nu)
\nonumber \\
&&
\Gamma_h( \tau_S + 2\nu) 
\Gamma_h\left( \omega-n \tau_A -\frac{\tau_S}{2}\right)
\prod_{j=0}^{n-2} 
\Gamma_h(2(j+2)\tau_S + 2(n-j-1) \tau_A+2\nu) \\
&&
\prod_{j=0}^{n} \Gamma_h((2j+1)\tau_S + 2(n-j) \tau_A)
\prod_{j=0}^{2n-2}
\Gamma_h\left( \!\! \left(j\!-\!\frac{1}{2}\!-\!2n\right)\tau_S \!-\! (j+1) \tau_A-\nu+\omega\right).
\nonumber
\end{eqnarray}

\subsubsection*{Field theory interpretation}

In order to provide a field theory interpretation of the identities (\ref{IIIBevenfinal}) and 
(\ref{IIIBoddfinal}) we organize the fields in the electric side by assigning their charges 
with respect to the global symmetries. Such charges are summarized in Table
\ref{TabIIIBele}.
\begin{table}[H]
	\centering
\begin{tabular}{c|c|cccc}
    &SU(N)&$U(1)_{\tilde Q}$ &$U(1)_S$ & $U(1)_A$ &$U(1)_R$ \\
\hline
$S$ &\symm&0&1&0&0 \\
$A$ &\asymm&0& 0&1&0\\
$\tilde Q_S$ &$\overline {\square}$&0&$-1/2$&0&1 \\
$\tilde Q$ &$\overline {\square}$&1&$0$&0&0 \\
\end{tabular}
	\caption{Charges for the matter fields in the electric of case III-B}
	\label{TabIIIBele}
\end{table}	
Then we study the gauge invariant combinations arising from these fields and the superpotential (\ref{WSQSQS}).
The analysis is different for $N=2n$ and $N=2n+1$.
In the case of $SU(2n)$ we have found the  combinations in Table \ref{Tab:IIIB_even}.
\begin{table}[H]
	\centering
	\begin{tabular}{l|c c c c}
		& $U(1)_S$ & $U(1)_A$ & $U(1)_{\tilde{Q}}$ & $U(1)_R$ \\ 
		\hline
		$\Upsilon_1 \equiv A^n $ & $0$ & $n$ & $0$ & $0$ \\
		$\Upsilon_2 \equiv S \tilde{Q}^2$ & $1$ & $0$ & $2$ & $0$ \\
		$\Upsilon_3^{(k)} \equiv S^{2k+2} A^{2n-2k-2}$ & $2k+2$ & $2n-2k-2$ & $0$ & $0$ \\
		$\Upsilon_4^{(j)} \equiv S^{2j+3} A^{2(n-j-1)} \tilde{Q}^2$ & $2j+3$ & $2(n-j-1)$ & $2$ & $0$ \\
		$\Upsilon_5 \equiv Y^{bare}_{SU(2n-2)} S^{2n-2} A^{n-1}$ & $-3/2$ & $1-n$ & $-1$ & $1$ \\
		$\Upsilon_6 \equiv Y^{bare}_{SU(2n-2)} S^{2n-1} A^{2n-2}$ & $-1/2$ & $0$ & $-1$ & $1$ \\
		$\Upsilon_7^{(m)} \equiv Y^{bare}_{SU(2n-2)} S^m A^{2n-3-m}$ & $m+1/2-2n$ & $-m-1$ & $-1$ & $1$ \\
		\hline
		$Y^{bare}_{SU(2n-2)}$ & $1/2-2n$ & $2-2n$ & $-1$ & $1$ \\
	\end{tabular}
	\caption{Gauge invariant combinations of the Case III-B even reproducing the singlets appearing in the RHS of the identity \eqref{IIIBevenfinal}. In the table $k=0,...,n-1$, $j=0,\dots,n-2$ and $m=0,...,2n-3$, and the bare monopole has $U(1)_2$-charge $-(2n-3)(2n-2)$.}
	\label{Tab:IIIB_even}
\end{table}	
The last two terms correspond to gauge invariant dressings of the bare  $Y^{bare}_{SU(2n-2)}$. This monopole indeed is not gauge invariant because it has charge 
$-(2n-3)(2n-2)$ under $U(1)_2$. 
On the other hand the three gauge invariant combinations $Y^{bare}_{SU(2n-2)} \\S^{2n-2} A^{n-1}$, $Y^{bare}_{SU(2n-2)} S^{2n-1} A^{2n-2}$ and $Y^{bare}_{SU(2n-2)} S^m A^{2n-3-m}$  are gauge invariant and they must be considered as singlets in the dual WZ model.
We observe that the three sphere partition function for the fields in Table \ref{Tab:IIIB_even}
is given by the RHS of (\ref{IIIBevenfinal}), corroborating the validity of the conjectured confining duality.

In the case of $SU(2n+1)$ we have found the gauge invariant  combinations in Table \ref{Tab:IIIB_odd}.
\begin{table}[H]
	\centering
	\begin{tabular}{l|c c c c}
		& $U(1)_S$ & $U(1)_A$ & $U(1)_{\tilde{Q}}$ & $U(1)_R$ \\ 
		\hline
		$\Upsilon_1 \equiv A^{n}(S \tilde{Q}) $ & $1$ & $n$ & $1$ & $0$ \\
		$\Upsilon_2 \equiv S \tilde{Q}^2$ & $1$ & $0$ & $2$ & $0$ \\
		$\Upsilon_3^{(j)} \equiv S^{2j+1} A^{2n-2j}$ & $2j+1$ & $2n-2k$ & $0$ & $0$ \\
		$\Upsilon_4^{(k)} \equiv S^{2k+4} A^{2n-2k-2} \tilde{Q}^2$ & $2k+4$ & $2n-2k-2$ & $2$ & $0$ \\
		$\Upsilon_5 \equiv Y^{bare}_{SU(2n-1)} S^{2n} A^{2n-1}$ & $-1/2$ & $0$ & $-1$ & $1$ \\
		$\Upsilon_6 \equiv Y^{bare}_{SU(2n-1)} S^{2n} A^{n-1} \tilde{Q}$ & $-1/2$ & $-n$ & $0$ & $1$ \\
		$\Upsilon_7^{(m)} \equiv Y^{bare}_{SU(2n-1)} S^m A^{2n-2-m}$ & $m-1/2-2n$ & $-m-1$ & $-1$ & $1$ \\
		\hline
		$Y^{bare}_{SU(2n-1)}$ & $-1/2-2n$ & $1-2n$ & $-1$ & $1$ \\
	\end{tabular}
	\caption{Gauge invariant combinations of the Case III-B odd reproducing the singlets appearing in the RHS of the identity \eqref{IIIBoddfinal}. In the table $j=0,...,n$, $k=0,...,n-2$ and $m=0,...,2n-2$, and the bare monopole has $U(1)_2$-charge $-(2n-2)(2n-1)$.}
	\label{Tab:IIIB_odd}
\end{table}
The last two terms in the table correspond to gauge invariant dressings of the bare  $Y^{bare}_{SU(2n-1)}$. This monopole indeed is not gauge invariant because it has charge 
$-(2n-2)(2n-1)$ under $U(1)_2$.
On the other hand the three gauge invariant combinations $Y^{bare}_{SU(2n-1)} S^{2n} A^{2n-1}$, $Y^{bare}_{SU(2n-1)} S^{2n} A^{n-1} \tilde{Q}$  and $Y^{bare}_{SU(2n-1)} S^m A^{2n-2-m}$  are gauge invariant and they must be considered as singlets in the dual WZ model.
We observe that the three sphere partition function for the fields in Table \ref{Tab:IIIB_odd} is given by the RHS of (\ref{IIIBoddfinal}), corroborating the validity of the conjectured confining duality.
Here we do not discuss the generic form of the superpotential but we will comment about
it in the next paragraphs.

\subsubsection*{A proof from tensor deconfinement}

\begin{figure}[H]
\begin{center}
  \includegraphics[width=14cm]{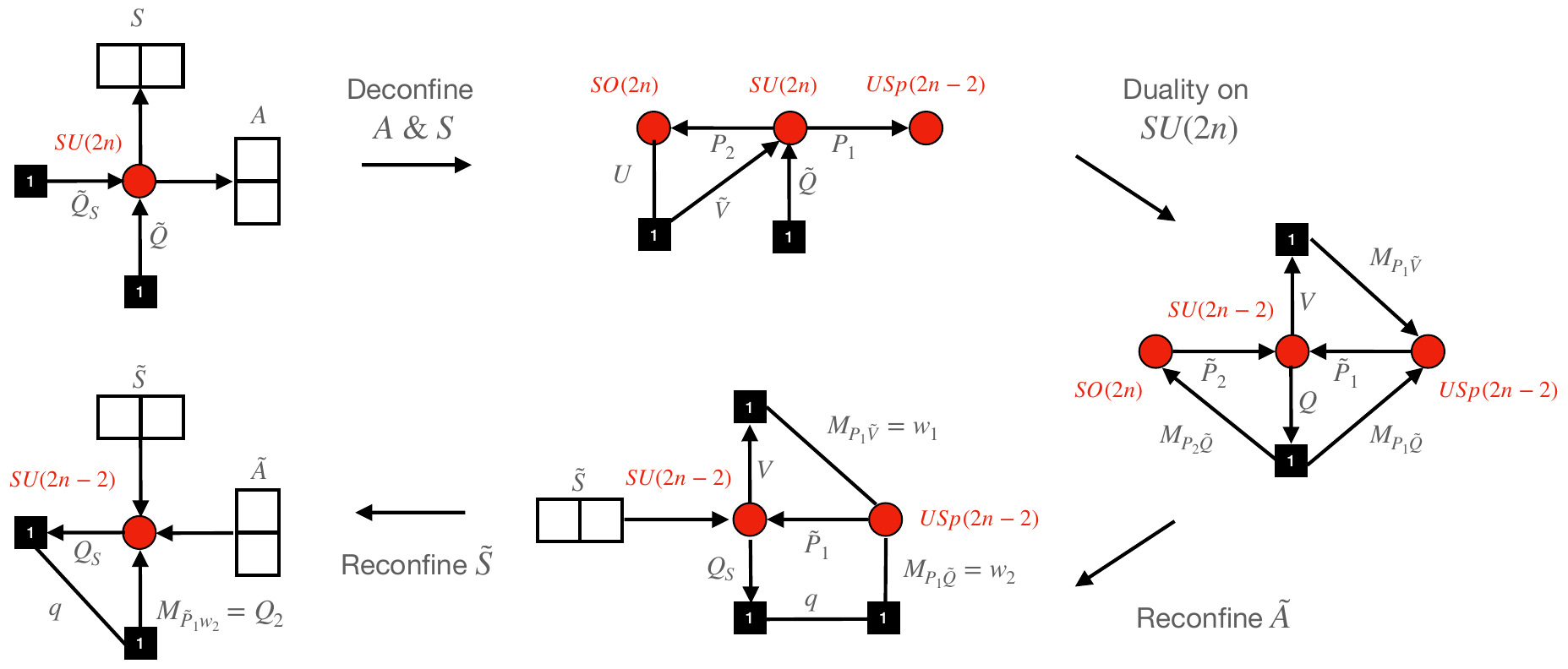}
  \end{center}
  \caption{In this figure we illustrate the various steps that we used in order to prove the confinement of case III-B for $SU(2n)$. Following the chain of  deconfinements and
  dualities in this figure we show how an $SU(2n)$ model is associated to an $SU(2n-2)$  
  model with the  charged matter content of case III-A.
The confinement of  case III-A with gauge group $SU(2n-2)$ has been proved above in subsection \ref{SAQsQ} and then the confinement of case III-B for $SU(2n)$ follows.
 }
  \label{decIIIB}
\end{figure}

In this case we can prove the confinement separately in the even and in the odd case.
In the even case we deconfine the symmetric and the antisymmetric as in Figure
\ref{decIIIB}.
We start considering the first quiver in Figure \ref{decIIIB} and for simplicity we flip some of the gauge invariant combinations. Denoting such  flippers $\beta$, $\rho$ and $\chi$ we turn on the superpotential
\begin{equation}
\label{WIIIB-I}
W = \beta S \tilde Q^2 + \rho \, \text{pf} A+Q_S^2 S+\chi \det S.
\end{equation}
Then we deconfine the antisymmetric $A$  and the symmetric $S$  
using an $USp(2n-2)$ and an $SO(2n)$ gauge group respectively.
The quiver for this phase is the second one in Figure
\ref{decIIIB}, where the superpotential is
\begin{equation}
\label{WIIIB-II}
W = Y_{SO(2n)}^+ +\beta  (P_2 \tilde Q)^2 +\gamma \, \epsilon \cdot P_2^{2n} +P_2 U \tilde V.
\end{equation}

Observe that the effect of the flipper $\chi$ in the superpotential (\ref{WIIIB-I})
is to remove the flipper of the $SO(2n)$ gauge invariant operator Tr$U^2$
in the superpotential (\ref{WIIIB-II}). Then the flipper $\chi$ (together with $\rho$) disappears from (\ref{WIIIB-II}), while there is a new singlet $\gamma$ 
interacting with the baryonic term $\epsilon \cdot P_2^{2n}$.
The $SU(2n)$ gauge group has $4n-2$ fundamentals and two antifundamentals and it can be dualized using the chiral duality reviewed in appendix (\ref{Niidual}).
The dual gauge group is then $SU(2n-2)$ and the model is represented schematically in the third quiver in Figure \ref{decIIIB}.
The dual superpotential is
\begin{equation}
\label{WIIIB-III}
W = \beta M_{P_2 \tilde Q}^2 +\gamma \tilde P_1^{2n-2} +M_{P_1 \tilde Q} \tilde P_1 Q+M_{P_2 \tilde Q} \tilde P_2 Q+M_{P_1 \tilde V} \tilde P_1 V,
\end{equation}
where the interaction $\gamma \, \epsilon \cdot \tilde P_1^{2n-2} $ descends from the duality map on the baryons.
Then we observe that the $SO(2n)$ gauge group has $2n-1$ vectors and it confines giving rise to a conjugate symmetric tensor $\tilde S$, the baryon monopoles $q$ and $Q_{\tilde S}$, a (massive) $SU(2n-2)$ fundamental, and an extra singlet. The baryon moopole $Q_{\tilde S}$ is in the fundamental representation of $SU(2n-2)$ and the last singlet acquires a mass term from the superpotential interaction with the singlet $\beta$.
The quiver in this case is the fourth one in Figure \ref{decIIIB} and  superpotential for this phase is
\begin{equation}
\label{WIIIB-IV}
W= \tilde S Q_S^2+M_{P_1 \tilde Q} \tilde P_1 q Q_S+  \,  \Sigma^2 \epsilon   \epsilon \cdot (S^{2n-3} M_{P_1 \tilde Q}^2 \tilde P_1^2)+\gamma \tilde P_1^{2n-2} +M_{P_1 \tilde V} \tilde P_1 V.
\end{equation}
The last quiver in figure Figure \ref{decIIIB} is obtained by confining the $USp(2n-2)$ gauge group, that has $2n$ fundamentals. The final superpotential is
\begin{equation}
\label{WIIIB-V}
W=\tilde S Q_S^2+Q_2 q Q_S+\Sigma^2  \,  \epsilon   \epsilon \cdot (S^{2n-3 } Q_2^2)+(\gamma+\eta M_{w_1 w_2}) \, \text{Pf} \tilde A,
\end{equation}
where the last term corresponds to the monopole of $USp(2n-2)$ acting as a singlet in this phase.
After this step we are not back to the original model, but we obtain the field content of the even III-A case  studied in subsection \ref{SAQsQ} (up to charge conjugation), that has already been proven to confine. Actually the model discussed here has a different superpotential, because of the presence of the singlets $q$ and $\Sigma$.

Before using the III-A even case we give the explicit map between the singlets obtained at this level with the ones appearing in Table \ref{Tab:IIIB_even}. We have
 	\begin{equation}
 		q \leftrightarrow \Upsilon_6, \quad \Sigma \leftrightarrow \Upsilon_5, \quad \gamma \leftrightarrow \text{flipper}(\Xi_1), \quad \eta \leftrightarrow \Upsilon_7^{(0)}, \quad M_{w_1 w_2} \leftrightarrow \text{flipper}(\Xi_4),
 	\end{equation}
 where the fields $\Xi_i$ are the ones appearing in Table \ref{Tab:IIIA_even} and the $\Upsilon_i$ are the fields in Table \ref{Tab:IIIB_even}. After confining the III-A case, the mapping between the $\Xi_i$ and the $\Upsilon_i$ is
 	\begin{equation}
 		\Xi_2^{(j))} \leftrightarrow \Upsilon_3^{(j)}, \quad \Xi_3^{(n-2-j))} \leftrightarrow \Upsilon_4^{(j)}, \quad \Xi_5^{(2n-k-3))} \leftrightarrow \Upsilon_7^{(k)},
 	\end{equation}
 where $j= 0,...,n-2$ and $k= 1,...,2n-3$. The missing fields $\Upsilon_2$, $\Upsilon_1$, $\Upsilon_3^{(n-1)}$ are flipped respectively by the flippers $\beta, \rho, \chi$.

\begin{figure}[H]
\begin{center}
  \includegraphics[width=14cm]{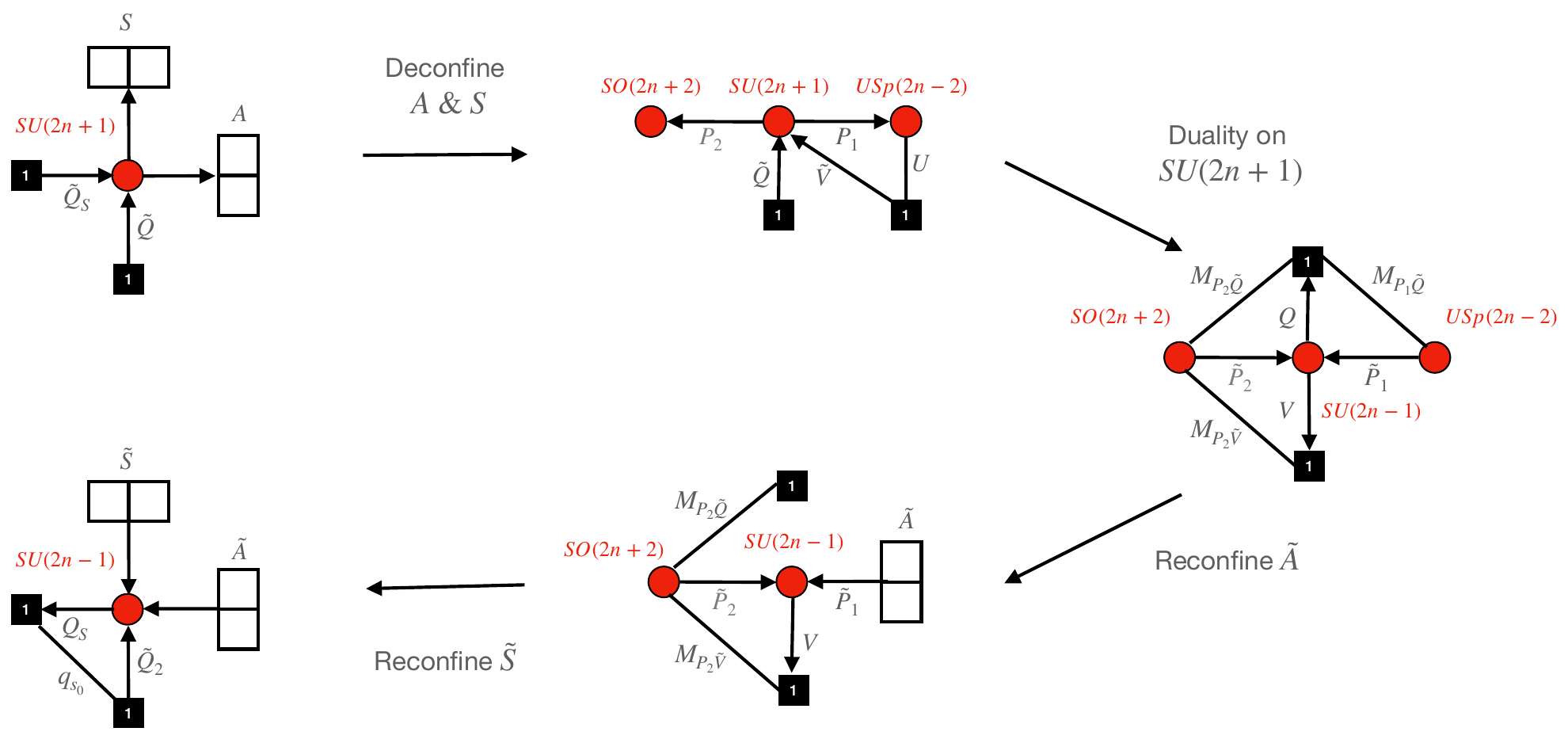}
  \end{center}
    \caption{In this figure we illustrate the various steps that we used in order to prove the confinement of case III-B for $SU(2n+1)$. Following the chain of  deconfinements and
  dualities in this figure we show how an $SU(2n+1)$ model is associated to an $SU(2n-1)$  
  model with the  charged matter content of case III-A.  
The confinement of  case III-A with gauge group $SU(2n-1)$ has been proved above in subsection \ref{SAQsQ} and then the confinement of case III-B for $SU(2n+1)$ follows.
 }
  \label{decIIIB2}
\end{figure}

In the odd case we deconfine the symmetric and the antisymmetric as in Figure
\ref{decIIIB2}.
 Then we dualize
the unitary node and reconfine the tensors.
After this step we are not back to the original model, but we obtain the odd case 
of subsection \ref{SAQsQ} (up to charge conjugation), that has already been proved.
We start considering the first quiver in  Figure \ref{decIIIB2} and for simplicity we  add the flippers 
$\beta$, $\Omega$ and $\zeta$, turning on the superpotential
\begin{equation}
\label{WIIIB2-I}
W = \beta S \tilde Q^2 +Q_S^2 S + \zeta S \tilde Q \epsilon_{2n+1} A^n+\Omega^2 \det S.
\end{equation}
Then we deconfine the antisymmetric $A$  and the symmetric $S$  
using an $USp(2n-2)$ and an $SO(2n+2)$ gauge group respectively.
The quiver for this phase is the second one in Figure
\ref{decIIIB2}, where the superpotential is 
\begin{equation}
\label{WIIIB2-II}
W = \beta P_2^2 \tilde Q^2 +Y_{USp(2n-2)}+P_1 U \tilde V + \zeta P_2^2 \tilde Q \tilde V.
\end{equation}
In this case we have turned on a linear monopole superpotential for the symplectic gauge group,
and two auxiliary fields $U$ and $\tilde V$.
The $SU(2n+1)$ gauge group has $4n$ fundamentals and two antifundamentals and it can be dualized using the chiral duality reviewed in appendix (\ref{Niidual}).
The dual gauge group is then $SU(2n-1)$ and the model is represented schematically in the third quiver in Figure \ref{decIIIB2}.
The dual superpotential is
\begin{equation}
\label{WIIIB2-III}
W = \beta M_{P_2 \tilde Q}^2 + M_{P_2 \tilde Q} \tilde P_2 Q+M_{P_1 \tilde Q} \tilde P_1 Q+M_{P_2\tilde V} \tilde P_2 V +\zeta M_{P_2 \tilde V}M_{P_2 \tilde Q}.
\end{equation}
The $USp(2n-2)$ gauge group has $2n$ fundamentals and it confines giving origin to an antisymmetric and an $SU(2n-1)$ fundamental. This last is massive because of the interaction
$M_{P_1 \tilde Q} \tilde P_1 Q$ in the superpotential (\ref{WIIIB2-III}).
There is also a singlet $\chi$ , corresponding to the $USp(2n-2)$ monopole interacting in the superpotential of this phase as
\begin{equation}
\label{WIIIB2-IV}
W = \beta M_{P_2 \tilde Q}^2 +M_{P_2\tilde V} \tilde P_2 V+\chi \epsilon_{2n-1}(\tilde A^{n-1}  M_{P_2 \tilde Q} \tilde P_2)+\zeta M_{P_2 \tilde V}M_{P_2 \tilde Q}.
\end{equation}
At this point we are left with the fourth quiver in Figure \ref{decIIIB2}, where we observe that the $SO(2n+2)$ gauge node has $2n+1$ vectors and it is confining. The analysis of its confining description requires the definition of various gauge invariant combinations.
We have an $SU(2n-1)$  symmetric tensor $\tilde S = \tilde P_2^2$ and other mesonic singlets
denoted as 
\begin{equation}
s_0 = M_{P_2 \tilde Q}^2,\,\, s_{01} = M_{P_2 \tilde Q}  M_{P_2 \tilde V}, \, \,
s_1 = M_{P_2 \tilde V}^2 ,\, \,  \tilde Q_2 = M_{P_2 \tilde Q} P_2, \, \,  \hat Q_2 = M_{P_2 \tilde V} P_2.
\end{equation}
There are also three baryon monopoles denoted as 
\begin{eqnarray}
q_{s_0}&=&Y_{SO(2n+2) \, \epsilon \cdot(\tilde P_2^{2n-1} M_{\tilde P_2 V})}^-,
\nonumber \\
q_{s_1}&=&Y_{SO(2n+2) \, \epsilon \cdot (\tilde P_2^{2n-1} M_{\tilde P_2 Q})}^-, \\
Q_S&=&Y_{SO(2n+2) \, \epsilon \cdot (\tilde P_2^{2n-2} M_{\tilde P_2 Q} M_{\tilde P_2 V})}^-. \nonumber 
\end{eqnarray}
These fields can be reorganized in a generalized symmetric matrix and in a generalized vector as
\begin{equation}
\mathcal{S}
=
\left(
\begin{array}{ccc}
\tilde S & \hat Q_2 &\tilde Q_2 \\
 \hat Q_2^T&s_0&s_{01} \\
\tilde Q_2^T&s_{01}&s_1
\end{array}
\right),
\quad
\mathcal{Q}_\mathcal{S}
=
\left(
\begin{array}{c}
Q_S \\
q_{s_0} \\
q_{s_1}
\end{array}
\right),
\end{equation}
such that the superpotential is
\begin{equation}
\label{WIIIB2-Va}
W =\mathcal{S}  \mathcal{Q}_\mathcal{S}^2+\beta s_0 +\hat Q_2 V+\chi \, \epsilon_{2n-1}(\tilde A^{n-1}  \tilde Q_2)+\Sigma^2 \det \mathcal{S} + \zeta s_{01},
\end{equation}
where $\Sigma$ corresponds to the fundamental monopole $Y_{SO(2n+2)}^+$ acting as a singlet in this phase.
Integrating out the massive fields we are left with the superpotential
\begin{equation}
\label{WIIIB2-Vb}
W = \tilde S \, Q_{\tilde S}^2+ \chi \, \epsilon_{2n-1}(\tilde A^{n-1}  \tilde Q_2)+s_1 q_{s_1}^2
+
\tilde Q_2 Q_S q_{s_0}+\Sigma^2\epsilon_{2n-1}\epsilon_{2n-1}(S^{2n-1}Q_2^2).
\end{equation}
Then we are left with an $SU(2n-1)$ gauge theory with one conjugate symmetric, one conjugate antisymmetric, one antifundamental $\tilde Q_2$ and one fundamental $Q_S$.
Up to conjugation, the field content of this model is identical to the one studied in case III-A.
Furthermore, the first two terms in the superpotential (\ref{WIIIB2-Vb}) corresponds to the ones in the superpotential (\ref{decoddIIIA-I}). The model is then confining and the other superpotential terms are just flippers
($\Sigma$ and $q_{s_0}$) or spectators ($s_1$ and $q_{s_1}$) of the III-A confining duality.

Before using the III-A odd case we give the explicit map between the singlets obtained at this level with the ones appearing in Table \ref{Tab:IIIB_odd}. We have
\begin{equation}
	\chi \leftrightarrow \text{flipper}(\Xi_1), \quad \Sigma \leftrightarrow \Upsilon_7^{(2n-2)}, \quad q_{s_0} \leftrightarrow \Upsilon_5, \quad q_{s_1} \leftrightarrow \Upsilon_6, \quad s_1 \leftrightarrow \Upsilon_3^{(0)},
\end{equation}
where this time the fields $\Xi_i$ are the ones appearing in Table \ref{Tab:IIIA_odd} and the $\Upsilon_i$ are the fields in Table \ref{Tab:IIIB_odd}. After confining the III-A case, the mapping between the $\Xi_i$ and the $\Upsilon_i$ is
\begin{equation}
	\Xi_2^{(n-j))} \leftrightarrow \Upsilon_3^{(j)}, \quad \Xi_3^{(n-2-l))} \leftrightarrow \Upsilon_4^{(l)}, \quad \Xi_4 \leftrightarrow \Omega, \quad \Xi_5^{(2n-k-3))} \leftrightarrow \Upsilon_7^{(k)},
\end{equation}
where $j= 1,...,n$, $l=0,...,n-2$ and $k= 0,...,2n-3$. The missing fields $\Upsilon_1$, $\Upsilon_2$ are flipped respectively by the flipper $\zeta, \beta$. 
Observe that the singlet $\Omega$ appears in both the initial superpotential (\ref{WIIIB2-I}) and in the final confining phase. It can be then flipped in both the phases consistently, and indeed it is not a singlet $\Upsilon$ in Table  \ref{Tab:IIIB_odd}. 
We already encountered an analogous behavior in the first deconfinement of case I-A.
This concludes the prove of the confinement for the case III-B.

\subsubsection*{A proof of (\ref{IIIBevenfinal}) and (\ref{IIIBoddfinal}) from field theory}

Here we derive the identity (\ref{IIIBevenfinal}) using the procedure  exposed at field theory level.
We consider the relation between the partition function of the first and the last quiver in Figure \ref{decIIIB}. Here we skip the intermediate steps, that can be derived by following chain of deconfinements and dualities as above.
The masses of the singlets in the III-B $SU(2n)$ phase are
\begin{eqnarray}
 m_{\beta }=2 \omega-2 \nu -\tau _S,  \quad
 m_{\rho } = 2 \omega -n \tau _A,  \quad
  m_{\chi } = 2 \omega -2 n \tau _S,
\end{eqnarray}
while  the masses of the singlets in the III-A $SU(2n-2)$ phase are
  \begin{eqnarray}
m_q &=&\omega -\nu -\frac{\tau _S}{2}, \quad
m_{\Sigma }=\omega -(n-1) \tau _A-\nu -\frac{3 \tau _S}{2} 
, \quad
m_{\gamma }=2 \omega -n \tau _S,
  \\
m_{\eta }&=&\omega-\tau _A-\nu +\left(\frac{1}{2}-2 n\right) \tau _S, \quad
m_{M_{w_1 w_2}}=\omega+\tau _A+\nu +\left(n-\frac{1}{2}\right) \tau _S. \nonumber
 \end{eqnarray}
The identity between such phases is given by the relation
\begin{eqnarray}
\label{almostIIIBeven}
&&
 \Gamma _h\left(m_{\beta },m_{\rho },m_{\chi }\right)
Z_{SU(2n)} \left(\omega-\frac{\tau_S}{2};\nu;\tau_S;-;\tau_A;-\right)
= 
 \Gamma _h\left(m_q, m_{\Sigma },m_{\gamma },m_{\eta },m_{M_{w_1 w_2}}\right)
\nonumber \\
\times
&&Z_{SU(2n-2)} \left(X+\nu;\omega-X+\frac{\tau_S}{2};-;2X-\tau_S;-;2X-\tau_A\right),
\end{eqnarray}
where the quantity $X$ can be read in the $SU(2n) \rightarrow  SU(2n-2)$ chiral 
duality and it is
\begin{eqnarray}
X=\frac{(n-1) \tau _A+n \tau _S}{2 (n-1)}.
\end{eqnarray}
The partition function on the RHS of (\ref{almostIIIBeven}) corresponds to the one studied in the III-A case for $SU(2n)$, but with the matter fields  in a conjugated representation.  In this case parity is unbroken due to the absence of a CS level 
and it allows us to consider such partition function with the conjugated representations, i.e. we can use the results from 
the identity (\ref{eq:Z2NAtSQQts}).
Then, evaluating the partition function in the RHS of (\ref{almostIIIBeven}) 
using  (\ref{eq:Z2NAtSQQts}), we obtain  the identity  (\ref{IIIBevenfinal}), up to the hyperbolic Gamma functions of the flippers $\beta, \rho$ and $\chi$  that can be moved on the other side of the equality with the help of the inversion formula (\ref{inversion})

We can also derive the identity (\ref{IIIBoddfinal}) using the procedure  exposed at field theory level.
We consider the relation between the partition function of the first and the last quiver in Figure \ref{decIIIB2}. Here we skip the intermediate steps, that can be derived by following chain of deconfinements and dualities as above.
The masses of the singlets in the III-B $SU(2n+1)$ phase are
\begin{eqnarray}
m_{\zeta }= 2\omega-\tau_S-n\tau_A-\nu,\quad
m_{\Omega}= \omega-\frac{2n+1}{2}\tau_S ,\quad
m_{\beta}=2 \omega-2\nu-\tau_S, 
\end{eqnarray}
while the masses of the singlets in the III-A $SU(2n-1)$ phase are
\begin{eqnarray}
m_{\chi }&=&2 \omega-\nu -(n+1) \tau _S, \quad
m_{\Sigma }=\omega-(2 n-1) \tau _A-\nu -\frac{5 \tau _S}{2}, 
 \nonumber \\
m_{q_{s_0}}&=&\omega-\nu -\frac{\tau _S}{2}, \quad 
m_{q_{s_1}}=\omega-n \tau _A-\frac{\tau _S}{2}, \quad
m_{s_1}=2 n \tau _A+\tau _S. 
\end{eqnarray}
The identity between such phases is given by the relation
\begin{eqnarray}
\label{almostIIIBodd}
&&
\Gamma_h(m_{\zeta },m_{\Omega},m_{\beta})
Z_{SU(2n+1)} \left(\omega-\frac{\tau_S}{2};\nu;\tau_S;\tau_A\right)
= 
\Gamma_h(m_{\chi },m_{\Sigma},m_{q_{s_0}},m_{q_{s_1}},m_{s_1}  )
\nonumber \\
\times
&&Z_{SU(2n-1)} \left(X+\nu,\omega-\frac{\tau_S}{2};-;2X-\tau_S;2X-\tau_A\right),
\end{eqnarray}
where the quantity $X$ can be read in the $SU(2n+1) \rightarrow  SU(2n-1)$ chiral 
duality and it is
\begin{eqnarray}
X = \frac{(n+1) \tau _S+(n-1) \tau _A}{2 n-1}.
\end{eqnarray}
Then we can evaluate the partition function in the RHS of (\ref{almostIIIBodd}) 
using the identity (\ref{eq:Z2NAtSQQts2}) derived above. After such evaluation we obtain  the identity  (\ref{IIIBoddfinal}), up to some flippers that can be moved on the other side of the equality with the help of the inversion formula (\ref{inversion}).
\subsection{Case III-C: $S\oplus A$}

The last model discussed in this section  corresponds  to an $SU(N)$ gauge theory with two antisymmetric tensors $A_{1,2}$ and four fundamentals.
This theory has already been discussed above, and here we apply the duplication formula in a different way, by freezing  the mass parameters for all the fundamentals.

\subsubsection*{Applying the duplication formula}

We start by freezing the mass parameters as in formula (\ref{freezing1}).
Then, by applying the duplication formula 
this assignation of parameters is compatible with the partition function of an $SU(N)$ gauge theory with one rank-two symmetric tensor $S$ and one rank-two antisymmetric tensor $A$.
We have two different identities for $N=2n$ and $N=2n+1$.
In the first case we obtain
\begin{eqnarray}
\label{IIICevenfinal}
&&Z_{SU(2n)}(-;-;\tau_S;\tau_A) 
=
\Gamma_h( n \tau_A) \Gamma_h(-\tau_S- (n-1) \tau_A)  \\
&&
\prod_{j=0}^{n-1} \Gamma_h(2(j+1)\tau_S + 2(n-1-j) \tau_A)
 \prod_{j=1}^{n-1}\Gamma_h(-(2j+1)\tau_S - (2n-1-2j) \tau_A),\nonumber
\end{eqnarray}
while in the second case we have
\begin{eqnarray}
\label{IIICoddfinal}
Z_{SU(2n+1)}(-;-;\tau_S;\tau_A) 
=&&
\prod_{j=0}^{n} \Gamma_h((2j+1)\tau_S + 2(n-j) \tau_A) \nonumber \\
\times &&
\prod_{j=0}^{n-1}\Gamma_h(-2(j+1)\tau_S - (2n-2j-1) \tau_A).
\end{eqnarray}

\subsubsection*{Field theory interpretation}
In order to provide a field theory interpretation of the identities (\ref{IIICevenfinal}) and (\ref{IIICoddfinal}) we organize the fields in the electric side by assigning their charges with respect to the global symmetries. Such charges are summarized in Table \ref{Tab:IIIC_ele}.
\begin{table}[H]
	\centering
\begin{tabular}{c|c|ccc}
    &SU(N)& $U(1)_S$ & $U(1)_A$ &$U(1)_R$ \\
\hline
$S$ &\symm&$1$&$0$&$0$ \\
$A$ &\asymm&$ 0$&$1$&$0$\\
	\end{tabular}
	\caption{Charged matter fields for the case III-C.}
	\label{Tab:IIIC_ele}
\end{table}
Then we study the gauge invariant combinations arising from these fields. The analysis is different for $N = 2n$ and $N = 2n + 1$. In the case of SU(2n) we have found the combinations in Table \ref{Tab:IIIC_even}.
\begin{table}[H]
	\centering
	\begin{tabular}{c|c c c}
		& $U(1)_S$ & $U(1)_A$ & $U(1)_R$ \\ 
		\hline
		$\Upsilon_1 \equiv A^{n} $ & $0$ & $n$ & $0$ \\
		$\Upsilon_2^{(j)} \equiv S^{2j+2} A^{2n-2-2j} $ & $2j+2$ & $2n-2-2j$ & $0$ \\
		$\Upsilon_3 \equiv Y^{bare}_{SU(2n-2)} S^{2n-1} A^{n-1}$ & $-1$ & $1-n$ & $0$ \\
		$\Upsilon_4^{(j)} \equiv Y^{bare}_{SU(2n-2)} S^{2n-2j-1} A^{2j-1}$ & $-(2j+1)$ & $2j+1-2n$ & $0$ \\
		\hline
		$Y^{bare}_{SU(2n-2)}$ & $-2n$ & $2-2n$ & $0$ \\
	\end{tabular}
	\caption{Gauge invariant combinations of the Case III-C even reproducing the singlets appearing in the RHS of the identity \eqref{IIICevenfinal}. In the table $j=1,...,n-1$ and the bare monopole has $U(1)_2$-charge $-(2n-2)^2$.}
	\label{Tab:IIIC_even}
\end{table}
The last two terms corresponds gauge invariant dressings of the bare monopole $Y^{bare}_{SU(2n-2)}$.
 This monopole indeed is not gauge invariant because it has charge $-(2n-2)^2$ under $U(1)_2$.
On the other hand we have found that the   combinations corresponding to $Y^{bare}_{SU(2n-2)} S^{2n-1} A^{n-1}$  and
$Y^{bare}_{SU(2n-2)} S^{2n-2j-1} A^{2j-1}$
are gauge invariant and they must be considered as singlets in the IR description. Nevertheless the singlets in this case do not admit any superpotential interactions and we expect that the IR dynamics is described by a quantum deformed moduli space. This is expected because there are neutral combinations of the singlets.
In the following we will study this case by flipping the Pfaffian of the antisymmetric. In this way the singlet $\Upsilon_1$ disappears from the low energy spectrum. The neutral combinations among the leftover fields $\Upsilon$ are $\Upsilon_3^2 \Upsilon_2^{(0)}$ and  $\Upsilon_2^{(j_1)}\Upsilon_2^{(j_2)}\Upsilon_4^{(j_3)}\Upsilon_4^{(j_4)} \delta_{j_1+j_2-j_3-j_4 +1}$ .

In the case of $SU(2n + 1)$ we have found the gauge invariant combinations in Table \ref{Tab:IIIC_odd}.
\begin{table}[H]
	\centering
	\begin{tabular}{l|c c c}
		& $U(1)_S$ & $U(1)_A$ & $U(1)_R$ \\ 
		\hline
		$\Upsilon_1^{(j)} S^{2j+1} A^{2n-2j} $ & $2j+1$ & $2n-2j$ & $0$ \\
		$\Upsilon_2^{(\ell)} \equiv Y^{bare}_{SU(2n-1)} S^{2n-2 \ell-1} A^{2 \ell}$ & $-2(\ell +1)$ & $2\ell+1-2n$ & $0$ \\
				\hline
		$Y^{bare}_{SU(2n-1)}$ & $-2n-1$ & $1-2n$ & $0$ \\
	\end{tabular}
	\caption{Gauge invariant combinations of the Case III-C odd reproducing the singlets appearing in the RHS of the identity \eqref{IIICoddfinal}. In the table $j=0,...,n$ and  $\ell =0,...,n-1$. The bare monopole has $U(1)_2$-charge $-(2n-1)^2$.}
	\label{Tab:IIIC_odd}
\end{table}
In this case we again observe that the field content does not allow a superpotential but the combinations  $\Upsilon_{1}^{(j_1)} \Upsilon_{1}^{(j_2)} \Upsilon_{2}^{(j_3)} \Upsilon_{2}^{(j_4)}  \delta _{j_1+j_2-j_3-j_4-1}$ are uncharged. We claim that in this case there is a quantum deformed moduli space 
as in family I-C and IIC.
In the following we will justify this claim using tensor deconfinement.
For completeness also in this case  adding a holomorphic mass $W = m Q \tilde Q_S$ to case III-A leads to the case III-C, for both $SU(2n)$ and $SU(2n+1)$. We 
leave the details of the analysis to the interested reader.

\subsubsection*{A proof from tensor deconfinement}

We can prove the confining duality by deconfining the symmetric and the antisymmetric tensors, using two confining dualities, for an orthogonal and for a symplectic gauge group respectively. Then we dualize the $SU(N)$ gauge group and reconfine the symmetric and the antisymmetric tensors.
This procedure gives another $SU(M<N)$ gauge theory with a conjugate symmetric and a conjugate antisymmetric tensor, interacting with some singlets.
By iterating the process we arrive to the confining case, where the various singlets generated at each step reconstruct the towers of gauge invariant operators of
the confining duality that we are looking for.
In the following we will discuss in the details the case of $SU(2n+1)$ and then we will show that the case of  $SU(2n)$ follows  from the $SU(2n+1)$  case.
\begin{figure}[H]
\begin{center}
  \includegraphics[width=12cm]{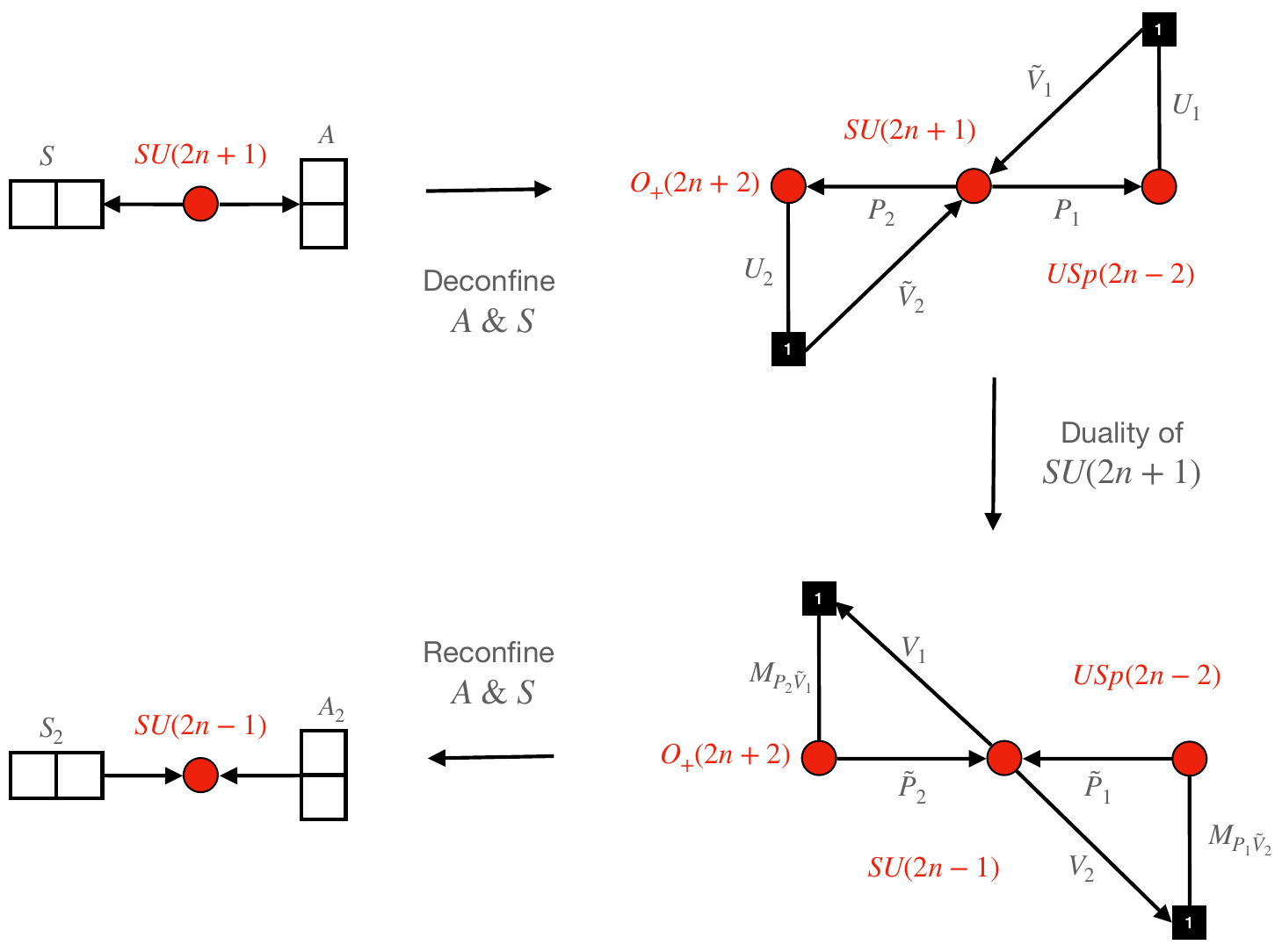}
  \end{center}
\caption{In this figure we illustrate the various steps that we used in order to prove the confinement of case III-C for $SU(2n+1)$. In this figure we show how an $SU(2n+1)$ model is associated to an $SU(2n-1)$   model with the same charged matter content in addition to further singlets that are clarified in the discussion. Using this relation we can iterate the discussion until we reach an $SU(3)$ gauge theory. The last step, i.e. the confinement of the $SU(3)$ gauge theory follows from the confinement discussed above in case I-C.}
    \label{oddAS}
\end{figure}
Following the discussion above we consider the case of $SU(2n+1)$ with a symmetric and an antisymmetric. The model and the various steps are illustrated in Figure \ref{oddAS}.
We then deconfine the symmetric and the antisymmetric in terms of an $O_+(2n+2)$ and an $USp(2n-2)$ gauge group respectively.
We turn on a linear superpotential for the fundamental monopoles of  $O_+(2n+2)$ and  $USp(2n-2)$.
The   superpotential for this deconfined theory is
\begin{equation}
W=Y_{O_+(2n+2)}+Y_{USp(2n-2)} + \tilde V_1 P_1 U_1+\tilde V_2 P_2 U_2+\alpha\, \text{Tr}\,( U_2^2).
\end{equation}
We then dualize $SU(2n+1)$ gauge theory, with $4n$ fundamentals  and 2 antifundamentals, using the chiral duality reviewed in appendix (\ref{Niidual}).
The dual theory is $SU(2n-1)$ and the dual superpotential is
\begin{equation}
\label{spotdecdual}
W=M_{P_2 \tilde V_1} \tilde P_2 V_1+M_{P_1 \tilde V_2} \tilde P_1 V_2+\alpha Y_{USp(2n-2)},
\end{equation}
where we claim that the last terms is dynamically generated.
Then we reconfine the two tensors, obtaining a conjugate symmetric $\tilde S_2= \tilde P_2^2$, a conjugate antisymmetric 
$\tilde A _2 = \tilde P_1^2$ and two singlets, $\sigma = M_{P_2 \tilde V_1}^2$ and $\rho$, that corresponds to the fundamental monopole of $O_+(2n+2)$. The orthogonal gauge group confines with a quantum modified moduli space, with the uncharged combination
\begin{equation}
\label{firstgiro}
\rho^2 \sigma \det \tilde S_2.
\end{equation}
Comparing with the singlets $\Upsilon_{1,2}$ in the tower of Table \ref{Tab:IIIC_odd}, the fields $\rho$ and $\sigma$ correspond to $\Upsilon_2^{(0)}$ and $\Upsilon_1^{(0)}$ respectively, while the operator  $\det \tilde S_2 $ corresponds to 
$\Upsilon_1^{(1)}$. By iterating the process we almost reconstruct the whole tower of singlets, until we reach $SU(3)$, with a symmetric and an antifundamental (or with the conjugate representations, depending on the parity of $n$).

Actually we can better study the process if we perform  a second iteration, along the lines Figure \ref{oddAS}, indeed this clarifies the duality map.
After two iterations we have an $SU(2n-3)$ gauge theory with a symmetric $S_3$ and an antisymmetric $A_3$ and there are two uncharged combinations, namely  
\begin{equation}
\label{twoconstr}
\big( \Upsilon_2^{(0)} \big)^2  \Upsilon_1^{(0)} S_3 A_3^{2n-4},\quad 
\big( \Upsilon_2^{(n-1)} \big)^2  \Upsilon_1^{(n)} \det S_3,
\end{equation}
where the first one is obtained from (\ref{firstgiro})  while the second one is generated by the confinement of $O_+(2n)$.
The operator $S_3 A_3^{2n-4}$ corresponds to $\det \tilde S_2$, and  it corresponds to the singlet 
$ \Upsilon_1^{(1)}$.
On the other hand, the operator $ \det S_3 $ corresponds to  $ \Upsilon_1^{(n-1)}$.

Depending on the parity of $n$ the process can be iterated until $SU(3)$ with a symmetric and an antisymmetric 
or with conjugated representations. Denoting $\mathbb{S}_n$ the $SU(3)$ (conjugate) symmetric,  corresponding  to ($\tilde S_{n=2m}$) $S_{n=2m+1}$  and analogously $\mathbb{A}_n$ the $SU(3)$ (anti)fundamental,  corresponding  to ($A_{n=2m+1}$ ) $\tilde A_{n=2m}$ we write the uncharged combinations in the  $SU(3)$ gauge theory as
\begin{eqnarray}
\label{SU3IIIC}
\Bigg( \big(\Upsilon_2^{(\ell)} \big)^2  \Upsilon_1^{(\ell)} \Upsilon_1^{(\ell+1)}\Bigg)_{\ell=0,\dots,\lfloor\frac{n}{2}\rfloor-2},
\quad
 \big(\Upsilon_2^{\left(\lfloor\frac{n}{2}\rfloor-1\right)} \big)^2  \Upsilon_1^{\left(\lfloor\frac{n}{2}\rfloor-1\right)} \mathbb{S}_n \mathbb{A}_n^2,
\nonumber \\
\Bigg( \big(\Upsilon_2^{(j-1)} \big)^2  \Upsilon_1^{(j)} \Upsilon_1^{(j-1)}\Bigg)_{j=n,\dots,\lceil \frac{n}{2} \rceil+2} 
,\quad
 \big(\Upsilon_2^{\left(\lceil \frac{n}{2} \rceil\right )} \big)^2  \Upsilon_1^{\left(\lceil \frac{n}{2} \rceil+1\right )} \det \mathbb{S}_n.
\end{eqnarray}
The last step consists of dualizing such $SU(3)$ gauge theory as discussed in Section \ref{SQt}, corresponding to case I-C. 
Indeed the three singlets of Table \ref{Tab:IC} are associated to the missing singlets in Table \ref{Tab:IIIC_odd}
by the relations
\begin{equation}
\Phi_1 \leftrightarrow  \Upsilon_1^{(\lfloor \frac{n}{2} \rfloor +1)}, \quad 
\Phi_2 \leftrightarrow  \Upsilon_1^{(\lfloor \frac{n}{2} \rfloor)}, \quad 
\Phi_3 \leftrightarrow  \Upsilon_2^{(\lfloor \frac{n}{2} \rfloor)}.
\end{equation}

\begin{figure}[H]
\begin{center}
  \includegraphics[width=12cm]{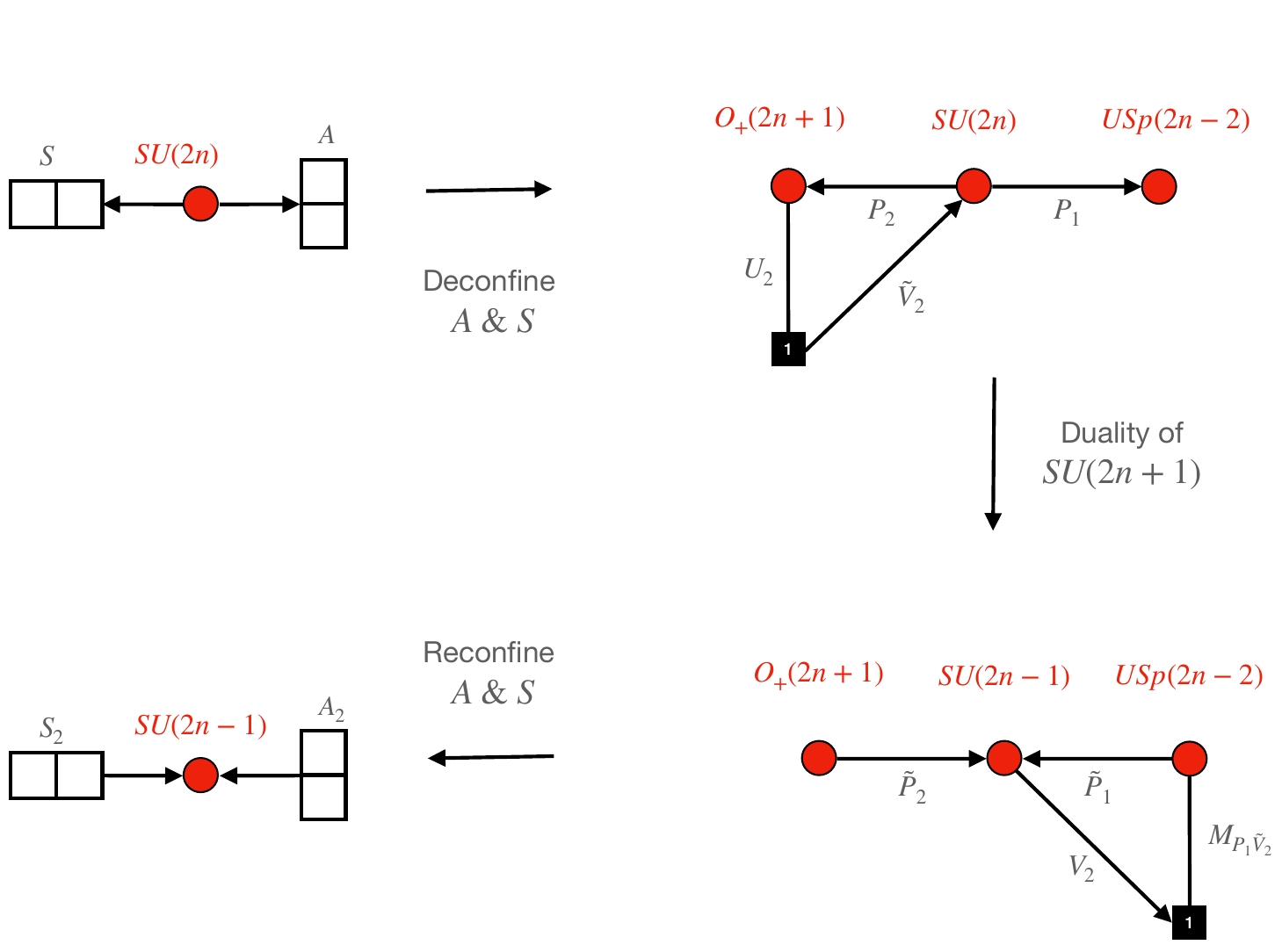}
  \end{center}
  \caption{In this figure we illustrate the various steps that we used in order to prove the confinement of case III-C for $SU(2n)$. In this figure we show how an $SU(2n)$ model is associated to an $SU(2n-1)$  
  model. Once we reach this model we observe that, up to conjugation and to a further superpotential term, this coincides with the case III-C with gauge group $SU(2n-1)$  discussed above. This proves the confining dynamics of this theory.}
    \label{evenAS}
\end{figure}

In the case of $SU(2n)$ we simplify the proof by flipping the Pfaffian of the antisymmetric $A$. We consider the superpotential 
\begin{equation}
\label{W1oddIIIC}
W = \sigma \text{Pf} A.
\end{equation}
Then we deconfine the antisymmetric and the symmetric, using an $USp(2n-2)$ and an $O_+(2n+1)$ gauge group.
The quiver for the deconfined theory is represented in the second picture in Figure \ref{evenAS}.
The superpotential for this theory is 
\begin{equation}
W=Y_{O_+(2n+1)}+\tilde V_2 P_2 U_2+\alpha\, \text{Tr}\,( U_2^2).
\end{equation}
Then we dualize the $SU(2n)$ gauge group, with $4n-1$ fundamentals and $1$ antifundamental using the duality discussed in appendix \ref{Niidual}. The charged field content can be read from the third quiver in Figure
\ref{evenAS}.
The superpotential for this theory is 
\begin{equation}
W=\tilde P_1 V_2 M_{P_1 \tilde V_2}+\alpha Y_{USp(2n-2)},
\end{equation}
where the last term is dynamically generated. 
The last quiver in Figure \ref{evenAS} is obtained by reconfining the antisymmetric and the symmetric tensor. The confinement for the $O_+(2n+1)$ gauge theory gives a quantum deformed moduli space, with the uncharged combination $Y_{O_{+}(2n+1)}^2\det S_2$.
At this point of the discussion we do not need to iterate the construction, because we transformed  the original $SU(2n)$ gauge group into $SU(2n-1)$ with a conjugate symmetric and a conjugate antisymmetric, that confines with a quantum moduli space, as discussed above.

\subsubsection*{Proving (\ref{IIICevenfinal}) and (\ref{IIICoddfinal}) from field theory}

Here we conclude our analysis on the case III-C by proving the relations 
(\ref{IIICevenfinal}) and (\ref{IIICoddfinal})  applying the 
iterative procedure discussed above at the level of the partition function.
For this reason we refer to the quivers in Figures \ref{oddAS} and \ref{evenAS} and at each step we discuss the main aspects of the identities generated by the various dualities.

We start from the $SU(2n+1)$ case, by fixing the parameters in the first quiver in Figure \ref{oddAS} as $\tau_S$ for the symmetric and  $\tau_A$ for the antisymmetric.
Then we deconfine the two tensors using the relations (\ref{dualityconfOpluswobc}) and   (\ref{USpfondmon})
and the new bifundamentals $P_1$ and $P_2$ in the second quiver 
have mass parameter $\frac{\tau_A}{2}$ and $\frac{\tau_S}{2}$ respectively. The 
new fields  $\tilde V_{1,2}$ and $U_{1,2}$ have mass parameters
\begin{equation}
m_{U_2} =  -n \tau _S-\frac{\tau _S}{2},\quad
m_{U_1} = 2 \omega-n \tau _A-\frac{\tau _A}{2},\quad 
m_{\tilde V_1} = n \tau _A,\quad 
m_{\tilde V_2} =2 \omega+ n \tau _S,
\end{equation}
while the singlet $\alpha$ has mass parameter $ m_{\alpha} = 2 \omega +2 n \tau _S+\tau _S $.
The we dualize $SU(2n+1)$ using the relation (\ref{chiralpqstarduality}). We obtain the third quiver in Figure 
\ref{oddAS}.
The new bifundamentals $\tilde P_1$ and $\tilde P_2$ have mass parameters $X-\frac{\tau_A}{2}$
and  $X-\frac{\tau_S}{2}$ respectively, with 
\begin{equation}
X = \frac{(n-1) \tau _A+(n+1) \tau _S}{2 n-1}.
\end{equation}
The $SU(2n-1)$ fundamentals  $V_{1,2}$ have mass parameters 
$m_{V_{1,2}} = 2\omega-X-m_{\tilde V_{1,2}}$. The two mesons that are not massive in this phase are   $M_{P_2 \tilde V_1}$ and their mass parameters are read from the ones of $P_{1,2}$ and $\tilde V_{1,2}$.
In the last step we reconfine the tensors using the identities (\ref{Opluswobc}) and (\ref{USpwomon}).
There are three singlets arising in this duality corresponding to the $USp(2n-2)$ monopole $\gamma$,
the $O_+(2n+1)$ monopole $Y$ and the singlet $\sigma= M_{P_2 \tilde V_1}^2$. 
The mass parameter of the monopole $\gamma$ is $m_{\gamma} = 2\omega-m_{\alpha}$ and this is consistent with the claim that there is an holomorphic mass for  these two singlets.
The other singlets have mass parameters 
$m_Y = (1-2 n) \tau_A -2\tau_S$ and $m_\sigma=\tau_S+ 2 n \tau_A$.
We have then obtained the relation 
 \begin{eqnarray}
Z_{SU(2n+1)} (-;-;\tau_S;-;\tau_A;-) &=&
\Gamma_h( (1-2 n) \tau_A -2\tau_S) 
  \Gamma_h(\tau_S+ 2 n \tau_A)\nonumber \\
  &\times &
Z_{SU(2n-1)} (-;-;- ;X-\tau_S;-;X-\tau_A). 
\end{eqnarray}
We can then iterate the analysis, using the same rules discussed above. Actually, in order to prove the identity it is better to iterate the process once. In this way we obtain the relation
\begin{eqnarray}
&&
Z_{SU(2n+1)} (-;-;\tau_S;-;\tau_A;-) =
\Gamma_h( (1-2 n) \tau_A -2\tau_S, \tau_S+ 2 n \tau_A, -  \tau_A -2 n \tau_S)
  \nonumber \\
 &&
\Gamma_h((2 n+1) \tau_S)
Z_{SU(2n-3)} \left(-;-;\tau _S+\frac{2 \left(\tau _A+\tau _S\right)}{2 n-3};-;\tau _A+\frac{2 \left(\tau _A+\tau _S\right)}{2 n-3};-\right). 
\end{eqnarray}
This relation can be iterated by observing that  the  mass parameters for the symmetric and the antisymmetric 
of the $SU(2n-4j+1)$ theory are  
\begin{eqnarray}
\tau_{A,S}^{(j+1)} =\tau _{A,S}+ \frac{2 j }{2 n-4j+1}\left(\tau _A+\tau _S\right), 
\end{eqnarray}
where $\tau_{A,S}^{(1)}  \equiv \tau_{A,S}$.
In this way we can iterate the process until $SU(3)$ for $n=2m+1$ or $SU(5)$ for $n=2m$.
In the first case we can collect all the singlets arriving to the identity between the initial partition function and the one for the $SU(3)$ model with a symmetric and an antifundamental, which are 
$\frac{1}{3} ((2m) \tau _A+(2m+3) \tau _S)$ and $\frac{1}{3} ((2m+3) \tau _A+(2m) \tau _S)$ respectively.
The identity at this level is 
\begin{eqnarray}
\label{almostevenIIIC}
Z_{SU(4m+3)} (-;-;\tau_S;-;\tau_A;-) &=&
\prod_{
\scriptsize
\begin{array}{c} 
j=0 \\
 j\neq m,m+1
 \end{array} }^{2m+1} \Gamma_h((2j+1)\tau_S + 2(2m+1-j) \tau_A) \nonumber \\
&\times &
\prod_{
\scriptsize
\begin{array}{c} 
j=0 \\
 j\neq m
 \end{array}}^{2m}\Gamma_h(-2(j+1)\tau_S - (2m-2j) \tau_A)
\\
&&\!\!\!\!\!\!\!\!\!\!\!\!\!\!\!\!\!\!\!\!
Z_{SU(3)} \big(-;-;\tau _S+ \frac{2m}{3} (\tau _A+\tau_S)  ;-;\tau _A+ \frac{2m}{3} (\tau _A+\tau_S)  ;-\big). \nonumber 
\end{eqnarray}
Plugging then (\ref{ZIC}) into (\ref{almostevenIIIC}) we arrive at (\ref{IIICevenfinal}) for $n=2m+1$.

The analysis for $n=2m+1$ is similar, but in this case once we reach the $SU(5)$ we 
need to implement a further iteration until we reach the $SU(3)$ case with conjugate tensors.
The mass parameter of the conjugate symmetric is
$\frac{1}{3} \left(2 m \left(\tau _A+\tau _S\right)+\tau _S\right)$
while the one of the conjugate antisymmetric is
$\frac{1}{3} \left((2 m-3) \tau _A+2 (m+2) \tau _S\right)$.
Again collecting all the singlets and evaluating (\ref{ZIC}) using these parameters we arrive ar
to (\ref{IIICevenfinal}) for $n=2m$.

The proof for $SU(2n)$ is simpler. In this case we start again defining $\tau_S$ and $\tau_A$ 
as the mass parameters for the symmetric and the antisymmetric respectively.
Then we add the contribution of the flipper $\sigma$ in (\ref{W1oddIIIC}), 
corresponding to $\Gamma_h(2\omega-n \tau_A)$.

Then we deconfine the two tensors using the relations (\ref{dualityconfOpluswobc}) and   (\ref{USpwomon})
and the new bifundamentals $P_1$ and $P_2$ in the second quiver in Figure \ref{evenAS}
have mass parameter $\frac{\tau_A}{2}$ and $\frac{\tau_S}{2}$ respectively. The 
new fields  $\tilde V_{2}$ and $U_{2}$ have mass parameters
\begin{equation}
m_{U_2} =  -n \tau _S,\quad
m_{\tilde V_2} =2 \omega+\frac{2 n+1}{2} \tau _S,
\end{equation}
while the singlet $\alpha$ has mass parameter $m_\alpha=2\omega+2n \tau_S$.

The we dualize $SU(2n+1)$ using the relation (\ref{chiralpqstarduality}). We obtain the third quiver in Figure 
\ref{evenAS}.
The new bifundamentals $\tilde P_1$ and $\tilde P_2$ have mass parameters $X-\frac{\tau_A}{2}$
and  $X-\frac{\tau_S}{2}$ respectively, with 
\begin{equation}
X = \frac{(2n+1) \tau _S+2(n-1) \tau _S}{2(2 n-1)}.
\end{equation}
The $SU(2n-1)$ fundamental  $V_{2}$ has mass parameters $m_{V_2} = 2\omega-X-m_{\tilde V_{2}}$. The mesons  $M_{P_{2} \tilde V_2}$  has mass parameter $2\omega-m_{U_2}$ 
and it disappears from the partition function, while $M_{P_{1} \tilde V_2}$ has mass parameter $m_{P_1}+m_{\tilde V_2}$.
In the last step we reconfine the tensors using the identities (\ref{Opluswobc}) and (\ref{USpwomon}).
There are two singlets arising in this duality corresponding to the $USp(2n-2)$ monopole $\gamma$ and
the $O_+(2n+1)$ monopole $Y$. 
The mass parameter of the monopole $\gamma$ is $m_{\gamma} = 2\omega-m_{\alpha}$ and this is consistent with the claim that there is an holomorphic mass for  these two singlets.
The other singlets has mass parameter
$m_Y = (1- n) \tau_A -\tau_S$.
We have then obtained the relation 
 \begin{eqnarray}
 \label{oneiterIIICeven}
 \Gamma_h(2\omega-n \tau_A)
Z_{SU(2n)} (-;-;\tau_S;-;\tau_A;-) &=&
  \Gamma_h((1- n) \tau_A -\tau_S)
 \\
 &\times&
 Z_{SU(2n-1)} (-;-;- ;X-\tau_S;-;X-\tau_A). 
 \nonumber
\end{eqnarray}
At this point it is not necessary to iterate the construction, indeed we can plug the RHS of (\ref{IIICevenfinal}) 
in (\ref{oneiterIIICeven}) obtaining (\ref{IIICoddfinal}).

\section{Comments on unitarity}
\label{unitarity}

In this section we comment on the existence of a conformal window in the dualities discussed in the paper. 
While confining dualities in four dimensions do not admit an interacting fixed point in the IR, in the 3d case the situation is different, because also in absence of gauge dynamics a WZ model with pure chiral  fields can be superconformal. The simplest example of a confining gauge theory with an interacting fixed point is represented by the SQED/XYZ duality.
A necessary condition in order to have a unitary duality is that the unitarity bound is satisfied by the operators in the chiral ring. This corresponds to the requirement that the scaling dimension $\Delta$ of the singlets that are not set to zero in the chiral ring by the F-term respects the constraint $\Delta \geq 1$, or equivalently that their exact R-charge is  $R_{ex} \geq 1/2$.
The exact R-charge can be obtained by maximizing the free energy on the three sphere, $F_{S^3} =\log |Z_{S^3}|$ in terms of the $U(1)_R$ R-symmetry. In the case of models with four supercharges, the ones analyzed here, the $R$-symmetry mixes with the other Abelian symmetries and the principle of $F$-maximization gives  the exact combination for such a mixing.
If there are singlets in the chiral ring that after $F$-maximization evade the unitarity bound, then the models does not correspond to a SCFT. It is nevertheless possible to modify the model, by coupling the UV gauge theory with a further singlet, a flipper, that sets the operator hitting the bound to zero in the chiral ring.
This prescription was originally provided in \cite{Barnes:2004jj} in order to keep unitarity and the 3d generalization of the mechanism was discussed in \cite{Morita:2011cs,Agarwal:2012wd,Benvenuti:2017lle}.
The  coupling with the flipper removes the problematic singlet from the dual superpotential, because it represents a mass term in a confining duality.  
Then one needs to perform $F$-maximization again, and check if the modified duality is unitary and compatible with the existence of an IR interacting fixed point.
In the discussion above we did not study the possibility of having a superconformal fixed point in the WZ models.

For the other cases a more detailed analysis is necessary and it goes beyond the scopes of this paper. 
Nevertheless, here we make some comments about such analysis.
We have observed that in general the unitarity bound is not satisfied for the models discussed in the paper. The presence of superpotential terms with high powers (higher than quartic) is indeed problematic because some of the singlets have necessarily R charge lower that the free value. It implies that some of the singlets need to be flipped in the electric description in order to have an interacting fixed point. We leave a general analysis to further studies and here we discuss the minimal possibility for case I-A
and I-B and discuss its consequence in the other models.

In case I-A we can flip all the singlets except $\Phi_1$ and $\Phi_5$. These two singlets are in non-trivial representations of the flavor symmetry and their interaction is crucial in the prove of Case II-A, as we discussed in the body of the paper. In such case F-maximization is equivalent to solve the equation
\begin{equation}
(n-1) \left(2 \Delta _{\Phi _1}-1\right) \cot \left(\pi  \left(2 \Delta _{\Phi _1}-1\right)\right)-\left(1-\Delta _{\Phi _1}\right) \cot \left(\pi  \left(1-\Delta _{\Phi _1}\right)\right).
\end{equation}
The solution of this equation is in the range $\left[\frac{2}{3},\frac{3}{4} \right)$, where the extremal values correspond to $n=2$ (on the left) and $n\rightarrow \infty$ (on the right).
Then this charge and the  charge of the other singlet $\Delta_{\Phi_5} = 2(1-\Delta_{\Phi_1} )$ are both above the unitarity bound.
The same situation holds in Case I-B by keeping the singlets $\Phi_1$ and $\Phi_3$.
For the other models studied in Family II and III the situation is similar. In such cases we have found unitary dualities by keeping only the singlets with cubic superpotential couplings.
For example, in case II-A we commented about unitarity below formulas (\ref{mapIIA}) and (\ref{IIASU(4)}). A similar discussion in Case II-B has been pursued below formula (\ref{WfinalIIBbello}) and formula (\ref{mapIIBodd}). 
We have also discussed issues related to the unitarity of case III-A below formula (\ref{map2}).

\section{Conclusions}
\label{conclusions}

In this paper we have studied the low energy dynamics of  3d $\mathcal{N}=2$ $SU(N)$ gauge theories with one symmetric tensor and, in addition, other matter fields in fundamental and/or antisymmetric representation.
Our analysis is motivated by the application of the duplication formula for the hyperbolic Gamma functions to confining dualities involving 3d $\mathcal{N}=2$ $SU(N)$ gauge groups with antisymmetric matter fields.
Indeed in this way we have obtained new identities involving on one side $SU(N)$ gauge theories with two index symmetric tensors and on the other side a set of singlets.
The natural guess and expectation that such models are confining needed a confirmatory analysis. We have provided such analysis in each case by investigating the structure of the Coulomb branch and deriving each duality by deconfining the tensorial matter. This last step required the use of   other well stated dualities and then, by applying other dualities, we arrived in each case to the confining duality expected from the analysis of the partition function. In this sense we have proved the dualities in terms of other dualities, corroborating the idea of the existence of few building blocks 
necessary to construct the whole spectrum of IR SUSY dualities.
Furthermore, the construction provided at field theory level, allowed us to find independent derivations of the identities that we started with, furnishing a consistency check of the analysis.
This last derivation is ``somehow" the mathematical counterpart of the observation made before, i.e. there are few identities that can be interpreted as the fundamental building blocks for the construction of other hyperbolic identities. 
We have provided also a classification of the various cases in three families, and in each family we have distinguished three different cases. The classification in families is motivated by the similarities between the various models in terms of the matter content, while the distinction between the cases is related to the procedure adopted to prove the dualities in terms of tensor deconfinement.
We have concluded our analysis commenting on the existence of SCFTs, discussing in general if and how the dualities treated here satisfy the bound of unitarity.

There are many possible developments and extensions of our results that require further investigations.
For example, it would be interesting to study $SU(N)$ confining gauge theories with symmetric tensors in presence of  linear monopole superpotentials, similarly to the ones studied in \cite{Benini:2017dud}. 
For the cases with a 4d origin this problem can be studied along the lines of the reduction of the 4d superconformal index to the 3d partition function on the three sphere, where the 4d balancing conditions enforcing a 4d non anomalous $R$-symmetry translates into the 3d balancing condition enforcing the presence of the KK monopole superpotential.
By applying the duplication formula in this case we expect 3d identities for $SU(N)$ gauge theories with a symmetric tensor and a linear superpotential for a (dressed) monopole. Such confining theories should then flow to the cases I-A, I-B and II-B studied in this paper through a real mass flow. We will study the details of such models and of the flows in a forthcoming publication.

Another interesting line of research regards the existence of  4d models with $SU(N)$ gauge groups, a two index symmetric tensor and non vanishing superpotential of the type (\ref{WSQSQS}) studied here. While such confining theories are forbidden in absence of superpotential, i.e. they do not appear in the classification of s-confining theories provided in \cite{Csaki:1996sm,Csaki:1996zb}, their ubiquity in 3d motivates the investigation is this direction.
Another possibility consists of performing a real mass flow starting from the 
dualities obtained here, generating a  non vanishing CS level, generalizing the appetizer of \cite{Jafferis:2011ns}.

Our classification is not conclusive, and we cannot exclude other confining dualities with symmetric tensors that cannot be obtained by applying the duplication formula starting from cases with antisymmetric matter. 
For example, one may start considering the cases II-A and II-C studied here with $SU(2n+1)$ gauge group and study the IR dynamics of such models. However in these cases we have not found evidences of a confining dynamics.
 
It would be also interesting to find a  brane realization of the confining dualities discussed here. Furthermore having a brane interpretation of the duplication formula itself is an intriguing possibility.
The construction of \cite{Kim:2023qwh,Hayashi:2023boy,Kim:2024vci}  for eight supercharges is promising in this direction.

Another issue that we did not investigate but that deserves further studies regards the existence of  3d $SU(N)$ dualities with symmetric tensors, such that our confining cases are limiting cases of them. The deconfinement techniques discussed here are promising starting points in this direction.  A related issue is related to the existence of RG flows from the $SU(N)/Spin(7)$ duality studied in \cite{Nii:2020eui} to the cases analyzed here.

 When the matter content allows the definition of a baryonic $U(1)_B$  symmetry it is in principle possible to make the background gauge field dynamical and in such case we have $U(N)$ gauge groups. The study of the consequence of such gauging is intriguing and deserves a separated analysis.

Confining dualities in 4d with $SU(N)$ gauge groups and symmetric tensors have been found in 4d in presence of  ADE type superpotentials as limiting cases of Seiberg-like dualities \cite{Intriligator:1995ax,Brodie:1996xm,Csaki:1998fm,Klein:1998uc,Hirayama:1998hu}.
By dimensional reduction such confining dualities can be extended to  3d $\mathcal{N}=2$ confining dualities as well (see for example \cite{Amariti:2015mva,Amariti:2016kat}). It would be interesting to connect such confining dualities to the ones discussed here. 

A last aspect that requires further studies regards the analysis performed here in appendix \ref{integralsfromdecforAS}, where we provided a derivation of the confining dualities proposed in \cite{Nii:2019ebv} in terms of other fundamental dualities using tensor deconfinement.
Here we focused on few models of the classification, because we were interested in the cases that can give rise to new confining dualities involving symmetric tensors. Nevertheless, it should be possible to derive the whole classifications along the same lines.

\section*{Acknowledgments}
We are grateful to Simone Rota for discussions. The work of the authors has been supported in part by the Italian Ministero dell'Istruzione, Universit\'a e Ricerca (MIUR), in part by Istituto Nazionale di Fisica Nucleare (INFN) through the “Gauge Theories, Strings, Supergravity” (GSS) research project.

\appendix
\section{Generalities on the three sphere partition function}
\label{appPZ}

In this appendix we collect some general definitions of the squashed three sphere partition and we fix the notations used in the body of the paper.

Localization on $S^3$ was first carried on in \cite{Kapustin:2009kz} for models with $\mathcal{N}>2$  and the case with four supercharges was first analyzed in \cite{Jafferis:2010un,Hama:2010av}. The partition function used in this paper was first obtained in \cite{Hama:2011ea}, where the round sphere was generalized to the case of a squashed one, with a $U(1)^2$ preserving metric, with squash parameters $\omega_1=i b$ and $\omega_2=i b^{-1}$. The real parameter $b$ represents the parameter in the defining equation of the ellipsoid $b(x_1^2+ x_2^2) + b^{-1}(x_3^2 +x_4^2)=1$.

For a gauge theory with gauge group $G$ and matter fields in the representation $\mathcal{R}$ of the gauge group the partition function 
is formally given by the expression
\begin{equation}
Z_G (\rho_{\mathcal{R}} (m)) = \frac{1}{|\text{Weyl}_G|}
\int [d\sigma] Z_{cl} \frac{\prod_\mathcal{R} \Gamma_h(\rho_\mathcal{R}(\sigma) +\rho_\mathcal{R}(m) ) }{\prod_\alpha \Gamma_h(\alpha(\sigma))}.
\end{equation}
Let us explain some salient features of the formula.
The integral is performed over the variable  $\sigma$ corresponding to the real scalar in the $\mathcal{N}=2$ vector multiplet in the Cartan of the gauge group. The weights $\rho_{\mathcal{R}}(\sigma)$ refer to the representations of the gauge group under which the fields transform, while the  weights $\rho_{\mathcal{R}}(m)$ refer to the representations of the flavor symmetry group under which the fields transform. In the denominator $\alpha(\sigma)$ are represented the non-zero roots, and indeed such term takes into account the contribution of the vector multiplets.
There is a further contribution from the classical action, collecting the CS and FI terms, that we have not turned on here, but that we report for completeness.

The one loop determinants in the integrand are hyperbolic Gamma functions. They can be defined as
\begin{equation}
\Gamma_h(x) \equiv \prod_{m,n=0}^{\infty} \frac{(m+1) \omega_1 +(n+1) \omega_2 -x}{m \omega_1 +n \omega_2 +x} 
\end{equation}
and there are two useful relations denoted as reflection and duplication formula that we have vastly used in the paper.
The reflection formula is
\begin{equation}
\label{inversion}
\Gamma_h(x) \Gamma_h(2\omega-x) =1
\end{equation}
and the duplication formula is
\begin{equation}
\label{duplication}
\Gamma_h(2x) = \Gamma_h(x) \Gamma_h\left(x+\frac{\omega_1}{2}\right) \Gamma_h\left(x+\frac{\omega_2}{2}\right) \Gamma_h\left(x+\omega \right). 
\end{equation}
The reflection formula  is very useful when flipping a field, indeed moving a singlet from the RHS to the LHS of an identity can be done by keeping the associated Gamma function in the numerator thanks to such formula, allowing for a clear field theory interpretation of the move.
Furthermore, the reflection formula represents the integration out of an holomorphic mass term in the partition function. On the other hand, the duplication formula has played a prominent role here, because, by freezing some mass parameters, it has allowed us to convert an antisymmetric into a symmetric tensor. Similarly this formula has been used to convert $USp(2N)$ gauge groups in $SO(2N)$ or $SO(2N+1)$, as done for example in \cite{Benini:2011mf}. 

In the bulk of the paper we have used various gauge groups and representation without giving the explicit wrights in the integrand. For this reason here we collect such weights for the representations of interest.

\begin{itemize}
\item Let us start from $SU(N)$: in this case we have considered matter fields in the fundamental, in the  symmetric and in the antisymmetric representation, and their conjugated. The weight of a fundamental in terms of $\sigma$ is given by $\rho_F(\sigma) = \sigma_i$, for $i=1,\dots,N$. For a symmetric tensor  we  have $\rho_S(\sigma) = \sigma_i + \sigma_j$, for  $1\leq i \leq j \leq N$ and for an antisymmetric one $\rho_A(\sigma) =  \sigma_i + \sigma_j$, for  $1\leq i < j \leq N$. For matter fields in conjugated  representations we need to consider the opposite signs. Furthermore the non-zero roots of  the  adjoint representation are $\alpha(\sigma) = \pm(\sigma_i-\sigma_j)$, with $1\leq i < j \leq N$.
 
\item We then consider orthogonal gauge groups with vectors. For $SO(2N)$ and $O_+(2N)$ the weights are  $\rho_V(\sigma) = \pm \sigma_i$, $i=1,\dots,N$, while for $SO(2N+1)$ and $O_+(2N+1)$ they are $\rho_V(\sigma) = \pm \sigma_i,0$, $i=1,\dots,N$.
The non-zero roots for the even case are $\alpha(\sigma) = \pm \sigma_i \pm \sigma_j$, for $1\leq i < j \leq N$, while in the odd case we have to consider also the contributions $\alpha(\sigma) =\pm \sigma_i$, for $i=1,\dots,N$.

\item If the gauge group is $USp(2N)$ we have  $\rho_F(\sigma) = \pm \sigma_i$, $i=1,\dots,N$ for the fundamental,  $\rho_A(\sigma)= \pm \sigma_i \pm \sigma_j$ (with $N-1$ zero weights if irreducible ) for the antisymmetric and $\rho_S(\sigma)= \pm \sigma_i \pm \sigma_j,\pm 2 \sigma_j$, for $1\leq i<j \leq N$ and $i=1,\dots,N$  for the adjoint (symmetric) (with $N$ zero roots).
\end{itemize}

For the $SU(N)$ gauge theories considered here we have encountered cases with fundamentals, symmetric, antisymmetric tensors and their conjugate. For this reason we refer to a generic partition using the formula
\begin{equation}
\label{ZSUgen}
Z_{SU(N)} (\vec\mu;\vec\nu; \tau_S,\tau_{\tilde S}; \tau_A;\tau_{\tilde A}),
\end{equation}
where $\vec \mu$ and $\vec \nu$ are vectors with $F$ and $\overline F$ components for the mass parameters of the fundamentals and the antifundamentals respectively.
$\tau_S$ and $\tau_{\tilde S}$ are mass parameters for a symmetric tensor and its conjugate,
$\tau_A$ and $\tau_{\tilde A}$ are mass parameters for a antisymmetric tensor and its conjugate.
We separate the various representations with a $;$ and if a representation is absent we use the $-$ symbol. 

In the case of $SO(N)$ and $O_+(N)$ we have encountered only cases with vectors, then we refer to a generic partition using the formula
\begin{equation}
\label{ZSOgen}
Z_{SO(N)/O_+(N)} (\vec \mu),
\end{equation}
where $\vec \mu$ is a vector of $F$ components corresponding to the mass parameters of the vectors. The difference between the $SO(N)$ and $O_+(N)$ cases stays is an extra factor of $2$ due to the gauging of the charge conjugation.

In the case of $USp(2N)$ we have encountered cases with fundamentals, an antisymmetric and a symmetric (adjoint). For this reason 
we refer to a generic partition function using the formula
\begin{equation}
\label{ZUSpgen}
Z_{USp(2N)}(\vec \mu;\tau_A;\tau_S),
\end{equation}
where $\vec \mu$ is a vector of $2F$ components, representing the mass parameters of the fundamentals, $\tau_A$ is the mass parameter of the antisymmetric and $\tau_S$ is the mass parameter of the adjoint.

\section{Integral identities for the confining dualities with antisymmetric matter from tensor deconfinement}
\label{integralsfromdecforAS}

In this appendix we prove the integral identities for the confining dualities 
with antisymmetric matter used in the body of the paper.
We restrict our attention to the dualities that cannot be derived from dimensional reduction from 4d parents. Indeed, in such case the integral identities can be derived by considering the hyperbolic limit of the elliptic integrals that 
represent the 4d s-confining dualities.
We start our analysis from the case discussed in Section \ref{sec:famII}, corresponding to 
$SU(2N)$ with an antisymmetric, a conjugate antisymmetric, four fundamentals 
and vanishing superpotential.
Then we move to the cases of Section \ref{sec:famIII}, with two antisymmetric tensors and either four fundamentals or three fundamentals and one antifundamental, both with $W=0$.

\subsection{$SU(2n)$ with $A$, $\tilde A$ and four fundamentals}
\label{appA1}
We start by assigning the mass parameters: we denote $\tau_A$ the one for the antisymmetric, $\tau_{\tilde A}$ the one of the  conjugate antisymmetric and $\mu_{1,2,3,4}$ the ones of the four fundamentals.
Then we deconfine the conjugate antisymmetric, using an $USp(2n-2)$ gauge group, by turning on a superpotential that flips the fundamental monopole for this
last group.
The mass parameter of the $SU(2n)\times USp(2n-2)$ bifundamental, that we denote here as $Q$, is $\frac{\tau_{\tilde A}}{2}$ and the one for the flipping field is $N \tau_{\tilde A}$.
The superpotential for this phase is 
\begin{equation}
W = \sigma Y_{USp(2n-2)}.
\end{equation}
The next step consists of confining the $SU(2n)$ gauge node. This node confines because it has an antisymmetric, four fundamentals and $2n-2$ antifundamentals.
Denoting the singlets as in (\ref{singlets1nii}) and in  (\ref{singlets2nii}), after the confinement of the $SU(2n)$ gauge group, the superpotential of the $USp(2n-2)$ gauge theory is
\begin{equation}
\label{spotappB1}
W = \sigma B_{n-2}\hat Y_{USp(2n-2)}^{(n-2)}+Y_A^{dressed}  
(\widetilde B_1 ^{n-2} M^2 B_{n-1} +T  \widetilde B_1 ^{n-3} M^4)
+
Y_{\tilde Q^{2n-2}}^{dressed} 
(B_{n-1}^2+T B_{n-2} ),
\end{equation}
where we refer to $\hat Y_{USp(2n-2)}$ as the fundamental monopole of the $USp(2n-2)$ 
gauge theory and the dressed monopole is $\hat Y_{USp(2n-2)}^{(j)}  = \hat Y_{USp(2n-2)} \tilde B_{1}^j$, where $ \tilde B_{1}$ is the $USp(2n-2)$ antisymmetric. 
The four $USp(2n-2)$  fundamentals, instead, correspond to the fields $M$.
The integral identity corresponding to the $SU(2n)$  confining duality is (\ref{idIAAS}) and using this identity we are left with the partition function of an $USp(2n-2)$ gauge theory  with an antisymmetric $ \tilde B_{1}$  and four fundamentals $M$, in addition to the singlets appearing in the superpotential (\ref{spotappB1}).
Such partition function is
\begin{eqnarray}
\label{afterdual}
&&
Z_{USp(2n-2)} \left(\mu+\frac{\tau_{\tilde A}}{2}; \tau_A +\tau_{\tilde A};- \right)
\Gamma_h(n \tau_{\tilde A}) \Gamma_h(n \tau_A)
 \Gamma_h \left((n-2)\tau_A + \sum_{a=1}^{4} \mu_a  \right)
\nonumber \\
&&
\prod_{1\leq a<b \leq 4}  \Gamma_h((n-1)\tau_A + \mu_a + \mu_b)
\Gamma_h\left(2 \omega  - \sum_{a=1}^{4} \mu_a-(2n-2) \tau_A\right)
 \nonumber
 \\
&& 
\Gamma_h\left(2 \omega - (n-1)\tau_{\tilde A}  - \sum_{a=1}^{4} \mu_a-(2n-3) \tau_A\right).
\end{eqnarray}
The $USp(2n-2)$ is confining as well. 
Such duality has been discussed in 
\cite{Amariti:2018wht,Benvenuti:2018bav}. The superpotential for this duality is rather complicated and there is a rapidly growing amount of terms by increasing the rank of the gauge theory. Nevertheless, there is a simple expression if the towers of singlets $\Tr (B_1)^{2,\dots,n-1}$ is flipped. In this case indeed the superpotential is a cubic functions of the dressed mesons and dressed monopoles.
On the partition function side, the confining duality for $USp(2n-2)$ with an antisymmetric and four fundamentals  corresponds to Theorem 5.6.6 of \cite{vanDeBult}, reviewed in formula (\ref{566vdb}).
We conclude then by plugging (\ref{566vdb}) inside (\ref{afterdual})
with $M=N-1$, $\tau= \tau_A +\tau_{\tilde A}$ and $m_a = \mu_a+\frac{\tau_{\tilde A}}{2}$.
We arrive at
\begin{eqnarray}
\label{finalidasa4fond}
&&
\prod_{j=0}^{n-2} 
\Gamma_h \left(2\omega -(2n-4-j)  (\tau_A +\tau_{\tilde A}) -\sum_{b=1}^4 \mu_a -2 \tau_{\tilde A} \right) \prod_{j=1}^{n-1} \Gamma_h(j (\tau_A +\tau_{\tilde A}) )
\nonumber \\
&&
 \Gamma_h \left((n-2)\tau_A + \sum_{a=1}^{4} \mu_a  \right)
 \prod_{j=0}^{n-2} \prod_{1 \leq a \leq b \leq  4} \Gamma_h(j  (\tau_A +\tau_{\tilde A})  + \mu_a+\mu_b +\tau_{\tilde A})
\nonumber \\
&&
\Gamma_h(n \tau_{\tilde A}) \Gamma_h( N \tau_A)
\Gamma_h\left(2 \omega - (n-1)\tau_{\tilde A}  - \sum_{a=1}^{4} \mu_a-(2n-3) \tau_A\right)
\nonumber \\
&&
\prod_{1\leq a<b \leq 4}  \Gamma_h((n-1)\tau_A + \mu_a + \mu_b)
\Gamma_h\left(2 \omega  - \sum_{a=1}^{4} \mu_a-(2n-2) \tau_A\right).
 \nonumber \\
\end{eqnarray}
This is exactly the expected RHS for the integral identity of the confining duality 
for the $SU(n)$ gauge theory with an antisymmetric, a conjugate antisymmetric and four fundamentals.

\subsection{Cases with two antisymmetric tensors }
\label{casesIIIas}

In the following we will prove the integral identities for the parent confining dualities discussed in family III.
In these cases the theories have an $SU(n)$ gauge group with an 
antisymmetric charged under an $SU(2)$  flavor symmetry, i.e. there are two antisymmetric that we need to consider. Here we are interested in the two cases 
discussed in the bulk of the paper. In the first case in addition to the antisymmetric there are four fundamentals.
In the second case there are three fundamentals and one antifundamental.

These two confining dualities do not have a 4d origin and the proof of the integral identities has to be performed through the deconfinement technique. Furthermore, similarly to the discussion performed in Section \ref{sec:famIII}, here we will see that the proof is iterative. Indeed, we find that for each of the four cases discussed below (where we have distinguished the parity of $n$), after deconfining the antisymmetric using two symplectic gauge groups, we will find a dual description of the unitary gauge group with lower rank. Then we will reconfine back the antisymmetric obtaining the initial version of the model, but with lower rank and with in addition new singlets. By iterating the process, we arrive at the confining description, finding the relation on the partition function that we expect from field theory. 
In this way we obtain the identities necessary in order to apply the duplication formula, that led us to conjecture the existence of new confining dualities 
with a symmetric and an antisymmetric tensor in Section \ref{sec:famIII}.

In the following analysis we will not reconstruct the full superpotentials of the WZ models, but we will limit ourself to the proofs of the integral identities. We leave such derivation to
future analysis.
\subsubsection*{Four fundamentals}

\begin{figure}[H]
\begin{center}
  \includegraphics[width=12cm]{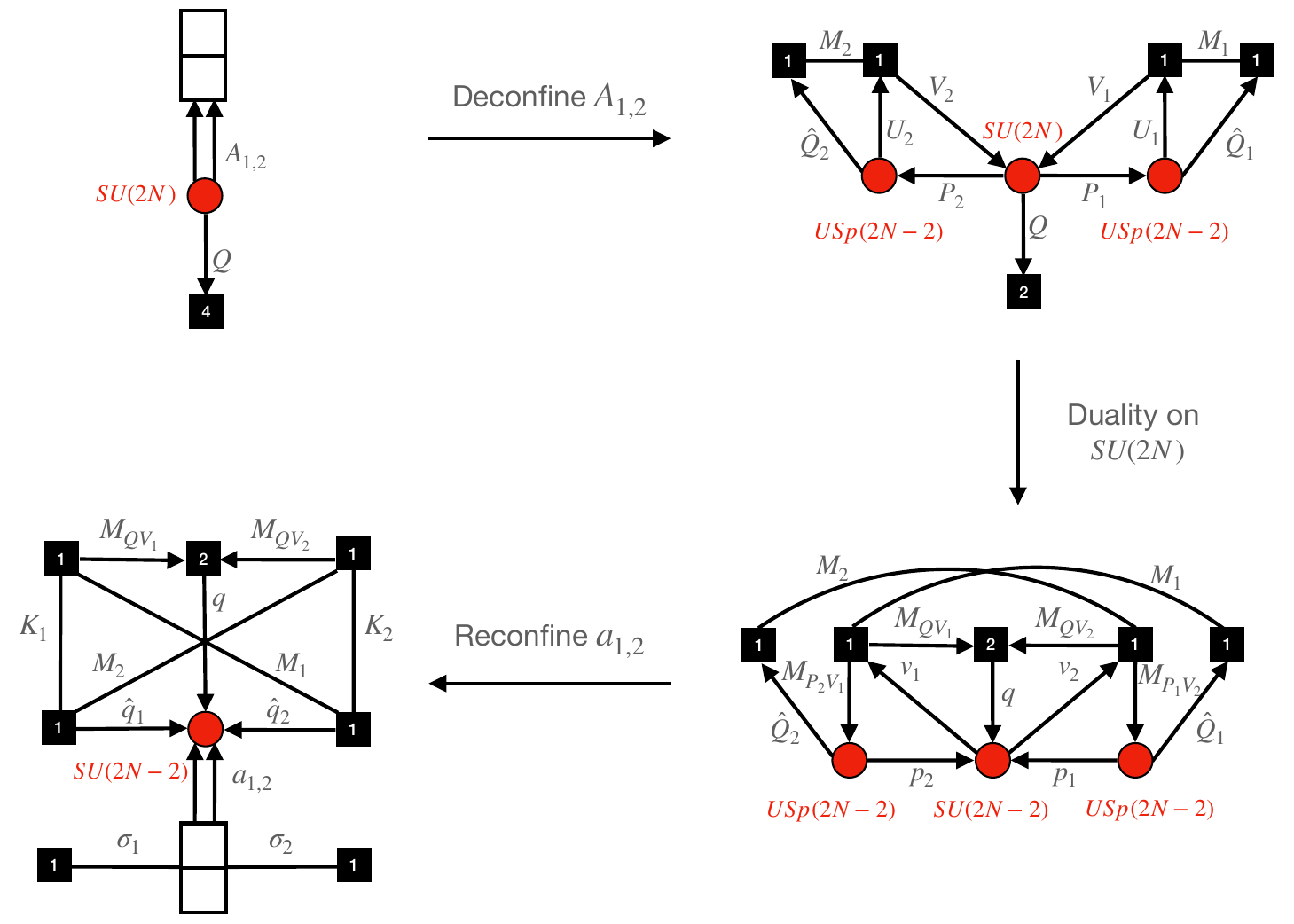}
  \end{center}
  \caption{Deconfinement of the two  antisymmetric tensors for $SU(2N)$ with four extra fundamentals and vanishing superpotential. The superpotential associated to the auxiliary  quiver 
  on the right is (\ref{WdecAA4e}). Two out of the four fundamentals $Q$ in the quiver on the right correspond to the composites $P_i \hat Q_i$ in the auxiliary quiver.  In the third quiver we have dualized the $SU(2N)$ gauge node while in the fourth quiver we have reconfined the two antisymmetrics.}
    \label{dec2Aev4}
\end{figure}

The analysis requires a different procedure for even and odd $N$.
In the case of $SU(2N)$ we start from the first quiver in Figure \ref{dec2Aev4} and  deconfine the two antisymmetric tensors 
using two $USp(2N-2)$ gauge groups. In this way we obtain the second quiver in Figure \ref{dec2Aev4}.
The superpotential for this phase is
\begin{equation}
\label{WdecAA4e}
W = P_1 U_1 V_1 +P_2 U_2 V_2 + U_1 M_1 \hat Q_1 +U_2 M_2 \hat Q_2 
+Y_{USp(2N-2)}^{(1)}+Y_{USp(2N-2)}^{(2)}.
\end{equation}
The next step consists of applying the duality of \cite{Nii:2018bgf} reviewed in appendix \ref{Niidual}
to the $SU(2N)$ gauge node. In this way we arrive to the third quiver in Figure \ref{dec2Aev4}.
The superpotential for this duality is of the form (\ref{spotchidu}) and there are massive interactions due to the superpotential (\ref{WdecAA4o}) that must  be  integrated out. 
At this point we can reconfine the $USp(2N-2)$ gauge nodes arriving at the fourth quiver in Figure \ref{dec2Aev4}, that up to conjugation corresponds to the original one, and in addition we have eight singlets. These fields have been denoted as $\sigma_{1,2}$, $M_{1,2}$, $M_{Q V_{1,2}}$ and $K_{1,2}$
in the final quiver.

In order to describe the confining dynamics iteratively we have found that it is simpler to consider the process of Figure \ref{dec2Aev4} twice. In such a way we reproduce the original quiver, with the same conjugation on the representation and the duality map of the global symmetries is simpler in order to iterate the procedure.  We observe indeed that, after two iterations, the $U(4)$ symmetry rotating the fundamentals is recovered.

After two steps we have the same quiver as the original, doubling the  singlets
discussed above.
We denote these new singlets as
$\tilde \sigma_{1,2}$, $\tilde M_{1,2}$, $\tilde M_{Q V_{1,2}}$ and $\tilde K_{1,2}$.

Depending on the parity of $N$, the process stops at either $SU(2)$ or $SU(4)$.
In the $SU(2)$ case we are left with $SU(2)$ with four fundamentals, in addition to singlets, while in the case of $SU(4)$ we have $SU(4)$ with four fundamentals and an $SU(2)$ antisymmetric flavors. Both these theories are confining. The first one corresponds to a limiting case of Aharony duality, and it can be dualized both in terms of an $USp(2)$ gauge theory with four fundamentals or as $SU(2)$ with two flavors.
In the second case we can use the results from the case studied above, i.e. $SU(2N)$ with an antisymmetric, a conjugate antisymmetric and four fundamentals because for $N=2$ the antisymmetric is self conjugate. 
In both cases, by collecting the fields arising from the iterative construction and the ones from the last confining duality one arrives to the singlets in (\ref{fourfund2ASevenmes}) and (\ref{fourfund2ASevenmon}).

This structure can be reproduced at the level of the partition function as well.
After deconfining the two antisymmetric the mass parameters of the fields in the second quiver of Figure \ref{dec2Aev4} are
\begin{eqnarray}
\label{firstdecev4f}
&&
m_{V_{1,2}} = \mu_{1,2} + (N-1) \tau_{1,2},\quad
m_{U_{1,2}} = 2\omega-\mu_{1,2}-N\tau_{1,2}+\frac{\tau_{1,2}}{2},
\nonumber \\
&&
m_{P_{1,2}}= \frac{\tau_{1,2}}{2},\quad
m_{\hat Q_{1,2}} =\mu_{1,2}-\frac{\tau_{1,2}}{2},\quad
m_{M_{1,2}} = N \tau_{1,2},\quad
m_{Q}= \mu_{3,4},
\end{eqnarray}
where only two of the masses of the fundamentals $Q$ are unchanged by the deconfinement and we referred to them as $\mu_{3,4}$.
The representations of these fields under the gauge groups can be read from the auxiliary quiver in the figure.

The next step consist of applying the duality on $SU(2N)$ on the partition function. This duality is performed by applying the dictionary spelled out in 
formula (\ref{dic}).
We define first the auxiliary quantity 
\begin{equation}
\label{Xdec4fund2Aeven}
X =\tau_1+\tau_2+\frac{\mu_3+\mu_4}{2(N-1)},
\end{equation}
such that the mass parameters in the third quiver in  Figure \ref{dec2Aev4}
are
\begin{eqnarray}
\label{afterdual4fundev}
&&
m_{M_{V_r P_s}} = \frac{\tau_r}{2}+\mu_s+(N-1) \tau_s, \quad
m_{M_{Q V_r }} = \mu_{3,4}+\mu_r+(N-1)\tau_r, \quad
m_{p_r} =  X-\frac{\tau_r}{2}, \nonumber \\
&&
m_{v_r} =  2\omega-\mu_r-(N-1) \tau_r-X, \quad 
m_{q_{3,4}}=  X-\mu_{3,4}, \quad
\end{eqnarray} 
with $r,s=1,2$ and $r \neq s$ in the first equality.
Then we confine the two symplectic gauge groups obtaining the fourth quiver in 
Figure \ref{dec2Aev4}. In this case the charged fields are the two conjugated antisymmetric $a_{1,2}$, two antifundamentals $q_{3,4}$ and two further fields denoted as $\hat q_{1,2}$ in the quiver, corresponding to the combinations
$M_{p_{1,2} \hat Q_{1,2} }$. 
Furthermore, the singlets $M_{Q V_r }$, $M_{1,2}$  survive this confinements and the other massless singlets arising in the spectrum are the two  combinations $K_{r} \equiv \hat Q_r M_{P_r V_s}$ (for $r \neq s$) and the two monopoles $\sigma_{1,2}$,
acting as singlets here. The mass of the fields arising in this phase are 
\begin{eqnarray}
\label{lastconfbackeven}
m_{\sigma_r} &=& 2\omega-(2N-2)\tau_r- \sum_{a=1}{4} \mu_{r},
\quad
m_{a_r} =  2X-\tau_r,
\nonumber \\
M_{\hat q_{r}}&=& \mu_{r}+ X-\tau_r,
\quad
M_{K_{r}}= \mu_{r}+\mu_s+(N-1) \tau_s,
\end{eqnarray}
while $m_{q_{3,4}}$ and $m_{M_{Q V_r }}$ are again given in (\ref{afterdual4fundev}) and $m_{M_r}$ is given in (\ref{firstdecev4f}).

At this point we iterate the construction deconfining the antisymmetrics, dualizing
the $SU(2N-2)$ gauge node and reconfining the two antisymmetric.
In this way we obtain the same model we started with, with gauge rank $2N-4$.
The charges of the charged fields and of the singlets in this phase can be found by iterating the process as well. We defined the new eight singlets as 
$\tilde \sigma_{1,2}$, $\tilde M_{1,2}$, $\tilde M_{Q V_{1,2}}$ and $\tilde K_{1,2}$.

By combining the singlets, we found the following three sets
\begin{eqnarray}
\label{summarymes}
T_r^{(1)} &\equiv& M_{r} \rightarrow N \tau_r, \nonumber \\
\Phi_{r,a,b}^{(1)}  &\equiv& \{ M_{Q V_r},M_{K_r},\tilde M_r, \tilde M_{Q V_r} \} \rightarrow 
\mu_a + \mu_b +(N-1) \tau_r,
\nonumber  \\
\Psi_r^{(1)} &\equiv&  \tilde M_{K_r} \rightarrow \sum_{a=1}^{4} \mu_a  +(N-2) \tau_r.
\end{eqnarray}
Furthermore, monopoles denoted as $\tilde \sigma_r$ have mass parameters
\begin{equation}
\label{sigmatildemon}
m_{\tilde \sigma_r} =
2 \omega-\sum_{a=1}^{4} \mu_a -(2 N-3) \tau _r-\tau _s,
\end{equation}
with $r=1$ and $s=2$ or viceversa.

In general the mass parameters for the matter fields after $j$ iterations can be expressed in terms of the mass parameters of the initial theory.
For this reason we define the mass parameters of the original model as
$\tau_{1,2}^{(1)}$ and $\mu _a^{(1)}$ and after $j$ iterations we have
\begin{equation}
\label{iteraeven}
\tau_r^{(j+1)} = \tau _r+\frac{j(\tau _1+\tau _2)}{N-2j},
\quad
\mu_a^{(j+1)} = \mu_a+\frac{j(\tau _1+\tau _2)}{2 (N-2j)},
\end{equation}
where $\tau_r^{(1)} \equiv \tau_r$ and  $\mu_a^{(1)} \equiv \mu_a$.
Observe that in (\ref{iteraeven}) the non Abelian flavor symmetry is 
explicitly realized for both the two antisymmetric tensors and the 
four fundamentals.
In this way we obtain at each step the mass  parameters for the singlets as
\begin{eqnarray}
m_{T_r^{(j+1)}}  &=& (N-2j+2)\tau_r^{(j)},\nonumber \\
m_{\Phi_{r,a,b}^{(j+1)}}  &=& (N-2j+1)\tau_r^{(j)}+\mu_a^{(j)}+\mu_b^{(j)}, \nonumber\\
m_{\Psi_r^{(j+1)}}&=& (N-2j)\tau_r^{(j)}+\sum_{a=1}^{4}\mu_a^{(j)},\\
m_{\sigma_r^{(j+1)}}&=& 2\omega-2(N-2j+1)\tau_r^{(j)}-\sum_{a=1}^{4}\mu_a^{(j)},
\nonumber\\
m_{\tilde \sigma_r^{(j+1)}}&=& 2\omega-(2N-4j+1)\tau_r^{(j)}-\tau_s^{(j)}-\sum_{a=1}^{4}\mu_a^{(j)}.\nonumber
\end{eqnarray}
We then iterate the process until dividing the case of even $N$ and the case of odd $N$. When  $N=2m$ we stop the process at $ j=m-1$ corresponding to an $SU(4)$ gauge group, while for $N=2m+1$ we reach  $j=m$, corresponding to an  $SU(2)$ gauge theory.  

For completeness here we report the final derivation of the identity for the case of $SU(4m)$. The analysis for $SU(4m+2)$ is analogous, and we do not work out the details explicitly. 
Fixing $N=2m$ we arrive at the following identity
\begin{eqnarray}
\label{ntl4}
Z_{SU(4m)}(\vec \mu;-;-;-;\vec \tau;-) &=& \prod _{r\neq s} \prod _{j=0}^{m-2} \Gamma _h\left((2 m-j) \tau _r+j \tau _s\right)
 \nonumber\\
&\times &
 \prod _{r\neq s} \prod _{j=0}^{m-2}
 \Gamma _h\left(\sum _{a=1}^4 \mu _a+(2 m-2-j) \tau _r+j \tau _s\right)
\nonumber\\
&\times &
\prod _{r\neq s} \prod _{j=0}^{m-2} 
\prod _{a<b} \Gamma _h\left(\mu _a+\mu _b+\tau _r (j+2 m)-(j+1) \tau _s\right)
\nonumber\\
&\times &
 \prod _{r\neq s} \prod _{j=0}^{m-2} 
 \Gamma _h\left(2 \omega-\sum _{a=1}^4 \mu _a+(2 j-4 m+2) \tau _r-2 j \tau _s \right)
 \nonumber\\
&\times &
 \prod _{r\neq s} \prod _{j=0}^{m-2}
  \Gamma _h\left(2 \omega-\sum _{a=1}^4 \mu _a+(2 j-4 m+3) \tau _r-(2 j-1) \tau _s \right)
 \nonumber\\
&\times &
Z_{SU(4)} (\vec \mu^{(m)};-;-;-;\tau_1^{(m)}; \tau_2^{(m)}),
\end{eqnarray}
where in the last line we distinguished the two representations of the antisymmetric tensors of $SU(4)$ for computational reasons, even if in this case this distinction is immanent.
Then we substitute to the last line of (\ref{ntl4}) the quantity appearing in formula (\ref{finalidasa4fond}), and by rearranging the fields we obtain the expected identity 
(\ref{4fund2ASeven}) for $SU(4m)$.

\begin{figure}[H]
\begin{center}
  \includegraphics[width=12cm]{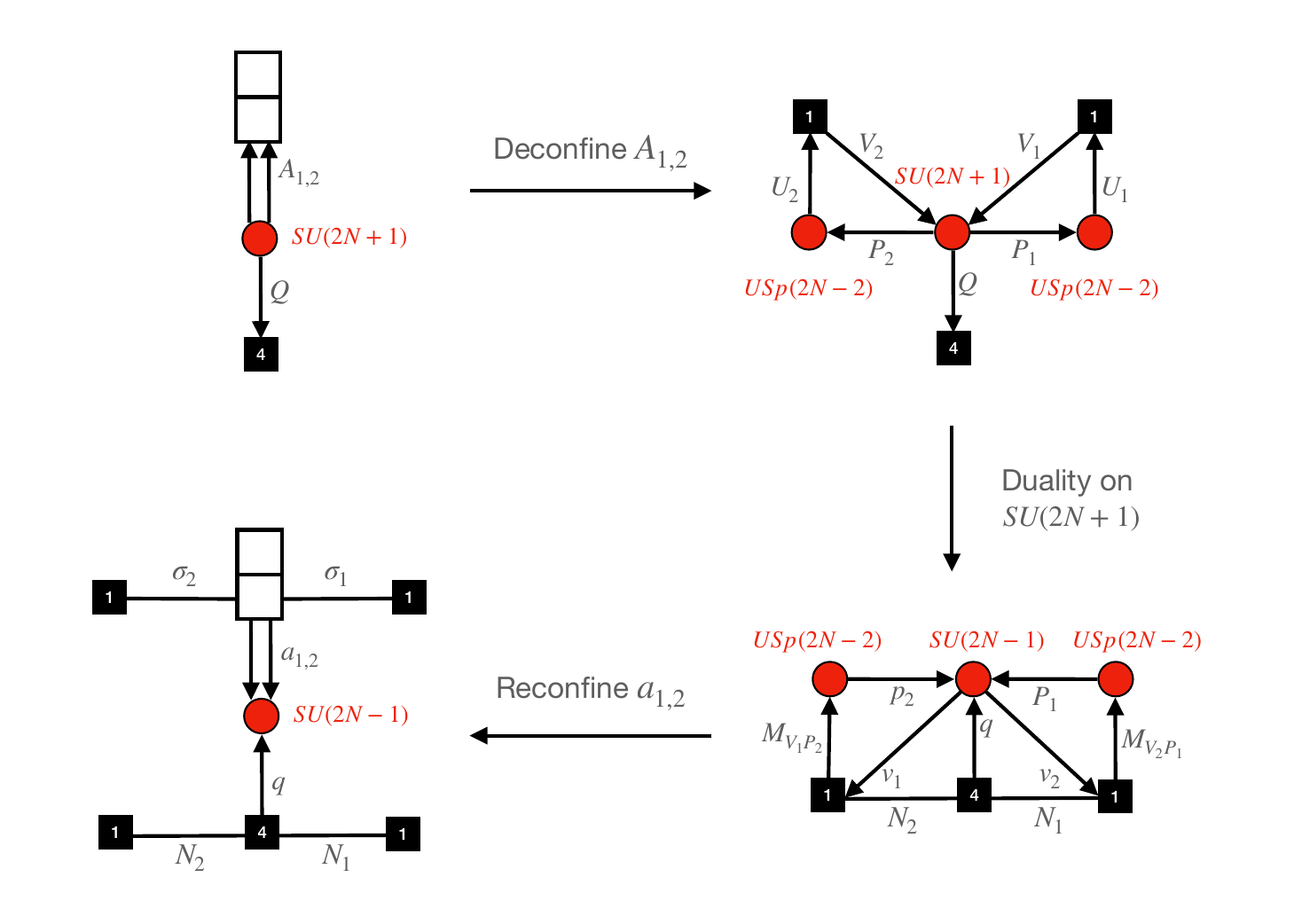}
  \end{center}
   \caption{Deconfinement of the two  antisymmetric tensors for $SU(2N+1)$ with four extra fundamentals and vanishing superpotential. The superpotential associated to the auxiliary  quiver  on the right is (\ref{WdecAA4o}).
   In the third quiver we have dualized the $SU(2N+1)$ gauge node while in the fourth quiver we have reconfined the two antisymmetrics.} 
  \label{dec2Aodd4}
\end{figure}

In the case of $SU(2N+1)$ we start from the first quiver in Figure 
\ref{dec2Aodd4} and  deconfine the two antisymmetric tensors 
using two $USp(2N-2)$ gauge groups. In this way we obtain the second quiver in Figure \ref{dec2Aodd4}.
The superpotential for the second quiver in  Figure \ref{dec2Aodd4} is 
\begin{equation}
\label{WdecAA4o}
W = P_1 U_1 V_1 +P_2 U_2 V_2 +Y_{USp(2N-2)}^{(1)}+Y_{USp(2N-2)}^{(2)}.
\end{equation}

The next step consists of applying the duality of \cite{Nii:2018bgf} reviewed in appendix \ref{Niidual}
to the $SU(2N+1)$ gauge node. In this way we arrive to the third quiver in Figure \ref{dec2Aodd4}.
The superpotential for this duality is of the form (\ref{spotchidu}) and there are massive interactions due to the superpotential (\ref{WdecAA4o}) that must be integrated out. 
At this point we can reconfine the $USp(2N-2)$ gauge nodes arriving at the fourth quiver in Figure  \ref{dec2Aodd4} , that up to conjugation corresponds to the original one, and in addition we have four singlets, two coming from the mesons of the $SU(2N+1)$ duality and two associated to the monopoles of the $USp(2N-2)$ 
gauge theories, acting as singlets in this phase. These fields have been denoted as $\sigma_{1,2}$ and $N_{1,2}$ in the final quiver.

In order to describe the confining dynamics iteratively we have found that it is simpler to consider the process of figure \ref{dec2Aodd4} twice. In such a way we reproduce the original quiver, with the same conjugation on the representation and furthemore the duality map of the global symmetries is simpler in order to iterate the procedure.
After two steps we have the same quiver as the original one with eight singlets, the four previous ones, $\sigma_{1,2}$ and $N_{1,2}$, and in addition four new singlets, that we denote as  $\rho_{1,2}$ and $M_{1,2}$.
At this point we can iterate the procedure until we reach an $SU(3)$ for odd $N$  or $SU(5)$ for even $N$.
In the $SU(5)$ case we can perform a further deconfiniment reducing to $SU(3)$
as well. The difference between the two $SU(3)$ models in the case of even and odd $N$ regards the conjugation of the representations. In the first case we have $SU(3)$ with four fundamentals and two antifundamentals, while the representations are conjugated in the second case. The final model is confining in both cases. In the first case  the confined degrees of freedom are the baryon, the meson and the monopoles $Y_{USp(2N-2)}^{(1,2)}$ dressed by the antifundamental, while in the second case we need to conjugate the representation of the baryon and of  the monopole. 
Summarizing, the various singlets created at each step reconstruct the singlets 
discussed in (\ref{singletsoddfour1}) and (\ref{singletsoddfour2}).

This structure can be reproduced at the level of the partition function as well.
After deconfining the two antisymmetric, the mass parameters of the fields in the second quiver of Figure \ref{dec2Aodd4} are
\begin{eqnarray}
m_{V_{1,2}} = N \tau_{1,2},\quad
m_{P_{1,2}}= \frac{\tau_{1,2}}{2},\quad
m_{U_{1,2}} = 2 \omega- N \tau_{1,2}-\frac{\tau_i}{2},
\end{eqnarray}
while the masses of the fundamentals $Q$ are unchanged by the deconfinement and they can be  be parameterized as $\mu_{a}$, with $a=1,\dots,4$.
The representations of these fields under the gauge groups can be read from the quiver.

The next step consist of applying the duality on $SU(2N+1)$ on the partition function. This duality is performed by applying the dictionary spelled out in 
formula (\ref{dic}).
We define first the auxiliary quantity 
\begin{equation}
\label{Xdec4fund2A}
X =\frac{(N-1)(\tau_1+\tau_2) + \sum_{a=1}^{4} \mu_a}{2N-1},
\end{equation}
such that the mass parameters in the third quiver in  Figure \ref{dec2Aodd4} are
\begin{eqnarray}
\label{afterdual4fund}
&&
m_{M_{V_r P_s}} = \frac{\tau_s}{2}+N \tau_r, \quad
m_{v_r} =  2 \omega- N \tau_r-X, \nonumber \\
&&
m_{p_r} =  X-\frac{\tau_r}{2}, \quad 
m_{q_a}=  X-\mu_a, \quad
m_{N_r} = \mu_a+N \tau_r,
\end{eqnarray} 
with $r,s=1,2$ and $r \neq s$ in the first equality.

Then we confine the two symplectic gauge groups obtaining the fourth quiver in 
Figure \ref{dec2Aodd4}. In this case, the charged fields are the two conjugated antisymmetric $a_{1,2}$ and the four antifundamentals $q$. 
Furthermore, the singlets $N_{1,2}$ survives this confinements and the other massless singlets arising in the spectrum are the two monopoles $\sigma_{1,2}$,
acting as singlets here. The mass parameters of $a_{1,2}$ and of $\sigma_{1,2}$
are
\begin{equation}
\label{lastconfback}
m_{\sigma_r} = 2 \omega-(2N-1)\tau_{r} -\sum_{a=1}^{4} \mu_a,
\quad
m_{a_r} =  2X-\tau_r,
\end{equation}
while $m_{N_r} $ and $m_{q_a}$ are again given in (\ref{afterdual4fund}).

At this point we iterate the construction deconfining the antisymmetric fields, dualizing the $SU(2N-1)$ gauge node and reconfining the two antisymmetric tensors.
In this way we obtain the same model we started with, with gauge rank $2N-3$.
The charges of the charged fields and of the singlets in this phase can be found by iterating the process as well. We defined the new four singlets as $M_{1,2}$
and $\rho_{1,2}$. The first two arise as mesons of the $SU(2N-1)$ duality
and the other two correspond to the monopoles of the two $USp(2N-4)$ factors. 
We have found that it is simpler to iterate the relations founds at this step.

In general the mass parameters for the matter fields after $j$ iterations can be expressed in terms of the mass parameters of the initial theory.
For this reason we define the mass parameters of the original model as
$\tau_{1,2}^{(1)}$ and $\mu _a^{(1)}$ and after $j$ iterations we have
\begin{equation}
\tau _r^{(j+1)}=\tau _r^{(1)}+\frac{2 j \left(\tau _1^{(1)}+\tau _2^{(1)}\right)}{2 N+1-4 j},
\quad
\mu _a^{(j+1)}=\mu _a^{(1)}+\frac{ j \left(\tau _1^{(1)}+\tau _2^{(1)}\right)}{2 N+1-4 j}.
\end{equation}
In this way we obtain at each step the mass  parameters for the singlets as
\begin{eqnarray}
m_{N_r^{(j+1)}} &=& \mu_a^{(j+1)}+(N-2j) \tau_r^{(j+1)}=
\mu _a^{(1)}+(N-j)\tau _r^{(1)}
+ j \tau _s^{(1)},
\nonumber \\
m_{M_r^{(j+1)}}&=&  \mu_a^{(j+1)}+\mu_b^{(j+1)}+\mu_c^{(j+1)}+(N-1-2j) \tau_r^{(j+1)}
\nonumber \\
&=&\mu _a^{(1)}+\mu _b^{(1)}+\mu _c^{(1)}+(N-j-1)\tau _r^{(1)}
+ j \tau _s^{(1)},
\nonumber \\
m_{\sigma_r^{(j+1)} }&=& 2 \omega-(2N-1-4j)\tau_{r}^{(j+1)} -\sum_{a=1}^{4} \mu_a^{(j+1)}
\nonumber \\
&=&2 \omega-(2N-1-2j)\tau_s-2j \tau_{r} -\sum_{a=1}^{4} \mu_a,
\nonumber \\
m_{\rho_r^{(j+1)}}&=&  2 \omega-\tau_s^{(j+1)}-2(N-1-2j) \tau _r^{(j+1)}
 -\sum_{a=1}^{4} \mu_a
 \nonumber \\
&=&2 \omega-2(N-j-1)\tau_s-(2j+1) \tau_{r} -\sum_{a=1}^{4} \mu_a.
 \end{eqnarray}

We then iterate the process until dividing the case of even $N$ and the case of odd $N$. When  $N=2m$ we stop the process at $ j=m-1$ corresponding to an $SU(5)$ gauge group, while for $N=2m+1$ we reach  $j=m$, corresponding to an  $SU(3)$ gauge theory.  
Here we do not report the final derivation of the formula, leaving the details to the interested reader, but we conclude with some further comments on it.
In the case of $N=2m+1$ the $SU(3)$ partition function has four fundamentals and two antisymmetrics. This field content corresponds to have four fundamentals and two antifundamentals, and in such case we can prove the formula (\ref{4fund2ASodd}) by using the identity (\ref{chiralpqstarconfining}), reconstructing all the singlets.
A similar derivation holds for the case $N=2m$. In this case we actually need one step further, corresponding to follow the steps of Figure (\ref{dec2Aodd4}) and reduce the model to $SU(3)$ with four antifundamentals and two conjugate antisymmetrics, equivalent to two fundamentals.  Up to charge conjugation, we apply the identity  (\ref{chiralpqstarconfining}), proving (\ref{4fund2ASodd}).

\subsubsection*{Three fundamentals and one antifundamental}

\begin{figure}[H]
\begin{center}
  \includegraphics[width=10cm]{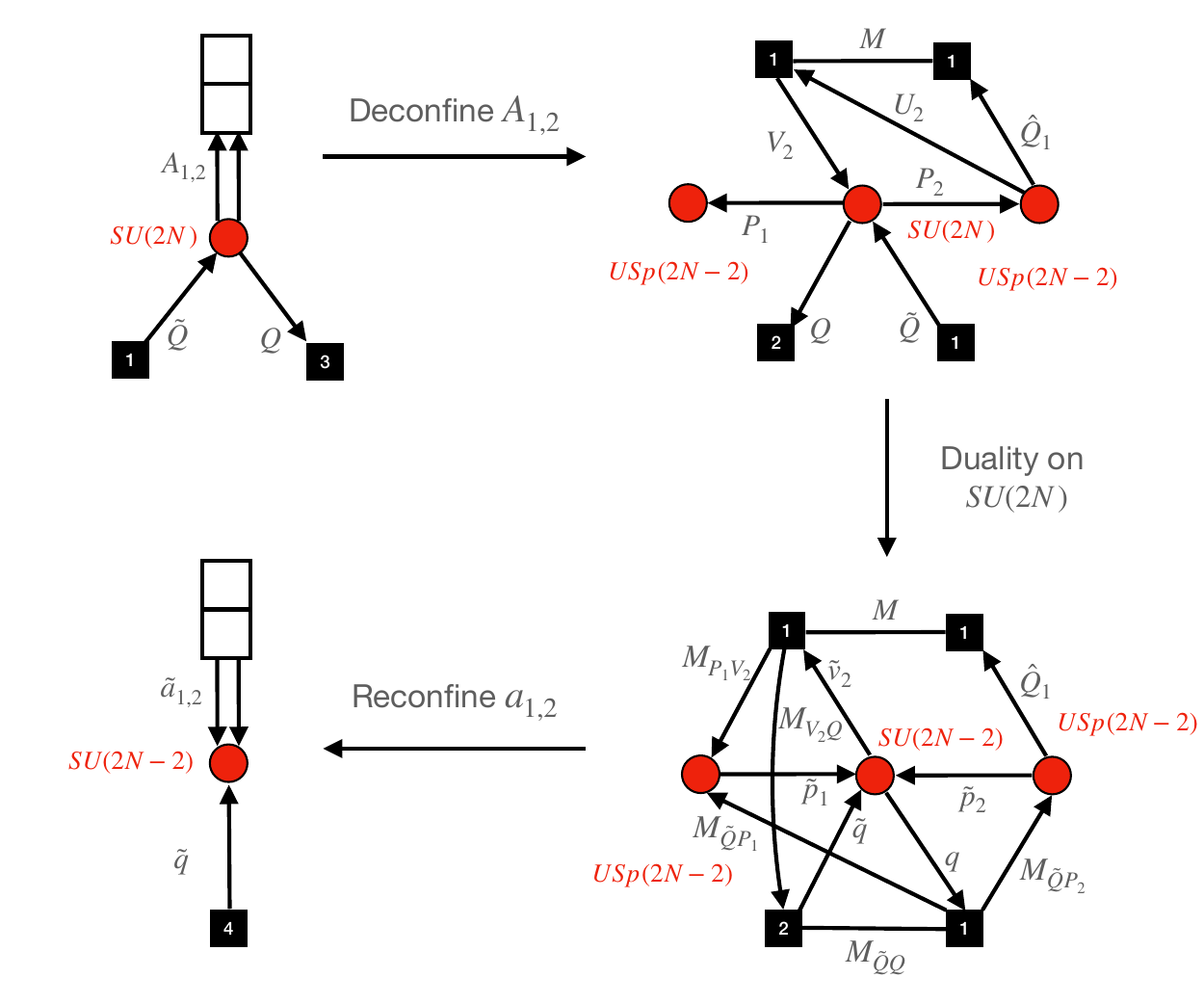}
  \end{center}
   \caption{Deconfinement of the two  antisymmetric tensors for $SU(2N)$ with three fundamentals and one antifundamental and vanishing superpotential. The superpotential associated to the auxiliary  quiver 
   on the right is (\ref{WdecAA31e}). There is a further singlet $\sigma$ flipping the monopole of the symplectic node on the left that we did not draw in the figure. Observe also that one of the fundamentals denoted as $Q$ in the quiver on the right corresponds to the composite $P_2 \hat Q_1$ in the auxiliary quiver.}     
    \label{dec2Aev31}
\end{figure}

In this case we start considering the case with even rank. We deconfine the two antisymmetric tensor $A_1$ and $A_2$ asymmetrically, using two $USp(2N-2)$ gauge groups. In one case we turn on a monopole superpotential while in the other case we flip the monopole. The quiver for this phase is the second one in Figure \ref{dec2Aev31}.
The superpotential for the deconfined phase is  
\begin{equation}
\label{WdecAA31e}
W = P_2 U_2 V_2+U_2 \hat Q_1 M  +Y_{USp}^{(2)}+\sigma Y_{USp}^{(1)}.
\end{equation}

In the partition function the fields in the quiver have the following real masses
\begin{eqnarray}
&&
m_{P_{1,2}} = \frac{\tau_{1,2}}{2},\quad
m_{Q_{2,3}} = \mu_{2,3},\quad
m_{\tilde Q} = \nu , \quad
m_{\hat Q_1} = \mu_1-\frac{\tau_2}{2}, \\
&&
m_{U_2} =2\omega +\frac{\tau_2}{2}-N \tau_2-\mu_1,\quad
m_{V_2} =(N-1) \tau_2+\mu_1 ,\quad
m_{\sigma} = N \tau_1 ,\quad
m_{M} = N \tau_2. \nonumber
\end{eqnarray}
Then we dualize the $SU(2N)$ gauge node and arrive to the third quiver in Figure \ref{dec2Aev31}.
The mass parameter for the fields in this phase are
\begin{eqnarray}
&&
m_{\tilde p_{1,2}} = X-\frac{\tau_{1,2}}{2},\quad
m_{\tilde q_{2,3}} = X-\mu_{2,3},\quad
m_{ q} =2\omega-\nu- X,\quad
m_{M_{\tilde Q P_r}}=\nu+\frac{\tau_{r}}{2},\nonumber \\
&&
m_{v_2} =2\omega-(N-1) \tau_2-\mu_1- X,\quad
m_{M_{P_1 V_2}}=\frac{\tau_1}{2}+(N-1) \tau_2+\mu_1,\nonumber \\
&&
m_{M_{\tilde Q Q}}=\nu + \mu_{2,3},\quad
m_{M_{V_2 Q}}=(N-1) \tau_2+\mu_1+\mu_{2,3},
\end{eqnarray}
where
\begin{equation}
X= \frac{\tau_1+\tau_2}{2}+\frac{\mu_2+\mu_3}{2(N-1)}.
\end{equation}
In the last step we reconfine the two antisymmetric fields and obtain the fourth quiver in Figure \ref{dec2Aev31}, where we have only depicted the charged matter fields.
Such matter fields have mass parameters
\begin{equation}
\label{step13fond}
m_{\tilde a_r} = 2X-\tau_r,\quad
\vec m_{\tilde q} = \{X-\tau_2+\mu_1,X-\mu_{2},X-\mu_3,X+\nu \},
\end{equation}
where the non Abelian flavor is explicitly broken by the various singlets and the interactions.
There are indeed new singlets arising in this phase. They are
the two mesons $M_{\tilde Q P_2}\hat Q_1$ and $M_{P_1 V_2}M_{\tilde Q P_1}$, with mass parameters
\begin{equation}
\label{mes3Q1Qt}
m_{M_{\tilde Q P_2}\hat Q_1}=\mu_1+\nu,\quad
m_{M_{P_1 V_2}M_{\tilde Q P_1}}=\mu _1+\nu +(n-1) \tau _2+\tau _1
\end{equation}
and the two fundamental monopoles $Y_{1,2}$ of the two $USp(2N-2)$ gauge groups, acting as singlets in this phase. They have mass parameters
\begin{equation}
\label{mon3Q1Qt}
m_{Y_1}=2 \omega-\sum_{a=1}^{3} \mu_a -\nu -2 (n-1) \tau _2-\tau _1
,\quad
m_{Y_2}=2 \omega-\sum_{a=1}^{3} \mu_a -\nu -(n-1)\tau _1.
\end{equation}
At this point we do not need to iterate the process, because we have obtained a theory with two conjugated antisymmetric tensors and four fundamentals.
Up to conjugation we proved the confining of this theory  above. At the level of the partition function we can simply plug in the values of the mass parameters in  (\ref{step13fond}) into formula (\ref{4fund2ASeven}), using $n=N-1$.
In this way we obtain formula (\ref{3fund2ASeven}) (taking into account the extra contributions from the hyperbolic Gamma functions in (\ref{mes3Q1Qt}) and (\ref{mon3Q1Qt}), i.e. proving the relation.

\begin{figure}[H]
\begin{center}
  \includegraphics[width=12cm]{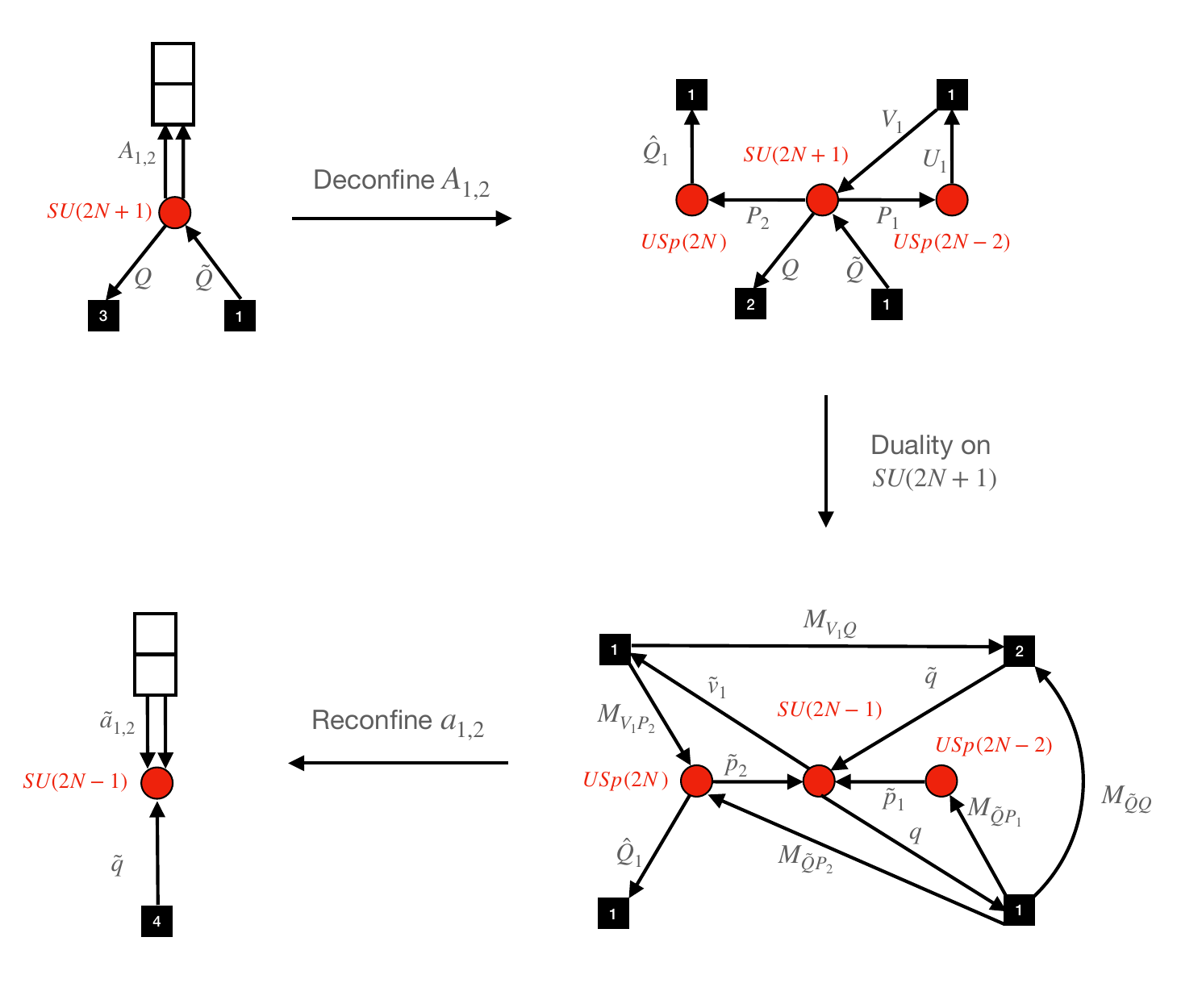}
  \end{center}
\caption{Deconfinement of the two  antisymmetric tensors for $SU(2N+1)$ with three fundamentals and one antifundamental and vanishing superpotential. The superpotential associated to the auxiliary  quiver 
  on the right is (\ref{WdecAA31o}). These is a further singlet $\sigma$ flipping the monopole of the symplectic node on the left that we did not draw in the figure. Observe also that one of the fundamentals denoted as $Q$ in the quiver on the right corresponds to the composite $P_2 \hat Q_1$ in the auxiliary quiver.}
      \label{dec2Aodd31}
\end{figure}

We conclude with  the $SU(2N+1)$ case. We deconfine the two antisymmetric tensor $A_1$ and $A_2$ asymmetrically, using an  $USp(2N-2)$ and an  $USp(2N)$ gauge group. In the first case we turn on a monopole superpotential while in the second case we flip the monopole. The quiver for this phase is the second one in Figure \ref{dec2Aodd31}.

The superpotential for the quiver in the RHS of Figure \ref{dec2Aodd31} is 
\begin{equation}
\label{WdecAA31o}
W = P_1 U_1 V_1  +Y_{USp}^{(1)}+\sigma. Y_{USp}^{(2)}
\end{equation}

In the partition function the fields in the quiver have the following real masses
\begin{eqnarray}
&&
m_{P_{1,2}} = \frac{\tau_{1,2}}{2},\quad
m_{Q_{2,3}} = \mu_{2,3},\quad
m_{\tilde Q} = \nu , \quad
m_{\hat Q_1} = \mu_1-\frac{\tau_2}{2}, \\
&&
m_{U_1} =2\omega -\frac{2N+1}{2}\tau_1,\quad
m_{V_1} =N \tau_1,\quad
m_{\sigma} =N \tau_2+\mu_1.
\end{eqnarray}
Then we dualize the $SU(2N+1)$ gauge node and arrive to the third quiver in the figure \ref{dec2Aodd31}.
The mass parameters for the fields in this phase are
\begin{eqnarray}
&&
m_{\tilde p_{1,2}} = X-\frac{\tau_{1,2}}{2},\quad
m_{\tilde q_{2,3}} = X-\mu_{2,3},\quad
m_{ q} =2\omega-\nu- X,\quad
m_{\tilde v_1} =2\omega-N \tau_1-X, \nonumber \\
&&
m_{M_{V_1 P_2}}  =N \tau_1+\frac{\tau_2}{2},\quad
m_{ M_{\tilde Q P_r}} =\nu + \frac{\tau_r}{2},\quad 
m_{M_{V_1 Q} } = N \tau_1+\mu_{2,3},\quad
m_{M_{\tilde Q Q}} =\nu + \mu_{2,3}, \nonumber \\
\end{eqnarray}
where
\begin{equation}
X= \frac{N \tau_1+(N-1) \tau_2}{2N-1}+\frac{\mu_2 +\mu_3}{2N-1}.
\end{equation}

In the last step we reconfine the two antisymmetric and obtain the fourth quiver 
in Figure \ref{dec2Aodd31}, where we have only depicted the charged matter fields.
Such matter fields have mass parameters
\begin{equation}
\label{step13fondbis}
m_{\tilde a_r} = 2X-\tau_r,\quad
\vec m_{\tilde q} = \{X-\tau_2+\mu_1,X-\mu_{2},X-\mu_3,X+\nu \},
\end{equation}
where the non Abelian flavor is explicitly broken by the various singlets and the interactions.
There are indeed new singlets arising in this phase. They are
the three mesons $K=\{M_{V_1 P_2} \hat Q_1, M_{V_1 P_2} M_{\tilde Q P_2}
M_{\tilde Q P_2} \hat Q_1\}$, with mass parameters
\begin{equation}
\label{mes3Q1Qtodd}
m_{K} = \{N \tau_1+\mu_1,N \tau_1+\tau_2+\nu,\nu + \mu_1 \},
\end{equation}
and the two fundamental monopoles $Y_{1,2}$ of the two $USp(2N-2)$ gauge groups, acting as singlets in this phase. They have mass parameters
\begin{equation}
\label{mon3Q1Qtodd}
m_{Y_1}=2 \omega-\sum_{a=1}^{3} \mu_a -\nu -2 N \tau _1
,\quad
m_{Y_2}=2 \omega-\sum_{a=1}^{3} \mu_a -\nu -(n-1)\tau _2 - \tau _1.
\end{equation}
As in the case above we do not need to iterate the process, because we have obtained an $SU(2N-1)$ confining theory with two conjugated antisymmetric and four fundamentals.
At the level of the partition function we  plug in the values of the mass parameters in  (\ref{step13fondbis}) into formula (\ref{4fund2ASodd}), using $n=N-1$.
In this way we obtain formula (\ref{3fund2ASodd}) (taking into account the extra contributions from the hyperbolic Gamma functions in (\ref{mes3Q1Qtodd}) and (\ref{mon3Q1Qtodd})),  i.e. proving the relation.

\section{Branching rules}
\label{Appendix_Branching_Rules}

In this appendix we list the branching rules for the matter and gauge fields under the breaking pattern $SU(N) \rightarrow SU(N-2) \times U(1)_1 \times U(1)_2$. For convenience we use different normalization in the Abelian charges of the decomposed fields, depending on the rank of the breaking group. Specifically, the normalization for the even $2n$ and odd $2n+1$ cases differs of a factor $2$ with respect to the generic $N$ case. We report them in the following.

	\begin{align}
		& \mathbf{\bullet \,\, SU(N) \longrightarrow SU(N-2) \times U(1)_1 \times U(1)_2} \nonumber \\
		& \square \longrightarrow 	\square_{(0,-2)} \oplus \textbf{1}_{(1,N-2)} \oplus \textbf{1}_{(-1,N-2)} \nonumber \\
		& \overline{\square} \longrightarrow 	\overline{\square}_{(0,2)} \oplus \textbf{1}_{(-1,-(N-2))} \oplus \textbf{1}_{(1,-(N-2))} \nonumber \\
		& \symmF \longrightarrow \symmF_{(0,-4)} 	\oplus \square_{(1,N-4)} \oplus \square_{(-1,N-4)} \oplus \textbf{1}_{(2,2N-4)} \oplus \textbf{1}_{(-2,2N-4)} \oplus \textbf{1}_{(0,2N-4)} \nonumber \\
		& \symmBF \longrightarrow  \symmBF_{(0,4)} \oplus \overline{\square}_{(-1,-(N-4))} \oplus \overline{\square}_{(1,-(N-4))} \oplus \textbf{1}_{(-2,-(2N-4))} \oplus \textbf{1}_{(2,-(2N-4))} \nonumber \\ & \phantom{\symmBF \longrightarrow} \oplus \textbf{1}_{(0,-(2N-4))} \nonumber \\
		& \asymmF \longrightarrow \asymmF_{(0,-4)} 	\oplus \square_{(1,N-4)} \oplus \square_{(-1,N-4)} \oplus \textbf{1}_{(0,2N-4)} \nonumber \\
		& \asymmBF \longrightarrow 	\asymmBF_{(0,4)} \oplus \overline{\square}_{(-1,-(N-4))} \oplus \overline{\square}_{(1,-(N-4))} \oplus \textbf{1}_{(0,-(2N-4))} \nonumber \\
		& \textbf{adj} \longrightarrow \textbf{adj}_{(0,0)} \oplus \overline{\square}_{(1,N)} \oplus \overline{\square}_{(-1,N)} \oplus \square_{(1,-N)} \oplus \square_{(-1,-N)} \oplus \textbf{1}_{(0,0)} \oplus \textbf{1}_{(0,0)} \oplus \textbf{1}_{(2,0)} \nonumber \\ & \phantom{\textbf{adj} \longrightarrow} \oplus \textbf{1}_{(-2,0)} \nonumber \\
		& \mathbf{ \bullet \,\, SU(2n) \longrightarrow SU(2n-2) \times U(1)_1 \times U(1)_2} \nonumber \\
		& \square \longrightarrow 	\square_{(0,-1)} \oplus \textbf{1}_{(1,n-1)} \oplus \textbf{1}_{(-1,n-1)} \nonumber \\
		& \overline{\square} \longrightarrow 	\overline{\square}_{(0,1)} \oplus \textbf{1}_{(-1,-(n-1))} \oplus \textbf{1}_{(1,-(n-1))} \nonumber \\
		& \symmF \longrightarrow \symmF_{(0,-2)} 	\oplus \square_{(1,n-2)} \oplus \square_{(-1,n-2)} \oplus \textbf{1}_{(2,2n-2)} \oplus \textbf{1}_{(-2,2n-2)} \oplus \textbf{1}_{(0,2n-2)} \nonumber \\
		& \symmBF \longrightarrow 		\symmBF_{(0,2)} \oplus \overline{\square}_{(-1,-(n-2))} \oplus \overline{\square}_{(1,-(n-2))} \oplus \textbf{1}_{(-2,-(2n-2))} \oplus \textbf{1}_{(2,-(2n-2))} \oplus \textbf{1}_{(0,-(2n-2))} \nonumber \\
		& \asymmF \longrightarrow \asymmF_{(0,-2)} 	\oplus \square_{(1,n-2)} \oplus \square_{(-1,n-2)} \oplus \textbf{1}_{(0,2n-2)} \nonumber \\
		& \asymmBF \longrightarrow 	\asymmBF_{(0,2)} \oplus \overline{\square}_{(-1,-(n-2))} \oplus \overline{\square}_{(1,-(n-2))} \oplus \textbf{1}_{(0,-(2n-2))} \nonumber \\
		& \textbf{adj} \longrightarrow  \textbf{adj}_{(0,0)} \oplus \overline{\square}_{(1,n)} \oplus \overline{\square}_{(-1,n)} \oplus \square_{(1,-n)} \oplus \square_{(-1,-n)} \oplus \textbf{1}_{(0,0)} \oplus \textbf{1}_{(0,0)} \oplus \textbf{1}_{(2,0)} \nonumber \\ & \phantom{\textbf{adj} \longrightarrow} \oplus \textbf{1}_{(-2,0)} \nonumber \\
		& \mathbf{\bullet \,\, SU(2n+1) \longrightarrow SU(2n-1) \times U(1)_1 \times U(1)_2} \nonumber \\
		& \square \longrightarrow 	\square_{(0,-1)} \oplus \textbf{1}_{(1,n-1/2)} \oplus \textbf{1}_{(-1,n-1/2)} \nonumber \\
		& \overline{\square} \longrightarrow 	\overline{\square}_{(0,1)} \oplus \textbf{1}_{(-1,-(n-1/2))} \oplus \textbf{1}_{(1,-(n-1/2))} \nonumber \\
		& \symmF \longrightarrow \symmF_{(0,-2)} 	\oplus \square_{(1,n-3/2)} \oplus \square_{(-1,n-3/2)} \oplus \textbf{1}_{(2,2n-1)} \oplus \textbf{1}_{(-2,2n-1)} \oplus \textbf{1}_{(0,2n-1)} \nonumber \\
		& \symmBF \longrightarrow \symmBF_{(0,2)} \oplus \overline{\square}_{(-1,-(n-3/2))} \oplus \overline{\square}_{(1,-(n-3/2))} \oplus \textbf{1}_{(-2,-(2n-1))} \oplus \textbf{1}_{(2,-(2n-1))} \nonumber \\ & \phantom{\symmBF \longrightarrow} \oplus \textbf{1}_{(0,-(2n-1))} \nonumber \\
		& \asymmF \longrightarrow \asymmF_{(0,-2)} 	\oplus \square_{(1,n-3/2)} \oplus \square_{(-1,n-3/2)} \oplus \textbf{1}_{(0,2n-1)} \nonumber \\
		& \asymmBF \longrightarrow 	\asymmBF_{(0,2)} \oplus \overline{\square}_{(-1,-(n-3/2))} \oplus \overline{\square}_{(1,-(n-3/2))} \oplus \textbf{1}_{(0,-(2n-1))} \nonumber \\
		& \textbf{adj} \longrightarrow \textbf{adj}_{(0,0)} \oplus \overline{\square}_{(1,n+1/2)} \oplus \overline{\square}_{(-1,n+1/2)} \oplus \square_{(1,-(n+1/2))} \oplus \square_{(-1,-(n+1/2))} \oplus \textbf{1}_{(0,0)} \nonumber \\ & \phantom{\textbf{adj} \longrightarrow} \oplus \textbf{1}_{(0,0)} \oplus \textbf{1}_{(2,0)} \oplus \textbf{1}_{(-2,0)} \nonumber \\
	\end{align} 

\section{3d dualities}
\label{3dreview}

In this appendix we review the  dualities used in the body of the paper in order to prove the nine $SU(N)$ confining 
dualities with symmetric tensors.
Some of these dualities  have been used to deconfine the symmetric and the antisymmetric tensors, while some others have been used in order to  dualize or confine the gauge groups in the auxiliary quivers.
We have organized the appendix in three main parts, reviewing the cases with an $SU(N)$ gauge group, the cases with an $SO(N)$ or $O_+(N)$ gauge group and the cases with a symplectic gauge group.
The $SU(N)$ dualities reviewed here have only $F$ fundamentals and $\overline F$ antifundamentals, with generically $F > N > \overline F$. These dualities have been extensively studied in \cite{Nii:2018bgf}, generalizing the construction of \cite{Benini:2011mf,Aharony:2014uya}. Here we will provide a derivation of them from 4d and we will obtain the matching of the three sphere partition function.
Then we move to the cases with orthogonal gauge groups, reviewing the confining dualities in presence of vectors. Such dualities have been proposed in \cite{Kapustin:2011gh,Benini:2011mf,Aharony:2011ci,Hwang:2011ht,Aharony:2013kma,Benvenuti:2021nwt}, and here we will provide the matching of the partition functions for the ones that have not been proved yet in the literature. 
Eventually, we will review the dualities for $USp(2N)$ distinguishing  the case with fundamentals, with  fundamentals and antisymmetric tensor and the case with fundamentals and an adjoint.

\subsection{$SU(N)$ gauge group}
\label{SUSUappD1}

In this section we review the duality involving $SU(N)$ SQCD with $F>N>\overline F$ antifundamentals. This duality has been extensively discussed in \cite{Nii:2018bgf}. We have used this duality in order to prove the confining dualities of type III. Furthermore, we have used the case with $F=N+2$ and $\overline F=N-2$ in subsection \ref{Niidual}. We have also used the confining limit of this duality with $F=N+1$ and  $\overline F=N-1$ in  subsection \ref{Niiconf}.
Here we provide a review of this duality and show how to derive it from 4d.

Before discussing the various possibilities we need to fix some notation. In the following $Y_0$ refers to the monopole associated to the breaking of $SU(2N) \rightarrow SU(N)^2 \times U(1)$ for the even ranks and 
$SU(2N+1) \rightarrow SU(N)^2 \times U(1)^2$ in the odd case.
In general, for $a\neq 0$ one considers the breaking pattern $SU(2N) \rightarrow SU(N-a)^2  \times SU(2a) \times U(1)^2$ and the relative monopole is denoted as $Y_a$.
In the odd case the $Y_a$ monopole is associated to the braking pattern
$SU(2N+1) \rightarrow SU(N-a)^2  \times SU(2a+1) \times U(1)^2$, which holds for $a=0$ as well.

\subsubsection{$SU(N)$ with $F>N$ fundamentals and $\overline F<N$ antifundamentals}
\label{Niidual}

The electric theory is $SU(N)$ SQCD with $F$ fundamentals and $\overline F$ antifundamentals, with $\overline F<N<F$ and $W=0$.
The moduli space of this theory is described by the meson $M = Q \tilde Q$ and the baryons $B=\epsilon_N Q^N$, while it is not possible to construct the antibaryons, because $\overline F < N$. Nevertheless, it is possible to construct gauge invariant monopole operators $Y_{dressed} = Y_1\dots Y_{N-1} \tilde Q^{N-2}$ dressing them with the antifundamentals.

While in general the dual theory has gauge group $SU(F-N)$ and superpotential
\begin{equation}
\label{spotchidu}
W = M q \tilde q 
\end{equation} 
the precise operator mapping is dependent on the parity of the gauge and flavor ranks, and  various cases have been explored in \cite{Nii:2018bgf}.

Here we review the main aspects of such dualities for the cases that have played a relevant role in our analysis.

\subsubsection*{$SU(2N)$ with $2F$ $\square$ and $2$ $\overline \square$}

We have used this case in the analysis of case III-B and  III-C.
In this case the gauge invariant monopoles are $Y_0$ and $Y_1 (\tilde Q)^{F-1}$. They are mapped to $\tilde Y_1 (\tilde q^2)^{F-1}$ and $\tilde Y_0$ respectively.
This identification is correct for $F>N+1$ while for $F=N+1$ the dressed monopole in the $SU(2)$ dual theory is not available. In such a case the $N$-th root of the  bare monopole of the electric theory is mapped to the anti-baryon $\tilde q^2$ in the dual phase.

\subsubsection*{$SU(2N+1)$ with $2F+1$ $\square$ and $1$ $\overline \square$}

We have used this case in the analysis of case III-A and  III-C.
In this case the operator mapping relates the dressed monopole operator of the electric phase $Y_0 \tilde Q^F$ to the bare monopole $\tilde Y_0$ of the dual phase.

\subsubsection*{$SU(2N+1)$ with $2F+2$ $\square$ and $2$ $\overline \square$}

We have used this case in the analysis of case III-B.
On the electric side the gauge invariant coulomb branch operator is $Y_0 \tilde Q^F$ and it is mapped in the dual to the dressed $\tilde Y_0 \tilde q^F$ dressed monopole.

\subsubsection*{$SU(2N)$ with $2F+1$ $\square$ and $1$ $\overline \square$}

We have used this case in the analysis of case III-A and  III-C.
In this case we have  only $Y_{0}$ in the Coulomb branch. This monopole is gauge invariant it is mapped in the dual to the dressed monopole 
$\tilde Y_0 \tilde q^F$. 

\subsubsection*{$SU(2N)$ with $2N+2$ $\square$ and $2 \overline F$ $\overline \square$}
We have used this case in the analysis of case I-A. In this case the dual gauge group is $SU(2)$.
The bare monopole of the dual gauge theory is gauge invariant.  In the dual description there are 
$2N+2$ fundamentals $q$, $2\overline F$ fundamentals $\tilde q$ and a meson $M = Q \tilde Q$. There is a  superpotential, $W = M q \tilde q$, that breaks the $U(2N + 2\overline F) \times U(1)$ flavor symmetry to  $SU(2N + 2) \times SU(\overline F)\times U(1)^2$.
The bare monopole operators of the electric gauge theory are $Y_{m}^{bare}$, with $m=\overline F,\overline F-1$ using the notation of formula (\ref{Eq:Sigma_m}) (in subsection \ref{genmondisc} we have given more details about these monopoles). They are not gauge invariant and they need to be dressed by $\tilde Q^{2 \overline F}$ and $\tilde Q^{2 \overline F-2}$ respectively. Actually, the two dressing are more sophisticated in this case. Indeed, in the case of $Y_{\overline F-1}^{bare}$ one need to defined a minimal bare monopole $(Y_{\overline F-1}^{bare})^{\frac{1}{N-\overline F+1}}$  and make it gauge invariant by dressing it by $\tilde Q^{2 \overline F-2}$. This gauge invariant combination is dual to the baryonic $\tilde q^2$ operator in the dual $SU(2)$ gauge theory. On the other hand, the $SU(2)$ gauge monopole of the dual theory is dual to the gauge invariant combination obtained by dressing the bare monopole operator $Y_{\overline F}^{bare}$ by $(\tilde Q^{2 \overline F})^{N+1-\overline F}$.

\subsubsection*{$SU(2N+1)$ with $2N+3$ $\square$ and $2 \overline F+1$ $\overline \square$}
We have used this case in the analysis of case I-A.
The situation is similar to the one discussed above, indeed in this case the dual gauge group is 
still $SU(2)$.
The bare monopole of the dual gauge theory is gauge invariant.  In the dual description there are 
$2N+2$ fundamentals $q$, $2\overline F$ fundamentals $\tilde q$ and a meson $M = Q \tilde Q$. There is a superpotential, $W = M q \tilde q$, that breaks the $U(2N + 2\overline F) \times U(1)$ flavor symmetry to  $SU(2N + 2) \times SU(\overline F)\times U(1)^2$.
The bare monopole operators of the electric gauge theory are $Y_{\overline F,\overline F-1}^{bare}$. They are not gauge invariant and they need to be dressed by $\tilde Q^{2 \overline F \pm 1}$. Again the two dressing are more sophisticated in this case. Indeed, in the case of $Y_{\overline F-1}^{bare}$ one need to defined a minimal bare monopole $(Y_{\overline F-1}^{bare})^{\frac{1}{N-\overline F+1}}$  and make it gauge invariant by dressing it by $\tilde Q^{2 \overline F-1}$. This gauge invariant combination is dual to the baryonic $\tilde q^2$ operator in the dual $SU(2)$ gauge theory. On the other hand, the $SU(2)$ gauge monopole of the dual theory is dual to the gauge invariant combination obtained by dressing the bare monopole operator $Y_{\overline F}^{bare}$ by $(\tilde Q^{2 \overline F+1})^{N+1-\overline F}$.

At the level of the partition function all these cases, independently from the parity of the ranks, can be studied starting from the reduction of ordinary $SU(N)$ Seiberg duality with $F>N$ pairs of fundamentals and antifundamentals on $S^1$, by triggering a real mass flow for some of  the antifundamentals. 
We give a positive real mass to $\frac{F-\overline F}{2}$ antifundamentals and a negative real mass to $\frac{F-\overline F}{2}$ antifundamentals, with even 
$F-\overline F$ (this is indeed the case in the various dualities reviewed above).
The original, duality, before performing such real mass flow, can be recasted in the following identity
\begin{eqnarray}
\label{beforerm}
Z_{SU(N)}(\vec \mu,\vec \nu) = \prod_{a=1}^F \prod_{b=1}^{ F} \Gamma_h(\mu_a +\nu_b) Z_{SU(F-N)}(\vec {\tilde \nu}, \vec {\tilde \mu}), 
\end{eqnarray}
with
\begin{eqnarray}
\tilde \mu_a =\frac{1}{F-N} \sum_{c=1}^{F}\mu_c-\mu_a,
\quad \tilde \nu_a =\frac{1}{F-N}  \sum_{c=1}^{F}\nu_c-\nu_a
\end{eqnarray}
and with the constraint
\begin{eqnarray}
\sum_{a=1}^{F} (\mu_a+\nu_a)=2 \omega.
\end{eqnarray}
This identity can be read from \cite{Aharony:2013dha} by rearranging the mass parameters.
Once we impose the real mass flow specified above we arrive at the identity 
\begin{eqnarray}
\label{chiralpqstarduality}
Z_{SU(N)}(\vec \mu,\vec \nu) = \prod_{a=1}^F \prod_{b=1}^{\tilde F} \Gamma_h(\mu_a +\nu_b) Z_{SU(F-N)}(\vec {\tilde \nu}, \vec {\tilde \mu}),
\end{eqnarray}
with
\begin{eqnarray}
\label{dic}
\tilde \mu_a =\frac{1}{F-N} \sum_{c=1}^{F}\mu_c-\mu_a,\quad \tilde \nu_a =2\omega - \frac{1}{F-N}  \sum_{c=1}^{F}\mu_c-\nu_a.
\end{eqnarray}
By inspection one can check that the dictionary imposed by the operator mapping  for each case discussed above implies the relation (\ref{dic}). Then the identity (\ref{chiralpqstarduality}) represents the matching of the partition function for the chiral duality proposed in \cite{Nii:2018bgf}.

\subsubsection{$SU(N)$ with $N+1$ fundamentals and $N-1$ antifundamentals}
\label{Niiconf}

In this case the duality is confining and we have often used it in the study of cases I and II.
Here we review the derivation of this confining limit of the duality starting from the 4d/3d reduction of the limiting case of 4d $SU(N)$ Seiberg duality with $F=N+1$ non-chiral flavors.
The effective duality on $S^1$ was studied originally in \cite{Aharony:2013dha}. On the electric side there is a non vanishing KK monopole superpotential, while the dual theory has the usual 
\begin{eqnarray}
\label{dualconf4dMbbt}
W = \det M - M B \tilde B
\end{eqnarray}
superpotential, enforcing the classical constraint among the baryons $B= \epsilon_N \cdot Q^N$ and $\tilde B =\epsilon_N \cdot \tilde Q^N$ and the  meson $M = Q \tilde Q$.

The KK monopole can be lifted by turning on a positive and a negative real mass for the $(N+1)$-th and the $N$-th components of 
the antifundamental $\tilde Q$. On the electric side we have 3d $SU(N)$ SQCD with $N+1$ fundamentals and $N-1$ antifundamentals
while on the dual side there are a \emph{rectangular} meson $M = Q \tilde Q$ and a baryon  $B= \epsilon_N \cdot Q^N$.
On the other hand two components of the antibaryon  $\tilde B$ are massive and the other $N-1$ are identified with the minimal dressed 
monopole $Y_d = Y_1 \dots Y_{N-1} \tilde Q^{N-2}$. The dual superpotential becomes
 \begin{eqnarray}
\label{dualconf3dMbbt}
W = M B Y_d.
\end{eqnarray}
We can study this flow on the three sphere partition function starting from the identity 
\begin{eqnarray}
Z_{SU(N)}(\vec \mu,\vec \nu) = \prod_{a,b} \Gamma_h(\mu_a +\nu_b) \prod_{a=1}^{N+1} \Gamma_h\left(\sum_{c=1}^{N+1}\mu_c-\mu_a\right)
\Gamma_h\left(\sum_{c=1}^{N+1}\nu_c-\nu_a\right),
\end{eqnarray}
that represents the effective duality on $S^1$, provided that the constraint
\begin{equation}
\sum_{a=1}^{N+1} \mu_a+\sum_{a=1}^{N+1} \nu_a=2 \omega
\end{equation}
is satisfied.
Fixing then $\nu_N = m+s$ and  $\nu_{N+1} = m-s$, we arrive at the final identity 
\begin{eqnarray}
\label{chiralpqstarconfining}
Z_{SU(N)}(\vec \mu,\vec \nu) = \prod_{a,b} \Gamma_h(\mu_a +\nu_b) \prod_{a=1}^{N+1} \Gamma_h\left(\sum_{c=1}^{N+1}\mu_c-\mu_a\right)
\prod_{a=1}^{N-1} 
\Gamma_h\left(2\omega -\sum_{c=1}^{N+1}\mu_c-\nu_a\right). \nonumber \\
\end{eqnarray}

\subsection{Orthogonal gauge groups}

\subsubsection{$SO(N)$}
\label{apppSO}
We have used in the paper some confining dualities for $SO(N)$ with vectors.
\begin{itemize}
\item
First we have used the duality between 
$SO(K+1)$ with $K$ vectors and flipped monopole, $W = \sigma Y^+$
and a WZ model with an $SU(K)$ symmetric $S$ and and antifundamental $q$, with $W= S q^2$. In this case the symmetric meson is built from the $SO(K+1)$ vectors and the antifundamental $q$ correspond to the baryon monopoles.

The identity between the partition functions  is 
\begin{equation}
\label{SOelemag}
\Gamma_h \left(\omega+\sum_{r=1}^{K} \mu_r\right) Z_{SO(K+1)} (\vec \mu)
= \prod_{r=1}^{K}  \Gamma_h(\omega-\mu_r)
\prod_{1\leq r\leq s\leq K } \Gamma_h(\mu_r+\mu_s).
\end{equation}

\item

Then by omitting the flip above one can also construct the duality between
$SO(K)$ with $K-1$ vectors and vanishing superpotential $W=0$ and a WZ model with an $SU(K-1)$ symmetric $S$, an antifundamental $q$ and a further singlet $\sigma$, with $W= S q^2 + \sigma^2 \det S$. In this case the singlet $\sigma$ corresponds to the $Y_+$ monopole.
The flip in this case has the effect to move the term $\Gamma_h \left(\omega+\sum_{r=1}^{K} \mu_r\right)$ from the LHS to the RHS of (\ref{SOelemag}), and, by using the inversion formula,  the net effect consists of having on the RHS of (\ref{SOelemag}) the term
$\Gamma_h \left(\omega-\sum_{r=1}^{K} \mu_r\right) $.
\item
Furthermore, another useful duality relates 
$SO(K)$ with $K+1$ vectors and linear monopole superpotential $W = Y^+$
to a WZ model with superpotential $W= S q^2 + \det S$.

The partition function of the electric theory is $ Z_{SO(K)} (\vec \mu)$, with balancing condition
\begin{equation}
\label{SOterzo}
\sum_{r=1}^{K+1} \mu_r= \omega.
\end{equation}
The  partition function of the dual confining theory  is 
\begin{equation}
\label{confsO}
Z_{conf} = \prod_{r=1}^{K+1} \Gamma_h(\omega-\mu_r)
\prod_{1\leq r\leq s\leq K+1} \Gamma_h(\mu_r+\mu_s).
\end{equation}
\end{itemize}

The integral identities for these dualities have been derived in \cite{Benini:2011mf} and \cite{Amariti:2022wae}.

\subsubsection{$O_+(N)$}
\label{apppO}
For the analysis of case I-C, II-C and III-C we have used a confining duality for the $O_+(N)$ gauge theory, that has been proposed in \cite{Benvenuti:2021nwt}.
Here we provide a derivation of the  integral identity that corresponds to the duality on the three sphere partition function.
Let's start by discussing the integral identity associated to the duality for $O_+(K)$ with $K-1$ vectors $P$, one vector $U$, a singlet $\alpha$, $K-1$ singlets $V$  and superpotential
\begin{equation}
\label{confOplus}
W = Y_{O_+} + a U^2 + P U V.
\end{equation}
This duality can be constructed starting from the duality for $SO(K+1)$ with $K+2$ vectors and monopole superpotential, by flipping some singlets and assigning an holomorphic mass to one vector.
Such mass assigns a vev in the dual confining phase that gives a mass to some of the singlets. The final result corresponds to a symmetric meson.
The details of the flow have been discussed in \cite{Benvenuti:2021nwt} and importantly the $\mathbb{Z}_2^{\mathcal{C}}$ discrete gauging leading to an $O_+(K+1)$ gauge group is 
a consequence of the presence of two disconnected branches for the vev.

Here we want to reproduce this behavior on the partition function, starting from the identity for $SO(K+1)$ with $K+2$ vectors.
We then assign the masses as
\begin{equation}
\mu_r=\{m_{1,\dots,K},m_0,\omega+i\epsilon\},
\end{equation}
where we have regulated the holomorphic mass of the $K+2$-th component. We are going to consider the $\epsilon \rightarrow 0$ limit 
in the following.
Let us also flip the singlets $a$ and $V$ in order to reproduce the duality discussed above.
The electric partition function becomes
\begin{equation}
Z_{ele}^{flipped} = \lim_{\epsilon \rightarrow 0} \prod_{r=1}^{K} \Gamma_h(2\omega-m_r-m_0) \Gamma_h(2\omega-2m_0) Z_{SO(K+1)}(\vec m,m_0,\omega+i \epsilon),
\end{equation}
where the $\epsilon \rightarrow 0$ limit can be safely taken by using the inversion formula on the $K+2$-th vector in the argument of the partition function.

On the other hand, the magnetic partition function becomes
\begin{eqnarray}
Z_{conf} = 
\lim_{\epsilon \rightarrow 0}&&
\prod_{1\leq r\leq s\leq K} \Gamma_h(m_r +m_s) \prod_{r=1}^{K} \left(\Gamma_h(m_r+\omega+i \epsilon) \Gamma_h(\omega- m_r) \right) \nonumber \\
\times 
&&
 \Gamma_h(2\omega+2i \epsilon) \Gamma_h(\omega+m_0+i \epsilon)  \Gamma_h(\omega-m_0) \Gamma_h(-i \epsilon). 
\end{eqnarray}

In this case before taking the $\epsilon \rightarrow 0$ limit we must use the relation
\begin{eqnarray}
\label{duplonSO}
\Gamma_h(-i \epsilon) =\frac{\Gamma_h(-2 i \epsilon)}{\Gamma_h(-i \epsilon + \omega)\Gamma_h\left(-i \epsilon + \frac{\omega_1}{2}\right)\Gamma_h\left(-i \epsilon + \frac{\omega_2}{2}\right)} =2 \Gamma_h(-2i \epsilon),
\end{eqnarray}
that corresponds to the application of the duplication formula.
By substituting this identity in the confining partition function, using the inversion formula, we obtain for $\epsilon \rightarrow 0$ the following identity 
\begin{equation}
\label{dualityconfOpluswobc}
 \Gamma_h(2\omega-2m_0) \prod_{r=1}^{K} \Gamma_h(2\omega-m_r-m_0)Z_{O_+(K+1)}(\vec m,m_0)=\prod_{1\leq r\leq s\leq K} \Gamma_h(m_r +m_s) ,
 \end{equation}
 with the balancing condition $m_0+\dots+m_K= 0$, reproducing the results expected from the confining duality obtained in \cite{Benvenuti:2021nwt}.
 Observe that the role  of the discrete gauging here has been played by the extra factor of $2$ in the last equality in (\ref{duplonSO}).

 There is a second confining duality  discussed in \cite{Benvenuti:2021nwt}, that is useful to deconfine a symmetric tensor and it corresponds to an $O(K+2)_+$ gauge theory with $K$ vectors and 
 superpotential $W = \sigma Y_{O_+}$. 
 This duality can be derived from the confining duality for $SO(K+2)$ with $K+1$ vectors in a similar fashion. After performing the same analysis we end up with the relation
  \begin{eqnarray}
  \label{Opluswobc}
 \Gamma_h\left(2\omega+\sum_{r=1}^{K}\mu_r \right)Z_{O_+(K+2)}(\vec \mu)=\prod_{1\leq r\leq s\leq K} \Gamma_h(\mu_r +\mu_s) ,
 \end{eqnarray}
 without balancing condition.

\subsection{Symplectic gauge groups}
We conclude the survey of the dualities used in the paper by discussing the case of $USp(2N)$. In such case we need to review the cases with fundamental matter, the case with four fundamentals and one antisymmetric and the case with two fundamentals and one adjoint.

\subsubsection{$USp(2N)$ with fundamentals}

Here we consider $USp(2N)$ SQCD with $2F$ fundamentals. In this case we have used two confining dualities in the body of the paper, necessary to deconfine antisymmetric tensors.

The two confining dualities are the following 
\begin{itemize}
\item
First we consider $USp(2N)$ with $2N+3$ fundamentals $Q$ and a fundamental $P$ with superpotential $W=Y_{USp}+ P Q U$. The singlet $U$ flips the mesonic term $P Q$, while $Y_{USp}$  corresponds to the fundamental monopole of the $USp(2N)$ theory. In this case the dual model consists of a free antisymmetric $M = Q^2$. This duality is a modified version of the 4d Intriligator Pouliot duality compactified on the circle and the relative matching of the three sphere partition function can be inferred from the matching of the superconformal index proved in \cite{rains2005transformations}.
The identity for such confining duality is
\begin{equation}
\label{USpfondmon}
\prod_{r=1}^{2N+3}\Gamma_h(2 \omega-m -\mu_r) Z_{USp(2N)} (\vec \mu,m) =
\prod_{1\leq r<s \leq 2N+3} \Gamma_h(\mu_r+\mu_s), 
\end{equation}
with the balancing condition enforced by the linear monopole superpotential
\begin{equation}
\sum_{r=1}^{2N+3} \mu_r + m = 2 \omega.
\end{equation}
\item
The other confining duality used in the paper to deconfine the antisymmetric matter corresponds to the limiting case of Aharony duality, where the gauge group is $USp(2N)$, there are $2N+2$  fundamentals $Q$ and the superpotential 
is $W = \sigma Y_{USp}$, where the singlet $\sigma$ flips the monopole.
The dual theory in this case is given just by a free antisymmetric field of the $SU(2N+2)$ flavor symmetry $M = Q^2$ and at the level of the three sphere partition function the identity is 
\begin{equation}
\label{USpwomon}
\Gamma_h \left(\sum_{r=1}^{2N+2} \mu_r\right) Z_{USp(2N)} (\vec \mu) =
\prod_{1\leq r<s \leq 2N+2} \Gamma_h(\mu_r+\mu_s) ,
\end{equation}
holding without balancing condition. The identity corresponds to {\bf Theorem 5.5.9} of \cite{vanDeBult} and it can be derived from 4d as well, by reducing the duality on the circle and then by performing a real mass flow \cite{Aharony:2013dha}.
\end{itemize}

\subsubsection{$USp(2N)$ with fundamentals and an antisymmetric}

In the body of the paper we have also used a confining duality for $USp(2N)$ with an antisymmetric tensor and four fundamentals.
The dualities for  $USp(2N)$ with antisymmetric matter have been extensively discussed in \cite{Amariti:2018wht,Benvenuti:2018bav} from dimensional reduction of the parent 4d cases studied in \cite{Razamat:2017hda}, generalizing the $E_7$ surprise of \cite{Dimofte:2012pd} (see also \cite{Spiridonov:2008zr}).

Here we have encountered in appendix \ref{appA1} the case with four fundamentals, in the proof of the confinement of $SU(2N)$ with four fundamentals and a conjugate pair of antisymmetrics.
On the partition function side the identity corresponds to {\bf Theorem 5.6.6} of \cite{vanDeBult}. Such identity is 
\begin{eqnarray}
\label{566vdb}
Z_{USp(2 N)} (m_{1,2,3,4};\tau) 
&=& \prod_{j=0}^{N-1} 
\Gamma_h \left(2\omega -(2N-2-j)\tau -\sum_{b=1}^4 m_a \right)
\nonumber \\
&\times&
\prod_{j=0}^{N-1}  \Gamma_h((j+1)\tau)\!\!\!
\prod_{1 \leq b \leq c \leq  4} \!\!\! \Gamma_h(j \tau + m_a+m_b), 
\end{eqnarray}
holding without balancing condition.

\subsubsection{$USp(2N)$ with fundamentals and an adjoint}
\label{app:BLMAdj}

Here we review the confining duality for $USp(2N)$, and adjoint $X$, a fundamental $q$ and a fundamental $p$, interacting with the adjoint through the
superpotential 
\begin{equation}
\label{BLMadj2fund}
W = X p^2 +\sum_{k=1}^{N} \sigma_k Tr X^{2k}.
\end{equation}
This duality was originally found in \cite{Benvenuti:2021nwt} and further discussed in \cite{Amariti:2022wae}.

In the superpotential above we have also added the flippers $\sigma_I$ that
are very useful to simplify the structure of the superpotential of the dual WZ model.
The theory confines and the confining dynamics is governed by the dressed mesons $M_\ell = q^2 X^{2\ell+1}$, with $\ell=0,\dots ,N-1$ and the dressed monopoles $Y_{\ell} = Y_{USp}^{bare} X^{\ell}$, with $\ell=0,\dots,2N-1$.

 The confining superpotential is
   \begin{equation}
  \label{confBLM}
W = \sum_{i,j,k} \, Y_i \, Y_j M_k \delta_{i+j+2k-4N+2}.
\end{equation}

The integral identity for such confinement was obtained in  \cite{Amariti:2022wae}. By assigning the mass parameter $\mu$ to the fundamental $q$ and $\tau$ to the adjoint $X$, the identity is
\begin{eqnarray}
\label{AR22}
\prod_{k=1}^{N} \Gamma_h(2\omega-2 k\tau) Z_{USp(2N)} (\mu,\omega - \frac{\tau}{2};-;\tau)
=
&&
\prod_{j=0}^{2N-1}
\Gamma_h\left(\omega-\left(j+\frac{1}{2} \right) \tau-\mu\right) \nonumber \\
\times &&
\prod_{\ell=0}^{N-1} \Gamma_h((2\ell+1) \tau+2\mu), \\
\end{eqnarray}
where the Gamma functions on the LHS of (\ref{AR22}) are the flippers $\sigma_k$. On the RHS of (\ref{AR22}) we observe the contribution of the dressed monopoles $Y_{j}$  in the first line and the contribution of the dressed mesons $M_\ell$ in the second line.

This duality has played a crucial role in the analysis of case II-A and II-B.
On the other hand, in the study of case II-C we have encountered another
duality that involved only $USp(2N)$ with an adjoint $X$ and a superpotential 
for the flippers
 \begin{equation}
 \label{confsoloADJ}
W =\sum_{k=1}^{N} \sigma_k Tr X^{2k}.
\end{equation}
The claim is that this theory confines and in the confined case there are only the fields 
$Y_{2\ell+1} = Y_{USp}^{bare} X^{2\ell+1}$, for $\ell=0,\dots,N-1$ with vanishing superpotential.

The duality can be derived from the one with two fundamentals   by adding the mass term $\Delta W = m \, Tr (q p)$ to  (\ref{BLMadj2fund}).
Indeed, as shown in \cite{Benvenuti:2021nwt}, this operator is mapped in the dual WZ model to the mass term
   \begin{equation}
  \label{confBLMmod}
W = m \sum_{k=1}^{N}  Y_{2N-2k} M_{k-1}. 
\end{equation}
By integrating out the massive fields, we are left then with the free $Y_{2\ell+1}$ dressed monopoles.
At the level of the partition function this mass corresponds to fix $\mu= \omega+\frac{\tau}{2}$. The final relation becomes
\begin{eqnarray}
\label{AM24}
\prod_{k=1}^{N} \Gamma_h(2\omega-2 k\tau) Z_{USp(2N)} (-;-;\tau)
=
\prod_{j=0}^{N-1}\Gamma_h(-(2j+1)\tau).
\end{eqnarray}
We conclude by observing that we can move the hyperbolic Gamma function 
from the LHS of (\ref{AM24}) to the RHS by using the inversion formula. 
This corresponds to consider only $USp(2N)$ with an adjoint, without any flip. In this case the model is dual to a set of singlets and dressed monopoles, that do not allow 
the existence of any superpotential term. 
Many combinations of the singlets are neutral and the claim is that such theory is confining with a quantum deformed moduli space. 
In general such neutral combinations are  $T_\ell T_k Y_i Y_j \delta_{\ell+k-i-j-1}$, where $T_k = Tr X^{2k}$ and $Y$ are the dressed monopole specified above.

We conclude focusing on the case of $USp(2)$:
in this case we do not turn on the flipper in the electric theory. The superpotential of the dual theory, including the operator dual to the mass deformation, is 
\begin{equation}
W = T_1 M_0 Y_0^2 + Y_1^2 M_0 + M_0 Y_0.
\end{equation}
Integrating out the massive fields, we remain with the uncharged combination $Y_1^2 T_1$.
Then we claim that in this case there is a quantum deformed moduli space.
The same conclusion can be reached by deconfining the $USp(2)$ adjoint using an $O_+(3)$ gauge group with superpotential (\ref{confOplus}). Then we confine the $USp(2)$ group using Aharony duality. 
The $O_+(3)$ gauge theory has a single vector and it has then a quantum deformed moduli space with the uncharged combination 
$Y_1^2 T_1$ as expected. 
For higher ranks, we expect a similar behavior, even if the number of terms  grows quite rapidly and working out the details is less straightforward.

 \bibliographystyle{JHEP}
\bibliography{ref.bib}

\end{document}